\documentclass[a4paper,11pt]{article}
\pdfoutput=1 

\usepackage{jcappub} 

\usepackage{natbib}
\usepackage{amsmath,amsfonts,amssymb}

\usepackage{mathrsfs}
\usepackage{eurosym}
\usepackage{bm}

\bibpunct{(}{)}{;}{a}{}{,}

\def\reff@jnl#1{{\rm#1\/}}

\def\aj{\reff@jnl{AJ}}                  
\def\araa{\reff@jnl{ARA\&A}}            
\def\apj{\reff@jnl{ApJ}}                
\def\apjl{\reff@jnl{ApJ}}               
\def\apjs{\reff@jnl{ApJS}}              
\def\ao{\reff@jnl{Appl.Optics}}         
\def\apss{\reff@jnl{Ap\&SS}}            
\def\aap{\reff@jnl{A\&A}}               
\def\aapr{\reff@jnl{A\&A~Rev.}}         
\def\aaps{\reff@jnl{A\&AS}}             
\def\azh{\reff@jnl{AZh}}                        
\def\baas{\reff@jnl{BAAS}}              
\def\jcap{\reff@jnl{JCAP}}              
\def\jrasc{\reff@jnl{JRASC}}            
\def\memras{\reff@jnl{MmRAS}}           
\def\mnras{\reff@jnl{MNRAS}}            
\def\pra{\reff@jnl{Phys.Rev.A}}         
\def\prb{\reff@jnl{Phys.Rev.B}}         
\def\prc{\reff@jnl{Phys.Rev.C}}         
\def\prd{\reff@jnl{Phys.Rev.D}}         
\def\prl{\reff@jnl{Phys.Rev.Lett}}      
\def\procspie{\reff@jnl{Proc.SPIE}}      
\def\physrep{\reff@jnl{Phys.Rep.}}      
\def\pasp{\reff@jnl{PASP}}              
\def\pasj{\reff@jnl{PASJ}}              
\def\qjras{\reff@jnl{QJRAS}}            
\def\skytel{\reff@jnl{S\&T}}            
\def\solphys{\reff@jnl{Solar~Phys.}}    
\def\sovast{\reff@jnl{Soviet~Ast.}}     
 \def\ssr{\reff@jnl{Space~Sci.Rev.}}    
\def\zap{\reff@jnl{ZAp}}                
\def\nat{\reff@jnl{Nature}}             

\newcommand{\core}{\textit{\negthinspace COrE\/}}
\newcommand{\coremfive}{\textit{\negthinspace CORE\/}}
\newcommand{\coreplus}{\textit{\negthinspace COrE+\/}}
\newcommand{\planck}{\textit{\negthinspace Planck\/}}
\newcommand{\Planck}{\planck}

\newcommand{\WMAP}{\negthinspace \textit{WMAP\/}}
\newcommand{\COBE}{\negthinspace \textit{COBE\/}}

\newcommand{\lsim}{\raise0.3ex\hbox{$<$\kern-0.75em\raise-1.1ex\hbox{$\sim$}}}
\newcommand{\gsim}{\raise0.3ex\hbox{$>$\kern-0.75em\raise-1.1ex\hbox{$\sim$}}}
\newcommand{\ltsima}{$\; \buildrel < \over \sim \;$}
\newcommand{\simlt}{\lower.5ex\hbox{\ltsima}}
\newcommand{\gtsima}{$\; \buildrel > \over \sim \;$}
\newcommand{\simgt}{\lower.5ex\hbox{\gtsima}}
\newcommand{\simprop}{\lower.5ex\hbox{$\; \buildrel \propto \over \sim \;$}}
\newcommand{\m}{\ifmmode $m$\else \,m\fi}

\def\st{\ifmmode ^{\mathrm{st}} \else $^{\mathrm{st}}$\fi}
\def\nd{\ifmmode ^{\mathrm{nd}} \else $^{\mathrm{nd}}$\fi}
\def\rd{\ifmmode ^{\mathrm{rd}} \else $^{\mathrm{rd}}$\fi}
\def\th{\ifmmode ^{\mathrm{th}} \else $^{\mathrm{th}}$\fi}

\title{Exploring Cosmic Origins with CORE: Survey requirements and mission design}

\author[1]{J.~Delabrouille,\note{Corresponding author.}}
\author[2]{P.~de~Bernardis,}
\author[3]{F.~R.~Bouchet,}
\author[]{A.~Ach\'ucarro,}
\author[]{P.~A.~R.~Ade,}
\author[]{R.~Allison,}
\author[]{F.~Arroja,}
\author[]{E.~Artal,}
\author[]{M.~Ashdown,}
\author[]{C.~Baccigalupi,}
% ...

\emailAdd{delabrouille@apc.in2p3.fr}

\abstract{

Future observations of cosmic microwave background (CMB) polarisation have the potential
to answer some of the most fundamental questions of modern physics and cosmology, including: What physical process gave
birth to the Universe we see today? What are the dark matter and dark energy that seem to constitute 95\%
of the energy density of the Universe? Do we need extensions to the standard model of particle physics and fundamental interactions?
Is the $\Lambda$CDM cosmological scenario correct, or are we missing an essential piece of the puzzle?
In this paper, we list the requirements for a future CMB polarisation survey addressing these scientific objectives, 
and discuss the design drivers of the \coremfive\ space mission
proposed to ESA in answer to the ``M5'' call for a medium-sized mission. The rationale and options, and the methodologies used to assess the mission's performance, are of interest to other future CMB mission design studies.
\coremfive\ has 19 frequency channels, distributed over a broad frequency
range, spanning the 60--600\,GHz interval, to control astrophysical foreground emission. 
The angular resolution ranges from $2^\prime$ to $18^\prime$, and the 
aggregate CMB sensitivity is about 2\,$\mu$K.arcmin.
The observations are made with a single integrated focal-plane instrument, consisting of an array of 2100 cryogenically-cooled,
linearly-polarised detectors at the focus of a 1.2-m aperture cross-Dragone telescope.
The mission is designed to minimise all sources of systematic effects, 
which must be controlled so that no more than $10^{-4}$ of the intensity leaks into polarisation
maps, and no more than about 1\% of $E$-type polarisation leaks into $B$-type modes.
\coremfive\  observes the sky from a large Lissajous orbit around the Sun-Earth L2
point on an orbit that offers stable observing conditions and avoids contamination from sidelobe pick-up of
stray radiation originating from the Sun, Earth, and Moon.
The entire sky is observed repeatedly during four years of continuous scanning, with
a combination of three rotations of the spacecraft over different timescales. With about 
50\% of the sky covered every few days, this scan strategy provides the mitigation
of systematic effects and the internal redundancy that are needed to convincingly extract the primordial 
$B$-mode signal on large angular scales, and check with adequate sensitivity the consistency of the observations
in several independent data subsets.
\coremfive\ is designed as a ``near-ultimate'' CMB polarisation mission which,
for optimal complementarity with ground-based observations,  
will perform the observations that are known to be essential to CMB polarisation science
and cannot be obtained by any other means than a dedicated space mission. 
It will provide well-characterised, highly-redundant multi-frequency observations
of polarisation at all the scales where foreground emission and cosmic variance dominate the final uncertainty 
for obtaining precision CMB science, as well as $2^\prime$ angular resolution maps of high-frequency foreground emission 
in the 300--600\,GHz frequency range, essential for complementarity with future ground-based observations 
with large telescopes that can observe the CMB with the same beamsize.}

\begin{document}
\maketitle
\flushbottom

\section{Introduction}
\label{sec:intro}

In the past few decades, the field of cosmology has undergone a period of dramatically rapid progress in which a standard model of cosmology has emerged, $\Lambda$CDM. The precision with which this model has been constrained has been largely driven by studies of the anisotropies in the cosmic microwave background (CMB). However, despite impressive advances, many open questions remain.  Did the very early Universe undergo a phase of inflation -- an accelerated expansion period in which macroscopic primordial inhomogeneities were seeded from local quantum fluctuations -- and if so, what are the physical mechanisms and the fields responsible for inflation? What is the nature of the elusive dark matter and dark energy that seem to constitute more than 95\% of the matter-energy density in our observable Universe? Are the apparent large-scale anomalies
observed in CMB temperature maps by the \WMAP\ and \Planck\ space missions a signature of deviation from isotropy and homogeneity, or a statistical fluke? Is there new physics at play in the Universe, beyond the standard model of particle physics and fundamental interactions? Is the overall $\Lambda$CDM cosmological scenario correct, or are we missing an essential piece of the puzzle?

Answers to these questions can be found in additional observations of the CMB, the relic radiation that was last scattered when the Universe was about 380{,}000 years old and became cold enough that the primordial plasma of light nuclei and electrons combined into neutral atoms, mainly hydrogen and helium. In the process, the Universe became transparent to radiation, so that CMB photons became free to propagate.  Hence when we observe them today they carry an image of the Universe at this recombination epoch, which encodes a wealth of information about the early Universe and about the interactions of CMB photons on their paths towards us. The \Planck\ space mission has extracted most of the information in the primordial CMB temperature anisotropy power spectrum \citep{2016A&A...594A...1P,2016A&A...594A..13P}. However, the sensitivity of \Planck\ to CMB polarisation -- about 50\,$\mu$K.arcmin (i.e., a noise level of $50\,\mu$K$_{\rm CMB}$\footnote{In CMB thermodynamic temperature units; we will drop the ``CMB'' subscript henceforth.} per pixel of 1 square arcminute solid angle) -- was not sufficient to extract all of the information that can be obtained from CMB polarisation. The near-optimal exploitation of CMB polarisation signals requires measurements at the level of a few $\mu$K.arcmin or better, i.e., at least an order of magnitude better than achieved by \Planck.

The scientific importance of measuring CMB polarisation has stimulated a huge amount of activity in the CMB community. A number of suborbital experiments have been or are being deployed, with the objective to either detect primordial CMB polarisation $B$ modes generically predicted in the framework of inflationary models, \citep[as recently reviewed in Ref.][]{2016ARA&A..54..227K}, or $B$ modes due to CMB lensing \citep{2006PhR...429....1L}, or both. However, it is widely accepted that a space mission will be necessary to fully exploit the scientific potential of CMB polarisation. 

Several concepts for next-generation space missions have already been presented in answer to calls for proposals by space agencies throughout the world. In Europe, \coreplus\ was proposed to ESA in January 2015, but was evaluated as incompatible with the technical and programmatic boundary conditions of the M4 call, which had an unusual schedule and tight budgetary constraints. \coreplus\ followed a previous proposal, COrE, submitted in December 2010 \citep{2011arXiv1102.2181T}, and the B-Pol concept \citep{2009ExA....23....5D}, proposed earlier within the same programme. A French small satellite mission, the SAMPAN satellite, was proposed to CNES and underwent a preliminary feasibility study with CNES and industry in around 2006 \citep{2005sf2a.conf..675B}. A Japanese satellite to study CMB polarisation, LiteBIRD, was proposed to JAXA in 2008 and is undergoing a study phase in Japan in collaboration with a team from the United States \citep{2014JLTP..176..733M,2016SPIE.9904E..0XI}. In the US, a mission concept study called EPIC/CMBpol was carried-out under a NASA contract in 2008--2009 \citep{2008arXiv0805.4207B,2009arXiv0906.1188B}, and an initial study is underway for a ``Probe-class'' mission currently called CMB-Probe. A different concept, PIXIE, using a Fourier transform spectrometer to observe in 400 narrow frequency bands between 30\,GHz and 6\,THz with only four bolometric detectors, has been proposed to observe not only CMB polarisation, but also measure spectral distortions of the background \citep{2011JCAP...07..025K,2016SPIE.9904E..0WK}. A comprehensive mission, PRISM, with a very broad science case, comprising both CMB polarisation and spectral distortions, was proposed to ESA in 2013 as a possible large mission, to be launched in 2028 or 2034 \citep{2013arXiv1306.2259P,2014JCAP...02..006A}. None of these proposals is selected yet, but the number of proposals testifies of the strong interest of the scientific community for a future CMB space mission.

These mission concepts all propose to observe the sky at millimetre to sub-millimetre wavelengths, but differ in sensitivity (by a factor of up to 10), angular resolution (by a factor of up to 20), frequency coverage (with $\nu_{\rm max}/\nu_{\rm min}$ ranging from 5 to 200), number of detectors (from 4 to more than 10{,}000), number of frequency bands (from 5 to 400) and orbit (from low-Earth orbit to the Sun-Earth L2 Lagrange point). These differences arise from: mission-specific science targets; varying assumptions about the plausible level and complexity of foreground astrophysical emission and about the range of frequency bands required to clean CMB maps from astrophysical contamination; and programmatic and budgetary constraints imposed by the calls for mission concepts by space agencies, which lead to inevitable compromises. 

In this paper, one of a series dedicated to the preparation of a post-\Planck\ CMB space mission, we discuss the performance requirements and the possible design of a future space mission concept that will observe CMB polarisation, in order to shed new light on cosmology, and that can be implemented as an ESA medium-size mission to be launched before 2030. This paper is part of the ``Exploring Cosmic Origins (ECO)'' collection of articles, each describing a different aspect of the Cosmic Origins Explorer (\coremfive), recently proposed to ESA in answer to the ``M5'' call for a medium-size space mission within the ESA Cosmic Vision Programme. 
We discuss the design drivers and the various options, and present the expected performance and scientific impact expected from the mission. We compare the \coremfive\ design with that of other proposals, and discuss the pros and cons of the various options.
A number of relevant questions are addressed in companion papers, which investigate in more detail: the scientific case for the mission \citep{2016arXiv161200021D,2016arXiv161208270C,ECO.lensing.paper,2017arXiv170310456M,2016arXiv160907263D,ECO.velocity.paper}; its ability to address contamination of the observations by astrophysical foreground
emission~\citep{ECO.foregrounds.paper}; data analysis techniques that can help mitigate systematic 
effects~\citep{ECO.systematics.paper}; and the design of the instrument~\citep{ECO.instrument.paper}.

\begin{figure}[bthp]
\begin{center}
\includegraphics[width=14cm]{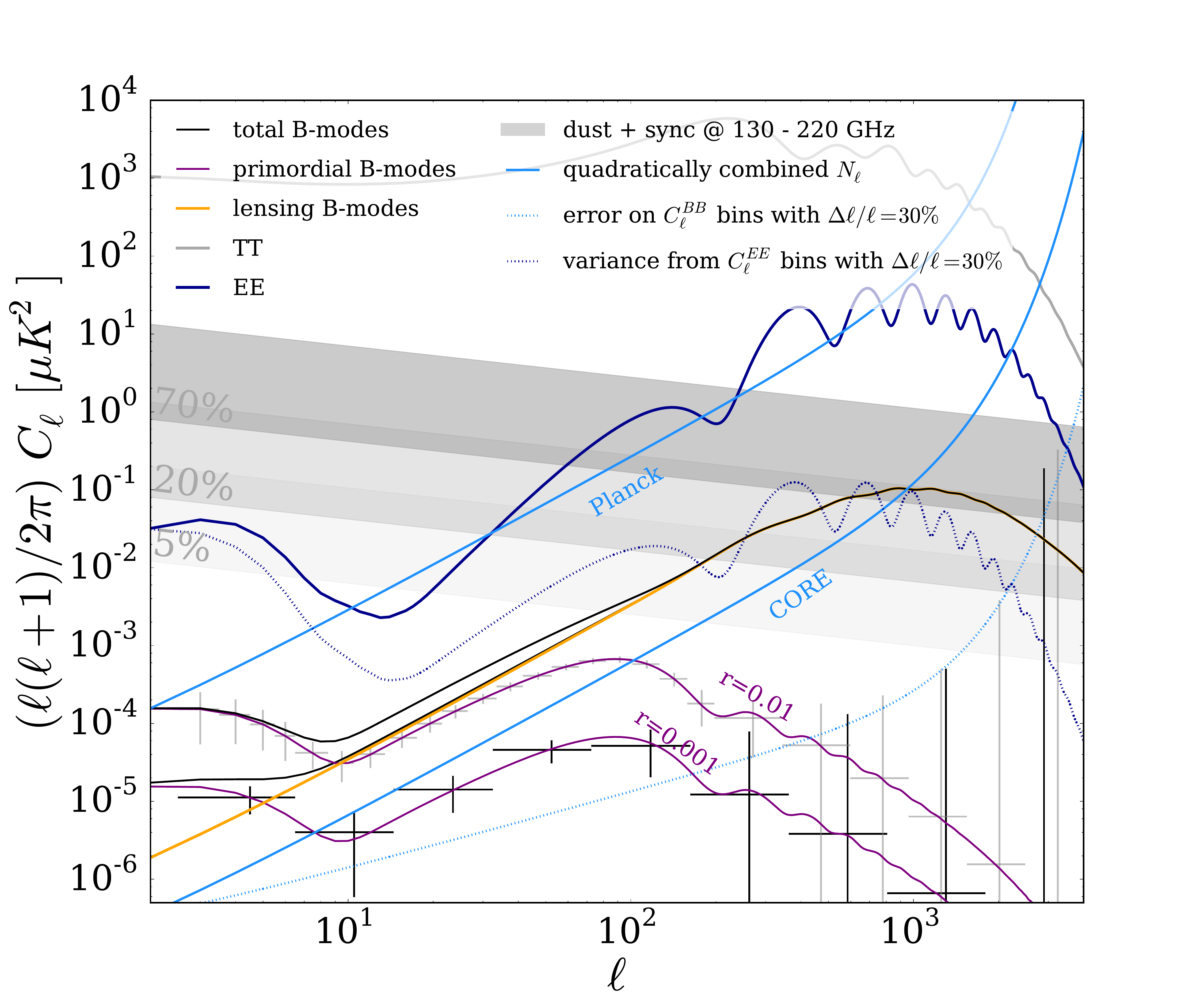}
\end{center}
\caption{
CMB polarisation angular power spectra $C_\ell^{EE}$ (dark blue), $C_\ell^{BB}$ from gravitational lensing of $E$ modes by large-scale structure (orange), $C_\ell^{BB}$ from inflationary gravitational waves $r$ (purple, for two values of the tensor-to-scalar ratio), and total $C_\ell^{BB}$ for $r=0.01$ (black). Two fundamental sources of error for measurements of these power spectra with \coremfive\ are shown for comparison: expected noise level (light blue); and average foreground emission over 70\%, 20\%, and 5\% of the sky (grey bands, from dark to light). Each of the grey bands shows the span of foreground contamination from 130 \,GHz (lower limit of the band) to 220\,GHz (upper limit). Uncertainties in power spectrum estimation over bands of $\Delta \ell/\ell=0.3$ coming from $E$ modes and noise sample variance (representative of the level at which errors must be understood to take full advantage of the survey raw sensitivity) are shown as dotted lines.
The error bars on the primordial $B$-mode spectra for $r=0.01$ and
$r=0.001$, corresponding to $1\sigma$ in bins ranging from $\Delta
\ell/\ell \simeq 0.2$ (for $r=0.01$, at low $\ell$) to 0.75 (for
$r=0.001$), illustrate the sensitivity that will be achieved for
inflationary science assuming perfect component separation over 70\%
of sky and reduction of the contamination by lensing using small-scale
CMB $E$ and $B$ modes measured by \coremfive.  }
\label{fig:B-sensitivity}
\end{figure}

\section{Overview of CORE}

The \coremfive\ mission concept proposed to ESA in answer to the
``M5'' call is a polarimetric imager that will observe the sky in 19
frequency bands between 60 and 600\,GHz, at an angular resolution
ranging from about $2^\prime$ at 600\,GHz to about $18^\prime$ at
60\,GHz. \coremfive\ is focussed on CMB polarisation, aiming at
exploiting the scientific information that can be extracted from CMB
polarisation $E$ and $B$ modes. One of the key science targets is the
detection, precise characterisation, and scientific exploitation of
CMB polarisation $B$ modes, both from inflationary gravitational waves
and from the gravitational lensing of last-scattering surface CMB $E$
modes by large-scale structure along the line of
sight~(section~\ref{sec:science}). Figure~\ref{fig:B-sensitivity}
gives a view of how well \coremfive\ will measure $E$ and $B$ modes,
and specifically primordial $B$ modes, for a tensor to scalar ratio
$r$ of 0.01 or 0.001. It also illustrates the relative importance of
various sources of error in polarisation measurements, and in
particular the need for accurate component separation on all angular
scales to fully exploit the CMB polarisation signals over a large
fraction of the sky. Indeed, over 70\% of sky, Galactic foreground
emission at 130\,GHz dominates over noise at all scales down to about
$12^\prime$ ($\ell\simeq1000$), and is larger than $E$-mode sample
variance in bins of $\Delta \ell/\ell=0.3$ at all scales. It also
dominates over lensing $B$ modes at all scales for large sky
fractions. The severity of foreground contamination would be reduced
if we restrict ourselves to exploiting only the cleanest part of the
sky: over 5\% of sky, the amplitude of foreground contamination is
reduced by an order of magnitude in amplitude, so that at 130\,GHz it
dominates over noise only on scales larger than about one degree. Over
such a smaller patch of sky, however, cosmic variance of $E$ modes or
$B$-modes is significantly increased.

{\small
\begin{table}[htb]
\begin{center}
\scalebox{0.95}{\scriptsize
\begin{tabular}{|c|c|c|c|c|c|c|c|c|c|c|c|}
\hline 
Channel& Beam& $N_{\rm det}$&  $\Delta T$& $\Delta P$& $\Delta I$& $\Delta I$& $\Delta y\times 10^6$&  PS ($5\,\sigma$)\\
{[}GHz{]}& {[}arcmin{]}& & {[}$\mu$K.arcmin{]}&  {[}$\mu$K.arcmin{]}& {[}$\mu K_{\rm RJ}$.arcmin{]}& {[}kJy/sr.arcmin{]}& {[}$y_{\rm SZ}$.arcmin{]}& {[}mJy{]}\\
\hline 
\hline 
60&  17.87&  48&   7.5&  10.6& 6.81& 0.75& $-1.5$& 5.0\\ 
70&  15.39&  48&   7.1&  10.0& 6.23& 0.94& $-1.5$& 5.4\\ 
80&  13.52&  48&   6.8&   9.6& 5.76& 1.13& $-1.5$& 5.7\\ 
90&  12.08&  78&   5.1&   7.3& 4.19& 1.04& $-1.2$& 4.7\\ 
100& 10.92&  78&   5.0&   7.1& 3.90& 1.20& $-1.2$& 4.9\\ 
115&  9.56&  76&   5.0&   7.0& 3.58& 1.45& $-1.3$& 5.2\\ 
130&  8.51& 124&   3.9&   5.5& 2.55& 1.32& $-1.2$& 4.2\\ 
145&  7.68& 144&   3.6&   5.1& 2.16& 1.39& $-1.3$& 4.0\\ 
160&  7.01& 144&   3.7&   5.2& 1.98& 1.55& $-1.6$& 4.1\\ 
175&  6.45& 160&   3.6&   5.1& 1.72& 1.62& $-2.1$& 3.9\\ 
195&  5.84& 192&   3.5&   4.9& 1.41& 1.65& $-3.8$& 3.6\\ 
220&  5.23& 192&   3.8&   5.4& 1.24& 1.85&  \dots& 3.6\\ 
255&  4.57& 128&   5.6&   7.9& 1.30& 2.59&    3.5& 4.4\\ 
295&  3.99& 128&   7.4&  10.5& 1.12& 3.01&    2.2& 4.5\\ 
340&  3.49& 128&  11.1&  15.7& 1.01& 3.57&    2.0& 4.7\\ 
390&  3.06&  96&  22.0&  31.1& 1.08& 5.05&    2.8& 5.8\\ 
450&  2.65&  96&  45.9&  64.9& 1.04& 6.48&    4.3& 6.5\\ 
520&  2.29&  96& 116.6& 164.8& 1.03& 8.56&    8.3& 7.4\\ 
600&  1.98&  96& 358.3& 506.7& 1.03& 11.4&   20.0& 8.5\\ 
\hline
\hline
Array& & 2100& 1.2& 1.7& & & 0.41& \\
\hline 
\end{tabular}
}
\end{center}
\vspace{-\baselineskip}
\caption{\small Proposed \coremfive\ frequency channels. 
The sensitivity is calculated for a 4-year mission, assuming $\Delta \nu/\nu=30\%$ bandwidth, 60\% optical efficiency, total noise of twice the expected photon noise from the sky and the optics of the instrument being cooled to 40$\,$K. This configuration has 2100 detectors, about 45\% of which are located in CMB channels between 130 and 220\,GHz. Those six CMB channels yield an aggregate CMB sensitivity in polarisation of $2\,\mu$K.arcmin ($1.7\,\mu$K.arcmin for the full array). Entries for the thermal SZ Comptonisation parameter $\Delta y$ are negative below 217\,GHz (negative part of the tSZ spectral signature).
}
\label{tab:CORE-bands}
\end{table}
}

The instrument uses an array of 2100 cryogenically cooled, broad-band, polarisation-sensitive Kinetic Inductance Detectors (KIDs) at the focus of a 1.2-m aperture crossed-Dragone telescope. 
The full array yields an aggregate CMB polarisation sensitivity of
about 1.7\,$\mu$K.arcmin (Table~\ref{tab:CORE-bands}). Frequency
channels are chosen to cover a frequency range sufficient to
disentangle the CMB from astrophysical foreground emission. Six
frequency channels ranging from 130\,GHz to 220\,GHz are dedicated
primarily to observing the CMB. The individual sensitivity of each of
these channels is comparable to the level of CMB lensing, of order
5\,$\mu$K.arcmin in polarisation. These sensitive observations at
different frequencies allow for cross-comparison and cross-correlation
of independent CMB maps to characterise foreground residuals and noise
properties. Six channels from 60 to 115\,GHz mostly serve to monitor
low-frequency and astrophysical foreground emission (polarised
synchrotron, but also free-free and spinning dust in intensity, and in
polarisation if required). In sky regions where synchrotron is faint
these channels can contribute to CMB sensitivity as well. Seven
channels ranging from 255 to 600\,GHz serve to monitor dust emission,
and to map cosmic infrared background (CIB) anisotropies that can
serve as a tracer of mass for ``de-lensing'' CMB polarisation $B$
modes \citep{2015PhRvD..92d3005S}. The telescope size (1.2-m aperture)
is such that the angular resolution is better than $18^\prime$ over
the whole frequency range, so that all the frequency channels can be
used for component separation down to this angular resolution. In the
cleanest regions of the sky, the CMB will be mapped in eight frequency
channels or more, with an angular resolution ranging from $\simeq
5^\prime$ to $10^\prime$ and a sensitivity to polarisation in the
5--8\,$\mu$K.arcmin range for each channel independently.

\begin{figure}[h]
\setlength{\belowcaptionskip}{-16pt}
\centering
\begin{minipage}{0.32\linewidth} 
\includegraphics[width=\textwidth, trim={3cm 0 2cm 3cm},clip]{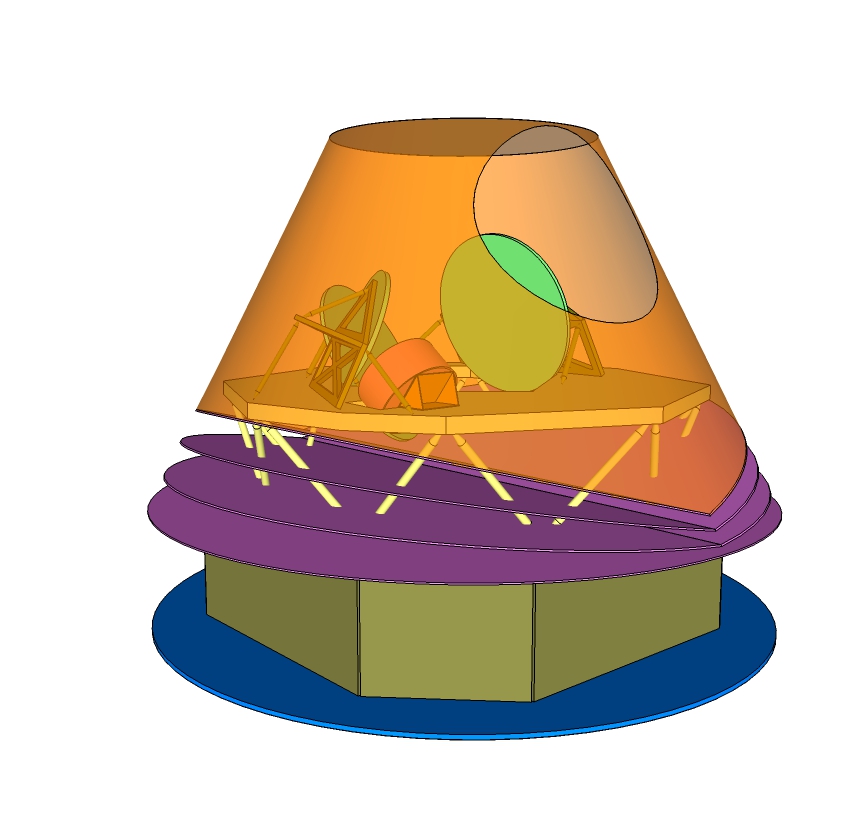}\\
\includegraphics[width=\textwidth]{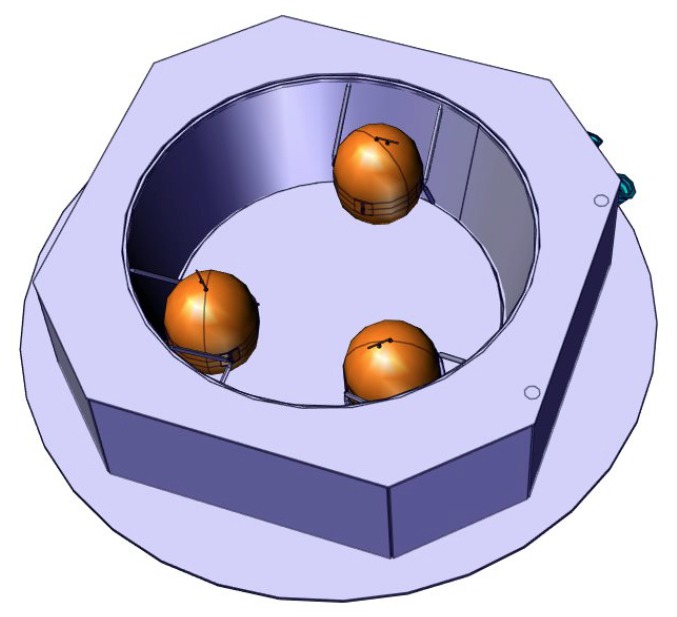}
\end{minipage}
\begin{minipage}{0.34\linewidth} 
\includegraphics[width=\textwidth, trim={5cm 1cm 7cm 0},clip]{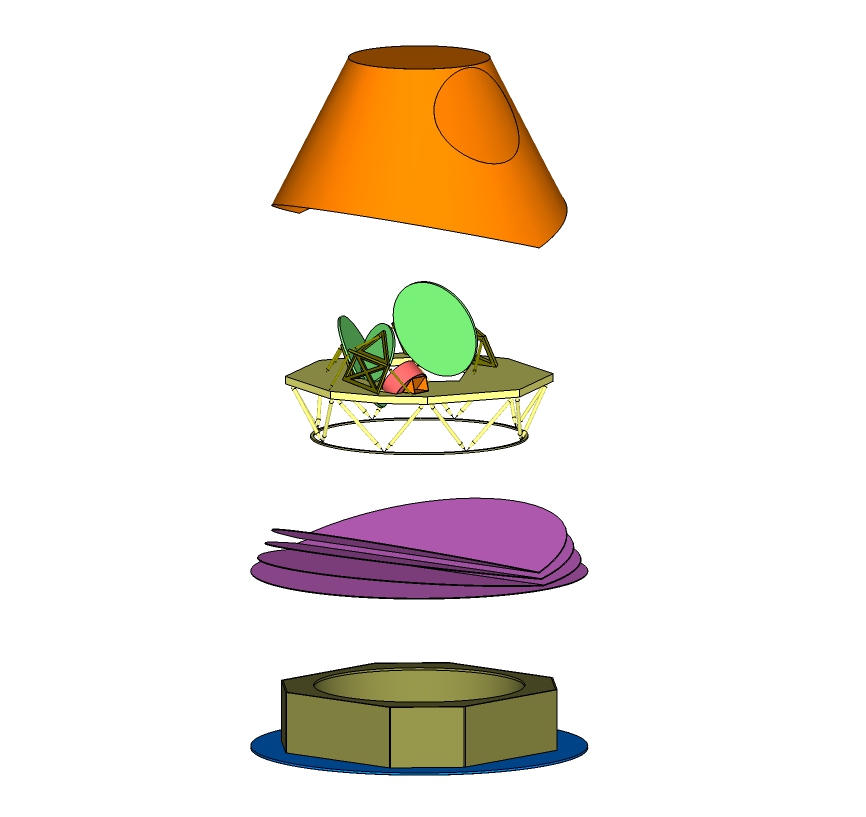}
{\vspace{0.5cm}}
\end{minipage}
\begin{minipage}{0.3\linewidth} 
\includegraphics[width=\textwidth, trim={8cm 0 6cm 3cm},clip]{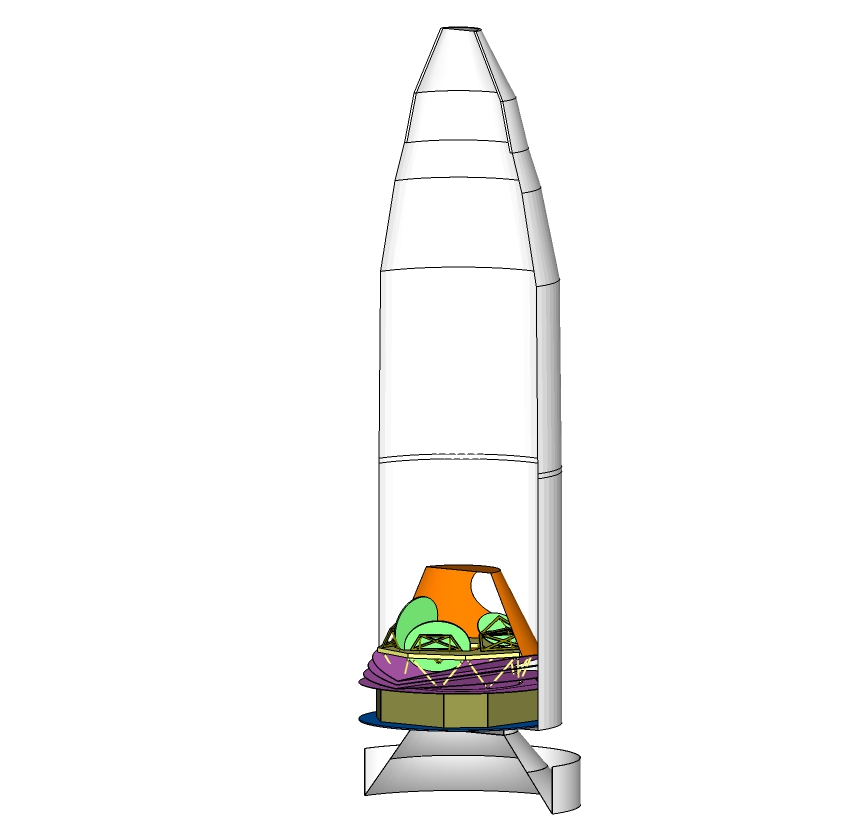}
\end{minipage}
\caption{\small 
Baseline \coremfive\ payload and service modules. {\it Top left}: Global view of the spacecraft.  {\it Bottom left}: View of the SVM following the preliminary design made by ESA in a short concurrent design facility study performed in March 2016 [\url{http://sci.esa.int/trs/57795-cmb-polarisation-mission-study}].
{\it Middle}: Global view of all spacecraft elements, showing the main shield (orange), the telescope (light green) on its optical bench (yellow), the focal-plane unit (FPU, red), the V-grooves (purple), and the SVM at the bottom. The FPU outer shield is not represented. {\it Right}: View of \coremfive\ in an Ariane-6.2 fairing.
}
\label{fig:spacecraft-views}
\vspace{\baselineskip}
\end{figure}

The geometry of the spacecraft, displayed in
figure~\ref{fig:spacecraft-views}, is as symmetric as possible to
avoid any thermal effect due to the modulation of the solar flux on
the spacecraft while it spins to scan the sky. The main elements of
the payload module (PLM), telescope, screens and baffles, will be kept
cold by passive cooling, to minimise the requirements on the active
cryogenic chain. Passive cooling of the PLM to approximately 40\,K
will be achieved by keeping the payload in the shadow of the service
module (SVM), and thermally decoupling the PLM from the SVM with a set
of highly reflective V-grooves \citep[a conceptual design similar to
that succesfully used on \Planck,][]{2010A&A...520A...1T}, while the
main payload conical screen radiates towards free space to compensate
for conductive heat inflow from the SVM.

Although the design and performance of the instrument do not
critically depend on the payload temperature actually achieved (which
could be as high as 90\,K or more with acceptable impact on the
mission performance), the low payload temperature that is achieved by
passive cooling also reduces the background on the detectors,
resulting in better sensitivity overall, in particular in the
frequency channels above 220\,GHz.

\begin{figure}[tbp]
\centering 
\includegraphics[width=.9\textwidth]{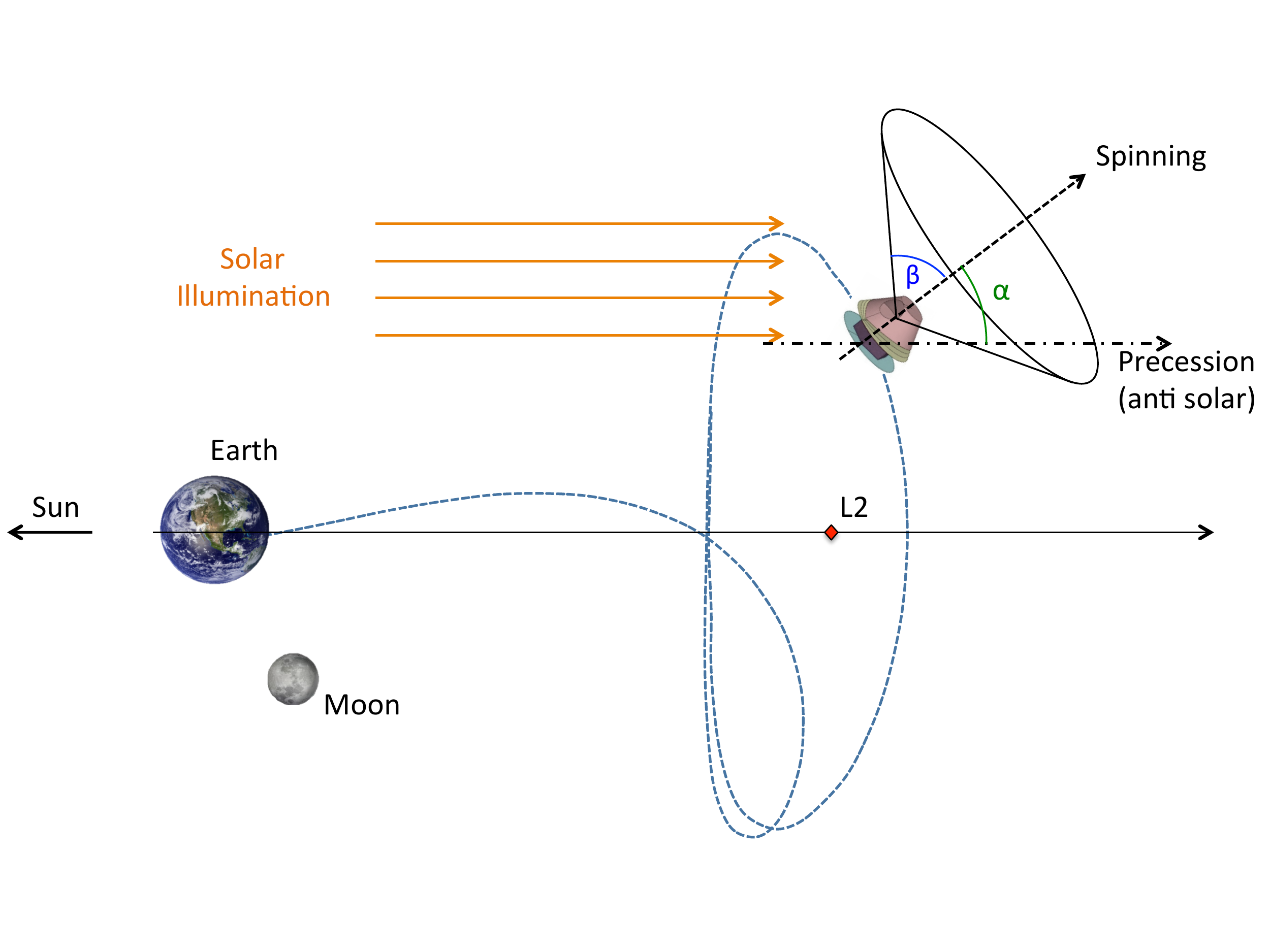}
\vspace{-2\baselineskip}
\caption{\label{fig:orbit} On an orbit around the Sun-Earth L2 Lagrange point, 1.5 million kilometre away from the Earth, the spacecraft scans the sky with three modulations of the pointing direction on various timescales. The spacecraft spins at a rate of order $f_{\rm spin} \simeq 0.5\,$RPM, so that the line of sight scans the sky on quasi-circles of opening angle $\beta$ with a period of about 2 minutes. The circles are not perfectly closed by reason of a slower precession, with a period of $T_{\rm prec} \simeq {\rm 4}$\,days, with precession angle $\alpha$. The precession axis is kept anti-solar, so that the symmetric spacecraft always receives the same amount of illumination from the Sun, ensuring hence the thermal stability of the payload. The last modulation is provided by the slow revolution of the whole system around the Sun with a period of one year.}
\end{figure}

\coremfive\ will be in orbit around the second Sun-Earth Lagrange
point (L2), and will scan the sky with a dedicated scanning strategy
combining a fast spin ($T_{\rm spin} \simeq 2$\,minutes) around the
spacecraft principal axis of symmetry, a slower precession ($T_{\rm
  prec} \simeq 4$\,days) around an axis that is kept anti-solar to
keep the solar flux on the spacecraft constant, and a slow revolution
of the whole system around the Sun with period 1\,year
(figure~\ref{fig:orbit}).  The precession angle is $\alpha=30^\circ$,
and the line of sight (LOS) is offset from the spin-axis by an angle
$\beta = 65^\circ$.
The baseline scan strategy guarantees that each sky pixel is seen by
each detector with a large number of different orientations, a
property that is crucial for measuring polarisation with a good
control of systematic effects. Contrarily to some mission concepts
proposed earlier, the baseline version of \coremfive\ does not make
use of an active polarisation modulator such as a rotating half-wave
plate (HWP). Systematic effects that generate confusion between all
Stokes parameters, and in particular those that result in a leakage of
intensity signals into much fainter polarisation, are controlled
through a combination of requirements on the instrument and on its
calibration, of a scanning strategy that provides polarisation
measurements and redundancies on a very large range of timescales, and
of carefully constructed data-processing pipelines. Systematics are
characterised and corrected for a posteriori, with a global
interpretation of the scientific data themselves, marginalising over
nuisance parameters that model instrument properties and sources of
systematic errors. The elementary tools of the data analysis pipeline
are outlined in section~\ref{sec:systematics-control} and discussed in
more detail in one of the companion papers
\citep{ECO.systematics.paper}.
%

\section{Scientific objectives}
\label{sec:science}

The baseline science programme of \coremfive\ focusses on understanding the fundamental processes that gave raise to our observable Universe. This science case can be addressed with precise observations of the polarisation of the CMB. 
The primary science programme aims to:
\begin{enumerate}
\itemsep0em
\item understand the mechanisms that gave raise to primordial inhomogeneities in the very early Universe, and in particular constrain scenarios of cosmic inflation;
\item test the standard $\Lambda$CDM model and look for possible missing pieces in our understanding of the cosmological picture;
\item look for cosmological signatures of extensions of the standard model of particles and interactions.
\end{enumerate}
Additional aspects of this science programme, achievable with the same data, must be considered in order to fully exploit CMB polarisation observations. These extensions, also of major scientific interest by themselves, are:
\begin{enumerate}
\itemsep0em
\setcounter{enumi}{3}
\item investigate and understand the cosmic structures that generate secondary CMB ani\-sot\-ropies superimposed on the primordial ones, in particular through the distortion of the CMB polarisation by gravitational lensing, which mixes polarisation $E$ and $B$ modes;
\item understand the astrophysical emission processes that are a source of foreground contamination for CMB polarisation observations;
\item understand the dust-obscured star-formation phase of galaxy evolution;
\item analyse cosmic dipoles in the microwave to test the isotropy and homogeneity of the Universe at the largest scales, and constrain energy dissipation processes from different cosmic epochs, including reionisation, through dipole spectrum distortions.
\end{enumerate}

We now further expand the main themes of this science case, concentrating on the transformational results that will be achieved with \coremfive. We split the science case into five main areas: inflation; testing and constraining in detail the standard hot big-bang $\Lambda$CDM cosmological model; constraining the standard model of particles and interactions; mapping structures in the Universe; and the legacy value of the \coremfive\ survey for other science goals.

\subsection{Inflation}

Cosmic inflation, postulated in the early 1980s to solve a number of
puzzles of the standard Big-Bang \citep[][and references
therein]{2014A&A...571A..22P,2016A&A...594A..20P}, is the current
baseline generic scenario for the generation of primordial
perturbations in the early Universe.  Inflationary models generically
predict the existence of primordial tensor perturbations at very early
times \citep{2016ARA&A..54..227K}.  The amplitude of these tensor
modes is parameterised with the tensor-to-scalar ratio, $r\equiv T/S$,
which specifies the power of tensor perturbations relatively to that
of scalar perturbations. Tensor modes (primordial gravitational waves)
contribute to the total CMB temperature anisotropies and to
polarisation $E$ and $B$ modes, while scalar modes (primordial density
perturbations) contribute only to $T$ and $E$ modes. The detection of
primordial CMB polarisation $B$ modes would provide direct evidence
for cosmic inflation and for quantum fluctuations of space time, as
well as determining the energy scale relevant for the inflationary
epoch.  Unambiguous detection of these primordial $B$ modes is hence
one of the primary targets of a future CMB space mission.

A few special cases of inflation deserve special attention. The
simplest models of single-field inflation with large fields ($\Delta
\phi > m_{\rm Planck}$) predict $r\gtrsim 0.002$--0.003 \citep[the
so-called Lyth bound,][]{1997PhRvL..78.1861L}. For a simple,
single-field slow-roll model, an expansion in terms of slow-roll
parameters, $\epsilon$ and $\eta$, gives $n_{\rm s}-1 =
2\eta-6\epsilon$, while $r=16\epsilon$. Taking \Planck's measurement
of $n_{\rm s} = 0.9655 \pm 0.0062$ for a standard $\Lambda$CDM model,
we infer that $6\epsilon-2\eta = 0.0345 \pm 0.0062$. If $\eta \, \lsim
\, \epsilon$, we have $\epsilon \simeq 0.005$--0.01 and $r \simeq
0.1$--0.2. This scenario is already in mild tension with the current
upper limit of $r\leq 0.07$ at 95\% CL, coming from BICEP/Keck after
foreground cleaning using \WMAP\ and \Planck\ data
\citep{2016PhRvL.116c1302B}, but is not completely ruled out. Taking
the 3$\,\sigma$ upper limit on $n_{\rm s}$, we get instead $r \simeq
0.05$, still compatible with present-day measurements; however, this
is likely to change in the coming years, with either a detection or a
clear rejection of this model.

In the case of the Starobinskii $R^2$ model
\citep{1980PhLB...91...99S}, the predicted level is $r = 3(n_{\rm
  s}-1)^2$ instead. The \Planck\ constraint on $n_{\rm s}$ suggests $r
\simeq 0.0035$, and $r \geq 0.0008$ for $n_{\rm s}$ at its 3$\,\sigma$
upper limit. This is a much bigger challenge for sub-orbital
observations. Section~\ref{sec:why-space} discusses why making a clear
detection of $B$ modes at this level must be done from space.

Lastly, we note that small field inflationary models, in general
easier to connect to fundamental physics, generically predict values
of $r$ smaller than the Lyth bound. Hence, $r \ll 0.001$ is as
plausible a scenario as anything else~\citep{1997PhRvL..78.1861L}. In
that case, detecting primordial inflationary $B$ modes is out of reach
for the foreseeable future, but nevertheless the precise and accurate
observation of CMB polarisation is still a scientific necessity ---
even if $r \ll 0.001$, ruling out large-field models is an essential
piece of information that cannot be obtained by any other means.

In the coming years, the sensitivity of ground-based CMB observatories
will substantially increase. Multi-frequency observations in the
atmospheric windows will improve the capability of controling
foreground contamination. This evolution is likely to result in
substantial improvement of the current upper limit of $r<0.07$,
perhaps down to $r \, \lsim \, 0.01$; although the contamination of
the observations by Galactic foreground emission is a challenge that
should not be underestimated. We require that \coremfive\ should
perform at least 10 times better then this, i.e., be able to
unambiguously detect or rule out $r \simeq 0.001$. The capability of
detecting CMB $B$ modes to that level of $r$ --- which is both well
motivated scientifically and plausibly out of reach of suborbital
experiments alone --- is a natural science objective for a future
space mission. A non-detection would rule out all large-field
inflationary models. A detection would be a major discovery, and also
make it possible to clearly decide between some of the currently
favoured inflationary models.

Constraining inflation is not exclusively the domain of the detection
of primordial $B$ modes and measurement of the value of
$r$. Inflationary models are also meaningfully constrained by
tightening the measurement of the spectral tilt $n_{\rm s}$ and on the
variations of $n_{\rm s}$ with scale, as well as on the level of
non-Gaussian signatures in the CMB maps. \coremfive\ is also designed
to dramatically improve on these other inflationary observables.  If
primordial $B$ modes are detected, the tensor spectral index $n_{\rm
  t}$ also becomes an observable of interest. We refer the reader to
the relevant companion paper \citep{2016arXiv161208270C} for further
details.

\subsection{The cosmological model}

Many of the main cosmological observations, such as the homogeneity
and isotropy on large scales, the expansion rate, the abundance of
light elements, the growth of structure, CMB temperature and
polarisation anisotropies, statistics of galaxy distributions, cosmic
shear measurements, supernova brightnesses, and cluster number counts,
are compatible (within current uncertainties) with a $\Lambda$CDM
model with just six main parameters. The remarkable agreement of this
disparate set of observations with a relatively simple model also
represents several big puzzles: it suggests the inflationary paradigm
for the original of the initial density perturbations (as discussed
above); it invokes the existence in the Universe of an unknown type of
dark matter, representing roughly 25\% of the total matter-energy
density; and it also requires the existence of the even more
mysterious dark energy, accounting for about 70\% of the energy
content in the Universe at present, and responsible for the observed
acceleration of the expansion and for the dilution of large-scale
structures at late times.  Clues about the exact nature of both
``dark'' components are lacking, leaving room for many possible
options, as well as for imaginative theoretical speculation.

In addition, in spite of a remarkable overall concordance, some
apparent tensions exist between the model and subsets of the
data. Although currently near the limits of statistical significance
that would be required to seriously challenge the standard
cosmological model, these tensions add to a sense of unease in
postulating that our Universe is filled at the 95\% level with forms
of matter and energy that are completely unknown. As an example, the
Hubble constant today inferred by \Planck\ from the CMB at the
last-scattering epoch \citep{2016A&A...594A..13P} is discrepant at
about the 2.4\,$\sigma$ level with Hubble Space Telescope
Cepheid+SNe-based estimates \citep{2011ApJ...730..119R}. These recent
values are consistent with the earlier tension noted in the first
\Planck\ cosmological parameters paper \citep{2014A&A...571A..16P},
which generated some debate in the community. Another intriguing
discrepancy is found between the value of the amplitude $\sigma_8$ of
density perturbations at the scale of 8$h^{-1}$\,Mpc inferred from
cluster number counts \citep{2016A&A...594A..24P} and the value
inferred from the CMB alone.  Additional discrepancies, at a lower
level of significance have been suggested through the inferred amount
of lensing in the CMB angular power spectrum, via differences in
sub-sets of CMB data, in the curvature $\Omega_k$ and in other
specific parameters
\citep{2016ApJ...818..132A,2017A&A...597A.126C,2016arXiv160802487P}.
So how can we determine if these tensions are more than just
statistical fluctuations?

The CMB is currently the key observable for quantifying this
global cosmological picture. CMB photons probe the Universe at the
earliest possible times and on the largest possible scales. The CMB is
also the unique backlight that shines on \emph{all} structures between
the last-scattering surface at $z\simeq 1100$ and observers on Earth
at $z=0$. The complete exploitation of the information it carries is a
scientific imperative for cosmology
\citep{2014PhRvD..90f3504G,2016JCAP...06..046S}. With high S/N maps of
$T$, $E$ and $B$, yielding cosmic-variance dominated measurements of
the temperature and polarisation angular power spectra $C_\ell^{TT}$,
$C_\ell^{TE}$, $C_\ell^{EE}$, and angular power spectrum $C_\ell^{\phi
  \phi}$ of the CMB lensing potential, errors on cosmological
parameters as currently best constrained with the CMB by \Planck\
\citep{2016A&A...594A..13P} can be reduced by factors that can reach
an order of magnitude or more \citep{2016arXiv161200021D}. Such a
drastic improvement will clarify whether existing tensions are an
indication of a departure from the standard cosmological scenario, a
statistical excursion, or a systematic error in one of the
measurements. New tensions that are undetectable as of now are 
also likely to be uncovered--- when considering
extensions to the standard cosmological scenario, the total volume of
the error, represented by the figure of merit
\begin{equation}
{\rm FoM} = \Big(\det\big[\mathrm{cov}\{\Omega_{\rm b}h^2,\, \Omega_{\rm c}h^2,\, \theta,\, \tau,\, A_{\rm s},\, n_{\rm s}, ...\}\big]\Big)^{-1/2},
\end{equation}
computed from the covariance matrix of the errors on a set of
cosmological parameters, can be improved by a factor as much as
$10^7$, depending on the extensions considered. This improvement in
constraining the cosmological scenario is essential for making
progress on the current puzzles.  We refer the reader to
Ref.~\citep{2016arXiv161200021D} for an in-depth discussion regarding
future constraints on cosmological parameters with \coremfive\ alone,
as well as in combination with other cosmological data sets.

Finally, improving the determination of the CMB dipole amplitude and 
direction and comparing it with analagous investigations in other 
wavebands, which exploit signals from different types of astrophysical 
sources, probing different shells in redshift, provide an important test of
 fundamental principles in cosmology. The extension of boosting effects 
 to polarization and cross-correlations with \coremfive\ will enable a more 
 robust determination of purely velocity-driven effects that are not degenerate 
 with the intrinsic CMB dipole, allowing us to achieve an overall 
 signal-to-noise ratio close to that of an ideal cosmic variance limited 
 experiment up to a multipole $l \simeq 2000$ significantly improving on the 
 Planck detection. We refer the reader to Ref.~\citep{ECO.velocity.paper} for further discussion.

\subsection{Fundamental particles and interactions}

The standard model of particles and interactions is remarkably
successful at describing the fundamental laws of nature. Families of
elementary particles, which constitute the building blocks for all of
the experimentally observed forms of matter, as well as the carriers
of the known interactions between them, have been identified and their
main characteristics have been determined. However, this model is
incomplete.

First and foremost, there exists at present no model that unifies the
force of gravity with the other known forces of nature. The coupling
constant for gravity is so small that the gravitational interaction
cannot be probed on the scale of individual particles (the ratio of
gravitational to electric interaction between an electron and a proton
is of the order of $10^{-40}$). Gravity can only be probed with
massive objects, for which all other interactions are effectively
screened by factors of at least $10^{40}$. Hence, the cosmos is an
essential laboratory for understanding the laws of physics when
gravity is taken into account.

Even if one ignores gravity, the standard model of particle physics is
still incomplete for a number of other reasons. For instance, the
standard model does not currently explain why neutrinos have mass,
while the observation of neutrino oscillations implies a non-vanishing
difference of squared mass for the different eigenstates, i.e.,
\begin{equation} 
\Delta m_{12}^2 \simeq 7.5 \times 10^{-5} \,{\rm eV}^2
\end{equation}
and
\begin{equation} 
\left | \Delta m_{13}^2 \right | \simeq 2.5 \times 10^{-3} \,{\rm eV}^2,
\end{equation}
but do not constrain the absolute mass scale of the neutrinos
\citep{NakamuraPetcov2016}.

Measuring CMB lensing, $C_\ell^{\phi\phi}$, breaks parameter degeneracies and
enables estimates to be made for the sum of the neutrino masses
\citep[e.g.,][]{2003PhRvL..91x1301K}.  The precise predictions 
depend on details of the neutrino sector (e.g., whether they have
the normal or inverted mass hierarchy) and on what other data are used in
combination.  However, one conclusion of Ref.~\citep{2016arXiv161200021D}
is that \coremfive, together with {\it Euclid\/} and DESI should provide
$\sigma(M_\nu)=16\,$meV, yielding a $4\,\sigma$ detection of the neutrino
mass sum.

Accurate measurements of CMB polarisation can also constrain additional 
neutrino species or other light relics.  This is parameterised by the quantity
$N_{\rm eff}$, which has the value 3.046 in the standard model (slightly higher
than 3 because of details of neutrino decoupling).  The expected uncertainty
is $\sigma(N_{\rm eff})=0.041$ from \coremfive\ alone, and
$\sigma(N_{\rm eff})=0.039$ in combination with future BAO data
\citep{2016arXiv161200021D}.

There are many other directions in which physics beyond the standard model
can be constrained with a sensitive CMB polarisation survey such as planned
with \coremfive.  This includes:
dark matter annihilation and decay;
variation of fundamental constants;
topological defects;
and signatures of stringy physics.

\subsection{Structures}

Much of cosmic history is probed by observations of the growth of
structures after the last scattering of CMB photons. A space mission
dedicated to precision CMB polarisation science will also trace the
growth of cosmic structures using three independent probes: CMB
lensing; galaxy clusters; and the cosmic infrared background.

\paragraph{Lensing:} Gravitational lensing by large-scale structures
along the path of the CMB photons slightly distorts the anisotropy and
polarisation patterns of the primordial CMB
\citep{2006PhR...429....1L}. This gravitational lensing effect mixes
$E$ and $B$ modes, giving rise to lensing $B$ modes on all angular
scales. $B$ modes due to the lensing of $E$ modes into $B$ modes peak
at $\ell \simeq 1000$, i.e., angular scales of order 10
arcminutes. Their amplitude on larger scales is similar to that of
white noise of amplitude $5\,\mu$K.arcmin. For $r<0.01$, lensing $B$
modes dominate $B$-mode polarisation at all scales except the very
largest ones ($\ell<10$), and they dominate over the noise in the
error budget for detecting primordial $B$ modes when maps reach a
noise level of order $5\,\mu$K.arcmin. Hence, a space mission
attempting to observe $r \, \lsim \,0.01$ will also inevitably observe
CMB lensing, and have to deal with the corresponding contamination,
which degrades the sensitivity to primordial $B$ modes.

CMB lensing, however, is not only a nuisance for measuring
inflationary $B$ modes; it also is a unique observable for probing the
full distribution of matter between us and the last-scattering surface
at $z \simeq 1100$, i.e., in the whole observable Universe, in a way
that does not rely on baryonic tracers and does not require us to
understand non-linear growth effects in detail. It is a way of
directly observing the distribution of dark matter and hence is a
primary goal for future CMB observations.
CMB lensing effects have already been detected by \Planck\ and by
several ground-based experiments
\citep[e.g.,][]{2016A&A...594A..15P}. These clear detections however,
still have limited signal-to-noise ratio per pixel, and/or limited
sky-coverage. A future $B$-mode survey can transform this area of
research, providing accurate maps that can be used for precision
cosmology and cross-correlation with large-scale structure surveys.

\paragraph{Clusters:} Galaxy clusters, detectable in the frequency
range of interest for CMB observations, distort the CMB spectrum via
the thermal Sunyaev-Zeldovich (tSZ) effect, which is interaction of
the hot intracluster gas with CMB photons through inverse Compton
scattering \citep{2002ARA&A..40..643C}. Clusters are a particularly
sensitive probe of the growth of cosmic structure
\citep[e.g.,][]{2014A&A...571A..20P}.  By measuring the abundance of
clusters as a function of redshift, we can tightly constrain the dark
energy equation of state and the neutrino mass scale, and look for
deviations to standard gravity theory.  Doing this requires accurate
and precise calibration of the cluster mass-observable scaling
relations, which in turn requires good lensing measurements of cluster
masses out to redshifts $z>1$. A CMB temperature and polarisation
survey can calibrate the normalisation of the SZ signal-to-mass
scaling relation using CMB halo lensing.  To obtain enough clusters
and calibrate their scaling relation to sufficient accuracy requires a
survey covering a large sky fraction with angular resolution
comparable to the scale of clusters, and high sensitivity in
temperature and polarisation.

For the baseline survey, we expect that \coremfive\ will detect tens
of thousands of galaxy clusters, with several hundred at redshifts
$z>1.5$.  The cluster sample will extend to higher redshifts than the
{\it eROSITA} catalogue and will be a critical resource for studies of
galaxy formation in dense environments, especially when coupled with
NIR surveys such as those from {\it Euclid\/} and {\it WFIRST}.  
Using CMB lensing measurements towards detected clusters, 
the normalisation of the SZ signal-to-mass relation can be calibrated to
the percent level at $z<1$, and to better than 10\% at redshifts
approaching $z=2$.  Under these conditions and in combination with
primary CMB constraints, a large cluster catalogue will tightly constraint 
the dark energy equation of state.
Moreover, with enough sensitivity and frequency coverage a cluster survey
will enable: studies of the relativistic SZ effect by stacking hundreds of
clusters; extraction of cluster pairwise momentum at signal-to-noise
$>70$; and measurement of the evolution of CMB temperature with
redshift to test the standard model.  Even if galaxy clusters are not 
considered a design-driver for \coremfive, joint analysis of \coremfive\
and CMB-Stage~4 (CMB-S4) data sets will push the detection mass limit
towards $3\times10^{13}\,{\rm M}_\odot$ and increase the cluster yield
by a factor of 4 over either experiment alone, including at the higher
redshifts \citep{2017arXiv170310456M}.

\paragraph{Gas, stars and dust:} One of the major research trusts in
modern cosmology is the understanding of the relative distributions of
luminous stars, diffuse gas and dark matter
\citep[e.g.,][]{2010MNRAS.404.1111G}.  In particular, we need to
understand how baryons cool to form stars and are reheated by feedback
in a cycle that must be finely tuned to allow less than 10\% of
baryons to end up in stars.  This is a central question in galaxy
formation studies and a critical element for interpreting stage~4 dark
energy programmes.  The stage~4 lensing surveys rely on percent-level
predictions for the total matter distribution, but feedback can modify
the matter distribution much more than this. New avenues of research 
in this area will be opened by observing the distribution of
the gas, through both the tSZ and the kinetic SZ (kSZ) effects, as
well as the total matter distribution through CMB lensing. Even if, again, 
this science topic is not a design driver for \coremfive, the wide frequency 
coverage that is needed for CMB polarisation science also is essential 
to extract an all-sky tSZ
map that accurately separates the signal from foregrounds, especially the CIB
anisotropies that limited the \Planck\ result.  In complement to the large galaxy
samples from planned imaging and spectroscopic surveys (e.g., {\it
Euclid}, {\it WFIRST}, LSST, DESI, and PFS), a space mission that maps
CMB lensing, the tSZ effect, and the CIB will measure for the
first time the relative distribution of galaxies, gas and total matter
out to redshifts beyond the peak of cosmic star formation at $z \simeq
2$.  The CIB measurements will also trace star-formation
activity and dust production at critical epochs around the peak epoch
of star formation.  A future survey such as proposed with \coremfive\ will 
substantially improve on Planck the characterisation of CIB fluctuations 
in both temperature and polarization and will use the frequency dependence 
of CIB dipole to reduce by at least one order of magnitude the uncertainty 
of absolute CIB spectrum currently provided by COBE/FIRAS.
Even if not design-drivers of the \coremfive\ mission, all these measurements 
represent unique capabilities of \coremfive\ to address key questions in the 
development of structure.

\subsection{Legacy}

\coremfive\ is a space mission with the ability to produce
well-characterised maps of the complete sky in the 60--600\,GHz
frequency range, with both very high sensitivity and good angular
resolution compared to existing data. As such, the mission's data set
can also be used to answer scientific questions beyond the primary CMB
science objectives described above, to an extent that depends on the
extent to which each of these areas can be considered as a design
driver.

In particular, the need to monitor Galactic foreground contamination
for CMB science is illustrated in figure~\ref{fig:B-sensitivity}. For
a sky coverage of $\simeq 70$\%, foreground emission is the dominant
source of error for all $\ell \,\lsim \, 1000$, i.e., at all angular
scales larger than about 12$^\prime$. It also is above the full-sky
cosmic variance of $E$ modes in multipole bins of $\Delta \ell/ \ell =
0.3$, on all scales. While observing in the cleanest few per cent of
sky for the first detection of primordial $B$ modes might be possible,
the full exploitation of CMB $E$ modes and of lensing $B$ modes
requires observations over a substantial fraction of the sky to avoid
loss of sensitivity (because of cosmic variance). Hence, observing
foreground emission on all relevant angular scales is required. This
opens up the opportunity to investigate the role of the Galactic
magnetic field in structuring the Galactic interstellar medium
\citep[as has started to be done using \Planck\
data,][]{2016A&A...586A.133P,2016A&A...586A.135P,2016A&A...586A.136P,2016A&A...586A.137P,2016A&A...586A.138P,2016A&A...586A.141P,2016A&A...596A.103P,2016A&A...596A.105P}.

Magnetism is a facet of our cosmic origins that observations have yet
to uncover in any detail.  Magnetic fields are not observable directly
but they may be studied by observing polarised radiation.  Early
attempts have shown the existence of coherent magnetic fields on all
observed scales from proto-planetary discs to clusters of galaxies,
but current data are too limited to reveal the processes that have
amplified and organised the much weaker primordial field, and to
unveil the role that magnetic fields play in the formation of
galaxies, stars and planets \citep[e.g.,][]{1979cmft.book.....P}.

Cosmic magnetism is a rapidly-advancing topic across astrophysics.
The \Planck\ all-sky dust polarisation map was a spectacular highlight
of that mission, which has revealed the fingerprints of the Galactic
magnetic field on interstellar matter \citep{2016A&A...594A...1P}.
ALMA is driving a new revolution where magnetic fields will be imaged
along the star-formation sequence from pre-stellar cores to
proto-stars and their proto-planetary disks.  Over the coming decade,
stellar polarisation combined with Gaia astrometry should yield a 3D
model of the magnetic field of the Milky Way on Galactic scales.
Further in the future, the SKA will extend our horizon further,
probing magnetic fields in distant galaxies, clusters, and the cosmic
web, while \coremfive\ will offer unprecedented statistics on dust
polarisation from the Galaxy to characterise the interplay between
gravity, magnetic fields and turbulence in cosmic space.

Polarisation observations provide an opportunity to study
magneto-hydrodynamical (MHD) turbulence and dynamo action in great
detail within our Galaxy.  What can be learned from CMB experiments on
dust polarisation will complement advances expected from Faraday
tomography measurements with lower frequency telescopes like LOFAR,
eVLA, ASKAP, and SKA \citep[e.g.,][]{2017A&A...597A..98V}.  The
detection potential for relevant plasma processes and their
characteristic scales, like those of turbulent energy injection and
dissipation, can be increased considerably via the sensitivity and
statistics expected from a future CMB polarisation space mission,
which, in order to monitor foreground contamination, must necessarily
map dust polarisation with an unprecedented combination of sensitivity
and angular resolution.

Dust and synchrotron radiation from the Galaxy provide complementary
views of interstellar magnetic fields.  Synchrotron radiation traces
magnetic fields over the whole volume of the Galaxy, while dust
polarisation traces them largely within the disk, where interstellar
matter is concentrated and stars form.  The statistical properties of
Galactic magnetic fields are imprinted on those observables and
methods to extract this information from observational data have
started to be developed.  Quantities highly relevant for an
understanding of Galactic turbulence and dynamo processes, such as the
energy, helicity, and tension force spectra, have been shown to be
encoded in synchrotron intensity, polarisation, and Faraday rotation
measures.  Likewise, the analysis of the \Planck\ data has prompted a
number of studies that are relating dust polarisation to the
magnetic-field structure and its interplay with the density structure
of matter \citep{2016A&A...594A...1P}.  Since dust sub-mm emission is
an optically-thin tracer of all ISM components (neutral, atomic and
molecular, and ionised), dust polarisation is best-suited to
investigate the magnetised interstellar matter, in particular the
formation of its filamentary structure, and within filaments the
initial conditions of star formation.

Due to our location within the Galaxy, the fluctuations at a given
angular scale in observables correspond to magnetic-field structures
of different physical sizes. Disentangling these in order to identify
physical scales and processes is a challenge, which calls for a
statistical approach.  The leap forward in statistics of \coremfive\
compared with \Planck\ (a factor of a few hundred in the number of
measured modes) will greatly enhance our ability to identify
signatures of the processes involved in MHD turbulence, in particular
coherent magnetic-field structures associated with localised
dissipation of turbulent energy.  What will be learned from these data
will complement what will be probed by ground-based telescopes (e.g.,
ALMA and SKA) observing dust polarisation from compact sources and
Faraday rotation. \emph{Together}, these projects will have a major
impact on our understanding of the role of magnetic fields in galaxy
and star formation.

Extragalactic sources are also a potential contaminant of CMB
observations.  High-redshift, dusty galaxies can be observed at
sub-millimetre wavelengths with angular resolution better than that of
\Planck, which did not have diffraction-limited angular resolution in
its three highest frequency channels.  A full-sky survey, such as that
of \coremfive, would detect thousands of strongly lensed (and hence
extremely bright) high-$z$ galaxies distributed over the full sky,
which can then be studied in extraordinary detail through follow-up
observations. Also, \coremfive\ can be used to detect high-redshift
proto-clusters beyond the reach of surveys in other frequency
bands. \coremfive\ will also detect the polarised emission from
thousands of individual radio sources and dusty galaxies. These
science objectives are further discussed in a companion paper
\citep{2016arXiv160907263D}.

The observation of the background of unresolved high-redshift
dusty galaxies that form the CIB, an essential tool for delensing CMB
$B$ modes and detecting low-level primordial $B$ modes, also open up
the possibility of further studying cosmological star formation, as
discussed by Ref.~\citep{2016arXiv161202474W}.
Finally, precise analyses of the dipole spectrum over a wide frequency 
range give us the chance to significantly improve with 
respect to COBE-FIRAS in the recovery of CMB spectral distortion 
parameters for both early and late dissipation processes, from a 
factor of several up to about 50 (or even much better for an ideal 
experiment with the \coremfive\ configuration), depending on the 
quality of foreground removal and relative calibration, allowing us to 
detect, for example, the energy release associated with cosmological 
reionisation.

\section{Survey requirements}
\label{sec:requirements}

Starting from main scientific objectives, we now discuss how the
mission design stems from the survey requirements and goals, in terms
of overall sensitivity, angular resolution, and channels of
observation. In this section, we assume that we want to extract
essentially all the cosmological information encoded in CMB
polarisation only, with the space survey alone. Down-scope options
stating the requirements for similar CMB polarisation performance in
combination with ground-based observations, as well as interesting
up-grade options for extra science in addition to CMB polarisation,
are discussed in section~\ref{sec:options}.

\subsection{The need for a space mission}
\label{sec:why-space}

It is reasonable to ask how much of the above CMB polarisation science
programme can plausibly be done from the ground? Plans for a very
sensitive ground-based CMB experiment, CMB-S4, are being actively
made, with a science case that covers many of the topics discussed
above \citep{2016arXiv161002743A}. The strawman design for an
ambitious ground-based CMB-S4 programme targets a CMB sensitivity of
the order of 1\,$\mu$K.arcmin and angular resolution of 1--3$^\prime$
at 150\,GHz, with a sky fraction of around 50\% (spread over regions
with various levels of Galactic foreground contamination).

With such sensitivity and angular resolution, in the absence of
additional sources of error, CMB-S4 would outperform a space mission
such as \coremfive, for which the sensitivity of the full array is
1.7\,$\mu$K.arcmin, for an angular resolution in the 5--10$^\prime$
range.
However, coming back to the sources of error displayed in
figure~\ref{fig:B-sensitivity}, we note the following important
issues.
\begin{itemize}
\item Over the cleanest 70\% of the sky, foreground emission dominates
  over noise for all $\ell \, \lsim \,1000$; hence, all scales larger
  than about $12^\prime$ should be observed at multiple frequencies in
  order to reduce foreground contamination and achieve noise-limited
  observations.
\item Over the same sky fraction, foreground emission dominates over
  $B$ modes for all multipoles; again, efficient component separation
  will be needed on all scales to observe $B$ modes (both primordial
  and lensing) with noise-dominated performance.
\item Foreground residuals after component separation will be
  difficult to characterise, and are hence a source of potential bias.
  For such residuals to be below noise and/or cosmic variance
  uncertainties in bins of $\Delta \ell/\ell = 0.3$, foreground
  contamination must be reduced by at least 3 orders of magnitude
  \emph{in amplitude} at $\ell \, \simeq 10$, 2 orders of magnitude at
  $\ell \simeq 100$, and 1 order of magnitude at $\ell \simeq 1000$;
  this is unlikely to be doable with ground-based experiments, which
  must thus exploit only significantly cleaner, and hence much smaller
  sky regions.  With the reasonable assumption that only half or less
  of the $50\%$ sky observed from the ground can be safely used for
  precision cosmology, a ground-based survey can at most exploit the
  CMB on $\lsim 25\%$ of the sky.
\item The cosmic variance of full-sky $E$ modes dominates over noise
  for all $\ell \, \lsim \, 2500$.  For cosmological constraints based
  on polarisation $E$ modes, it is hence preferable to increase the
  size of the survey, rather than to observe smaller patches deeper;
  this is best done from space, with enough channels for accurate
  monitoring of the foreground emission.
\item The cosmic variance of full-sky lensing $B$ modes dominates over
  the noise for all $\ell \, \lsim \, 1000$.  Hence, again, for
  cosmological constraints based on polarisation lensing $B$ modes, it
  is preferable to increase the size of the survey, rather than to
  observe smaller patches deeper; in addition, the confusion between
  primordial and lensing $B$ modes dominates the error on primordial
  $B$ modes for all scales below $\ell \simeq 1000$.  Space offers the
  opportunity to accurately map the CIB, for $B$-mode delensing by
  a factor of 2--3 over a large fraction of the sky; for a quick
  comparison, CIB-based delensing by a factor of 2--3 over 70\% of sky
  is as efficient at reducing the cosmic variance of residual lensing
  as CMB-based delensing by a factor of 5--8 over 10\% of sky (which
  requires $\sqrt{7}$ times better delensing to compensate for the
  reduced sky fraction).
\end{itemize}
For all these reasons, when considering multipoles up to $\ell \simeq
1000$--2000, the performance of future CMB observations for exploiting
CMB polarisation power spectra will be limited not by raw detector
sensitivity, but \emph{by the capability of removing foreground
  contamination and by the capability to separate lensing $B$ modes from primordial
  $B$ modes over the largest solid angle}. A space mission with
sufficient sensitivity and angular resolution is vastly superior to
ground-based observatories for controlling these main sources of error
over a large fraction of the sky.

We now discuss in more detail some of the key issues with ground-based
observations, so that we can make a realistic assessment of the
capability of ground-based programmes to reach the \coremfive\ science
targets.

\subsubsection{Atmosphere}

CMB temperature and polarisation anisotropies are best observed in the
frequency range extending from a few tens to a few hundreds of GHz,
i.e., at wavelengths ranging from about 1\,mm to about 1\,cm, around
the peak of the CMB 2.725-K blackbody emission. In this frequency
range, ground-based observations are possible in a set of windows
through which the atmosphere is sufficiently transparent. The main
atmospheric windows are centred around minima of atmospheric emission
at about 30, 90, 150, and 220\,GHz. The transmission at $60^\circ$
elevation is of order 99\% at $30\,$GHz, 98\% at 90 and $150\,$GHz and
of order 96\% at 220\,GHz from the Atacama plateau, when the amount of
precipitable water vapour is at the level of $0.5\,$mm (at Llano de
Chajnantor in Chile, the observing conditions are better than that
about 25\% of the time).
Even in these atmospheric windows, the atmosphere contributes to the
total photon background and hence the photon noise, so that the
mapping speed of a space-borne instrument is at least 100 times better
than on the ground, for an identical number of detectors (see appendix 
\ref{appendix:atmosphere}).

Even more problematic than background loading, fluctuations of
atmospheric emission due to inhomogeneities in temperature or water
vapour content generate strong parasitic signals, and are a source of
unstable calibration (because of varying airmass and opacity), i.e.,
from the ground, the CMB is observed through a shiny and fluctuating
curtain of atmospheric absorption and emission.
In the best CMB channels for ground-based observations (90 and
150\,GHz), about 75\% of the time the atmosphere above one of the best
observing sites on Earth (the Atacama plateau) is more than 2\%
emissive, i.e., contributes a background of more than 6\,K. As the
telescope scans the sky, it scans through inhomogeneities of this
emission. Even at a level as low as 0.1\% (easily achieved with
fluctuations of air temperature and/or water content at about the same
order of magnitude), one gets 6\,mK of spurious large-scale signal or more,
correlated between focal-plane detectors. A more detailed model of
atmospheric turbulence gives fluctuations in the 15--30\,mK range for
the best 25\% of the observing time \citep{2015ApJ...809...63E}.  At a
scale of around $2^\circ$ this atmospheric signal is about 6 orders of
magnitude larger than the 8-nK raw sensitivity of a 1\,$\mu$K.arcmin
survey. Since scanning the same patch $10^{12}$ times is not a
realistic option, this signal must be removed by a combination of
processing, e.g., filtering, exploitation of multi-frequency or
multi-detector observations with analysis methods such as those
discussed in Refs.~\citep{2002MNRAS.330..807D} or
\citep{2008ApJ...681..708P}, and polarisation modulation with a
rotating half-wave-plate. Current observations demonstrate that
polarisation modulation can reduce this signal by 2--3 orders of
magnitude in amplitude. Even then, residuals are still at a
challenging 3 orders of magnitude above the target sensitivity on
$2^\circ$ angular scales. The situation is even worse at larger scales
and/or at higher frequencies.

A space mission completely avoids the complexity of atmospheric
absorption, emission, and fluctuations. More details about the atmosphere
can be found in Appendix~\ref{appendix:atmosphere}.

\subsubsection{Astrophysical foregrounds}
\label{sec:foregrounds}

CMB observations must address the problem of astrophysical foreground
emission. At frequencies below about 100\,GHz CMB observations are
contaminated by a complex mixture of low-frequency astrophysical
sources of electromagnetic radiation that include Galactic
synchrotron, free-free, and anomalous dust emission (presumably
spinning dust, or possibly magnetic dust, or both), as well as
numerous extragalactic radio sources, while at frequencies above about
100\,GHz thermal dust emission is the dominant foreground. At
frequencies above approximately 200\,GHz, in addition to thermal dust
emission, anisotropies of the cosmic infrared background (CIB), and to
a lesser extent zodiacal-light emission dominate the fluctuations of
the observed sky brightness over most of the sky. Molecular lines,
notably those of carbon monoxide at multiples of 115\,GHz, clearly
seen in \Planck\ data \citep{2014A&A...571A..13P}, must also be taken
into account (as well as those of isotopologues at nearby frequencies, 
$^{13}$CO and C$^{17}$O near multiples of 110\,GHz, C$^{18}$O near 
multiples of 112\,GHz). Several lines of CO emission, at about 220, 225 and
230\,GHz, are located in one of the main atmospheric windows.

Of all of those foregrounds, synchrotron is the one that is known to
be the most polarised (in theory, up to 75\%). At high frequency, only
thermal dust is known to be very clearly polarised
($\gsim$10\%). Other sources of emission can be somewhat polarised (at
a level $\lsim$1\%).

While for temperature fluctuations there are regions where CMB signals
strongly dominate over astrophysical foregrounds, this is not the case
for polarisation (see figure~\ref{fig:B-sensitivity}). To remove
foreground contamination, it is necessary to observe the sky at
several frequencies and exploit the fact that the emission law of the
CMB is substantially different from that of most foreground emission
processes \citep{2009LNP...665..159D}. Exploiting these colour
differences is best done by combining observations in a set of well
chosen different frequency bands. \COBE-DMR observed the sky in three
different frequency bands, \WMAP\ in five, and \Planck\ in nine. To
exploit the largest possible fraction of the sky at a sensitivity
level at least an order of magnitude better than \Planck, even more
frequency bands will be required.

Although the details will be known only with observations at the
appropriate level of sensitivity, i.e., with future CMB data
themselves, a simple accounting argument suggests that no less than
ten channels are required, and preferably more.
The two main known polarised Galactic emission sources in the
frequency range of interest are synchrotron and thermal dust.  To
model the synchrotron (parameterised in each sky pixel by intensity,
spectral index, and possible curvature of spectral index), at least
three low-frequency channels are required, four to provide some
redundancy, or even more if one has to model synchrotron emission with
more than one simple emission law per pixel. The same is true for
thermal dust at high frequency, for which at least three parameters
per pixel are needed to adjust a model with a single modified
blackbody emission law. Four channels at least are needed to make this
measurement, with an extra channel for a consistency check.  This
means that a total of eight channels are needed to model the
foreground emission if only synchrotron and dust must be taken into
account. On top of this, thermal dust emits with more than one
population of grains so it is possible that a single modified
blackbody is not sufficient for a model that is accurate at the $\lsim
1$\% level.
Finally, the CMB itself must be observed in at least two frequency
bands that are sufficiently distant in frequency for a useful
cross-check, which is essential to detect possible residual foreground
emission, and preferably more bands to understand the origin of any
discrepancy (\Planck\ effectively used comparisons between the CMB
seen in four channels, at 70, 100, 143, and 217\,GHz, to investigate
foreground residuals and systematic errors). The conclusions is that
ten channels is the absolute minimum to monitor foreground emission in
polarisation.

There is no way for this number of frequency channels, which must be
well spread over the useful frequency range, to be accommodated in
only four atmospheric windows. While synchrotron can in principle be
observed from the ground at a few frequencies $\nu \, \lsim \,
30\,$GHz, thermal dust emission, which dominates at frequencies where
the observing conditions from the ground are poor, must be mapped from
space (or, as an intermediate solution, from stratospheric balloons,
which, however, have significant residual atmospheric noise, less
flexibility for choosing the observing strategy, and much reduced
observing time compared with a space mission).

\subsubsection{Systematic effects}

Very precisely controlling systematic effects due to non-idealities in
the instrument, which are a potential major source of error for future
sensitive CMB observations, is mandatory for achieving the science
goals of future CMB polarisation observations. Instrumental
imperfections include complex response of the instrument to external
radiation or stimuli. Such non-idealities impact the shape of the
response in time, in space (beams), or in frequency (bandpass); in
practice these can be different from the design specifications and can
be mismatched between detectors.  There can also be gain fluctuations,
susceptibility to events that are not related to the observations
(such as cosmic ray hits), magnetic susceptibility, variations of the
observing environment that impact the detector response, etc.
To minimise such effects, space offers unmatched observing conditions,
in an extremely stable environment.  This allows for:
\begin{itemize}
\itemsep0em
\item minimising sidelobe pickup of emission from the Earth, Sun, and
  Moon;
\item minimising thermal fluctuations of parts of the instrument that
  are optically coupled to the detectors;
\item avoiding fluctuations of the response of the instrument, which
  is essential for enabling the calibration of instrumental
  imperfections at the level of accuracy required to correct for their
  impact on the CMB science;
  \item redundancy on different timescales without need for specific re-pointing, 
  nor changing the observing conditions;
  \item maximal sky coverage, allowing for redundant analyses exploiting
   independently several large regions of sky. 
  \end{itemize}
This is essential because the impact of systematics can be 
minimised and accurately assessed only through analyses which require
enough redundancy and a stable instrument.

\subsubsection{Why space -- summary}

Sub-orbital CMB observations have been key pathfinders in CMB science,
from the initial detection of the CMB more than fifty years ago to
now. However, it also is true that all the major steps forward have
been achieved by space missions. \COBE-DMR, \WMAP, and \Planck\ have
all been transformational for cosmology: \COBE\ confirmed the
blackbody spectrum of the CMB, ruling-out alternatives to the hot
Big-Bang scenario, and detected the first temperature anisotropies
that were required to explain the origin of structures; \WMAP\ set the
stage for precision cosmology; and \Planck, in turn extracted
essentially all cosmological information available in the CMB TT
spectrum, provided today's reference maps of CMB temperature and
polarisation, of CMB lensing on large scales, and of polarised
emission of astrophysical foregrounds that contaminate the CMB signal
\citep{2016A&A...594A...1P}.

Data from CMB space missions have also been essential for planning and
analysing CMB observations made with ground-based instruments. The
sensitive polarisation observations on degree-scales made from the
ground with BICEP2 and the Keck array have used the high-quality
\Planck\ CMB temperature map for calibration, and for systematic
effect corrections.  \Planck's observations have also been essential
to assess the level of dust in the BICEP2 detection of $B$-mode
polarisation. Thinking a decade into the future, high-quality CMB
$E$-mode polarisation maps, as well as $E$- and $B$-mode foreground
maps from space missions, will also be essential for the
interpretation of deep observations of CMB polarisation from the
ground on selected patches of the sky.

There are good reasons why the best observations of CMB temperature
anisotropies have been carried out from space, starting with the first
detections with \COBE-DMR to the precision cosmological picture
brought by \Planck. Polarisation is more challenging, and must be done
from space. Ground-based observations will serve as a technological
roadmap, and for observing the small scales that are too costly from
space. Along the way, the sub-orbital programme may bring some
breakthroughs, perhaps even including a first tentative detection of
primordial $B$ modes. Irrespective of what happens, however, the next
polarisation space mission should be designed to wrap up everything
and to deliver CMB polarisation data sets that will be the reference
for decades to come. Just as \Planck\ did for temperature anisotropies
before, \coremfive\ is designed with this objective in mind for
polarisation.

\subsection{What survey?}

The range of options for a future space mission is quite wide, as
illustrated by the diversity of existing proposals. The main drivers
of the design of the survey we propose are: (i)~guaranteed scientific
breakthrough for the science targets discussed above (at the time the
mission delivers its scientific results); (ii)~maximum science versus
cost, within the programmatic and budgetary constraints of an ESA
M-class mission; (iii)~focus on the science objectives and
observations that are out of reach from the ground, with a priority on
precision CMB polarisation science; (iv)~within practical constraints,
achieve a performance that is limited by fundamental and astrophysical
constraints (e.g., the CMB cosmic variance, the possibility to
separate foregrounds in practice, and the possibility to separate
inflationary $B$ modes from lensing $B$ modes) rather than by the
instrument, so that the mission is ``near ultimate'' (in the sense
that another CMB polarisation mission will not be required after
\coremfive).
The main parameters defining the mission, further discussed in the
next subsections, are:
\begin{itemize}
\itemsep0em
\item sky coverage (section~\ref{sec:coverage}), for best sensitivity
  on the largest scales and for reducing the part due to cosmic
  variance in the determination of $E$- and $B$-mode power spectra;
\item sensitivity and angular resolution
  (section~\ref{sec:detperformance}), which determine the capability
  of the mission to detect $E$ and $B$ modes with sufficient S/N over
  the required range of angular scales;
\item frequency range of observations and number of frequency channels
  (section~\ref{sec:channels}), which govern the ability of the
  experiment to separate CMB observations from other sources of
  astrophysical emission;
\item strategy for modulating polarisation and controlling systematic
  effects (section~\ref{sec:systematics}), in particular those arising
  from the confusion between CMB temperature anisotropies and CMB
  polarisation.
\end{itemize}

\subsection{Sky coverage}
\label{sec:coverage}

\coremfive\ must observe the CMB over the largest possible sky
fraction to accurately measure the largest angular scale $E$ and $B$
modes, where the reionisation bumps in the CMB polarisation power
spectra are located ($\ell \, \lsim \, 10$). It is essential for the
mission to detect both the reionisation and recombination bumps of the
primary CMB in order to confirm the inflationary origin of any
detected $B$ modes. The largest scales are also essential for
measuring the reionisation optical depth $\tau$ from $E$-type
polarisation, and to lift the degeneracy of $\tau$ with the sum of
neutrino masses. On the largest angular scales, one can also check
whether anomalies detected in temperature maps by \WMAP\ and \Planck\
\citep{2004ApJ...605...14E,2011ApJS..192...17B,2014A&A...571A..23P}
are also visible in polarisation maps.  Since measuring these large
scales from the ground is very challenging (if possible at all), a
space mission is the only way to obtain this essential piece of
information.

Considering also that at all angular scales the cosmic variance
depends on the inverse of the sky fraction, we must seek to observe a
large solid angle (e.g., about 50\% of clean CMB sky or more) so that
error bars on the measured spectra are not increased much (e.g., not
by more than $\sqrt{2}$) by reason of cosmic variance. Given this
constraint, and considering the legacy value of the observations, it
is logical (but not strictly required) to plan that a CMB space
mission should observe the complete sky, as was the case for all other
CMB space missions before. The sampling variance being minimal for
homogeneous survey depth, we also initially require the sky coverage
to be as uniform as possible (the same observation time in all
pixels), although this last requirement can be relaxed with better
overall mission sensitivity.

However, if $r \, \lsim \, 10^{-3}$, a tentative detection of
primordial $B$ modes will require very accurate delensing and
foreground subtraction. If the errors turn out to be dominated by
foreground residuals it would make sense (for the particular objective
a detecting very low primordial $B$ modes) to seek the best possible
sensitivity on the cleanest 10--20\% of sky that will also be observed
from the ground, to ultimately combine observations with both the many
frequency channels provided by space (for foreground monitoring), and
the high angular resolution in atmospheric windows provided by the
ground (for optimal delensing using the CMB only).  We should hence
envisage the capability to concentrate observing time on selected
patches at the end of the mission in the case that primordial $B$
modes still escape detection by then. Although not a strict
requirement, this capability could also be used for non-CMB science on
targets of interest, such as Galactic dust filaments, interesting
galaxy clusters, or deep fields observed by other instruments, for an
extension of the mission's science harvest.

\subsection{Sensitivity and angular resolution}
\label{sec:detperformance}

\begin{figure}[tbp]
\centering 
\includegraphics[width=\textwidth]{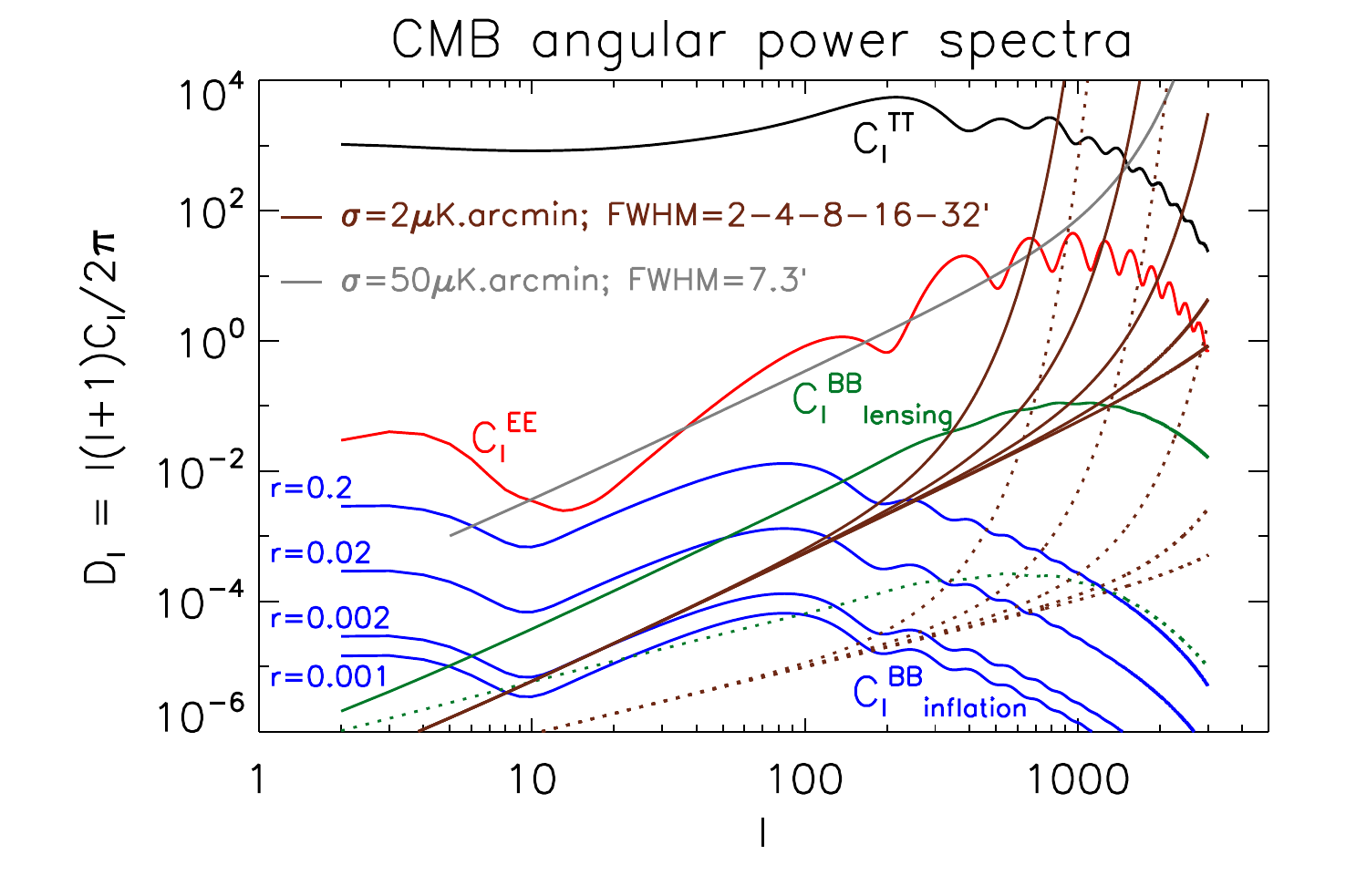}
\caption{\label{fig:plotcmb} CMB temperature and polarisation angular
  power spectra. The grey line corresponds to the spectrum of
  instrumental noise representative of the polarisation sensitivity
  achieved with the \Planck\ mission, at the level of 50 $\mu$K.arcmin
  for a beam of $7.3^\prime$. While \Planck\ has observed most of the
  CMB temperature anisotropies with a high signal-to-noise ratio
  (better than inferred from the figure, which shows the {\it
    polarisation\/} sensitivity), $E$ modes have been observed with
  $S/N \sim 1$, and the level of lensing $B$ modes has been detected
  only statistically.  Solid brown lines show the level of noise for a
  map sensitivity of 2\,$\mu$K.arcmin and a beam of $2^\prime$,
  $4^\prime$, $8^\prime$, $16^\prime$, or $32^\prime$ (from bottom to
  top), and dotted brown lines correspond to the error on $C_\ell$ for
  the same noise level and angular resolution, after smoothing by
  $\Delta \ell = 0.3 \ell$. A full-sky observation with angular
  resolution of $8^\prime$ or better would detect lensing $B$ modes
  (green line) with $S/N=2.5$ per individual mode up to $\ell \simeq
  1000$. For $r> 0.002$, both the recombination bump (peaking at $\ell
  \simeq 80$) and the large-scale reionisation bump can be detected in
  bins of $\Delta \ell = 0.3 \ell$. The error, however, is dominated
  by the cosmic variance of the $B$-mode lensing spectrum in such bins
  (dotted green). Unambiguously detecting the recombination bump at
  the level of 0.001 requires reducing by a factor of $\simeq 3$ the
  contamination by lensing $B$ modes (delensing).}
\end{figure}

To obtain all of the information from CMB polarisation, it is
necessary to measure with good S/N all of the CMB $E$ and $B$
modes. Figure~\ref{fig:plotcmb} compares the final error on
polarisation $C_\ell$ for a noise level of 2\,$\mu$K.arcmin
(sufficient for measuring lensing $B$ modes with S/N $\simeq 2.5$ per
mode in amplitude) with a beam ranging from $2^\prime$ to $32^\prime$.
This illustrates the modes that are lost when the angular resolution
is reduced.

\paragraph{$E$ modes:}
For a beam of $4^\prime$ or smaller, $E$-mode observations are signal
dominated almost up to $\ell \simeq 3000$. When the beam gets bigger,
the $E$-mode measurement becomes progressively more degraded, so that
for a $32^\prime$ beam, the future mission would not do better than
\Planck\ for $\ell \, \gsim \, 700$ (the exponential cut-off of the
Gaussian beam yields a very rapidly raising error at high
$\ell$). Hence, the ultimate $E$-mode measurement requires a noise
level $\lsim \, 2\,\mu$K.arcmin and angular resolution $\lsim \,
4^\prime$.

\begin{table}[htb]
\begin{center}
\begin{tabular}{|c|c|c|c|c|c|c|}
\hline 
Beam& 2$^\prime$& 4$^\prime$& 8$^\prime$& 16$^\prime$& 32$^\prime$& 64$^\prime$\\
\hline 
\hline 
  $b_\ell^2$ for $\ell=80$& 1.000& 0.998& 0.994& 0.975& 0.904& 0.667\\ 
 $b_\ell^2$ for $\ell=200$& 0.998& 0.990& 0.961& 0.855& 0.534& 0.081\\ 
$b_\ell^2$ for $\ell=1000$& 0.941& 0.783& 0.378& 0.020& 0.000& 0.000\\ 
\hline
\end{tabular}
\end{center}
\vspace{-\baselineskip}
\caption{\small Impact of the beam on $C_\ell^{BB}$ over the first $B$-mode recombination bump, at $\ell = 80$ and $\ell = 200$, and at the peak of the lensing $B$ modes ($\ell = 1000$), for various angular resolutions.}
\label{tab:beam-for-primordial}
\end{table}

\paragraph{Inflationary $B$ modes:}
For primordial $B$ modes, under the assumption that the lensing
contamination can be completely removed by other means (external data
sets such as the CIB, futuristic matter surveys, or lensing maps from
very accurate small-scale CMB polarisation observations), dotted brown
lines in figure~\ref{fig:plotcmb} show that 2\,$\mu$K.arcmin seems
adequate to detect r=0.001. However, we also require that the mission
provides a means of delensing $B$-mode maps, so that the lensing
residual does not exceed much the noise level of the mission.

The angular resolution of the final CMB map should be such as not to
reduce the effective sensitivity to the recombination bump, which
peaks at $\ell \simeq 80$ and extends up to $\ell \simeq 200$
(higher-order bumps are well below the noise level for
2\,$\mu$K.arcmin noise, irrespective of the angular resolution).
Table~\ref{tab:beam-for-primordial} gives the effective beam damping,
$b_\ell^2$ at $\ell=80$ and $\ell=200$ for various beam
sizes. Covering that range of multipoles is important to confirm the
inflationary origin of the observed signal. A beam of $64^\prime$
clearly is very sub-optimal, since it degrades the sensitivity to
primordial $B$ modes across the recombination peak, from a factor of
2/3 at $\ell=80$ to a factor of more than 10 at $\ell=200$. With a
loss of power ranging from 2.5\% to 14.5\% only between $\ell=80$ and
$\ell=200$, a beam of 16$^\prime$ is completely adequate. A beam of
32$^\prime$ is adequate at $\ell=80$ (10\% power loss), but marginal
at $\ell=200$ (47\% power loss).

\begin{figure}[tbp]
\centering 
\includegraphics[height=0.39\textwidth]{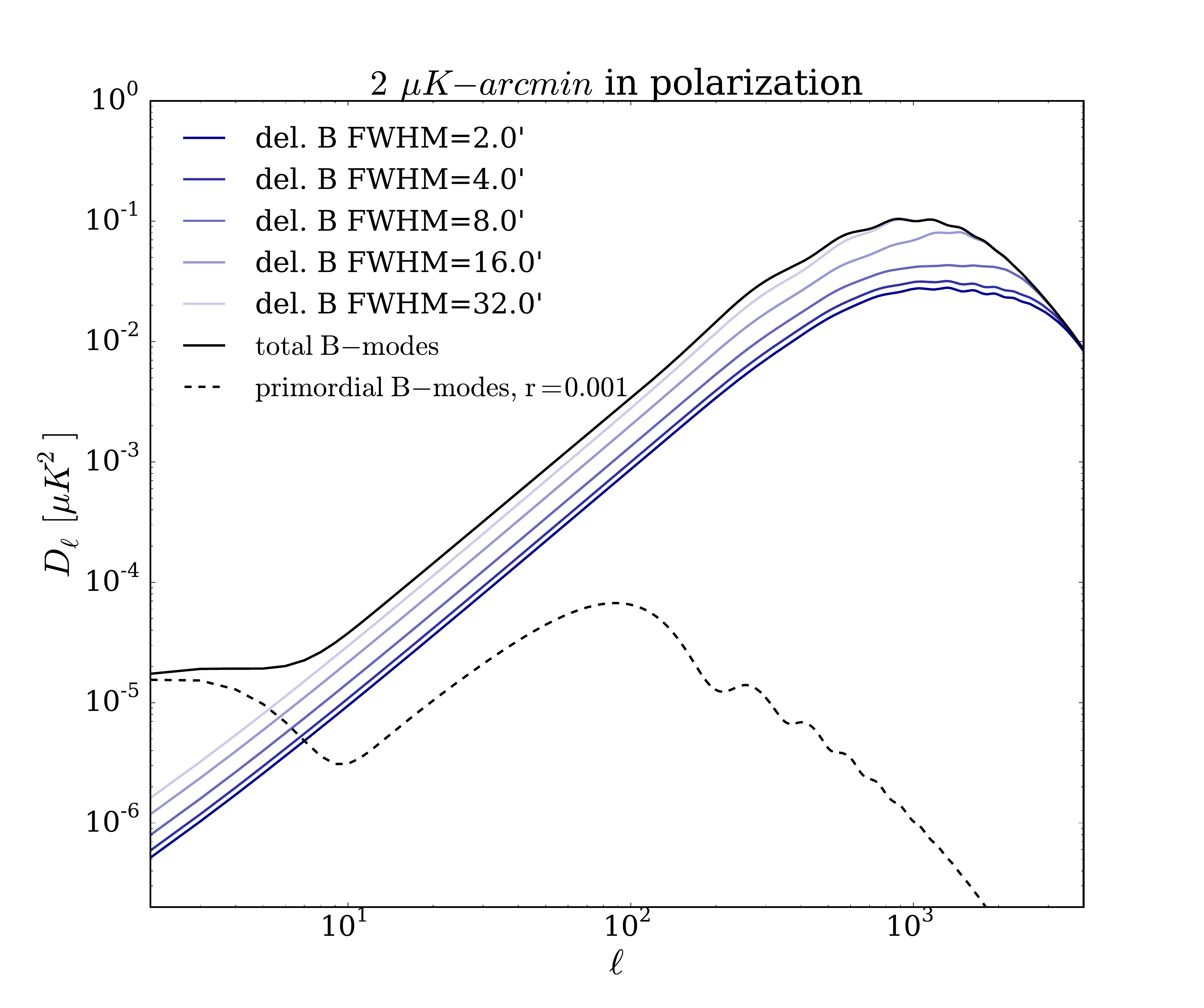}
\includegraphics[height=0.39\textwidth]{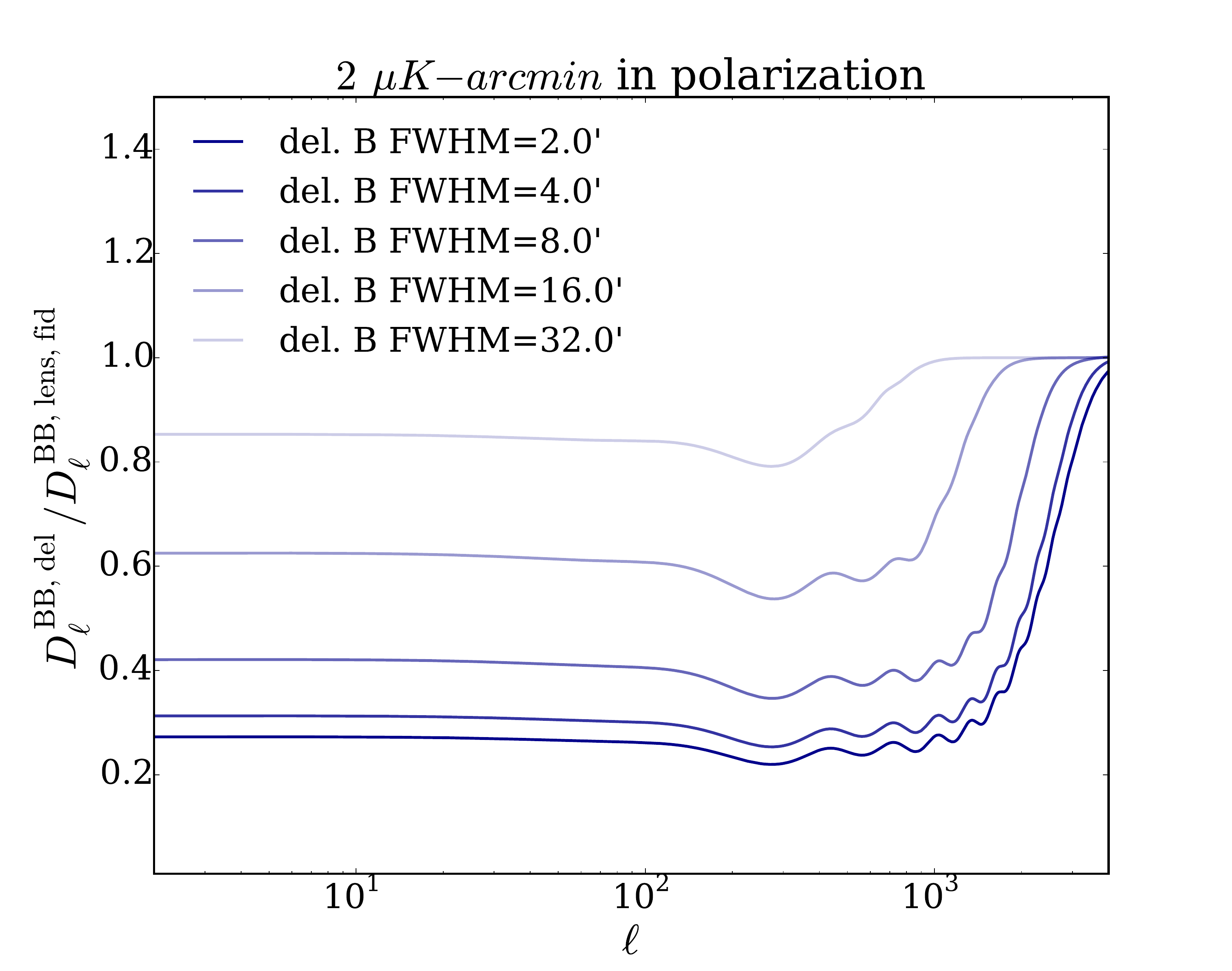}
\includegraphics[height=0.38\textwidth]{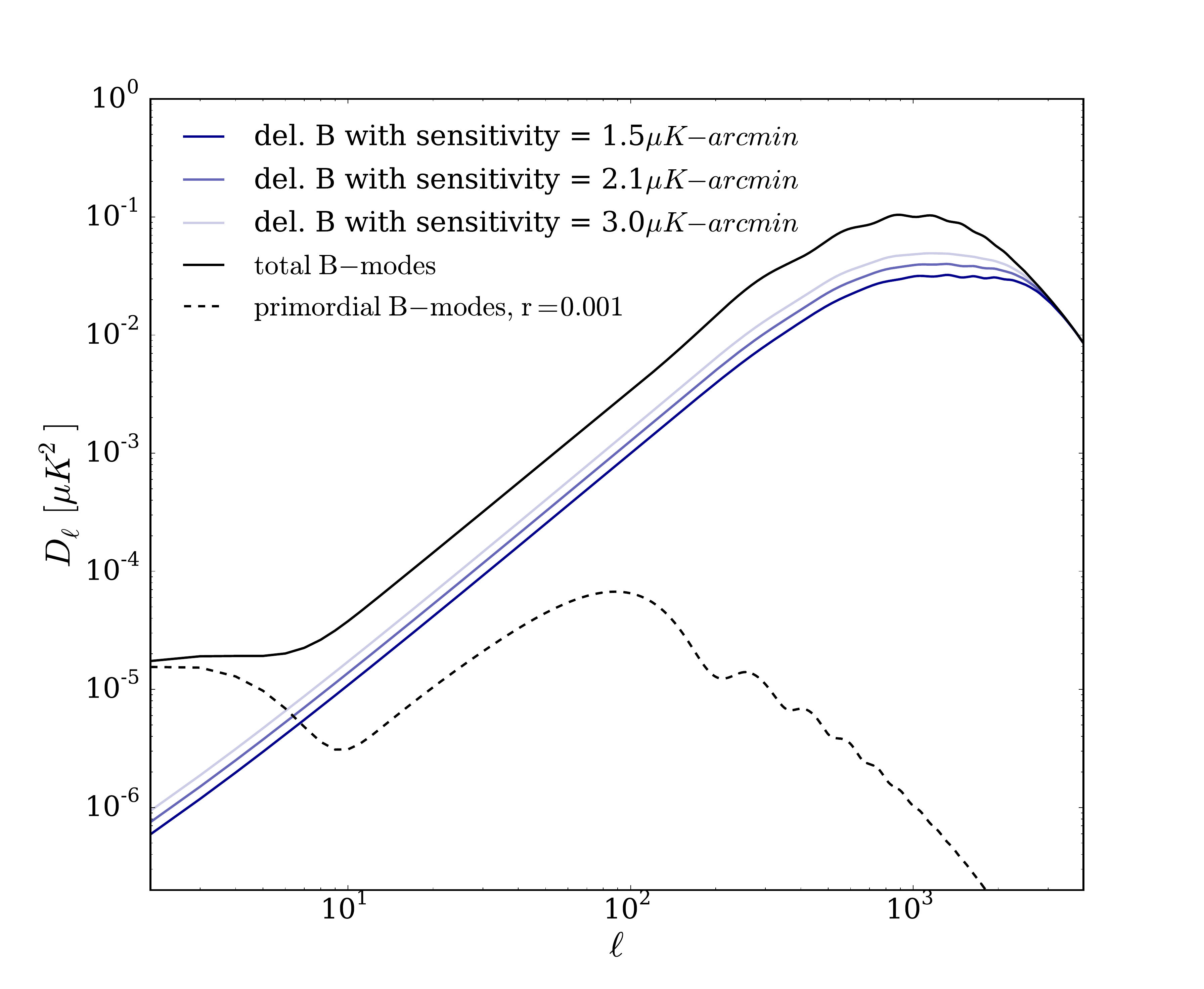}
\includegraphics[height=0.37\textwidth]{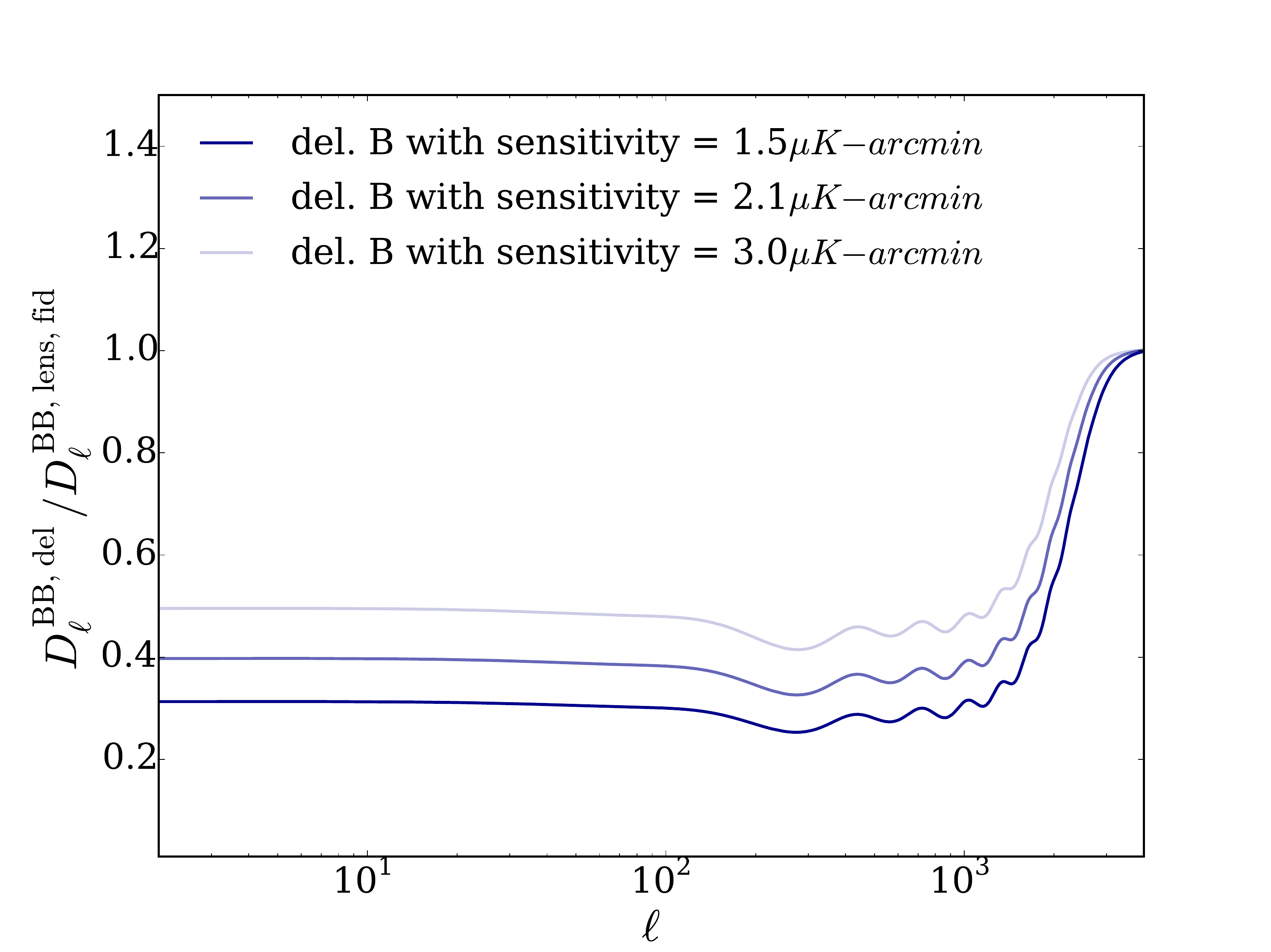}
\caption{
  \label{fig:plotdelensing} Effectiveness of CMB $B$-mode polarisation
  delensing for varying sensitivity and angular resolution
  assumptions.  {\it Top}: Delensing effectiveness for
  2\,$\mu$K.arcmin noise and various beam sizes. The left panel shows
  the level of residual lensing, compared to primordial $B$ modes for
  $r=0.001$, and the right panel shows the ratio of residual lensing
  power over initial lensing power; a beam of $4^\prime$ is requested
  for removing 2/3 of the lensing contamination. {\it Bottom}:
  Delensing effectiveness for the \coremfive\ optical configuration,
  and three different sensitivity assumptions.  }
\end{figure}

\paragraph{Lensing $B$ modes:} The shape of the lensing $B$-mode
spectrum is such that the lensing power is measured almost equally
well with 4$^\prime$ or $8^\prime$ angular resolution, for a detector
noise level of 2\,$\mu$K.arcmin, so that to improve the measurement of
the lensing power spectrum, better sensitivity is more important than
better angular resolution. A signal-to-noise ratio per mode of 2--3
(in amplitude) is achieved with a sensitivity in the
1.7--2.5\,$\mu$K.arcmin range. Table~\ref{tab:beam-for-primordial},
however, also gives the impact of the beam on the $B$-mode lensing
power at $\ell = 1000$, and shows that for fully exploiting the
lensing $B$ modes, an angular resolution $\lsim \, 4^\prime$ is
desirable. An 8$^\prime$ beam reduces the CMB power at $\ell=1000$ by
a factor close to 3, and a beam significantly larger than about
$8^\prime$ does not resolve the lensing $B$ modes anymore. This is in
line with the capability to map the lensing potential as a function of
the sensitivity and angular resolution of the survey, discussed in
Ref.~\citep{2002ApJ...574..566H}. Their figure~4 shows that the level
of the noise in the reconstructed lensing map decreases until the
angular resolution is about 2$^\prime$--4$^\prime$ and the noise level
in polarisation maps is about 0.1--$0.3 \times
\sqrt{2}$\,$\mu$K.arcmin.

Noting that ground-based observations can make accurate small-scale
CMB observations to complement the space mission data in the
atmospheric windows, we can assume that they might measure the
smallest CMB scales on at least a fraction of sky if the space mission
fails to do so. It is necessary, however, that the space mission
provide a matching angular resolution of the order of
2$^\prime$--4$^\prime$ at frequencies $\gsim\,300$\,GHz, which are
hard to observe from the ground, to monitor dust contamination at all
scales. A space mission that fulfills this requirement also provides
an angular resolution of order $6^\prime$--$12^\prime$ in CMB channels
between 100 and 200\,GHz.

\paragraph{Other science:} Improving the angular resolution of the
space mission would be beneficial for non-CMB science, in particular
for cosmology with clusters \citep{2017arXiv170310456M}, for observing
extragalactic sources~\citep{2016arXiv160907263D}, and for
investigating Galactic magnetism with polarised dust emission on small
scales. However, complementarity with sub-orbital observations can
also be used for some of these science goals, which are not primary
drivers of the \coremfive\ mission concept.

\paragraph{Summary of sensitivity and angular resolution
  requirements:} In summary, the space survey should be designed to
provide, after component separation, a clean CMB map over more than
50\% of the sky with at least:
\begin{itemize}
\itemsep0em
\item sensitivity in the 1.7--2.5\,$\mu$K.arcmin range for a
  signal-dominated $B$-mode map (with a signal-to-noise ratio of 2--3
  on the lensing $B$ modes);
\item CMB angular resolution better than about 30$^\prime$ for
  detecting primordial $B$ modes;
\item CMB angular resolution $\lsim\, 8^\prime$ for exploiting most of
  the detectable lensing $B$-mode spectrum;
\item CMB angular resolution $\lsim\, 4^\prime$ for optimal
  measurement of polarisation $E$ modes and lensing $B$ modes, as well
  as for separating lensing from primordial $B$ modes.
\end{itemize} 
The actual performance of different surveys for CMB polarisation
science is discussed in more detail in companion papers
\citep{2016arXiv161208270C,2016arXiv161200021D,ECO.lensing.paper}.

We also note that to provide a foreground-cleaned CMB map at a given
resolution, all the channels used for foreground cleaning must have at
least that resolution. For primordial $B$-mode science at the level of
$r\simeq 10^{-3}$, both the synchrotron and the dust should be mapped
with an angular resolution of about $30^\prime$ or better. For
primordial $B$-mode science with $r \, \lsim \, 0.01$, signal
dominated primordial $B$-mode maps at $\ell \, \gsim \, 10$ also
require delensing, by a factor of about 3 in power for $r =0.003$, and
about 10 for $r=0.001$. In addition, for $r=0.001$,
figure~\ref{fig:plotcmb} shows that the cosmic variance of the lensing
signal in bins $\Delta \ell/\ell = 0.3$ is at about the level of the
$B$-mode recombination bump, so that delensing is required to achieve a
detection. For a noise level of 2\,$\mu$K.arcmin, the required
delensing by a factor of 3 in power can be achieved with CMB
polarisation if the survey has an angular resolution of about
$4^\prime$ (see figure~\ref{fig:plotdelensing}).

Alternatively, delensing can be achieved with precise maps of the CIB,
and the capability to delens with both small-scale CMB polarisation
and the CIB would provide a useful cross-check, as well as better
delensing overall.
Hence, in addition to the above CMB surveys, for enabling ``ultimate''
CMB polarisation science, possibly in combination with ground-based
observations reaching an angular resolution of about $2^\prime$, the
space survey must also provide maps of high-frequency foregrounds at a
matching angular resolution.  This leads to the following additional
requirements:
\begin{itemize}
\itemsep0em
\item dust polarisation maps with an angular resolution of about
  $2^\prime$, for component separation on small scales;
\item CIB intensity maps for an alternative way of $B$-mode delensing,
  essential for detecting primordial $B$ modes below $r \simeq 0.01$.
\end{itemize}
These high-frequency observations must be done from space.  On the
other hand, small-scale synchrotron polarisation maps, which are too
challenging to obtain from space because of the required telescope
size, can be obtained with large ground-based radio telescopes if
necessary.

\subsection{Frequency channels}
\label{sec:channels}

We distinguish requirements for the main CMB science (addressed with
the observation of CMB polarisation $E$ and $B$ modes) and for other
science goals.

\paragraph{CMB polarisation:} The minimum of foreground emission
relative to the CMB is located between 60 and 100\,GHz. One might then
think that the best frequency to observe CMB polarisation would be
located precisely at this minimum. However, there are practical
advantages to observing at higher frequencies.

\begin{itemize}
\itemsep0em
\item Since the beam size of a diffraction-limited telescope scales as
  the inverse of the frequency, the angular resolution that can be
  achieved at 120 and 180\,GHz is, respectively, 2 and 3 times better
  than at 60\,GHz.
\item When CMB observations are made at 150$\,$GHz or above, most of
  the complex low-frequency foreground emission signals are low enough
  that it suffices to reduce their contamination by a factor $\lsim \,
  10$. Only dust contamination needs to be accurately subtracted (at
  the sub-percent level) from the CMB polarisation observations.
\item This dominant astrophysical foreground can be monitored at
  $\nu>200$\,GHz with better angular resolution than that of the CMB
  channels, while for a single dish multi-frequency instrument,
  low-frequency foregrounds are observed at lower angular resolution,
  and hence cannot be monitored over the full range of useful angular
  scales.
\item Finally, as shown in Appendix~\ref{appendix:atmosphere}, the CMB
  sensitivity per focal-plane area in space is maximal in the
  150--250\,GHz frequency range.
\end{itemize}

Taking this into account, we should observe the CMB mostly in bands
centred between about 130 and 200\,GHz, avoiding the $\nu=115\,$GHz
and $\nu=230\,$GHz CO $(J\,{=}\,1\,{\rightarrow}\,0$) and
($J\,{=}\,2\,{\rightarrow}\,1$) lines. We also need channels below and
above this frequency range for monitoring foreground emission. A
factor of around 2 in frequency provides proper leverage for
distinguishing the various emission processes, i.e., we need frequency
bands extending from about 65 to 400\,GHz. Ground-based telescopes can
provide synchrotron observations over large patches of sky at $\nu \,
\lsim \, 40\,$GHz, so observing those frequencies with a space-based
mission is not a priority. High frequencies, however, must be observed
from space.

As discussed above in section~\ref{sec:foregrounds}, at least 10
different channels in that frequency range, and preferably more, are
required to separate the different foreground emission signals from
the CMB over a substantial fraction of sky. Additional bands are
required to extend the fraction of sky that can be used for CMB
science. Spreading frequency channels logarithmically in the
65--400\,GHz frequency range with frequency ratios such that
$\nu_{n+1} \simeq 1.15 \, \nu_{n}$ (a sampling in frequency well
matched with a bandwidth $\Delta \nu/\nu$ of approximately 30\% per
channel) yields a set of 15 frequency channels that is adequate for
CMB science, with some safety margin. Actual \coremfive\ channels are 
obtained by a similar process (starting at $\nu_0=60$\,GHz instead of 65\,GHz, 
and making the channels a bit closer in the main CMB frequency range).

Further optimisation depends on assumptions about polarised foreground
emission properties. To be on the safe side, it is preferable to pick
a baseline with more channels than strictly required, in case
polarised foreground emission turns out to be more complicated than
current models might suggest.

\paragraph{Other science:} We now consider options to further optimize
the legacy value of the survey, in particular in combination with
complementary ground-based CMB observations with CMB-S4.

Accurate science with galaxy clusters requires a frequency range
covering both the minimum and the maximum of the thermal SZ
distortion, i.e., ranging from around 120\,GHz to 400\,GHz. A few
channels in that frequency range (at least three, but preferably more
for redundancy) are required to separate the thermal SZ from the
kinetic SZ and also to monitor temperature effects, which, for hot
clusters, can modify the thermal SZ spectrum by 5--10\%. These
requirements are fulfilled with the channels selected on the basis of
CMB polarisation science.

Additional channels above and below this frequency range are required
to monitor contamination by radio and/or infrared sources in clusters,
and to avoid confusion with small-scale foreground emission, in
particular Galactic dust and the CIB. Again, a factor of around 2 in
frequency below 120\,GHz and above 400\,GHz seems adequate, with at
least four channels on both sides. This suggests extending the
frequency range to span 60--800\,GHz, with four channels above
400\,GHz and four channels below 120\,GHz. A 220-GHz channel, close to
the zero of the thermal SZ effect, separates CMB anisotropies and the
kinetic SZ from the thermal SZ (assuming that the CO lines are
controlled in some way, e.g., with notch filters). For the best
synergy with the future CMB-S4 programme, the angular resolution of
the high-frequency channels above $\nu \simeq 300$\,GHz on the space
mission must match that of the ground-based survey at 150 and
220\,GHz.

CIB anisotropies are useful as a delensing tool for primary CMB
science. As demonstrated with \Planck, mapping the CIB
anis\-ot\-ropies is relatively easy in patches with low dust emission,
using channels at 350, 550 and 850\,GHz. At higher frequency,
low-redshift infrared galaxies, observable by other means, start to
dominate the CIB emission, and dust emission from our own galaxy is
relatively stronger. With at least five channels above 300\,GHz (but
preferably more for increased accuracy), it will be possible to use a
generalised internal linear combination method
\citep{2011MNRAS.418..467R} to separate CIB emission from thermal dust
over a significant fraction of sky, as demonstrated using \Planck\ and
{\it IRAS\/} observations \citep{2016A&A...596A.109P}. The CIB map
obtained in this way, in addition to its intrinsic scientific
interest, can be used to delens CMB $B$ modes and improve constraints
on the primordial CMB $B$-mode spectrum, independently of what can be
done with CMB delensing only. This is an essential tool for
cross-checking and validating delensing based on CMB data only.

A principal component analysis can be used on multi-channel CIB
observations to infer the star-formation rate (SFR) as a function of
redshift. A study of how well different future experiments would
perform \citep{2016arXiv161202474W}, showed that the constraints
steadily improve with the number of channels between 220 and 850\,GHz,
provided sufficient angular resolution is available. From their
figure~3, a factor of 30 is gained on the SFR figure of merit they
define by increasing the number of channels above 200\,GHz from 5 to
10, and yet another similar factor is gained with 25 channels above
200\,GHz. Although not designed for CIB science, with eight channels
above 200\,GHz, \coremfive\ outperforms other experiments such as
LiteBIRD (which in addition suffers from reduced angular resolution)
and CMB-S4 (with frequency bands limited to atmospheric
windows). \coremfive\ could be further optimised for CIB science; this
could be considered at a later stage if compatible with budgetary and
programmatic constraints of the implementation of the mission.

\subsection{Systematic effects}
\label{sec:systematics}

Polarising bolometric detectors integrate the electromagnetic power
along one polarisation axis. They measure a combination of intensity
and linear polarisation Stokes parameters. Assuming an infinitely
small beam and frequency band, the signal on the detector, once it is
calibrated in units of CMB temperature fluctuations, can be modelled
as
\begin{equation}
x = I + \eta (Q\cos(2\psi) + U\sin(2\psi)) + n,
\end{equation}
where $I$ is the sky brightness and $Q$ and $U$ are the two Stokes
parameters describing linear polarisation in the observed sky
direction and at the relevant frequency, $\eta$ is polarisation
efficiency (ideally equal to unity), $n$ is the noise term, and $\psi$
is an angle of observation with respect to a set of reference
axes. $E$ and $B$ polarisation are obtained by non-local linear
combinations of $Q$ and
$U$~\citep{1997PhRvD..55.1830Z,1997PhRvD..55.7368K}.
In order to measure $Q$ and $U$, it is necessary to invert a linear
system of measurements at different orientations of the form
\begin{equation}
x_i = I + \eta (Q\cos(2\psi_i) + U\sin(2\psi_i)) + n_i,
\label{eq:simple-polarimeter-measurement}
\end{equation}
which can be rewritten in vector-matrix format
\begin{equation}
{\bm x} = {\sf A} {\bm s} + {\bm n},
\label{eq:matrix-polarimeter-measurement}
\end{equation}
where ${\bm s} = [I,Q,U]$ is the vector of sky Stokes parameters, and
{\sf A} is an operator that integrates the pixel and
frequency-dependent sky signal ${\bm s}(p,\nu)$, and can be thought of
as a generalised mixing or pointing matrix that depends on detector
responses (calibration, polarisation efficiency, beams, and frequency
bands) and on the scanning and observing strategy (pointing,
orientation, and polarisation modulation).
If an estimate $\widehat {\sf A}$ of the mixing matrix ${\sf A}$ is
known, an estimate of ${\bm s}$ is obtained by linear inversion as
\begin{equation}
{\widehat {\bm s}} = \left [ {\widehat {\sf A}}^{\sf T} {\sf W} {\widehat {\sf A}} \right ] ^{-1} {\widehat {\sf A}}^{\sf T} {\sf W} {\bm x},
\label{Eq:inversion}
\end{equation}
and the best estimate of ${\bm s}$ in the least squares sense is
obtained when the ``weighting matrix'' ${\sf W}$ is the inverse of the
noise covariance matrix of the observations.

Inserting ${\bm x} = {\sf A} {\bm s} + {\bm n}$ in
eq.~(\ref{Eq:inversion}) we immediately see that if ${\widehat {\sf
    A}} = {\sf A}$, and if there are enough measurements in each sky
pixel (with different scanning angles) for the matrix $ {\widehat {\sf
    A}}^{\sf T} {\sf W} {\widehat {\sf A}} $ to be invertible, the
estimate of $\bm s$ is unbiased, since ${\widehat {\bm s}} = {\bm s}$
up to an additive noise term that does not depend on $\bm s$. If,
however, $\widehat {\sf A} \neq {\sf A}$, we obtain
\begin{equation}
{\widehat {\bm s}} = \left. \left [ {\widehat {\sf A}}^{\sf T} {\sf W} {\widehat {\sf A}} \right ] ^{-1} \left [ {\widehat {\sf A}}^{\sf T} {\sf W}
{\sf A} \right ] \right. {\bm s} \, + \, {\rm noise},
\label{Eq:inversion2}
\end{equation}
where the product of the first two matrices (in square brackets) is
different from the identity. Non-vanishing off-diagonal terms generate
mixing between $I$, $Q$, and $U$. Since $I \gg E \gg B$, the most
problematic effects are leakage of intensity into polarisation, and
leakage of $E$ into $B$.

The survey should allow for the inversion of the system in
eq.~(\ref{eq:matrix-polarimeter-measurement}) accurately enough that
the total uncertainty on the final CMB data products is dominated by
detector noise at the required level of $2\,\mu$K.arcmin. Instrumental
imperfections that generate systematic errors, and ways to mitigate
them by design and during the data analysis phase, are further
discussed in section~\ref{sec:systematics-control}, as well as in a
dedicated companion paper \citep{ECO.systematics.paper}.

\subsection{Flexibility, safety margins, and redundancy}
\label{sec:margins}

The survey requirements above are somewhat approximate (by
choice). Indeed, the scientific capabilities of \coremfive\ are
continuous functions of the main parameters defining the survey, so
there is no reason to be too prescriptive in the specific value of a
particular parameter. For example, although full-sky coverage is
preferable, the main goals of the mission are not sacrificed if a
fraction of sky is missing, up to a few tens of percent if the missing
part is in regions of high Galactic emission. Additionally, although
it is highly desirable to map the CMB with a sensitivity sufficient
for signal-dominated $B$-mode mapping, the ability of the mission to
reach its science goals is not radically different if the
signal-to-noise ratio per mode (or pixel) is only $\simeq 2$ or
$\simeq 3$ (even if, of course, the higher the better). Similarly,
there is some flexibility in the exact angular resolution that needs
to be achieved --- whether primordial $B$ modes are mapped with
30$^\prime$ angular resolution or 20$^\prime$ is not absolutely
critical, nor whether lensing $B$ modes are mapped with 4$^\prime$ or
6$^\prime$ angular resolution. The impact of the mission
characteristics on the scientific performance of \coremfive\ is
studied more rigorously in the companion papers
\citep{2016arXiv161208270C,2016arXiv161200021D,ECO.lensing.paper}.

In spite of this flexibility, we must realise that the task is
challenging. As can be seen in figure~\ref{fig:B-sensitivity}, the
primordial $B$-mode power spectrum, $C_\ell^{BB}$, must be detected at
a level that is 4--6 orders of magnitude below the foreground
contamination $C_\ell$, an order of magnitude below the lensing
$B$-mode spectrum, 8 orders of magnitude below $C_\ell^{TT}$ and 3
orders of magnitude below $C_\ell^{EE}$. It is not sufficient to
design the mission to just barely be able to detect a small excess of
power that could be primordial $B$ modes (assuming that foreground and
lensing contamination as well as systematic effects will be under
control). Even with a design that is optimised to avoid such
contamination, the ability must be built in for ascertaining that the
desired signal is what is being measured.  This calls for safety
margins and redundancy in the capacity to detect primordial $B$ modes.

The lensing $B$-mode power spectrum, $C_\ell^{BB}$, although larger
and hence easier to detect than primordial $B$ modes for $r \lsim
0.01$, as already seen in existing results
\citep{2014ApJ...794..171T,2015ApJ...807..151K}, is nevertheless
challenging to accurately measure.  The power spectrum at its peak is
4 orders of magnitude below $C_\ell^{TT}$, 2 orders of magnitude below
$C_\ell^{EE}$, and at about the level of the foreground contamination
over most of the sky. Detecting these $B$ modes is one thing, but
measuring them precisely enough that the uncertainty on $C_\ell^{BB}$
is dominated by the sample variance of Gaussian noise, or by their own
cosmic variance, is another thing entirely. Systematic effects and
foregrounds must be controlled at the proper level.

The survey's ability to detect and scientifically exploit $B$ modes
must have built-in redundancy. After the analysis of the data from the
mission, the only way to convince oneself that an excess $B$-mode
signal al low $\ell$ is indeed a detection of primordial $B$ modes is
to show that the signal persists when the data are split into parts.
One can ask if the signal is seen equally well in each of the first
and second halves of the mission? In the north and south parts of the
sky? With different Galactic cuts? In separate frequency channels,
after component separation with different methods? For such tests to
be implemented, the survey must have extra frequency channels, and be
capable of detecting $B$ modes in such data cuts, each of which will
have sensitivity reduced by a factor that ranges from $\sqrt{2}$ to a
few.

\subsection{Survey requirements and goals -- summary}

Although there is some flexibility, there are nevertheless stringent
specifications that the survey must meet to reach its main science
objectives. These {\it requirements\/} can be summarised as follows.
\begin{itemize}
\itemsep0em
\item The survey must map a large fraction of sky, allowing us to
  produce foreground-cleaned maps of CMB polarisation over at least
  $50\%$ of the sky in a least two (but preferably three or more)
  different frequency channels in the 100--200\,GHz range.
\item The survey must be sufficient to unambiguously confirm the
  inflationary origin of any detected large-scale $B$-mode signal for
  the targetted value of $r$. For this purpose, both the reionisation
  bump below $\ell=10$ and the recombination bump up to $\ell \simeq
  200$ must be detected, with the ability to see the turnover above
  $\ell=80$. For $r=0.001$, a foreground-cleaned, delensed CMB
  $B$-mode map is required to have noise level
  ${\lsim}\,3\,\mu$K.arcmin at angular resolution $30^\prime$ or
  better.
\item All foregrounds must be mapped at several frequencies, with
  sufficient angular resolution for cleaning their emission in the CMB
  channels to below the noise sample variance at all $\ell$ between 2
  and 200, and for characterising the level of the residuals. In
  particular, polarised synchrotron and dust must each be mapped in at
  least four different frequency channels, with 30$^\prime$ angular
  resolution or better in each channel.
\item Lensing $B$-mode science requires a survey sensitivity $\lsim \,
  2.5 \, \mu$K.arcmin and CMB angular resolution $\lsim \,
  8^\prime$. For foreground cleaning, the dominating dust foreground
  emission must be mapped at a matching angular resolution in at least
  four frequency channels. These requirements also guarantee
  near-ultimate $E$-mode science.
\item The survey must allow for delensing the $B$ modes down to a
  residual level of about $3 \, \mu$K.arcmin over the 50\% of sky used
  for CMB science, either using the CIB, which must then be mapped up
  to $\ell =2000$ or better in several frequency channels above
  200\,GHz, or using small-scale CMB polarisation observations, or
  (preferably) both.
\item The survey must allow for all the necessary mitigation
  strategies for systematic effects.
\item Adequate margins and redundancy strategies must be implemented
  for fulfilling the above requirements.
\end{itemize}
To maximise the scientific reach of the mission, we design the mission
to also fulfill the following {\it goals}.
\begin{itemize}
\itemsep0em
\item For the legacy value of the survey, full-sky maps at all
  frequencies should be obtained by the space mission.
\item For exploiting CMB polarisation to its full potential, $E$ and
  $B$ modes must be mapped
  with an angular resolution $\lsim \, 4^\prime$.
\item For complementarity with future ground-based CMB observations,
  dust emission should be mapped from space with an angular resolution
  of about $2^\prime$.
\end{itemize}

\section{Mission design}

\subsection{Practical constraints}

The mission concept must be designed so that it can be implemented in
the framework of an ESA M-class mission, for a total budget $\lsim
{\rm EUR}\,700$\,million for the project (funded mostly by ESA with a
${\rm EUR}\,550\,$million cost cap for an M-class mission, the
remainder being funded by European national funding agencies and
possible international partners). This is also the typical cost envisaged 
for future NASA Probe-class missions. The spacecraft must fit in the
fairing of the future Ariane 6.2 launcher. The spacecraft mass and
size, and the overall programme cost cannot much exceed what has been
implemented for \Planck, i.e., a total mass of order 2 (metric)
tonnes, a 1.5-m class telescope (for which the diffraction limit at
150\,GHz is $\simeq 5.6^\prime$), and a power consumption of order
2\,kW. The drastic performance improvement as compared with \Planck\
will primarily come from an increased number of photon-noise-limited
detectors (a few thousand instead of a few tens), from an increased
number of frequency channels (for better component separation
performance in foreground-dominated regions), and from a
highly-redundant observational strategy better suited to the 
measurement of polarisation, with precise control of all systematic 
effects that could impact the performance of the mission.

\subsection{Orbit}

The choice of an orbit impacts all aspects of the mission: launch
requirements; payload and spacecraft geometry; and size of the
telecommunications system.  As for \WMAP\ and \Planck\ before,
\coremfive\ will be in orbit around the second Lagrange point (L2) of
the Sun-Earth system, to ensure that the Sun, Earth, and Moon are well
away from the line of sight at all times. This has been the baseline
for EPIC, for previous versions of \coremfive\ (\core, \coreplus), and for
PRISM. Alternative proposed missions have considered a circular
Sun-synchronous low-Earth orbit (LEO), similar to that of \COBE, which
relaxes the requirements on the launcher, as well as on the
communications system for telemetry; this was the case for early
versions of LiteBIRD and PIXIE.

The main drawback of an LEO is the contamination of the measurement by
sidelobe pickup from celestial bodies. The Moon and the Sun each
subtend a solid angle of about 700\,arcmin$^2$, for an emission
temperature of approximately 300\,K and 6000\,K, respectively, in the
frequency range of interest. The target polarisation sensitivity on
the same angular scale is about 0.07\,$\mu$K. The level of sidelobe
rejection towards the Moon must thus be of the order of $4\times 10^9$
or better, while that of the Sun must be of the order of $8 \times
10^{10}$. From an LEO, the Earth subtends a very large fraction of sky
(tens of percent). Inhomogeneities in surface temperature and
emissivity, modulated in the far sidelobe pickup of the telescope
radiation pattern, are therefore a concern for a sensitive CMB space
mission.

The impact of this sidelobe stray light is difficult to assess, and
demonstrating prior to launch that it does not degrade the performance
of the mission would require comprehensive studies (the far-field
integrated sidelobe response under normal instrument observing
conditions cannot be reliably measured on the ground). It is hence
safer to minimise the risk of sidelobe stray light by design, and
select the L2 point as a baseline to keep all of the Sun, Earth, and
Moon well away from the main beam of the instrument, as well as masked
by a number of screens.

A possible orbit around the Sun-Earth L2 point is shown in
figure~\ref{fig:orbit}. The size of the orbit impacts the required
amount of propellant for injection into orbit, and the maximum
elongation of the Earth and Moon with respect to the Sun as seen from
the spacecraft.
Table~\ref{tab:orbit-injection} lists the typical $\Delta v_{\rm orb}$
required to inject a 2-tonne spacecraft into orbit for three possible
Lissajous orbits around L2. The maximum Earth elongation with respect
to the Sun is also listed; this is relevant for the design of the
telecommunications system and for sidelobe rejection.

\begin{table}[h]
\begin{center}
\begin{tabular}{|l|c|c|c|}
\hline 
Orbit around L2& $A_x \times A_y$ {[}km{]}& $\Delta v_{\rm orb}$& $\theta_{\rm Earth}$ at maximum\\
\hline 
\hline 
Large Lissajous& $250{,}000\times673{,}000$& $90\,{\rm m}\,{\rm s}^{-1}$& $24^\circ$\\ 
Medium Lissajous& $100{,}000\times320{,}000$& $270\,{\rm m}\,{\rm s}^{-1}$& $12^\circ$ \\ 
Small Lissajous& $50{,}000\times160{,}000$& $330\,{\rm m}\,{\rm s}^{-1}$& $6^\circ$\\ 
\hline 
\end{tabular}
\end{center}
\caption{Typical required $\Delta v_{\rm orb}$ and maximum Earth elongation 
for various Lissajous orbits around the Sun-Earth L2 Lagrange point.}
\label{tab:orbit-injection}
\end{table}

\subsection{Observing strategy}
\label{sec:scanning}

Following the survey requirements, the observing strategy is selected
to cover the whole sky in temperature and polarisation, with
redundancy of the measurement over many different periods of time.
As discussed in section~\ref{sec:systematics}, the observational plan
should be designed for an adequate control of residuals of $I$ into
$Q$ and $U$ (temperature-to-polarisation leakage) after inversion of
the linear system of observations, and for adequate control of errors
in the polarisation angle (mixing of $Q$ and $U$, which generate
leakage of $E$ into $B$). To that effect, we place the following
requirements and preferences on the observing strategy.

\begin{enumerate}
\itemsep0em
\item Each pixel must be observed by the same detector with many
  different polarisation angles over the course of the mission,
  preferably evenly spread in $[0,2\pi]$ for best polarisation
  sensitivity \citep{1999A&AS..135..579C}.
\item Each pixel must be observed with different polarisation angles
  on relatively short timescales (timescales on which the calibration
  parameters are not expected to vary significantly).
\item Each pixel must be observed at very different times during the
  course of the mission.
\item The line of sight must never come close to the direction of the
  Sun, Earth, or Moon (this specification must be quantified using
  models of the 2-dimensional radiation pattern of the telescope).
\item To avoid strong fluctuations of the temperature of the payload,
  the solar flux absorbed by the spacecraft should be as constant as
  possible.  The cooling chain should preferably provide stable
  instrument temperature.  Any variation of the payload or instrument
  temperature should be slow compared to the timescale of polarisation
  measurements.
\item For negligible impact of low-frequency (additive) noise, most of
  the pixels should be revisited on timescales of the order of the
  inverse of the knee frequency of the low-frequency noise (i.e., the
  frequency at which the power spectrum of low-frequency noise equals
  that of the white noise). This makes it possible to remove the
  low-frequency noise using appropriate data-processing techniques
  \citep{2000A&AS..142..499R,2005A&A...436.1159D,2009A&A...506.1511K,2011A&A...534A..88T}.
\item Distant pixels must be observed at time-lags smaller that the
  typical timescale of any instrumental long-term instability, e.g.,
  low-frequency noise, gain drifts, or payload temperature drifts.
\item For minimising the sample variance (and for the legacy value of
  the observations), the complete sky must be observed, with
  integration time being evenly spread over the sky.
\item It may be useful to implement a scanning strategy that can
  concentrate observations in the cleanest regions of the sky during a
  mission extension, after identification of such regions during the
  main survey. A mission design compatible with this option is
  preferred.
\end{enumerate}
The choice to implement an active polarisation modulation system
onboard the spacecraft, using a spinning or stepped HWP, is an
important element for the definition of the observing strategy. The
use of an HWP simplifies the implementation of requirements 1 and 2
(above) by adding the extra flexibility to rotate the polarisation
without rotating the whole spacecraft. However, an HWP also has
several drawbacks, discussed in section~\ref{sec:hwp}, that impact the
performance and feasibility of the mission. As a baseline, we favour
an observing strategy with no HWP. As a consequence, all the
requirements above must be fulfilled by scanning only.

\begin{figure}[ht]
\centering
\includegraphics[width=0.9\textwidth]{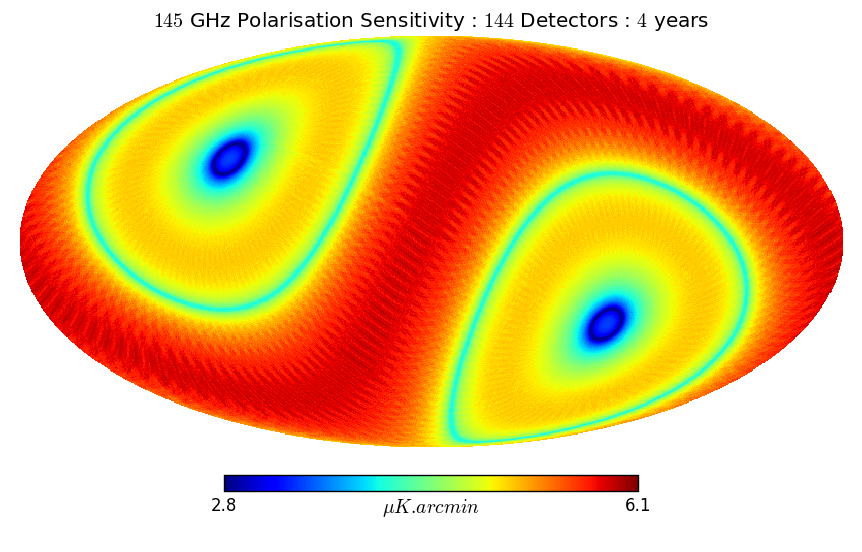}
\includegraphics[width=0.49\textwidth]{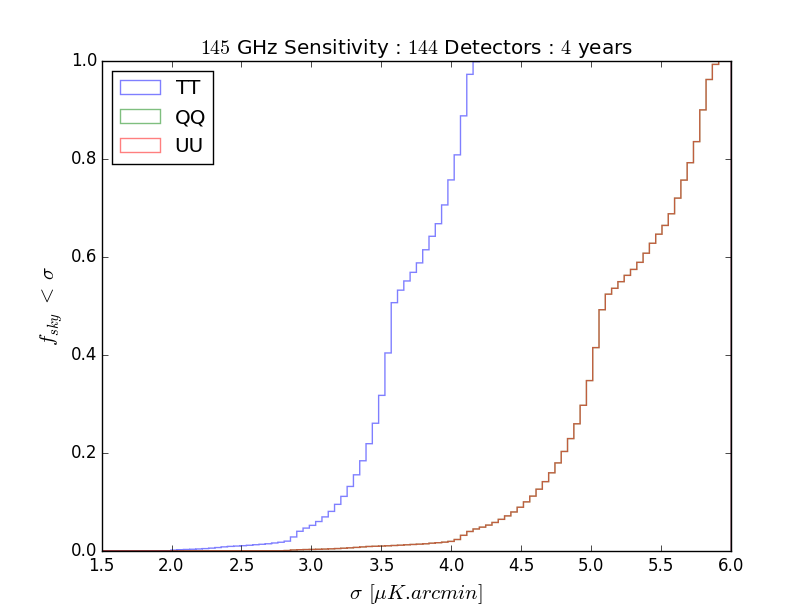}
\includegraphics[width=0.49\textwidth]{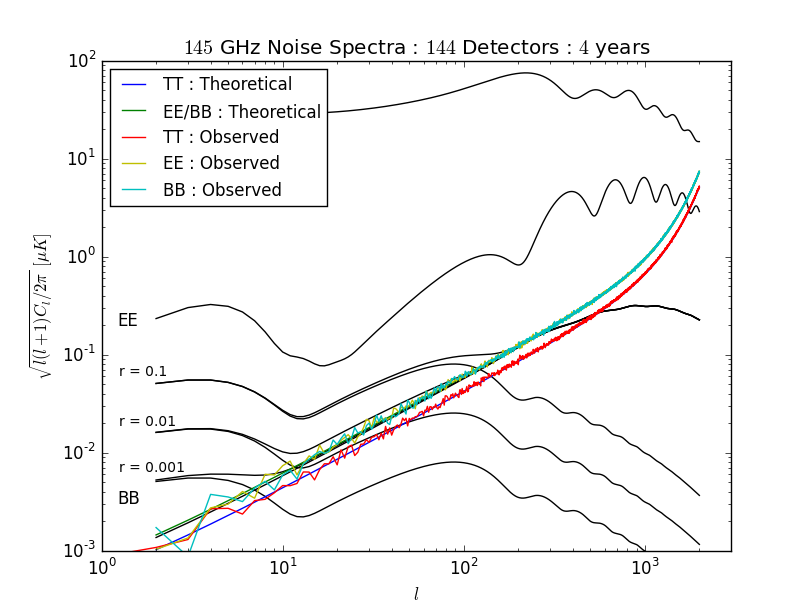}
\caption{{\it Top}: Distribution of the polarisation sensitivity over
  the sky for the 145\,GHz detectors of the \coremfive\ baseline
  design. For this single channel, the sensitivity ranges from about
  3\,$\mu$K.arcmin in two deep patches located at the ecliptic poles
  (blue), to about 6\,$\mu$K.arcmin near the ecliptic plane, with a
  median of about 5\,$\mu$K.arcmin. {\it Bottom left}: Cumulative
  distribution function of the sensitivity over the sky; most of the
  pixels are seen with a polarisation sensitivity ranging from 4 to
  6\,$\mu$K.arcmin (with the $Q$ and $U$ distributions overlapping in
  the plot). {\it Bottom right}: Power spectra of the noise maps for
  the 145-GHz channel (4 years of observation with 144 detectors
  distributed in 9 rows of 16), compared to the theoretical estimate
  from Table~\ref{tab:CORE-bands} for homogeneous coverage.}
\label{fig:coverage}
\end{figure}

The selected strategy impacts the choice between a spin-stabilised or
a 3-axis stabilised spacecraft. With spin-stabilisation, the whole
spacecraft rotates around a pre-defined axis that is fixed in the
frame of the spacecraft. When the telescope line of sight is offset
from this spin axis by some angle $\beta$, the line of sight scans the
sky along a circle of radius $\beta$. A large value of $\beta$
combined with fast spinning connects distant points on short
timescales (requirement 7 above); this was the case for \Planck, for
which $\beta=85^\circ$. On the other extreme, $\beta=0$ results in a
sky pixel being observed with all possible orientations on a short
timescale (requirement 2); this is the baseline scanning strategy for
PIXIE.

To observe the complete sky, the direction of the spin axis must be
changed during the mission lifetime. For \Planck, this was achieved
using small thrusters that corrected the direction of the spin axis
every 40 minutes or so. If motions must be made on smaller timescales,
attitude corrections require a significant amount of propellant (see
Ref.~\citep{2017MNRAS.466..425W} for a more complete discussion of the
corresponding constraints on the mission).
With a 3-axis stabilised spacecraft, reaction wheels rotate to keep
the satellite in the desired orientation as a function of time. This
solution is more flexible, but the scanning speed is limited by the
maximum momentum of the reaction wheels, which must compensate for the
momentum of the whole spacecraft.

\begin{figure}[ht]
\centering
\includegraphics[width=0.49\textwidth]{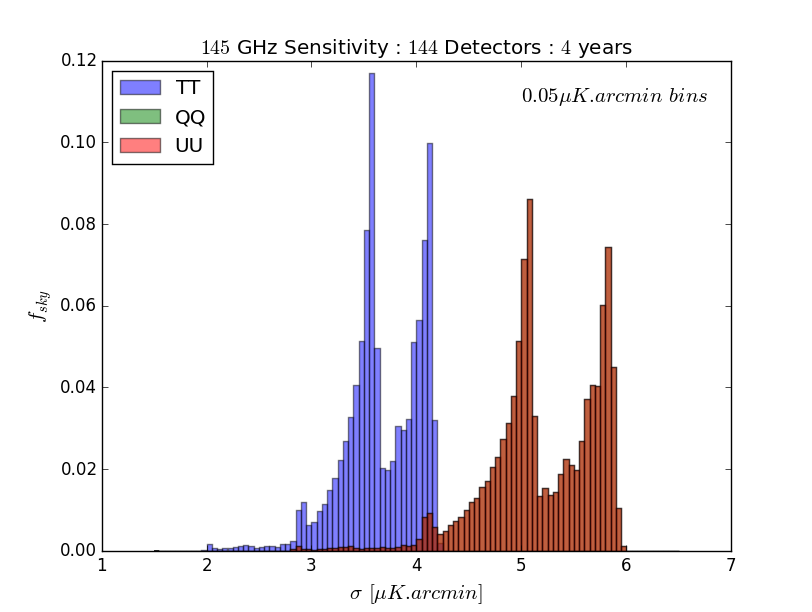}
\includegraphics[width=0.49\textwidth]{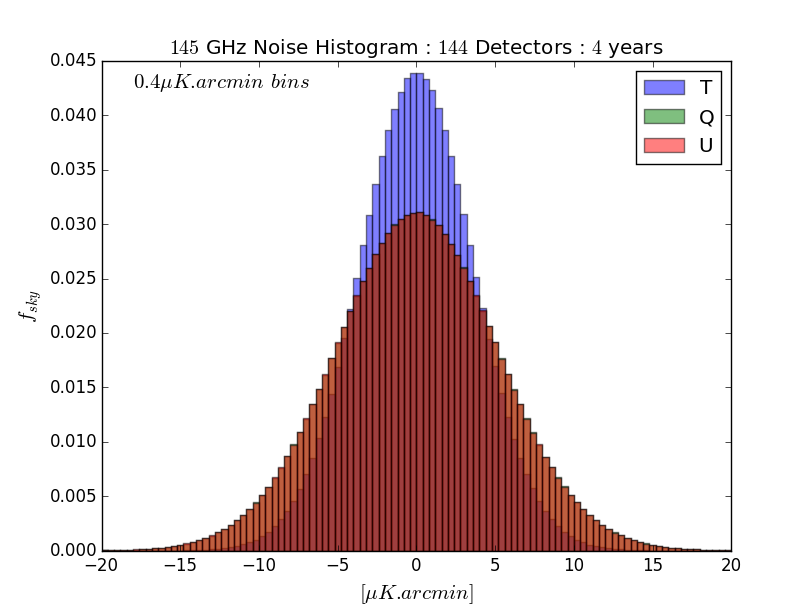}
\caption{{\it Left}: Histogram of noise level for temperature and
  polarisation, in bins of $\delta\sigma = 0.1$; the two peaks of the
  distribution, lying for polarisation slightly below 5\,$\mu$K.arcmin and
  6\,$\mu$K.arcmin, correspond to the dominating colours in the top
  panel of figure~\ref{fig:coverage}. {\it Right}: Histogram of the
  noise realisation on the map in bins of 0.4\,$\mu$K.arcmin. The
  noise is generated by projection of 4-year timelines for a patch of
  36 detectors each sampled at 85\,Hz onto an $N_{\rm side}=1024$ {\tt
    HEALPix} \citep{2005ApJ...622..759G} map. The noise values are
  then rescaled to 144 detectors (four patches of 36 following each
  other in the scan) and to an equivalent $1\,{\rm arcmin}^2$ pixel
  size, for easier comparison with the expected noise level. For both
  panels, the $Q$ and $U$ histograms overlap.}
\label{fig:coverage2}
\end{figure}

We select a baseline in which the satellite is spun 
with a short period $T_{\rm spin} \simeq 2$\,min (frequency $f_{\rm
  spin} \simeq 8\,$mHz). The spin axis itself will precess with a
frequency $f_{\rm prec}$ (with a longer period $T_{\rm prec} \simeq
4$\, days), around a direction that will be maintained roughly
anti-solar (and thus rotate around the Sun with a period of about 1
year to follow the annual motion of the Earth). The spin axis will be
offset from the precession axis by an angle $\alpha \simeq 30^\circ$,
and the optical axis will be offset from the spin axis by an angle
$\beta \simeq 65^\circ$ (figure~\ref{fig:orbit}). The resulting
temperature and polarisation sensitivity distribution over the sky for
a set of 144 detectors at 145\,GHz, and the resulting noise power
spectrum, are shown in figures~\ref{fig:coverage} and
\ref{fig:coverage2}.

The optimisation of $\alpha$ and $\beta$ is discussed in
Ref.~\citep{2017MNRAS.466..425W}. In addition to the distribution of
observing time over the sky and the distribution of scanning angles
for each pixel, these choices impact the design of the payload
itself. Further discussion on the trade-offs that lead to the
particular selection of $\alpha$ and $\beta$ values, and of the spin
and precession periods, can be found in section~\ref{sec:payload} and
Appendix~\ref{appendix:scanning}.

\subsection{Mission phases and operations}

The main mission phases are listed below.
\begin{itemize}
\itemsep0em

\item {\bf Launch and Early Operations phase:} The spacecraft will be
  launched warm, coolers will be turned on shortly after launch. The
  launcher will place \coremfive\ on an orbit towards the Sun-Earth L2
  point. Small corrections to the orbit as deemed necessary will take
  place in the first few days after launch.

\item {\bf Decontamination phase:} The temperature will be allowed to
  progressively descend to around 170\,K, and will be stabilised for a
  period of about 2 weeks for out-gassing and decontamination of the
  optical surfaces and of the focal plane.

\item {\bf Commissioning and transfer phase:} For about 2.5 months,
  the spacecraft will cruise towards L2. Checks of the basic
  functionality of the spacecraft and of the payload (commands, AOC,
  telemetry, and payload basic functionality) will also be performed.

\item {\bf Calibration and Performance Verification phase:} This
  phase, which will start immediately after injection of \coremfive\
  on its orbit around L2, and last for about 2 months, will consist of
  instrument tuning, verification of the sensitivity, initial
  measurement of instrument key parameters, and initial
  characterisation of systematic effects.

\item {\bf Nominal observation phase:} For about 4 years, \coremfive\
  will perform routine scanning of the sky following the selected
  scanning strategy.

\item {\bf Extended observation phase:} An extension of the mission
  for deeper integration on selected small patches of sky will be
  considered if justified by the scientific results obtained from the
  nominal survey data.

\item {\bf Decommissioning phase:} At the end of operations, the
  spacecraft will be removed from its nominal orbit around L2,
  injected into an heliocentric orbit, and passivated.
\end{itemize}

\subsection{Telemetry}

For \Planck, a small Lissajous orbit around L2 was selected to keep
the Earth elongation compatible with telecommunication with a fixed,
large-beam antenna at the bottom of the spacecraft. With about 30
times more detectors, \coremfive\ needs a continuous data rate of
about $1.15\,{\rm Mbit}\,{\rm s}^{-1}$ (for the baseline
configuration). A small (roughly 30\,cm) steerable antenna operating
in the Ka band, communicating with large ground-based antennas
currently in use, is compatible with this telemetry requirement.

\begin{figure}[ht]
\centering
\includegraphics[width=\textwidth]{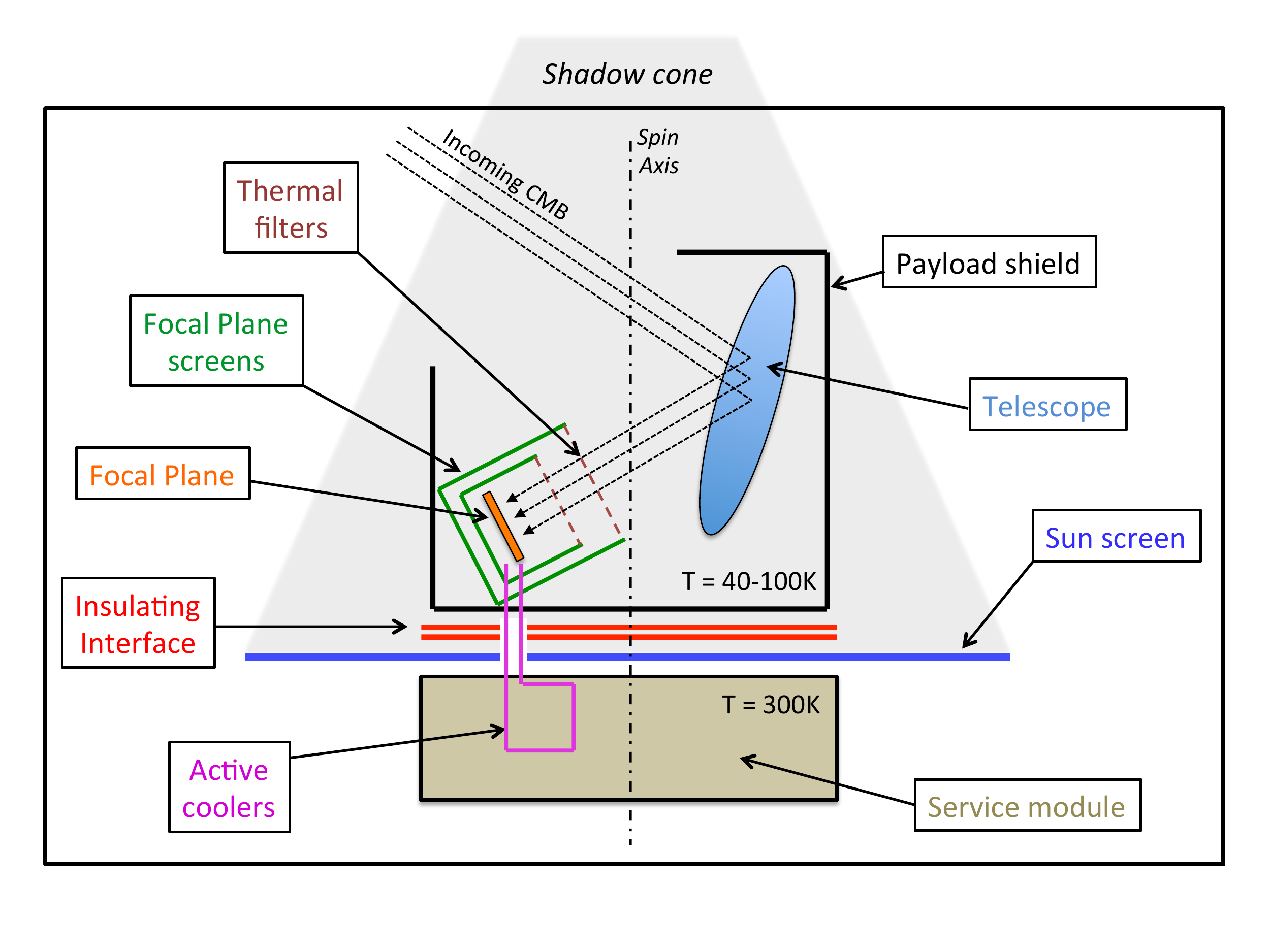}
\caption{Conceptual sketch of the main elements of the \coremfive\
  payload; The Sun screen and insulating interface keep the payload in
  the shade and cold; The telescope focuses the light onto the focal
  plane; The main payload shield reduces stray light from telescope
  sidelobes; Active coolers that can be located in the service module,
  or in a interface element between the service module and the cold
  payload, cool the focal plane to 100$\,$mK, with intermediate stages
  at 1.7$\,$K, 4$\,$K and 15--20$\,$K.}
\label{fig:payload-sketch}
\end{figure}

\section{Payload}
\label{sec:payload}

Figure~\ref{fig:payload-sketch} illustrates the conceptual design of
the \coremfive\ spacecraft. The passively cooled payload module (PLM)
is separated from the warm service module (SVM) by a main Sun
screen. Throughout the mission, the Sun, Earth, and Moon will always
be on the same side of this Sun screen, the PLM being on the other
side.

To avoid modulation of the solar flux during the scan, the spacecraft
is designed with a general axis of symmetry that coincides with the
spin axis. The Sun itself remains, throughout the scanning, on a cone
of opening angle $\beta$ with respect to the spacecraft, i.e., always
at the same angular distance from the spin axis. This defines a shadow
cone on the PLM side, in which all PLM elements must be accommodated
to avoid any direct solar illumination during the course of the
observations. The Sun screen does not necessarily have to be a flat
disc. It is possible, for example, to extend it vertically with a
cylinder or a cone to increase the available payload volume.

The payload must be designed to accommodate a telescope that focusses
incoming radiation onto a large focal plane.  The size of the
telescope is driven by the scientific requirements on angular
resolution and on the level of sidelobe rejection. The angular
diameter of the main beam (between the zeros of the Airy function) of
a disc of diameter $D$ is $\theta_{\rm beam} \simeq 70 \lambda/D$ (in
degrees), where $\lambda$ is the wavelength of observation. 
The FWHM is about half of that for full illumination of the aperture. However, 
for an aperture illumination with an edge taper of $\lsim 20$\,dB, the 
effective aperture size is close to half of that of the telescope, so that
the actual FWHM of the beam is close to the size of the Airy disc. At
150\,GHz (2\,mm), a 1-m aperture corresponds to a main beam full width
of $\simeq 8.4^\prime$, while a 2-m aperture corresponds to half of that,
i.e., $\simeq4.2^\prime$ (actual FWHM beams are slightly different, since
they also depend on the exact illumination pattern of the aperture by horns
or lenslets in the focal plane, or cold stops if any). To fit in the
4.5-m fairing of an Ariane 6.2 launcher, the focal number of a 1.5-m
telescope should remain $\lsim\,$3 (less for a larger telescope).

The focal plane must accommodate a few thousand background-limited
detectors. Hence, the typical diameter of a focal plane observing with
thousands of detectors of size $\gsim\,\lambda$ in the 100--200\,GHz
frequency range is tens of centimetres. Detector coupling using
lenslets or horn feeds require more space per pixel, typically of the
order of 3--6$\lambda$ diameter per focal-plane pixel. The larger the
focal plane, the more pixels it can accommodate, and hence the better
the sensitivity of the instrument. The \coremfive\ baseline focal
plane is 50\,cm in diameter. For an increased number of detectors,
dual-polarisation, multi-frequency detectors, such as those deployed
on some ground-based instruments, are an appealing technological
solution.

To reach background-limited performance, the detectors must be
cryogenically cooled to sub-kelvin temperatures. The space environment
makes it possible to passively cool the whole payload itself to a
temperature $\lsim$\,100\,K, to reduce the radiative background from
the telescope mirrors and from the payload on the detectors, as well
as the thermal load on the active cooling stages. Such passive cooling
is an advantage, but not a strict necessity -- the CMB can be observed
through a warm telescope. As long as the optical coupling of the
payload with the detectors is kept below about 1\%, the load from a
300-K payload onto the detectors is $\lsim 3\,$K$_{\rm RJ}$, i.e.,
comparable to the load from the CMB itself.

A set of reflecting and absorbing shields and baffles prevent stray
light from reaching the focal plane.  The main payload shield itself
has two roles. In addition to being a protection against stray light
originating from the sky itself (blocking in particular all spillover
radiation around the telescope reflectors), it also contributes to the
passive cooling by radiating towards cold space. The same concept was
used on \Planck, for which the payload temperature was $\simeq45$\,K.

\subsection{Instrument design}

The \coremfive\ instrument is designed to observe the sky in the
frequency range 60--600\,GHz with a multi-beam, multi-band
polarimeter.  As a baseline, the sky emission is collected with a
1.2-m projected aperture telescope that feeds a large focal plane
populated with an array of 2100 background-limited detectors.
The focal plane is actively cooled to 0.1\,K using a continuum-cycle
dilution refrigerator.  A set of reflecting, low-pass filters reduces
the radiative loading reaching the focal plane.  Detectors are
distributed among 19 frequency bands, each of which have an
approximate fractional band-width of 30\%. The bands, which monitor
foregrounds at the lower and higher end of the spectrum, with the CMB
in the middle, are defined by plastic-embedded metal-mesh filters.

Sensitivity to polarisation is obtained by means of plastic-embedded
metal grids for all single-polarisation detectors ($\nu > $ 115\,GHz),
and by means of planar ortho-mode transducers for the
dual-polarisation detectors ($\nu \leq$ 115\,GHz).  Radiation is
coupled to LEKID resonators by means of embedded-mesh lenslets and
short waveguide sections for $\nu \leq$ 220\,GHz, and to MKIDs via
standard silicon lenslets for $\nu \geq$ 255\,GHz. For simplicity,
most of the \coremfive\ detectors are baselined to be single
frequency, single polarisation, but this could be revised in the
future, for improved sensitivity. More details can be found in the
companion instrument paper \citep{ECO.instrument.paper}.

\subsection{Telescope}

Two main telescope options have been considered: a Gregorian telescope similar to that of \Planck\ \citep{2010A&A...520A...2T}; or a crossed-Dragone design, similar to that considered for EPIC \citep{2009arXiv0906.1188B}, LiteBIRD \citep{2014JLTP..176..733M,2016SPIE.9904E..0XI}, or for future ground-based experiments \citep{2016ApOpt..55.1686N}.

The Gregorian telescope has the advantage of compactness for a given aperture. However, the focal surface, neither flat nor telecentric, is not optimally suited to big arrays of detectors. This drawback could be corrected with the use of tertiary optics, e.g., either a refractive lens, or a set of lenses. As lenses are emissive (typical emissivity $\lsim10$\%), they must be cooled to cryogenic temperatures, typically below 20\,K to contribute negligible background loading. For a good optical efficiency of the whole system, they must also be covered with anti-reflection coating, which is hard to achieve for a very broad band. 

\begin{figure}[htb]
\centering
\includegraphics[height=6cm]{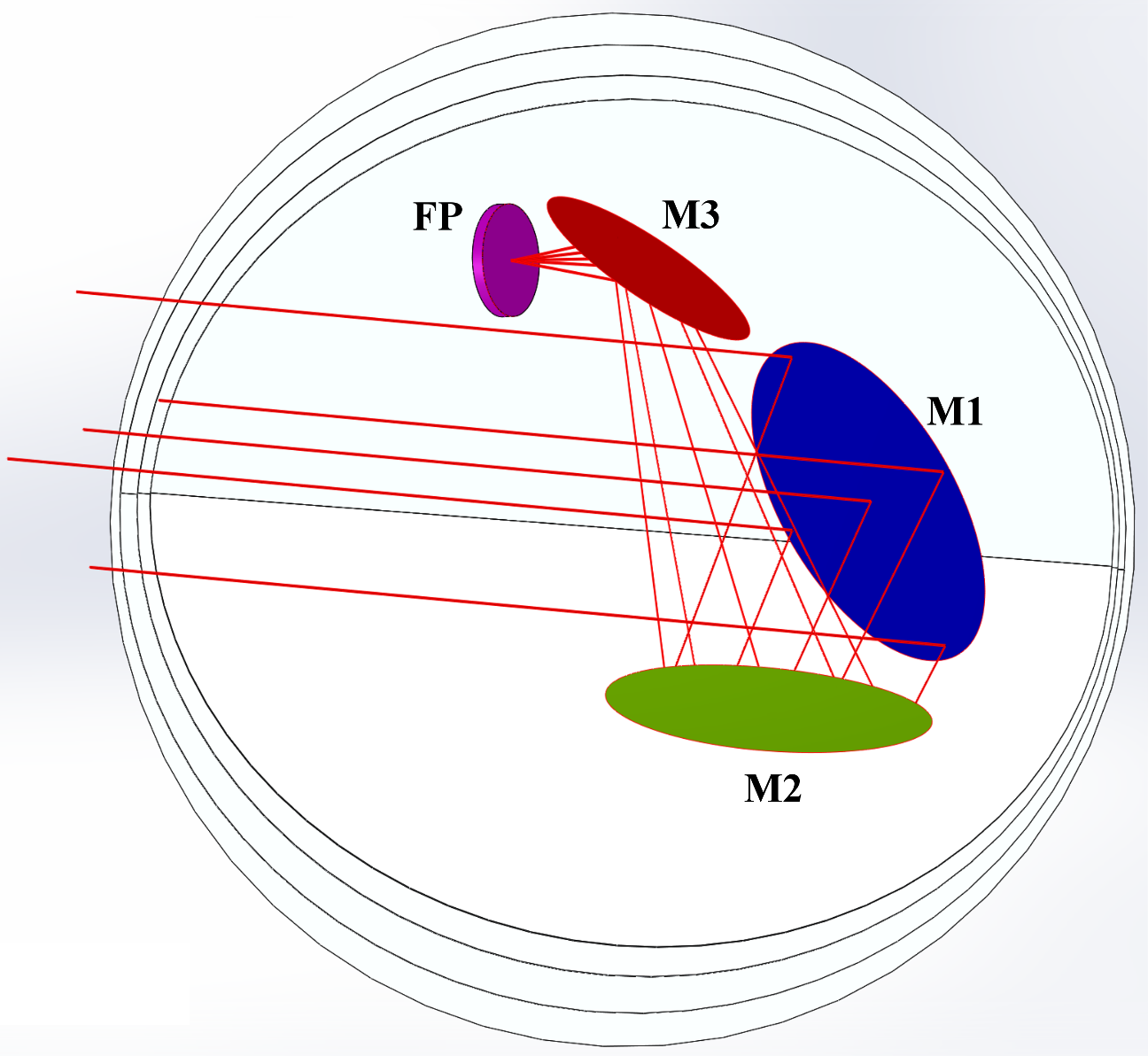}
\includegraphics[height=6cm]{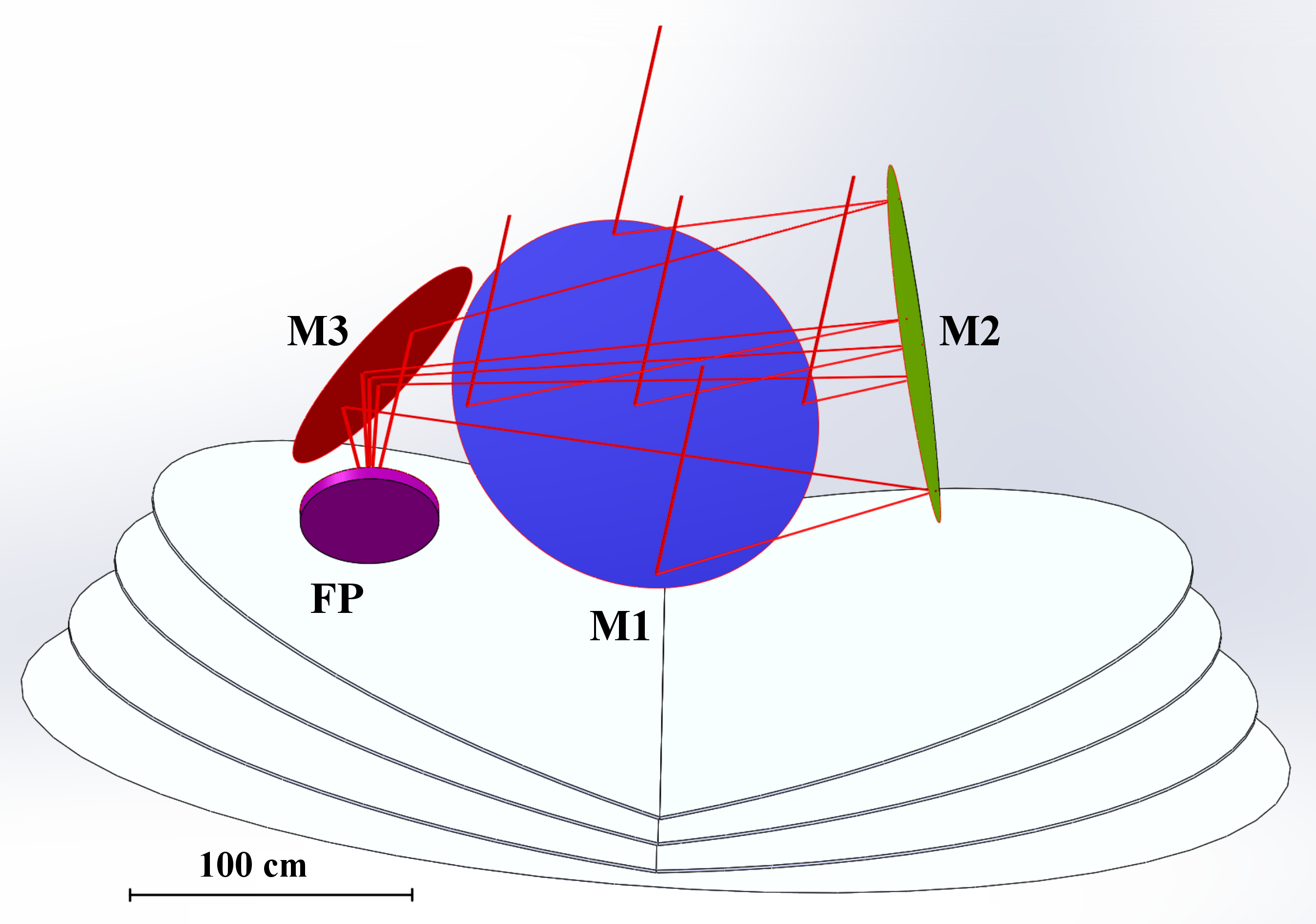}
\caption{View of the \coremfive\ optical system, on top of the V-grooves. 
Shields and structures are not represented. The primary mirror is shown 
in blue, the secondary mirror in green, the flat tertiary mirror in red, 
and the focal-plane array in purple.}
\label{fig:optics}
\end{figure}

The crossed-Dragone design offers a larger, flat, near-telecentric
focal plane, but is more cumbersome, and hence harder to fit into the
fairing for a 1.2-m aperture. In addition, the focal plane is near the
incoming beam, and is thus exposed to direct illumination from the
sky, which is a source of sidelobe contamination. The baseline we
consider, shown in figure~\ref{fig:optics}, solves both problems with
a relatively large focal number ($F \simeq 2.5$) and a flat tertiary
mirror, which re-locates the focal surface so that it fits in the
payload and can be shielded from direct illumination from the
sky. Further details of the optics are given in the companion
instrument paper \citep{ECO.instrument.paper}.

\subsection{Shielding against sidelobe stray light}

Sidelobe rejection is essential to avoid contaminating CMB
measurements with stray light emission originating from bright sky
regions away from the line of sight. Difference maps of observations
made with \Planck\ in different surveys highlight the variations in
the signal observed in two different orientations. Patterns due to
differential integrated sidelobe emission are clearly seen in those
difference maps \citep{2014A&A...571A..14P}.

A model of the \Planck\ 2-D radiation pattern identifies the main
features responsible for the observed sidelobe pickup as spillover
around the edge of the secondary mirror and the primary mirror, and
reflections on the sides of the main baffle. The level of these
features ranges typically from $-75$ to $-85$\,dB, while the majority
of the remainder of the radiation pattern is at $-95$\,dB or
below. The total integrated spillover around the primary or the
secondary typically is at the level of 0.1--0.2\%
\citep{2010A&A...520A...2T}.

While the satellite spins to scan the sky, each of these sidelobe
patterns also sweeps the sky with a large elongated ``beam,'' of
typical size tens of ${\rm deg}^2$. When such spillover patterns
(which can be though of as elongated structures scanning the sky)
cross large structures with brightness of a few mK amplitude, such as
the Galactic ridge, they contribute a signal of the order of a few
$\mu$K equivalent amplitude brightness, extending over few-degree
scales. For comparison, the average noise of \coremfive\ integrated
over a $10\,{\rm deg}^2$ sky patch is about 0.01\,$\mu$K, i.e., 100 to
1000 times smaller.

\begin{figure}[htb]
\centering
\includegraphics[height=6.5cm]{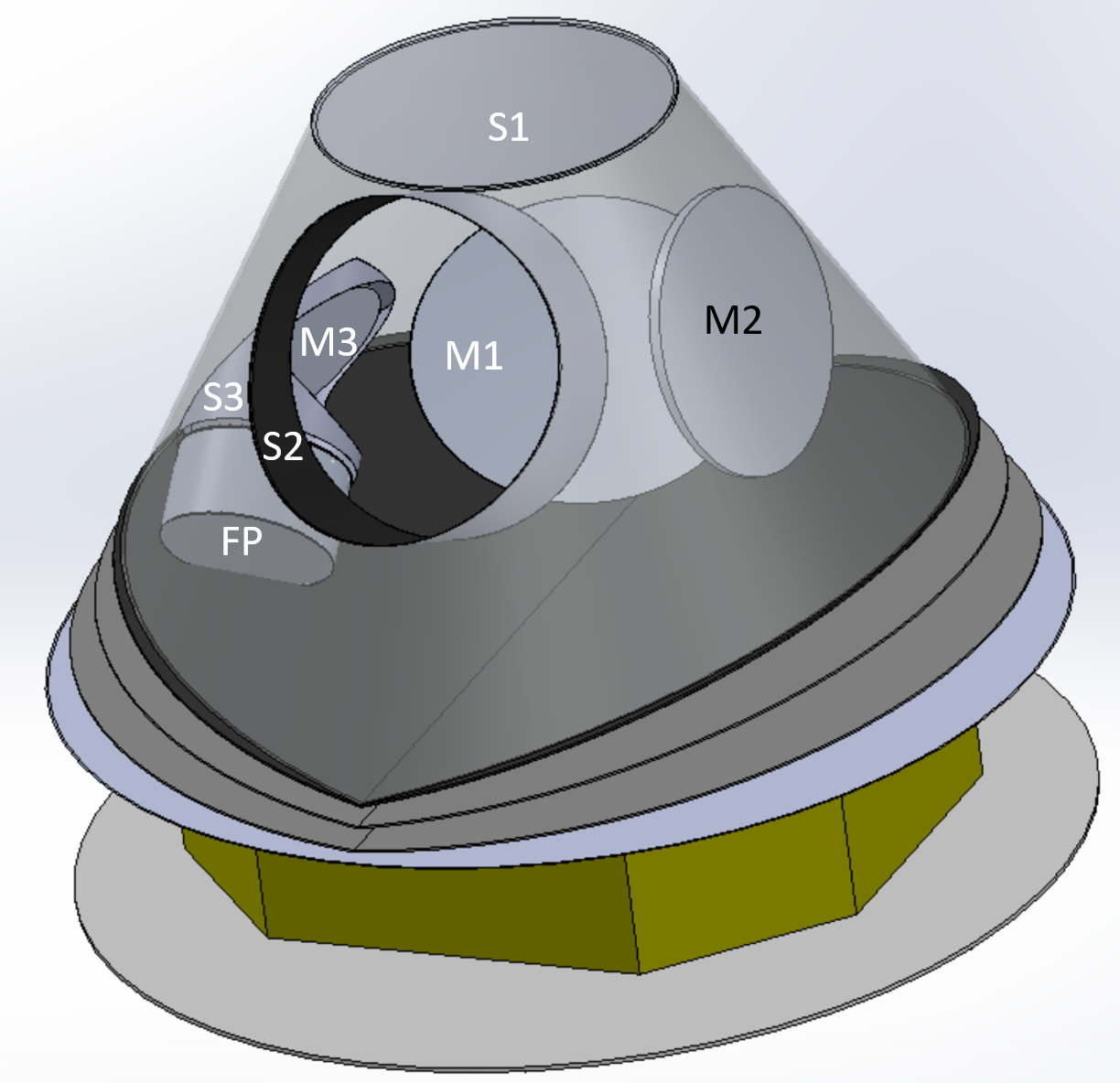}
\includegraphics[trim = 0mm 0mm 0mm 25mm, clip, height=6.5cm]{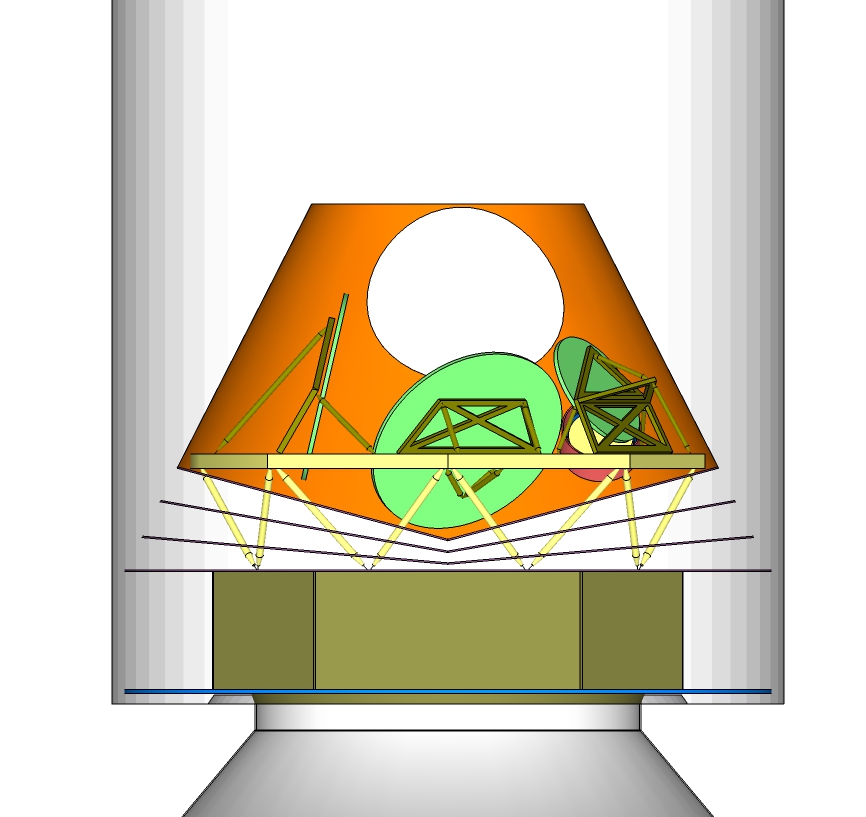}
\caption{Right: View of the spacecraft, showing the configuration of
  the three telescope mirrors, M1, M2,
  and M3, the focal plane, FP, and the various screens. The screens
  S1, S2, and S3 are designed so that the focal plane is not in direct
  view of the sky, and hence receives radiation from the sky only
  through the telescope, or after several reflections on the (mostly
  absorptive) screens. Left: Side view of the spacecraft in the fairing of an
  Ariane~6 launcher, illustrating the ``book-shaped'' geometry of the V-grooves.}
\label{fig:shielding}
\end{figure}

Since sidelobe patterns are very polarised (diffraction around
conductive edges polarise the diffracted light), the sidelobe signal
is not reduced much by differencing detectors with orthogonal
polarisation sensitivity, nor by a rotating HWP. We must control
sidelobes down to a level 100 to 1000 times better than was achieved
for \Planck. For \coremfive, this is achieved with the combination of:
\begin{itemize}
\itemsep0em
\item under-illumination of the secondary and primary mirrors, so as
  reduce direct spillover around telescope reflectors (we assume a
  typical required edge taper similar to that of \Planck, i.e., of the
  order of 20--25\,dB);
\item an approximately 90\% absorptive payload main screen, which
  reduces the level of stray light by a factor of about 10 at each
  (partial) reflection;
\item a main baffle geometry such that the 10\% reflected power off
  each absorptive surface is primarily reflected to another absorptive
  surface, away from the focal plane;
\item additional absorptive baffles that avoid any direct view of the 
sky from the focal-plane array.
\end{itemize}
This design results in additional rejection of sidelobe pickup by as
many orders of magnitude as the number of (partial) reflections that
are required for incoming stray light to reach the focal plane, i.e.,
at least 1 order of magnitude, but likely more. The \coremfive\
baffling/shielding design is shown in figure~\ref{fig:shielding}.

In addition to being designed for the minimisation of the sidelobe 
stray light contamination, \coremfive\ is a highly redundant mission. 
Each 4-detector set observes a connected map of 45\% of sky every
4 days. This provides thousands of independant maps of each sky 
region, obtained at different times, with different spacecraft orientation,
and different detector sets, that can be compared to find signatures
of any detectable residual of stray light contamination.

Absorptive screens have the drawback of increased load on the
detectors and on the active cooling chain, and thus should preferably
be as cold as possible. The other drawback is the impact of
temperature fluctuations of the payload on the detected signals. A
1-mK temperature fluctuation of a payload 1\% coupled to the detectors
generates a 10-$\mu$K signal. Such internal stray light emission is
usually not correlated to the spacecraft pointing on the sky
(contrarily to sidelobe stray light), and thus generates spurious
signals that are additive (and uncorrelated with the sky emission),
rather than multiplicative. They can be removed by processing in a
map-making step, along with other low-frequency noise terms; they
should nonetheless be kept small by design.

\subsection{Polarisation modulation}
\label{sec:hwp}

Some CMB polarisation experiments make use of active modulation of
the polarisation with a rotating half-wave plate (HWP). Polarisation
modulation with an ideal HWP has the advantage of uniformly spreading
the angles of polarisation sensitivity in each observed pixel (for
optimal polarisation sensitivity), and of mitigating some systematic
effects, such as those due to beam asymmetry (when a single detector
observing sky pixels with various orientations is used to make
polarisation maps). When continuously rotated at a frequency $f_{\rm
  HWP}$, it also shifts the frequency of the useful signal to wings of
a carrier located at $4f_{\rm HWP}$ in the frequency domain, while the
signal of a spinning telescope is concentrated in narrow frequencies around
multiples of $f_{\rm spin}$. A fast modulation moves the signal away
from the frequency range impacted by any long-term instabilities,
such as low-frequency noise gain drifts, and more generally
time-evolution of the instrumental response. These desirable properties
make the use of HWPs appealing for measuring CMB polarisation.
However, there also are several drawbacks to the use of an HWP (in
particular in a satellite mission). Perfect modulation is
technologically hard to implement, and modulation by an imperfect HWP
is likely to generate systematic effects of its own, which might be
hard to correct in the data analysis processing.

Whether or not to use an HWP is not a simple question -- it must be
answered in a difficult trade-off between the advantages, the impact on
science, and the risks. The drawbacks of an HWP can be serious, so an HWP 
should be used \emph{only if it clearly helps mitigate critical systematic 
effects that cannot be avoided by any other means.} The \coremfive\ baseline 
mission does \emph{not} make use of an HWP. We explain why in more 
detail below.

\subsubsection{Technical complexity}

A rotating HWP in space is a (possibly large) cryogenic moving part on
the payload. If continuously rotating on a superconducting magnetic
bearing, it is not straightforwardly thermalised, and is also
susceptible to being a source of magnetic fields close to the focal
plane. Since detectors and their readouts are susceptible to magnetic
fields, enhanced magnetic shielding will be required.
The angular position (orientation of the principal axes of the
birefringent material) must be accurately known at all times to avoid
time-dependent $E$--$B$ mixing. Any such error impacts all focal-plane
detectors in a correlated way.

A stepped HWP can be clamped to be thermalised more easily. However,
it does not modulate the signal towards wings of $4f_{\rm HWP}$ in the
frequency domain anymore, and hence does not help with low-frequency
noise and long-term instability. lt also requires an accurate clamping
mechanism to precisely know the HWP orientation each time it is fixed (in
a reproducible way).

These technical issues can probably be solved (half-wave plates are
operated in many sub-orbital experiments with no obvious show-stopper
so far). In space however, they generate instrumental complexity (and
hence cost and risk). This must be taken into account in the decision
process.

\subsubsection{Impact on science performance}

Besides the technical challenges, the first major drawback of an HWP
is the impact on the scientific capabilities of the mission. 
\begin{itemize}
\itemsep0em
\item An HWP reduces the sensitivity of the focal-plane instrument, as
  compared to the broad-band photon-noise limit, either by reason of
  extra loading due to radiation emitted by the HWP material and/or
  (as was the case for the \core\ M3 proposal) by constraining the width
  of the spectral bands. To avoid extra loading, the HWP must be cold,
  so that the product $\epsilon \times T$ (emissivity times
  temperature) is a fraction of the sky load (dominated by the 3-K CMB
  blackbody).
\item A single HWP is efficient only in a restricted range of
  available frequencies of observation, so that covering the full
  frequency range required for proper monitoring of foreground
  emission is not guaranteed. As a result, either several instruments
  covering different frequency ranges must be implemented (with an
  independent HWP each), or the capability to monitor foreground
  emission is potentially reduced.
\item Used as a first optical element, practical considerations of
  technical feasibility result in a restriction in the size of the
  optical aperture, and hence of the angular resolution and of the
  throughput of the optics (large half-wave plates are
  challenging). This impacts negatively the scientific reach of the
  mission, through the ability to de-lens $B$-modes, component separation, 
  and sidelobe rejection,
  since a smaller aperture has more far-sidelobe pickup.
\end{itemize}
A reflective HWP was proposed as a baseline for the ``M3'' version of
the \core. The original proposed design used a polarising grid in front
of a mirror, a design that allows for a reasonably large aperture (a
baseline of 1.2\,m in the \core\ proposal) with relatively low
emissivity. It also avoided the limited frequency range of operation
of current transmissive HWPs. This concept had several appealing
properties, its main drawback (besides technological readiness) being
a polarisation-modulation efficiency that strongly depended on the
frequency of observation. With such a design, optimal polarisation
modulation is achieved at frequencies of $(2N\!+\!1)\nu_0$, where
$\nu_0$ is a fundamental frequency set by the distance between the
grid and the mirror (and was 15\,GHz for \core), and $N$ is an
integer. As a result, the experiment could only observe with high
efficiency at frequencies centred around 45, 75, 105, 135, 165, 195,
\dots\,GHz, in bands of typical width 15\,GHz. This constraint
generates two difficulties. Firstly, the set of frequencies available
for observation cannot be completely optimised for foreground
subtraction. The \core\ concept had only three channels below 130\,GHz
to monitor synchrotron emission, which was presumed to be good enough
but without much margin for surprises. Secondly, at high frequency,
the width of the bands is small, of the order of only 10\% at 135 and
165\,GHz (which are the main frequency channels for mapping the
CMB). To compensate for the narrow band, many detectors must be
deployed, more than 6300 for \core\ to reach a final CMB sensitivity not
better than what is achieved with \coremfive\ as proposed for ``M5,''
with only 2100 detectors.

A novel type of reflective half-wave plate (R-HWP) was manufactured
and tested in recent years, as a result of an ESA funded project,
``Large radii half-wave plate development,'' aimed at the development
of this technology for future CMB satellite missions. The new design,
based on embedded mesh technology, has high polarisation-modulation
efficiency across a 150\% bandwidth at incidence angles up to
45$^\circ$. The design could be further improved to achieve the 164\%
bandwidth required for \coremfive\ if necessary. In that case, one
could consider it either for a payload design similar to that of \core\
(M3 version), or for an M5-type concept (baseline design presented in
the present paper), for replacing the tertiary mirror of the telescope
with such an R-HWP. Both options would preserve the angular resolution
and sensitivity of the baseline \coremfive\ design. However, an
embedded mesh HWP is still more emissive than a simple reflecting
surface, and hence would negatively impact the sensitivity of the
mission, unless it is cooled down to about 10\,K (depending on the
exact emissivity). The reflective HWP has the advantage of easier
thermalisation, since it has a reflective side that can be thermally
conductive.
 
Transmissive HWPs are an alternative to the reflective concept
proposed for \core. Sapphire HWPs are used on balloons or in
ground-based experiments
\citep{2014RScI...85b4501K,2016SPIE.9914E..2UH,2017arXiv170303847T}. However,
sapphire HWPs are currently limited in size (to $\lsim 50$\,cm).
As an HWP modulates all the incoming polarisation, it is best to use
it as a first optical element, so that instrumental polarisation, such
as induced by the telescope, is not modulated. This has a consequence
on the size of the entrance aperture.

To give a concrete example, let us assume that we select a design with
a 30-cm transmissive HWP as the first element in the optical
chain. Such an aperture means that the angular resolution would be a
factor of approximately 4 times poorer than that of the \coremfive\
baseline. The resulting angular resolution in CMB channels (of the
order of $30^\prime$ at 150\,GHz instead of 7.5$^\prime$), is
sufficient in principle to search for primordial $B$ modes. However,
component separation is compromised by the fact that in the
\coremfive\ lowest frequency channel (60\,GHz), the angular resolution
would be about $75^\prime$ instead, 2.5 times poorer than at
150\,GHz. This does not allow for component separation with the full
set of frequencies down to the $30^\prime$ angular scale. In addition,
at $75^\prime$ angular resolution, masking strong, polarised, and
mostly variable radio sources that contaminate the polarisation maps
will leave many missing pixels.  Such a small aperture would also have
an impact on the sensitivity of the mission, since a smaller telescope
necessarily means less throughput and a smaller available focal-plane
area for a given acceptable Strehl ratio. For a similar detector
technology and implementation design, the loss of sensitivity is
directly proportional to the decrease in aperture diameter, i.e., a
factor of 4, for a final sensitivity of 6.8\,$\mu$K.arcmin for the
full array (instead of 1.7), clearly insufficient to reach the
scientific objectives of \coremfive.

\subsubsection{Mitigation or generation of systematics?}

An HWP mitigates some systematics. A perfect HWP mitigates the impact
of beam asymmetry (assuming that the beam does not change with the
rotation). It relaxes the need to know the satellite attitude and
pointing accurately. It allows for polarisation measurements with
single detectors, relaxing the need to know their spectral response
very accurately. It also helps with the impact of long-term
instabilities that generate low-frequency noise, which are modulated
out of the main sky signal frequencies.
An HWP, however, does not solve everything; sidelobe signals, for
instance, are modulated by the HWP at the same rate as the scientific
signal incoming from the main beam of the instrument.

A rotating HWP will also generate systematic effects of its own, which
must be dealt with in the analysis of the observational data
\citep{2014MNRAS.437.2772M,2016RScI...87i4503E}. Chromatic effects on
polarisation rotation angles and polarisation efficiency must be
accounted for. Also, if detectors do not ``illuminate'' the HWP
homogeneously (i.e., if the full HWP aperture is not homogeneously
coupled to the detectors by the downstream optics), which is usually
the case, any inhomogenities in the HWP thermal emission will generate
spurious signals at the HWP spin frequency and harmonics. To give the
rough size, assuming illumination inhomogeneities of the order of
10\%, and 0.5\% temperature inhomogeneities of a 5-K HWP with 1\%
emissivity, we obtain a spurious signal of the order of 2.5\,$\mu$K, 3
orders of magnitude larger than targeted CMB $B$ modes (note that in
reality, illumination inhomogeneity is larger than that for edge
tapering the aperture against sidelobes, but the tapering is
approximately symmetric, and the symmetric part of the illumination
does not generate this type of systematic effect).

Even for an achromatic, perfectly thermalised HWP, systematics due to
transmission inhomogeneities will be present. Inhomogeneities of the
HWP transmission of the order of 0.1\%, modulating a 3-K CMB signal,
generate fluctuations of the order of 3\,mK across the HWP
aperture. Even if the impact of these inhomogeneities averages out by
integration over the aperture, a small fraction will subsist,
generating spurious signals at the HWP rotation frequency and
harmonics. Whether or not this signal can be kept below the noise must
be demonstrated with further work, since cancellation by several
orders of magnitude is needed.

The impact of many of these imperfections can presumably be addressed
in the data analysis procedures. However, an HWP in the payload makes
the response and properties of the instrument explicitly time
dependent (by design). This time dependence exists not only for the
instrument's required polarisation response, but also for many
instrumental imperfections. Instead of calibrating these instrumental
imperfections for one single stable instrument, with a rotating HWP
they would have to be characterised for each polarisation angle (for
instance, one sidelobe pattern for each HWP angle). If so, in-flight
calibration of instrument properties and data analysis would be
significantly more challenging than with a single, stable instrument
with no rotating HWP.

Weighing the pros and cons, a design with no HWP seems both to be
easier and to allow for improved scientific performance. Hence it is
the baseline for \coremfive. The mitigation of systematic effects
without an HWP is further discussed in
section~\ref{sec:systematics-control}, as well as in the companion
paper on systematic effects \citep{ECO.systematics.paper}, in which  
simulation based analyses show how polarization can be recovered 
without making use of a HWP.

\subsection{Cooling chain}

The performance of \coremfive\ (and more generally of any future CMB
space mission) is critically dependent on operating a (potentially
large) array of sub-kelvin detectors in space. Such low temperature
instruments have already been operated in previous space missions such
as \Planck\ \citep{2011A&A...536A...2P}, {\it
  Herschel} \citep{2010AIPC.1218.1510C}, and {\it Hitomi}
\citep{2016SPIE.9905E..3NS}, and are also planned for future missions
such as {\it Athena} \citep{2014SPIE.9144E..5VB,2016SPIE.9905E..2JC}. Nonetheless,
cryogenic detectors in space are a challenge, and an element of risk
for the performance of the mission.

Following the strategy adopted for the \Planck\ satellite, \coremfive\
uses the environment of space to achieve the lowest possible payload
temperature by passive cooling. Active coolers provide stages at
around 15--20\,K, 4\,K, 2\,K, and 100\,mK, with shields and thermal
filters at each stage that screen the coldest stages from the power
radiated by the hotter ones. The baseline active coolers, all
European-made, are pulse tubes for the 15--20-K stage, Joule-Thomson
4-K and 1.7-K coolers, and a continuous $^3$He-$^4$He dilution fridge
for the sub-kelvin stage.

The development of the active cooling chain depends on the question of
the redundancy strategy to safeguard the mission against cooler
failures. The cooling chain indeed is a single point failure system,
and thus it is in general a good philosophy to implement redundancy
here. There is, however, a drawback to this, each cooler that is kept
off nonetheless thermally connects the cold stages to the hot stages,
increasing the conductive load. For \Planck, the 20-K sorption cooler
was redundant (which turned out to be useful to extend the mission
lifetime, after faster than expected aging of the first sorption
cooler unit), while the lower temperature stages were not duplicated
\citep{2011A&A...536A...2P}. For \coremfive\, we assume redundant
cooling down to 1.7\,K as a baseline, but with only one sub-kelvin
cooler. This can be reconsidered at later stages, in a trade-off
between integration complexity (and hence cost and schedule risk)
versus in-flight failure risk.
Details of the cooling chain design can be found in the companion
instrument paper \citep{ECO.instrument.paper}.

\subsection{Mass and power budgets}

The mass of the SVM is estimated to be about 1150\,kg, and that of the
PLM to be about 390\,kg (telescope reflectors, assuming Silicon
Carbide, 100\,kg; structures, 120\,kg; V-grooves, 90\,kg; main payload
shield, 40\,kg; focal-plane instrument and thermal control, 40\,kg),
for a total dry mass of 1540\,kg. The mass of propellant needed for
orbit injection and operations is 150\,kg (assuming the use of fly
wheels for attitude control). The total wet mass is 1690\,kg
(amounting to about 2 tonnes when margins are included).

The required on-board power is about 1700\,W (2100\,W with margins
included), dominated by a cooling chain requirement of 1290\,W. The
main power consumption is taken by the two pulse tubes (450\,W
each). Solar panels, for a total effective area of 14.2\,m$^2$,
populated with the latest generation of triple junction 3G30\% solar
cells, are illuminated by the Sun under a constant solar incidence
angle of 30$^\circ$. This provides a worst case end of life electrical
power slightly above 2300\,W (based on $190\,{\rm W}/{\rm m}^2$ at
normal incidence).

\subsection{Scanning strategy and payload design}
\label{sec:payload-geometry}

The payload described above is designed for a precession angle $\alpha
= 30^\circ$ between the spin axis and the direction of the Sun, and
scanning angle $\beta = 65^\circ$ between the line of sight and the
spin axis (see section~\ref{sec:scanning}).
As discussed in Ref.~\citep{2017MNRAS.466..425W} however,
temperature-to-polarisation leakage effects are lowered when the
precession angle $\alpha$ is increased while the scanning angle
$\beta$ is decreased. This is due to the better distribution of
polarisation angles over the sky for a pair of detectors. In particular, 
all pixels can be seen with all possible orientations only when 
$\alpha \geq \beta$, and with an appropriate choice of the spinning 
and precession periods..

The choice of the precession angle $\alpha$ impacts the design of
several subsystems of the spacecraft. As $\alpha$ is the incidence
angle of solar illumination on the bottom panel of the spacecraft, the
power on board for a fixed area of solar panels scales as
$\cos{\alpha}$. The total area that is available is $A = \eta \pi
D^2/4$, where $D$ is the diameter of the bottom disc, restricted to be
less than the diameter of the fairing, and $\eta$ is the fraction of
the area that can effectively be used taking into account the space
needed for structural elements and telecommunication
antenna(s). Assuming $D \simeq 4.5\,$m and $\eta=0.9$, the area
available for solar panels is $A \simeq 14\,$m$^2$. For an on-board
power of 2.1\,kW, assuming 190\,W/m$^2$ from solar panels at normal
incidence, the precession angle is constrained to $\alpha \leq
37.8^\circ$.

The precession angle also impacts the geometry of the
payload. Increasing $\alpha$ reduces the volume of the shadow cone in
which the cold payload must fit, unless one accepts that the Sun
illuminate the payload screen, or one changes completely the geometry
of the payload. 

The spinning period $T_{\rm spin}$ also impacts several sub-systems of
the spacecraft. For a 3-axis stabilised system, the reaction wheels
must compensate the total angular momentum of the spacecraft. Assuming
a moment of inertia similar to that of Planck, the total angular
momentum of the spacecraft spinning at 0.5 RPM is $\simeq
105$\,Nms. As large reaction wheels can store $\simeq 70$-100\,Nms,
the spacecraft requires two such reaction wheels, which allows for a
maximum spin rate of 0.66-0.95\,RPM.
Pointing reconstruction accuracy also depends on the spin rate, as
star sensors used for attitude reconstruction become less accurate
when they scan fast. The combination of reaction wheel dimensioning
and attitude reconstruction limits the typical allowed spin rate to
$\lsim 2$\,RPM.

The \coremfive\ scan-strategy and payload design result of a
compromise between these different constraints.

\section{Controlling systematic effects}
\label{sec:systematics-control}

The strategy to control systematics with \coremfive\ is based on
in-flight calibration and deprojection of intensity leakage from
polarisation maps in the data analysis process.  Such deprojection
requires an accurate model of the instrument, which is not immediately
available before launch at the required level of accuracy. A first
model of the instrument is obtained from a combination of theoretical
modelling and ground-based calibration. This first model comprises
estimates of: the beams for each detector (shape and pointing
direction with respect to the spacecraft reference frame);
polarisation parameters (polarisation efficiency and orientation of
the polarisation sensitivity with respect to the spacecraft reference
frame); spectral response parameters (models and measurements of the
spectral bands); and models of radiation patterns, including 4$\pi$
sidelobe patterns. However, none of these ground-based measurements or
models can be expected to match the accuracy that is required to
invert the system of eq.~(\ref{eq:matrix-polarimeter-measurement})
accurately enough for measuring
polarisation $B$ modes with errors dominated by the nominal detector
sensitivity. Additional instrumental knowledge must be obtained in
flight, either with dedicated measurements during a payload
calibration phase, or with the scientific data themselves. We sketch
in the next subsections the general strategy to achieve this.

\subsection{Systematic-correction mapmaking}

A linearly polarised detector scanning the sky along a (quasi-circular) path $p(t)$
ideally observes
\begin{equation}
s(p) = I(p) + \eta \left ( Q_\parallel(p) \cos 2\psi + U_\parallel(p) \sin 2\psi \right ) ,
\label{eq:ideal-scan}
\end{equation}
where $Q_\parallel$ and $U_\parallel$ stand for linear polarisation
Stokes parameters in the frame where the $x$-axis is along the scan
and the $y$-axis perpendicular to it, and $\psi$ is the angle of
orientation of the polarimeter with respect to the scanning direction,
which is fixed by construction of the payload and by the definition of
a fixed spin-axis in the spacecraft frame. Here $I$, $Q_\parallel$,
and $U_\parallel$ should be understood as Stokes parameters of
the sky emission smoothed with some ideal symmetric beam. 
The polarisation efficiency, $\eta$,
is ideally equal to unity, but can be somewhat lower in practice with
no major impact on the measurement as long as $\eta$ is large enough
(closer to 1 than to 0) and that its value is known. The fact that the
scanning is quasi-circular (up to the slow precession of the
spin-axis) defines at each pixel along the scanning trajectory a
natural frame in which the beams (for intensity and polarisation) and
the polarimeter orientation are fixed. If the scanning is at constant
angular speed $\Omega_{\rm spin}$ (which we assume), any time constant
of the detectors (or more generally the whole impulse response of the
detectors and readout system) can also be included into an effective
shape of a \emph{scanning beam} that does not change with time.

To first order in polarisation and second order in intensity,
systematic effects transform the ideal signal of
eq.~(\ref{eq:ideal-scan}) into the following:
\begin{eqnarray}
s(p) & \simeq & I(p) + \eta \left ( Q_\parallel(p) \cos 2\psi + U_\parallel(p) \sin 2\psi \right ) \nonumber \\
 & + & a_\parallel \nabla_\parallel^2 I(p) + a_\perp \nabla_\perp^2 I(p) + a_\times \nabla_\perp \nabla_\parallel I(p) \nonumber \\
 & + & b_\parallel \nabla_\parallel \left [ I(p) + \eta \left ( Q_\parallel(p) \cos 2\psi + U_\parallel(p) \sin 2\psi \nonumber \right ) \right ] \nonumber \\
 & + & b_\perp \nabla_\perp \left [ I(p) + \eta \left ( Q_\parallel(p) \cos 2\psi + U_\parallel(p) \sin 2\psi \nonumber \right ) \right ] \nonumber \\
 & + & 2\delta \, \eta \left [ - Q_\parallel(p) \sin 2\psi + U_\parallel(p) \cos 2\psi \right ] \nonumber \\
 & + & \epsilon I(p) + \xi \left [ Q_\parallel(p) \cos 2\psi + U_\parallel(p) \sin 2\psi \right ],
\label{eq:real-scan}
\end{eqnarray}
where $\nabla_\parallel$ and $\nabla_\perp$, denote gradients along
the scan or perpendicular to the scan, respectively. When several
measurements with different orientations (i.e., along different scans)
are combined to reconstruct the three Stokes parameters $I$, $Q$, and
$U$, the second line generates a leakage of $I$ into polarisation by
reason of beam ellipticity.\footnote{In fact, pointing error also contributes 
to this term, but if the pointing error is much smaller than the beam size,
this is a small correction only.} The coefficients $a_\parallel$, $a_\perp$,
and $a_\times$ measure the amplitude of each of the terms and depend
on the amplitude and direction of the $I$-beam ellipticity with
respect to the scanning direction. The third and fourth lines
represent the pointing error, and depend on the depointing of the
centre of the beam with respect to the nominal direction. This
depointing, assumed to be fixed through the duration of the mission,
generates in particular a leakage of gradients of $I$ into $E$ and
$B$, and of gradients of $E$ into $B$.  Note that in
eq.~(\ref{eq:real-scan}) it is assumed that the displacement is the
same for the $I$-beam as it is for the polarisation beams, but this
assumption could be relaxed. The third line corresponds to a pointing
error along the scan (and could include the impact of an error on the
time constant), while the fourth line corresponds to the effect a
pointing error across the scan.  The fifth line arises from the
first-order expansion of the sines and cosines when we replace angle
$\psi$ with angle $\psi-\delta$, i.e., describes the impact of a small
misalignment of the polarimeter direction in the focal plane. Finally,
the sixth line measures the impact of photometric calibration errors
and depolarisation due to incorrectly calibrated cross-polarisation
leakage.

We assume here that these effects do not vary with time, so that all
the parameters, $a_\parallel$, $a_\perp$, $a_\times$, $b_\parallel$,
$b_\perp$, $\delta$, $\epsilon$, and $\xi$, are fixed and constant for
any given detector (at least for a long-enough period of time to make
a map of a substantial fraction of sky). If all of these parameters
are known a priori with near-perfect precision (i.e., the instrument
is perfectly calibrated), then $b_\parallel$, $b_\perp$, $\delta$, and
$\epsilon$ can be made to vanish (by correcting the pointing solution)
and can be ignored. Similarly, the polarisation efficiency correction
term (for $\xi \neq 0$) can be taken into account immediately in the
map-making step by simply changing the value of $\eta$.  Only the
corrective term from the second line of eq.~(\ref{eq:real-scan}), due
to beam ellipticity, remains.  It is then possible to correct the
observations from these remaining systematic leakage effects as
follows. First, construct a map of $I$ that ignores them. Use that map
to compute the beam-asymmetry terms (the second line of
eq.~\ref{eq:real-scan}) and subtract them from the timelines. Then use
those timelines to obtain a new map of $I$, $E$, and $B$ corrected for
beam ellipticity. The leakage of $E$ and $B$ into $I$ in the first
map-making process for $I$ is a small higher-order correction that can
be ignored, although it is possible to iterate the correction if
necessary. This method has been investigated and shown in
Ref.~\citep{2007A&A...464..405R} to perform well on simulations in the
framework of preparation for the \Planck\ mission.  A new
implementation has been developed specifically for \coremfive, and
demonstrated to reduce the impact of beam asymmetry well below
\coremfive's sensitivity target
\citep{ECO.systematics.paper,ECO.instrument.paper}.

It is theoretically possible to do even better and correct the maps in
the case where the calibration is not perfect and the beam shape not
exactly known. Instead of computing the correction terms assuming that
all of $a_\parallel$, $a_\perp$, $a_\times$, $b_\parallel$, $b_\perp$,
$\delta$, $\epsilon$, and $\xi$ are known, we instead \emph{calibrate}
$a_\parallel$, $a_\perp$, $a_\times$, $b_\parallel$, $b_\perp$ by
fitting for the five unknown parameters in a map-making step (i.e.,
build the least-square map solution for $I$, $Q$, $U$, and all of
$a_\parallel$, $a_\perp$, $a_\times$, $b_\parallel$, $b_\perp$)
assuming that all $\nabla^2 I(p)$ and $\nabla I(p)$ terms are known to
first order from the first iteration of the reconstruction of the
intensity map, so that the system to be solved is linearised. We then
iterate once again, injecting in the system $Q_\parallel(p)$ and
$U_\parallel(p)$ from the map of $E$ to fit for the terms $\epsilon$
and $\xi$ that govern the leakage of $E$ into $B$.

\subsection{Bandpass leakage correction}

The previous paragraph deals with all of the angular response mismatch
of a single detector. When mapmaking requires differencing detectors
that have different frequency bands, an additional source of
intensity-to-polarisation leakage arises.
Each detector $i$ observes the integral of the sky emission over a
frequency band $h_i(\nu)$, so that the intensity detected by each
detector can be written as
\begin{equation}
d_i = \int {\rm d}\nu \, h_i(\nu) \left [ I_{\nu} + Q_{\nu} \cos 2\psi_i + U_{\nu} \sin 2 \psi_i \right ],
\end{equation}
where $I_{\nu}$, $Q_{\nu}$, and $U_{\nu}$ now are the Stokes
parameters of the sky emission brightness as a function of frequency
$\nu$, and $h_i(\nu)$ is the frequency band of detector $i$.

The total sky brightness arises from the superposition of emission
signals from different astrophysical processes. In a given pixel, the
total sky intensity is
\begin{equation}
I_{\nu} = \sum_c f_{\nu c} \, I_c,
\end{equation}
where $f_{\nu c}$ is the spectral emission law of component $c$, and
$I_c$ is the amplitude of component $c$ at some reference frequency
(typically near the centre of the spectral band defined by
$h_i(\nu)$). Similar equations hold for $Q_{\nu}$ and $U_{\nu}$.

In principle, the spectral emission laws for all of the three Stokes
parameters $I$, $Q$, and $U$ can be different for a given component of
sky emission, but here we are primarily concerned with the frequency
band mismatch for $I$, which is the dominant term.  The total signal
observed by detector $i$ is
\begin{equation}
d_i = \sum_c a_{ic} \left [ I_c + Q_c \cos 2\psi_i + U_c \sin 2\psi_i \right ],
\label{eq:multicomponent-matrix-polarimeter-measurement}
\end{equation}
where
\begin{equation}
a_{ic} = \int {\rm d}\nu \, h_i(\nu) f_{\nu c} \, .
\end{equation}
This multi-component model replaces the single component model of
eq.~(\ref{eq:matrix-polarimeter-measurement}). Calibrating 
the observations on one given component (e.g., the CMB)
amounts to rescaling the observations so that for all $i, \; a_{ic}=1$
for that particular component $c$. There is, however, no guarantee
that all of the coefficients $a_{ic}$ will be the same for all
components contributing significantly to the observed emission; hence
in the general multi-component case, it is not possible to calibrate
the data so that all of the $a_{ic}$ coefficients are equal to
unity. A set of different detectors measures in each pixel a mixture
of components of the form
\begin{equation}
{\bm d} = \sum_c {\sf A}_c {\bm s}_c + {\bm n},
\label{eq:multicomp-linear-system1}
\end{equation}
with, for each component $c$, a component-specific ``mixing matrix'' for $I$, $Q$, and $U$:
\begin{equation}
{\sf A}_c = \begin{pmatrix}
a_{1c} & a_{1c} \cos 2\psi_1 & a_{1c} \sin 2\psi_1\\
a_{2c} & a_{2c} \cos 2\psi_2 & a_{2c} \sin 2\psi_2\\
\vdots & \vdots  & \vdots \\
a_{Nc} & a_{Nc} \cos 2\psi_N & a_{Nc} \sin 2\psi_N
\end{pmatrix}.
\label{eq:matrix-Ac}
\end{equation}
The consequence is that there is no immediate way to invert the
observations to recover the values of $I=\sum I_c$, $Q=\sum Q_c$, and
$U=\sum U_c$ from multi-detector observations. Unless all of the
coefficients $a_{ic}$ are equal, any direct inversion assuming a
matrix ${\sf A}$ calibrated on one of the components will inevitably
result in a leakage of $I$ into $Q$ and $U$ for the other components,
compromising the interpretation of the observed polarisation. The
problem is further complicated by the fact that the emission laws of
some of the components vary across the sky.

The band-mismatch problem can potentially be severe for detecting
low-level primordial $B$ modes. In units of ${\rm MJy}\,{\rm
  sr}^{-1}$, the emission law of synchrotron scales roughly as
$\nu^{-1}$, that of thermal dust emission roughly as $\nu^{3.6}$, and
that of the CMB as the derivative with respect to temperature of a
2.725-K blackbody. The very different colours of these various
emission processess result in differences of a few percent between the
various $a_{ic}$ coefficients.  For CO emission, concentrated in thin
spectral lines centred at frequencies that are multiples of
$\nu\simeq115\,$GHz (and nearby frequencies for isotopologues), the
exact spectral response may vary significantly between detectors
(e.g., by factors of a few).

Similarly to the impact of the angular response, the problem can be
solved iteratively as follows.  In a first step, maps of intensity are
obtained (neglecting the bandpass mismatch) in several frequency
bands. These maps are used to obtain maps of intensity for all
components in each of the average frequency bands, in a
component-separation step. Estimated component maps $\widehat {\bm
  s}_c$ (with vanishing polarisation at this stage) are then plugged
in
eq.~(\ref{eq:multicomponent-matrix-polarimeter-measurement}). Expanding
each mixing matrix as
${\sf A}_c = {\sf A} +\delta{\sf A}_c$, that equation can be recast as
\begin{equation}
{\bm d} = {\sf A} {\bm s} + \sum_{c \neq {\rm CMB}} \delta {\sf A}_c \, \widehat{\bm s}_c + {\bm n} .
\label{eq:multicomp-linear-system2}
\end{equation}
This is a linear system in the unknowns ${\bm s}$ and $\delta {\sf
  A}_c$, which can be solved by standard linear inversion.

This method has been implemented on simulations of \coremfive,
demonstrating that the bandpass-mismatch effect can be reduced to a
level compatible with the required mission sensitivity with one
iteration, provided that the mismatch between the bands is no worse
than was the case for \Planck\
\citep{ECO.systematics.paper}. Additional technical details and
results, as well as a discussion of a second correction method, can be
found in Ref.~\citep{Ranajoy.bandpass.paper}.

\section{Options}
\label{sec:options}

\subsection{Descoping options}

\coremfive\ is an ambitious mission. If a drastic descope were deemed
necessary, one could consider reducing the ambitions and concentrating
on the observations that cannot be obtained by any other means and are
crucial for achieving the science goals of \coremfive, either with the
mission alone, or in combination with other observations that could be
obtained independently (even if not as well as with \coremfive). Using
the name ``MiniCORE'' to refer to this descoped mission, we would
require MiniCORE to provide at least the following capabilities.
\begin{itemize}
\itemsep0em
\item Clean, multi-frequency, full-sky CMB maps at large and medium
  angular scales, i.e., at all scales where foreground emission and
  cosmic variance dominate the errors for measuring $E$ modes and
  lensing $B$ modes when the noise is below 5\,$\mu$K.arcmin, i.e.,
  all scales larger than about 10--15$^\prime$.
\item Full-sky maps of high-frequency foregrounds at all useful scales
  (i.e., down to a few arc-minutes), to complement ground-based CMB
  observations at the same angular scale.
\item CIB maps, for delensing the $B$ modes (independently of methods based on CMB polarisation itself).
\end{itemize}
A downsized version of \coremfive, with aperture reduced to 80\,cm, 900
detectors instead of 2100, mission duration of 3 years instead of 4
years, and reduced frequency range, would fulfill these
requirements. A possible distribution of frequency channels is
outlined in Table~\ref{tab:MiniCORE-bands}.

{\small
\begin{table}[htb]
\begin{center}
\scalebox{0.95}{\scriptsize
\begin{tabular}{|c|c|c|c|c|c|c|c|c|}
\hline 
Channel& Beam& $N_{\rm det}$& $\Delta T$& $\Delta P$& $\Delta I$& $\Delta I$& $\Delta y\times 10^6$& PS ($5\,\sigma$)\\
{[}GHz{]}& {[}arcmin{]}& & {[}$\mu$K.arcmin{]}& {[}$\mu$K.arcmin{]}& {[}$\mu K_{\rm RJ}$.arcmin{]}& {[}kJy/sr.arcmin{]}& {[}$y_{\rm SZ}$.arcmin{]}& {[}mJy{]}\\
\hline 
\hline 
100& 16.93& 40& 8.4& 11.8& 6.51& 2.00& $-2.0$& 12.7\\ 
115& 14.81& 40& 8.2& 11.7& 5.92& 2.41& $-2.2$& 13.4\\ 
130& 13.18& 40& 8.3& 11.7& 5.43& 2.82& $-2.5$& 13.9\\ 
145& 11.89& 90& 5.6& 7.9& 3.34& 2.16& $-2.0$& 9.6\\ 
160& 10.84& 90& 5.8& 8.1& 3.09& 2.43& $-2.6$& 9.9\\ 
175& 9.96& 90& 6.0& 8.5& 2.88& 2.71& $-3.5$& 10.1\\ 
195& 9.00& 90& 6.5& 9.2& 2.63& 3.07& $-7.0$& 10.4\\ 
220& 8.04& 90& 7.3& 10.4& 2.39& 3.55& \dots& 10.7\\ 
255& 6.99& 90& 9.3& 13.1& 2.15& 4.29& 5.7& 11.2\\ 
295& 6.08& 40& 19.6& 27.6& 2.96& 7.91& 5.7& 18.0\\ 
340& 5.3& 40& 31.1& 43.9& 2.81& 9.98& 5.6& 19.8\\ 
390& 4.62& 40& 55.9& 79& 2.75& 12.85& 7.1& 22.3\\ 
450& 4.00& 40& 120.9& 171& 2.75& 17.11& 11.3& 25.7\\ 
520& 3.46& 40& 315.2& 445.8& 2.79& 23.18& 22.4& 30.0\\ 
600& 2.99& 40& 987.9& 1397.1& 2.84& 31.47& 55.2& 35.3\\ 
\hline
\hline
Array& & 900& 2.3& 3.2& & & 0.90& \\
\hline 
\end{tabular}
}
\end{center}
\vspace{-\baselineskip}
\caption{ \label{tab:MiniCORE-bands} 
  \small Possible MiniCORE frequency channels. 
  The sensitivity is calculated for a 3-year mission, assuming 
  $\Delta \nu/\nu=30\%$ bandwidth, 60\% optical efficiency, total 
  noise of twice the expected photon noise from the sky, and the optics 
  of the instrument being cooled to 85$\,$K. This configuration has 
  900 detectors, about 55\% of which are located in CMB channels 
  between 130 and 220\,GHz. Those six CMB channels yield an 
  aggregate CMB sensitivity in polarisation of $3.6\,\mu$K.arcmin 
  ($3.2\,\mu$K.arcmin for the full array). Entries for the thermal SZ 
  Comptonisation parameter $\Delta y$ are negative below 217\,GHz
  (negative part of the tSZ spectral signature).
}

\end{table}
}

This downsized option still has angular resolution better than
$17^\prime$ in all channels, in order to observe both bumps of
inflationary $B$ modes. With eight channels between 100 and 220\,GHz
it can check for foreground contamination in CMB maps, but
complementary ground-based observations at 90\,GHz (and below 40\,GHz
for monitoring the synchrotron) would be useful. Dust and CIB are
mapped between 255 and 600\,GHz, with angular resolution ranging from
3 to 7$^\prime$, which is about adequate for complementarity with
future ground-based observations for CMB polarisation science. The
aggregate CMB sensitivity of 3.2\,$\mu$K.arcmin with the full array,
slightly less than 2 times worse than the \coremfive\ baseline (4
times worse in power), is still good enough for the lensing $B$ modes
to be mapped with S/N $\simeq 1.5$. The primordial $B$-mode
recombination bump is below the noise for $r \, \lsim \, 0.006$. This
is not optimal, as it fails to clearly satisfy the ``margins and
redundancy'' requirement. While the final sensitivity can be improved
a posteriori with an extension of the mission duration, more descoping
would not be adequate for addressing the science goals of \coremfive\ and
guaranteeing scientific breakthroughs -- including for the detection of
primordial $B$ modes.
MiniCORE as defined here can hence be considered as the ``minimal''
next-generation CMB polarisation space mission.

This downsizing does not require any major redesign of the
spacecraft. The smaller telescope and payload would fit in an
Ariane~6.2 launcher, without the need of a tertiary mirror (provided
the focal plane can still be shielded from stray light). The mass of
the focal plane would be drastically reduced, to about 3\,kg instead
of 8\,kg, allowing for reduced conductive losses and hence more margin
on the cooling power. Additional details about the instrumental
configuration of this descoped option are available in the companion
instrument paper in this series \citep{ECO.instrument.paper}.

\subsection{Upgrades}

Going in the other direction, \coremfive\ could be improved in several
ways for better scientific performance. The most straightforward
improvement would be to increase the sensitivity by making all
detectors dual-polarisation. This improves the noise level in all
channels centred at $\nu \ge 115\,$GHz by a factor $\sqrt{2}$, for a
total CMB sensitivity of 1.5\,$\mu$K.arcmin using channels from 130 to
220\,GHz (1.3\,$\mu$K.arcmin for the full array).

Another simple improvement with little impact on the overall design is
to add frequency channels above 600$\,$GHz, for better addressing the
Galactic and extragalactic science goals. For instance, 96
photon-noise-limited detectors at 1200$\,$GHz would increase the
sensitivity to dust and IR sources, at an improved angular resolution
of $1^\prime$ instead of $2^\prime$ (which would also provide improved
pointing reconstruction using science data). This would require very
limited additional resources (negligible focal-plane area and small
increase in telemetry).

Similarly, the use of multi-chroic detectors can potentially improve
the sensitivity by increasing the number of detectors by a factor of
2--3, potentially allowing a final map sensitivity of 1\,$\mu$K.arcmin
or better. This increased sensitivity would be useful for CMB science,
however, only if the foreground emission residuals in CMB maps can be
reduced by a matching amount.

Finally, one could consider increasing the telescope size for better
angular resolution, and hence improved lensing science and Galactic
and extragalactic astrophysics studies. A small aperture increase
(e.g., aperture diameter $D \simeq 1.5\,$m) could possibly be achieved
simply by optimisation of the proposed geometry. A more significant
aperture increase (e.g., $D \simeq 1.8\,$m) would require both a
revision of the baseline design, and an increase in the overall size
of the payload, which probably could only be considered within a large
international collaboration. With the large optical system, the
available focal-plane area would also be increased, allowing for more
detectors and extra sensitivity.

While they do not drastically improve the performance for inflationary
science or for investigations of the cosmological model, the added
value of increased angular resolution and sensitivity is substantial
for cluster science, extragalactic sources, and for exploiting CIB
maps to constrain star formation at distant redshifts. Some of the
companion science papers to this one investigate the added value of
such upgrades.

\section{Discussion}

\begin{table}[bt]
\begin{center}
\scalebox{0.84}{\scriptsize
\begin{tabular}{|c|c|c|c|c|c|c|c|c|c|c|}
\hline 
Mission & Year & $N_{\rm det}$ & $\Delta P$ & Aperture & Beam & $N_{\nu}$ & $\nu$ range & $N_{\rm rec.}$ & $t_{\rm map}$ & HWP \\
 & & & {[}$\mu$K.arcmin{]} & & size & & {[}GHz{]} & &  & \\
\hline 
\hline 
SAMPAN & 2006 & ${\sim}\,10\,000$ & 2.7 & 30\,cm & $40^\prime$--$20^\prime$ & 6 & 70--545 & 6 & months & (yes) \\
EPIC-LC (NTD) & 2007 & 830 & 3.0 & 30\,cm & $155^\prime$--$16^\prime$ & 7 & 30--300 & 6 & months & yes \\
EPIC-LC (TES) & 2007 & 2\,366 & 1.8 & 30\,cm & $155^\prime$--$16^\prime$ & 7 & 30--300 & 6 & months & yes \\
BPOL & 2007 & 1\,006 & 3.4 & 4.5 to 26.5\,cm & $15^\circ$, $68^\prime$--$40^\prime$ & 6 & 45--345 & 8 & months & yes \\
LiteBIRD & 2010 & 2\,622 & 2.5 & 20 and 40\,cm & $69^\prime$--$17^\prime$ & 6 $\rightarrow$ 15 & 40--400 & 2 & months & yes \\
\hline
\hline
EPIC-IM (4K) & 2009 & 11\,094 & 0.9 & 1.4\,m & $28^\prime$--$1^\prime$ & 9 & 30--850 & 1 & months & no \\
EPIC-IM (30K) & 2009 & 2\,022 & 2.3 & 1.4\,m & $28^\prime$--$1^\prime$ & 9 & 30--850 & 1 & months & no \\
\core\ (M3) & 2010 & 6\,384 & 1.8 & 1.2\,m & $23^\prime$--$1.3^\prime$ & 15 & 45--795 & 1 & few days & yes \\
\coreplus\ (M4) & 2015 & 2\,410 & 2.0 & 1.5\,m & {\bf $14^\prime$--$1.4^\prime$} & 19 & 60--600 & 1 & few days & no \\
{\bf \coremfive\ (M5)} & {\bf 2016} & {\bf 2\,100} & {\bf 1.7} & {\bf 1.2\,m} & {\bf 18$^\prime$--2$^\prime$} & {\bf 19} & {\bf 60--600} & {\bf 1} & {\bf few days} & {\bf no} \\
\hline
\hline
EPIC-CS  & 2007 & 1\,520 & 1.8 & 3.0\,m & $15^\prime$--$1^\prime$ & 8 & 30--500 & 1 & months & TBD \\
PIXIE  & 2010 & 4 & 4.2 & 55\,cm & $2.6^\circ$ tophat & 400 & 30--6000 & 2 & months & no \\
PRISM (imager)  & 2013 & 7\,600 & 1.1 & 3.5\,m & $17^\prime$--$5^{\prime \prime}$ & 32+300 & 30--6000 & 1 & few days & no \\
PRISM (spectro)  & 2013 & a few & 2.9 & 50\,cm & 1.4$^\circ$ & 400 & 30--6000 & 2 & months & no \\
\hline 
\end{tabular}
}
\end{center}
\caption{ \label{tab:compare-perf} Main characteristics of proposed CMB space missions. Columns are, from left to right: mission name (with possible options); year of initial conceptual design; number of detectors; aggregated CMB sensitivity from all channels; aperture size; beam size; number of frequency channels; frequency range; number of receivers (optical systems with a focal plane); typical time required to observe, with a single 4-detector set, a sizeable, well sampled map (e.g., tens of percent of sky, connected, with no holes); and whether or not the mission uses a rotating HWP. The three main vertical sections identify missions focused on large-scale CMB polarisation (category 1, top), missions targeting most of CMB polarisation science (category 2, middle), and missions specifically designed to also address science objectives beyond CMB polarisation (category 3, bottom). In each category, missions are ordered according to the year of design.
Entries for LiteBIRD correspond to a recent version with 3 years of observation, in which the initial number of six frequency bands in a single receiver evolved to a new baseline of 15 bands with two receivers.  The PIXIE sensitivity assumes a polarisation sensitivity of 70\,nK for the full instrument \citep{2011JCAP...07..025K}. PRISM has two independent instruments, specified on two different rows, but both are part of the proposed baseline mission; the PRISM imager has 32 broad-band channels, and also a narrow-band spectrometer (with $R = \Delta\nu/\nu \simeq 100$, for about 300 narrow spectral bands).}
\end{table}

We now turn to a discussion of our choices for \coremfive\ and put those choices in the context 
of other CMB polarisation space missions. 
Table~\ref{tab:compare-perf} gives the main characteristics of space missions that have been 
proposed or studied since 2007. The Table has three parts: (i) missions with telescope aperture $\simeq 30$\,cm, with relatively coarse angular resolution and 
noise level  $\gsim 2\,\mu$K.arcmin, targetting large scale CMB polarisation only; (ii) missions with a telescope size in the 1.2-1.5\,m range, with angular resolution of a few arcminutes and a noise level in the 1-2\,$\mu$K.armin range typically, designed for comprehensive CMB polarisation science; 
(iii) missions specifically designed for addressing a scientific program that extends beyond CMB polarisation, with either high angular resolution, or including CMB spectroscopy as a main science goal. 
Among the various options, \coremfive\ is the right choice for an ESA M-class mission for the following reasons: 
\begin{itemize}
\item
It has the combination of resolution and low noise to give $\sigma_r = 0.0004$, thus clearly distinguishing between 
inflationary models with $r \ll 0.001$ or $r=0.003$, avoiding in particular the risk of a possible ambiguous hint
of $r = 0.002 \pm 0.001$, which would neither be a clear detection, nor rule-out
a Starobinsky-type inflationary model for which $r \simeq 0.003$. We designed \coremfive\ with this capability because 
current theory gives strong motivation for such inflationary models, and we 
believe that to be relevant in the 2020's {\it any} space mission must be able to provide this discrimination;
Delensing capability is essential to reach this level of sensitivity: with lensing reconstructed by \coremfive, 
one can reduce the lensing B-mode power by 70\%,
leading to an improvement of a factor of 2.5 in the error on the amplitude of primordial
gravitational waves \citep{ECO.lensing.paper}.
\item
It has the necessary number of frequency bands, with noise per band sufficient to measure
sources of galactic emission. {\it Any} mission that aspires to measure primordial $B$ modes at $\ell<20 $ must contend with 
foreground levels that are orders of magnitude stronger than the $B$ mode, and must contend
with yet unknown potential foreground complexities. Some of the early proposed polarisation missions do not have 
sufficient foreground determination capabilities;
\item
A mission that targets only $r$ may have a null-result as its main science output. While setting limits on $r$ has 
important consequences for the physics at ultra-high energies, 
the resolution of the mission strengthens the \coremfive\ constraints on the physics of the inflation, 
even if $r$ is not detected, through high fidelity measurements of $n_{\rm s}$, of the scale-dependent
running of $n_{\rm s}$, and of non-Gaussianity. This requires the 
measurement of the E-mode power spectrum over a broad range of $\ell$'s.
We also believe a next generation CMB space mission should 
provide a broader range of cosmological and astrophysical results, serve a broader community of astrophysicists, and give a legacy dataset 
to be mined for more than just inflationary physics. 
We designed \coremfive\ to provide cosmic variance limited observation of the $E$ modes up to $\ell \simeq 2500$
and of the lensing $B$ modes up to $\ell \simeq 1000$, enabling investigation
of possible extensions with parameters describing curvature, neutrino physics, extra light relics, 
primordial helium abundance, dark matter annihilation, recombination physics, variation of fundamental constants, 
dark energy, modified gravity, reionisation, cosmic birefringence. The ground-breaking post-CORE overall reduction of the allowed parameter 
space will be as much as $\sim 10^7$ as compared to Planck 2015, and $10^5$ with respect to Planck 2015 + future BAO measurements \citep{2016arXiv161200021D}.
The angular resolution of \coremfive\ is optimised to probe a broad range of science goals while fitting within an ESA M-class mission budget. 
\item
We designed the scan strategy of \coremfive\ to give strong discrimination of polarimetric systematic errors. 
We considered an alternative approach to mitigate low frequency noise and polarimetric systematic effects - the use of a continuously rotating 
half-wave plate - a technical risk that (i) would therefore increase costs, and (ii) is not necessary given the mitigation provided
by the scan strategy. Experience with sub-orbital instruments suggests that a half-wave plate should be the first element 
in the optical path. A half-wave plate, whether rotated continuously or in steps, with an entrance aperture diameter larger 
than 0.5~m and compatible with the broad frequency coverage required for foreground cleaning is a technical challenge with consequences on costs and schedule. 
\end{itemize}

\subsection{Complementarity with sub-orbital experiments}

Past experience with CMB temperature anisotropies shows that precision
CMB science requires a space mission when the dominating sources of
error are foreground contamination, cosmic variance, and systematic
effects, rather than raw CMB sensitivity. 

The same will be true for
polarisation. While the noise limit for $E$-mode and lensing $B$-mode
detection was first overcome by deploying increasing numbers of
detectors observing from the South Pole and the Atacama Plateau (not
forgetting circumpolar balloon flights), a comprehensive, precise and
accurate cosmological exploitation of CMB polarisation (including $E$
modes and lensing $B$ modes), cannot be made without a space mission.

This, however, does not preclude exploiting the best ground-space
complementarity. A diffraction limit of $2^\prime$ requires a 4.2-m
telescope at 150\,GHz (6.3-m at 100\,GHz). While such telescopes can
be deployed on the ground, they cannot be envisaged in orbit within
the budget of an M-class mission. As demonstrated with SPT and ACT,
even for mapping temperature, such large telescopes perform well for
measurements at $\ell \, \gsim \, 500$. They are a perfect complement to a
space mission that observes up to $\ell=1000$ (a $12^\prime$ beam) in the
same frequency range, providing a good overlap in sensitivity for
cross-calibration of gains and beams in the $500\leq \ell \leq 1000$
angular scale range.

The combination of a spectrometer mission such as PIXIE, an imager 
such as \coremfive, and a high resolution ground-based observatory
would be very powerful for the best observation and scientific exploitation of the 
CMB in the next decade.

\section{Conclusion}

Cosmological observations support a concordance inflationary
$\Lambda$CDM cosmological scenario, in which seeds for density
perturbations are generated in the very early Universe during a phase
of cosmic inflation, by stretching to macroscopic scales quantum
fluctuations of the spacetime metric. These perturbations then evolve
in the primordial plasma until baryons decouple from radiation,
releasing the CMB, and become free to collapse under the force of
gravitation to generate the large-scale structures observed in the
present Universe. But fundamental questions still remain: did cosmic
inflation really happen, and if so what is the physics that drives it?
What are the dark matter and dark energy required by this scenario,
which appear to represent 96\% of the total energy density in the
Universe? Is there something essential still missing in our
understanding of our cosmos?

The CMB is a crucial tool for further investigating this global
picture.  Three space missions have already scrutinised the CMB to
exploit the scientific information encoded in its tiny fluctuations of
intensity and polarisation, largely contributing to the adoption of
the standard $\Lambda$CDM model. However, only temperature
anisotropies have been mapped with good signal-to-noise ratio over
most of the sky and for most of the useful angular scales. Much can
still be learnt from detailed observations of CMB polarisation on all
scales larger than about two arcminutes: polarisation $E$ modes are
yet a largely unexploited probe of the cosmological model; lensing $B$
modes offer the opportunity to map all the dark matter structures
between the last scattering surface at $z \simeq 1080$ and present-day
observers, to further understand how it clusters and how it interacts;
the detection of primordial $B$~modes on scales larger than about
thirty arcminutes is essential to confirm the inflationary scenario
and obtain clues about the physics at work in the early Universe, on
grand unification energy scales $10^{12}$ times higher that those
probed by the largest human-made particle accelerator to date. These
new observations of CMB polarisation still have the potential to
revolutionise our understanding of our Universe. Their optimal
exploitation is a must.

Ideally, CMB polarisation should be observed accurately over the full
sky, at all scales down to about $2^\prime$, and with a sensitivity of
the order of a fraction of a $\mu$K.arcmin. This is a challenging
task. While such a sensitivity can theoretically be reached observing
continuously for a few years from the ground with several hundred
thousand detectors, or from space with several thousand detectors,
controlling foreground contamination or systematic effects with a
matching level of accuracy is the major challenge to overcome in order
to reach the science goals of a future CMB polarisation
survey. Space-borne observations of the CMB offer unmatched precision
and accuracy when astrophysical foreground emission, cosmic variance,
or instrumental systematic effects are the dominant sources of error
in the interpretation of CMB observations. This is the case for all
angular scales down to about 10 arcminutes; smaller scales, which
require telescopes of size several metres, are best observed from
ground-based observatories.

The most effective scientific exploitation of CMB polarisation must
hence make use of the complementarity between the ground and
space. The space mission must map the polarised sky in more than 10
frequency channels with an angular resolution sufficient to cover the
range of scales where foregrounds or cosmic variance dominate the
errors in CMB maps, i.e., all scales $\gsim 10^\prime$. Such a space
survey can be complemented with ground-based observations to extend
the angular resolution down to $2^\prime$ in a few atmospheric
windows; such information is hard to gather from space by reason of
the size of the telescope required to reach this angular resolution at
those frequencies. A space mission is, however, ideally suited to map
the high-frequency foreground emission (i.e., dust and CIB) above
300\,GHz, with a matching angular resolution of the order of
$2^\prime$. This can be achieved with a metre-class telescope in
space, while the atmosphere precludes a wide and sensitive survey from
the ground at those frequencies. Dust must be mapped to accurately
subtract its contribution to CMB $E$ and $B$ modes and to characterise
any possible residuals that could bias the extraction of those
signals. The CIB must be mapped as a useful tracer of cosmic
structure, which is essential to disentangle primordial $B$ modes from
lensing.

\coremfive\ reaches the sensitivity and angular resolution
requirements of such a future space mission, across a frequency range
that extends from 60 to 600\,GHz, with an array of 2100
cryogenically-cooled detectors at the focus of a 1.2-m aperture
telescope. The sky is mapped in 19 frequency channels with angular
resolution ranging from 2 to 18$^\prime$, for an aggregate CMB survey
sensitivity of 1.7\,$\mu$K.arcmin in polarisation after 4 years of
continuous observations. The observing strategy is such that 45\% of
the sky is mapped every 4 days, allowing for many cross-checks of the
measurement over the course of the mission, essential to control
systematic effects. Additionally, the broad frequency coverage, with
six independent channels between 130 and 220\,GHz, and CMB
polarisation sensitivity of order 5\,$\mu$K.arcmin each, allows for
cross-checking to ensure that the measured CMB spectra are not
contaminated by residual foregrounds after subtraction of a model of
polarised astrophysical emission. This built-in redundancy is
essential to ascertain the accuracy of the observed CMB, and avoid any
false interpretation of the observed sky emission.

\coremfive\ is designed to minimise systematic effects that could
potentially be generated by thermal instability or side lobe pickup of
strong astrophysical emission: The spacecraft and the scan-strategy
are designed in such a way that the solar power absorbed remains
constant throughout the scientific observations. Sun, Earth and Moon
are kept well away from the line of sight at all times, and masked by
absorptive screens that avoid illumination of the focal plane, either
directly or by reflection. With no moving part in the optical path of
the instrument, and with continuous observations in a very stable
configuration, \coremfive\ is designed for a maximally stable
instrumental response, allowing to calibrate the instrument in flight
with an accuracy matching the stringent requirements of the
measurement. This allows for correcting potential systematic effects
in a data-processing step that jointly measures the sky emission and
relevant parameters of the instrumental response.

Should descoping be deemed necessary for technological or programmatic
reasons, it would be possible to reduce the aperture size to 80\,cm
instead of 120\,cm, divide the number of detectors by 2, and drop the
lowest frequency channels below 100\,GHz (which are the most
challenging from space by reason of required volume and mass). This
would increase the dependence of the mission upon high-quality
measurements from the ground in all atmospheric windows, but would
preserve most of the essential CMB polarisation
science. Alternatively, the quality of the survey could be improved
with dual-polarisation and/or multichroic detector technology, as well
as with a somewhat increased telescope aperture.

Although optimised for accurate CMB polarisation observations,
\coremfive\ also has the potential to tackle additional science
goals. Its sensitivity to foreground polarisation will help understand
the role of magnetic fields in the structuring the interstellar medium
and in Galactic star formation. By detecting tens of thousands of
galaxy clusters (and hundreds of thousands in combination with a
high-resolution ground-based CMB survey), it will provide a means to
constrain the nature of dark matter and dark energy or to check for
modified gravity, independently of the primary CMB or of any other
probe. By observing distant, strongly lensed dusty galaxies and
protoclusters of galaxies, it will help us to understand the history
of cosmic star formation and the role of baryons in the formation of
matter structures. Finally, the legacy value of the 19 accurate
intensity and polarisation maps it will deliver to the scientific
community will make the \coremfive\ survey a lasting resource for many
additional astrophysical and cosmological investigations, the impact
of which cannot yet be foreseen.

\acknowledgments

The \coremfive\ collaboration thanks CNES, Thales Alenia Space, 
and Air Liquide Advanced Technologies for advice and technical 
support during the preparation of the \coremfive\ proposal. We also 
thank the ESA CDF team for the CMB Polarisation CDF study performed 
in March 2016, the results of which were extensively used to define the 
mission concept presented in this paper.
J.G.N. acknowledges financial support from the Spanish MINECO for a ÔRamon y CajalÕ fellowship (RYC-2013-13256) and the I+D 2015 project AYA2015-65887-P (MINECO/FEDER).
CJM is supported by an FCT Research Professorship, contract reference
IF/00064/2012, funded by FCT/MCTES (Portugal) and POPH/FSE.
F.J.C., R.F.-C., E.M.-G. and P.V. acknowledge support from the Spanish Ministerio de Econom'a y Competitividad project ESP2015-70646-C2-1-R 
(cofinanced with EU FEDER funds), Consolider-Ingenio 2010 project CSD2010-00064 and from the CSIC "Proyecto Intramural Especial"
project 201550E091.
FA is supported by the National Taiwan University (NTU) under Project No. 103R4000 and by the NTU Leung Center for Cosmology and Particle Astrophysics (LeCosPA) under Project No. FI121.
 BFR acknowledges support from the National Science Centre, Poland,
 under grant 2014/13/B/ST9/00845.


\newpage
\appendix
\section{Impact of atmosphere on ground-based CMB observations}
\label{appendix:atmosphere}

\subsection{Atmosphere and detector sensitivity}
\label{sec:sensitivity}

Space-borne detectors benefit from a potentially very cold and quiet
environment, allowing for the sensitivity of broad-band detectors
below $200\, $GHz to be limited by CMB photon noise. From the ground,
typical atmospheric emissivity in atmospheric windows is at the level of
a few percent, depending on the frequency and the amount of
precipitable water vapour. This extra radiation generates additional
loading on the detectors, at the level of a few percent of the atmospheric 
temperature of $230-290\,$K
(i.e., 10--$20\,$K of background), rapidly increasing at frequencies
above $200\,$GHz. The actual loading observed with the BICEP2
instrument is 22\,K$_{\rm RJ}$ \citep{2014ApJ...792...62B}, partly due to the 
atmosphere, the rest being due to the instrument itself (forebaffle, window and filters).

\begin{figure}[htbp]
\centering 
\includegraphics[width=.49\textwidth]{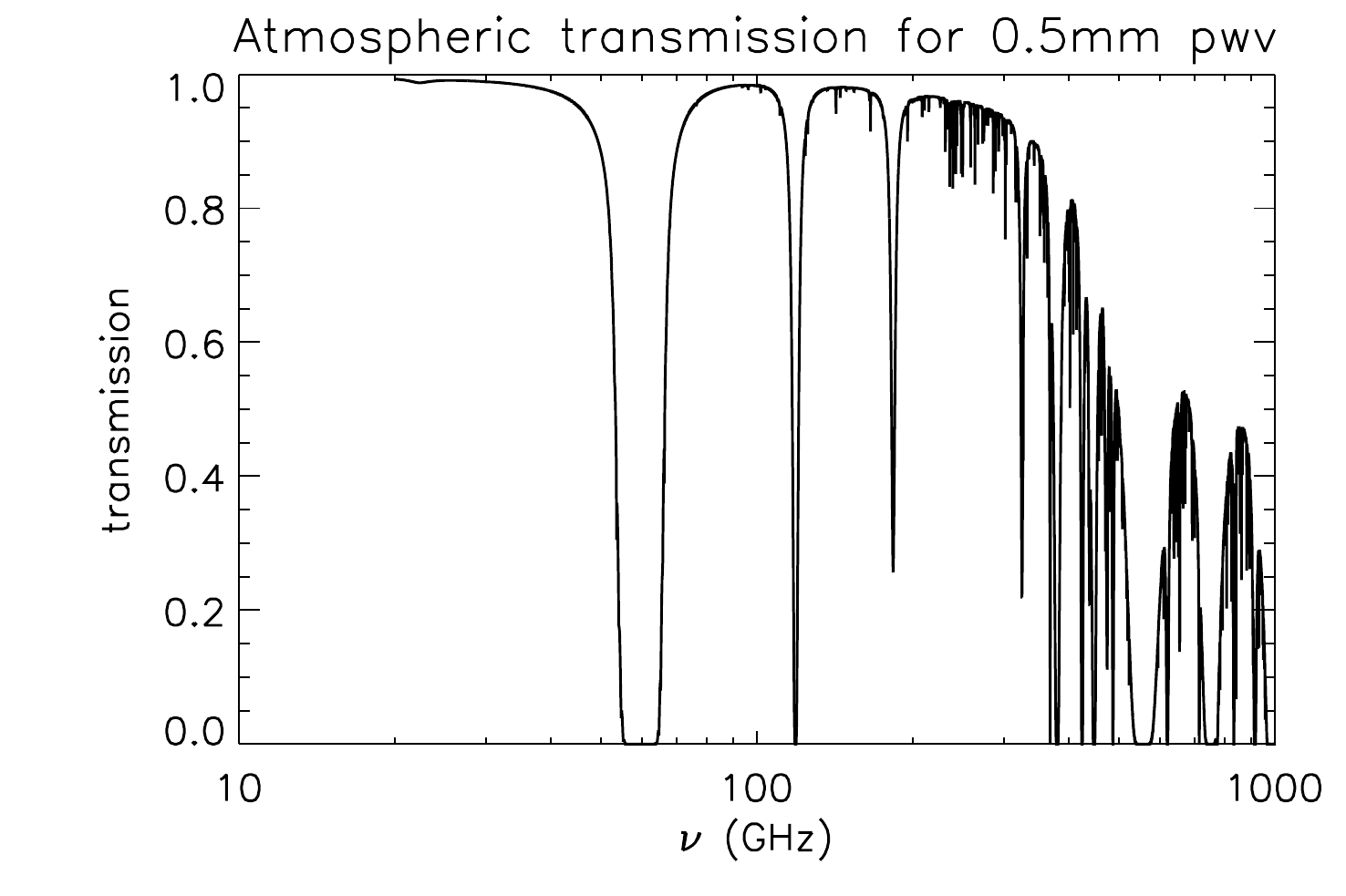}
\hfill
\includegraphics[width=.49\textwidth]{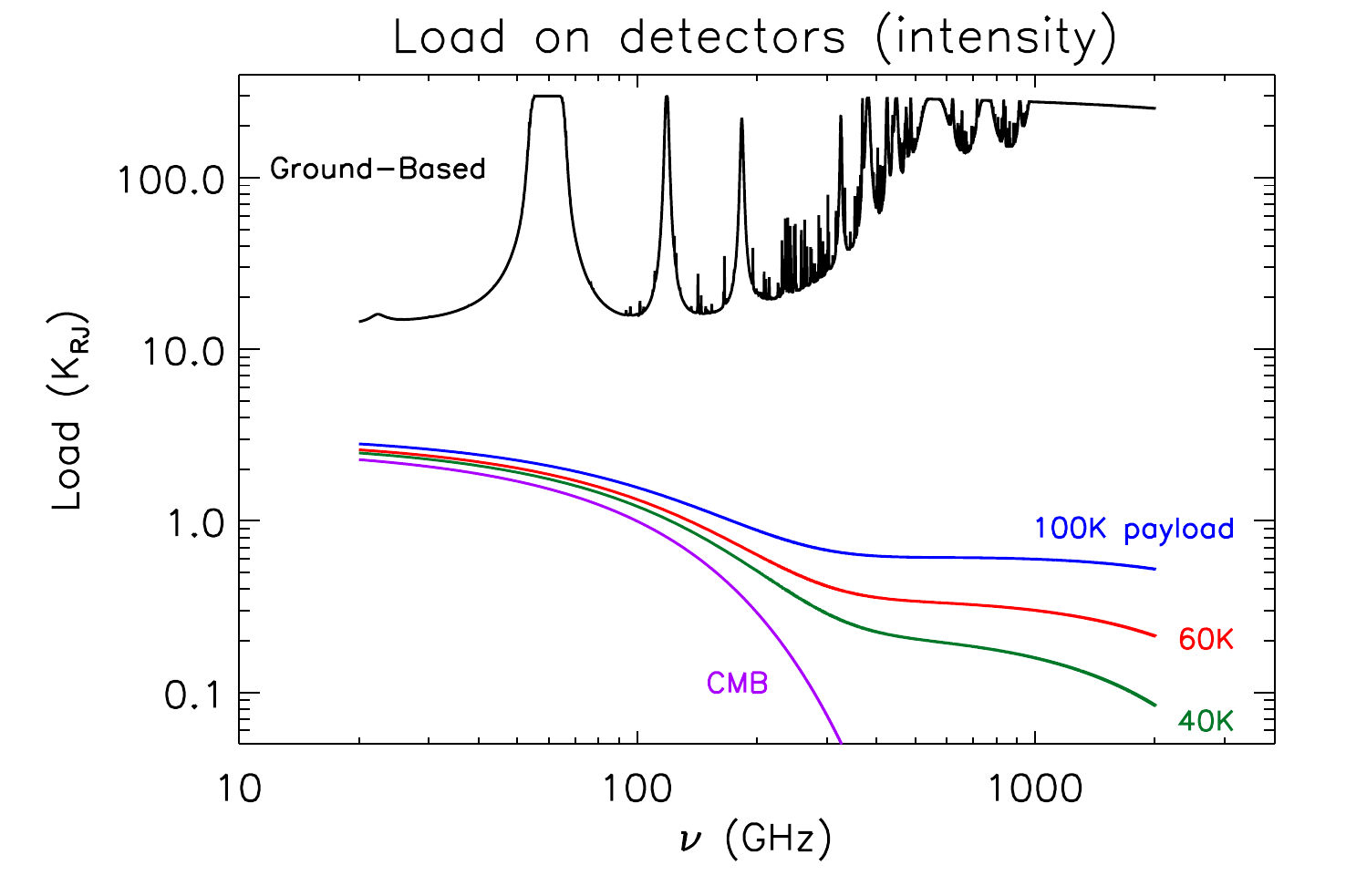}
\hfill
\includegraphics[width=.49\textwidth]{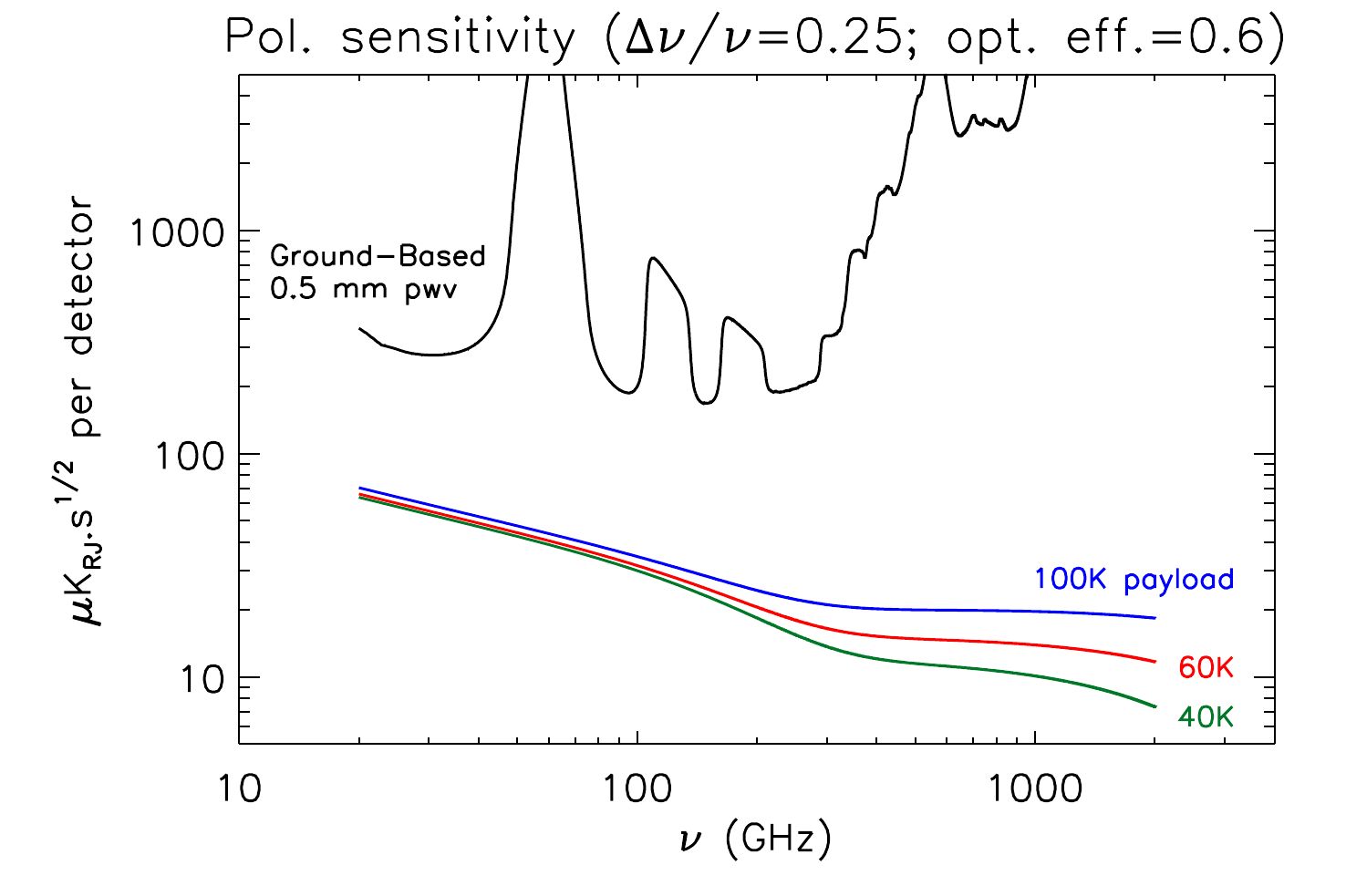}
\hfill
\includegraphics[width=.49\textwidth]{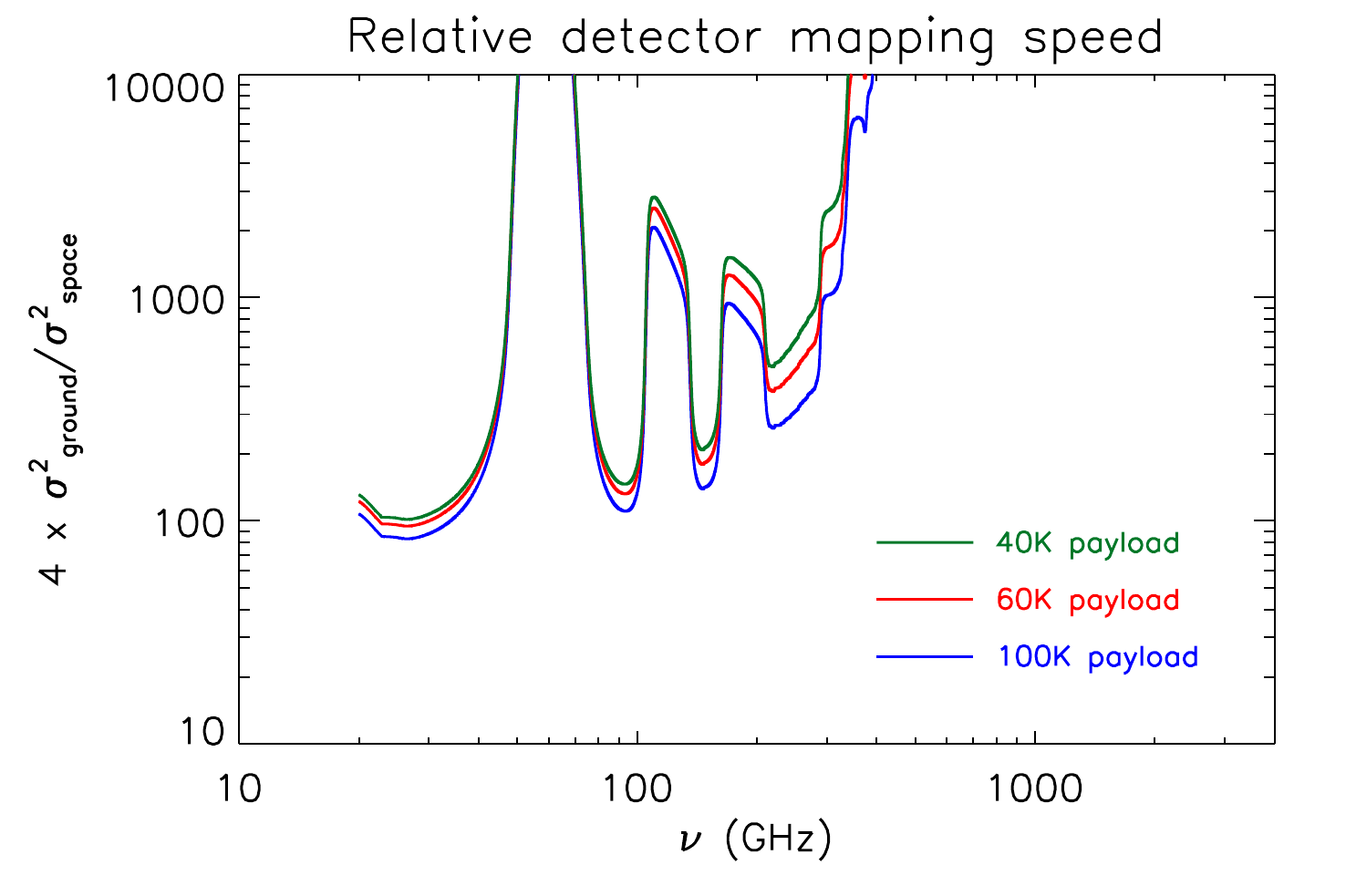}
\caption{\label{fig:det_sensitivity} {\it Top left}: Typical
  atmospheric transmission from the Atacama plateau at $60^\circ$
  elevation, for an average of half a millimetre of integrated
  precipitable water vapour. {\it Top right}: Load on a
  detector for a ground-based instrument (black) and for a space-borne
  instrument with various payload temperatures. In the ground-based
  case, we assume 3\% total emissivity for the optics (telescope
  reflectors and entrance window), and assume that all the environment
  is at $290\,$K, while for the space mission we assume a telescope
  emissivity similar to that of \Planck, and 0.5\% stray light from a
  black payload. {\it Bottom left}: Corresponding detector sensitivity
  (noise level of a single polarised detector calibrated in intensity)
  as a function of band central frequency; we assume square bands with
  $\Delta\nu/\nu=0.25$, an optical efficiency of 60\%, and consider
  single-moded detectors with a throughput of $\lambda^2$. {\it Bottom
    right}: Relative mapping speed of one single space-borne versus
  one single ground-based detector with the same assumptions, also
  considering an observing efficiency of 25\% from the ground as
  compared to space.}
\end{figure}

Figure~\ref{fig:det_sensitivity} compares the background load and
noise level for a detector in the focal plane of a 40-K, 60-K, or
100-K space-borne telescope with emissivity similar to that of
\Planck, with those of a ground-based detector with 3\% emissive
optics (total emissivity for the telescope and the cryostat window if
any), and 0.5\,mm of precipitable water vapour, observing at
$60^\circ$ elevation from the Atacama plateau. The emissivity of the
space telescope is modelled following the measured performance of the
\Planck\ reflectors \citep{2010A&A...520A...2T}.
We assume square frequency bands with 25\% bandwidth, 60\% optical
efficiency, and assume that each detector integrates incoming
radiation over a throughput of $\lambda^2$, where $\lambda$ is the
central wavelength of the band. We also assume that the total noise is
$\sqrt{2}$ times the photon noise, i.e., intrinsic detector noise at
the same level as the photon noise, the two adding-up in quadrature.
The mapping speed comparison assumes an illustrative time efficiency
from the ground of 25\% (due to maintenance, calibration, and discarding
of data taken in bad weather conditions). For instance, BICEP2 at the south pole had a
total observing efficiency of about 30\% \citep{2014ApJ...792...62B}
in the period extending from February~2010 to November~2012, while
that of other ground-based experiments has typically been lower. By
comparison, the observing efficiency in space is expected to be close
to 100\%; the \Planck-HFI data loss due to glitches ranges from 6\% to
20\% depending on the detector \citep{2014A&A...571A..10P}. Note,
however, that such a good observing efficiency requires a continuous
sub-kelvin cooler.


Figure~\ref{fig:det_sensitivity} shows that, within the atmospheric
windows, a single space-borne detector can reach a sensitivity
equivalent to 100--200 ground-based detectors (depending on frequency,
and ignoring fluctuations of atmospheric emission, ground pickup and
other systematics, which further degrade the performance of
ground-based observations relatively to space). Outside of the
atmospheric windows, i.e., close to the main O$_2$ and H$_2$O lines,
and above 300$\,$GHz, ground-based observations are extremely
challenging, and a space mission seems the only option to observe
large patches of sky with good sensitivity and with an angular
resolution of a few arcminutes or better (i.e., without resorting to
massively multi-moded observations).

The sensitivity of space-borne detectors weakly depends on the
temperature of the payload. For instance, at $\nu=150\,$GHz, the
respective noise levels of a detector in the conditions described
above are 49, 43, and 40$\, \mu$K$_{\rm CMB}.\sqrt{{\rm s}}$,
respectively. This estimate is slightly better than (but generally in
good agreement with) the observed sensitivity of \Planck\ $143\,$GHz
polarisation-sensitive detectors, which range from 50 to 53$\,
\mu$K$_{\rm CMB}.\sqrt{{\rm s}}$ (except for one, which is at the level of
59$\, \mu$K$_{\rm CMB}.\sqrt{{\rm s}}$ \citep{2011A&A...536A...4P}).

\subsection{Required observing time and focal-plane area}
\label{sec:obs_time_and_area}

\begin{figure}[tbp]
\centering 
\includegraphics[width=.49\textwidth]{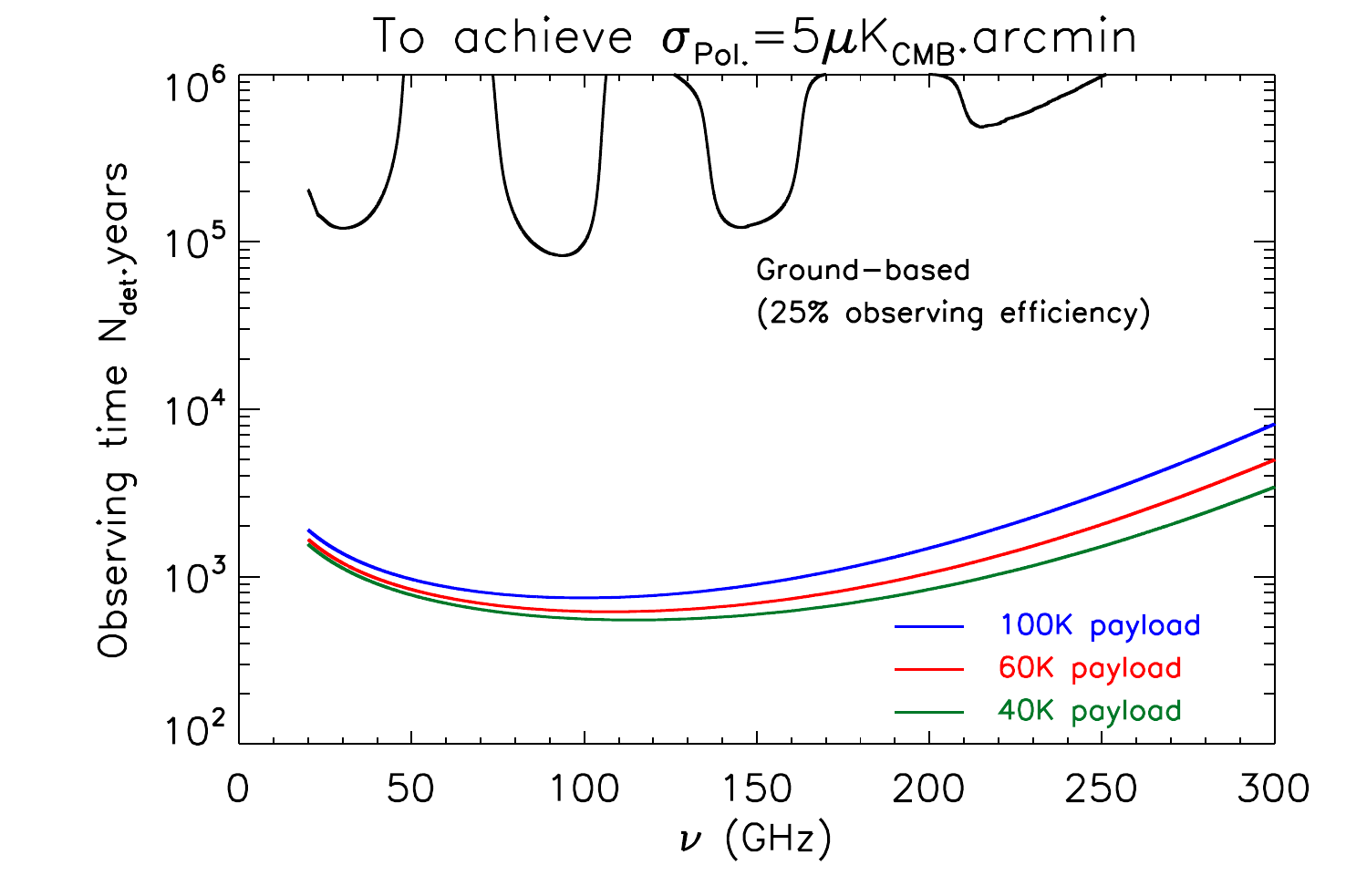}
\hfill
\includegraphics[width=.49\textwidth]{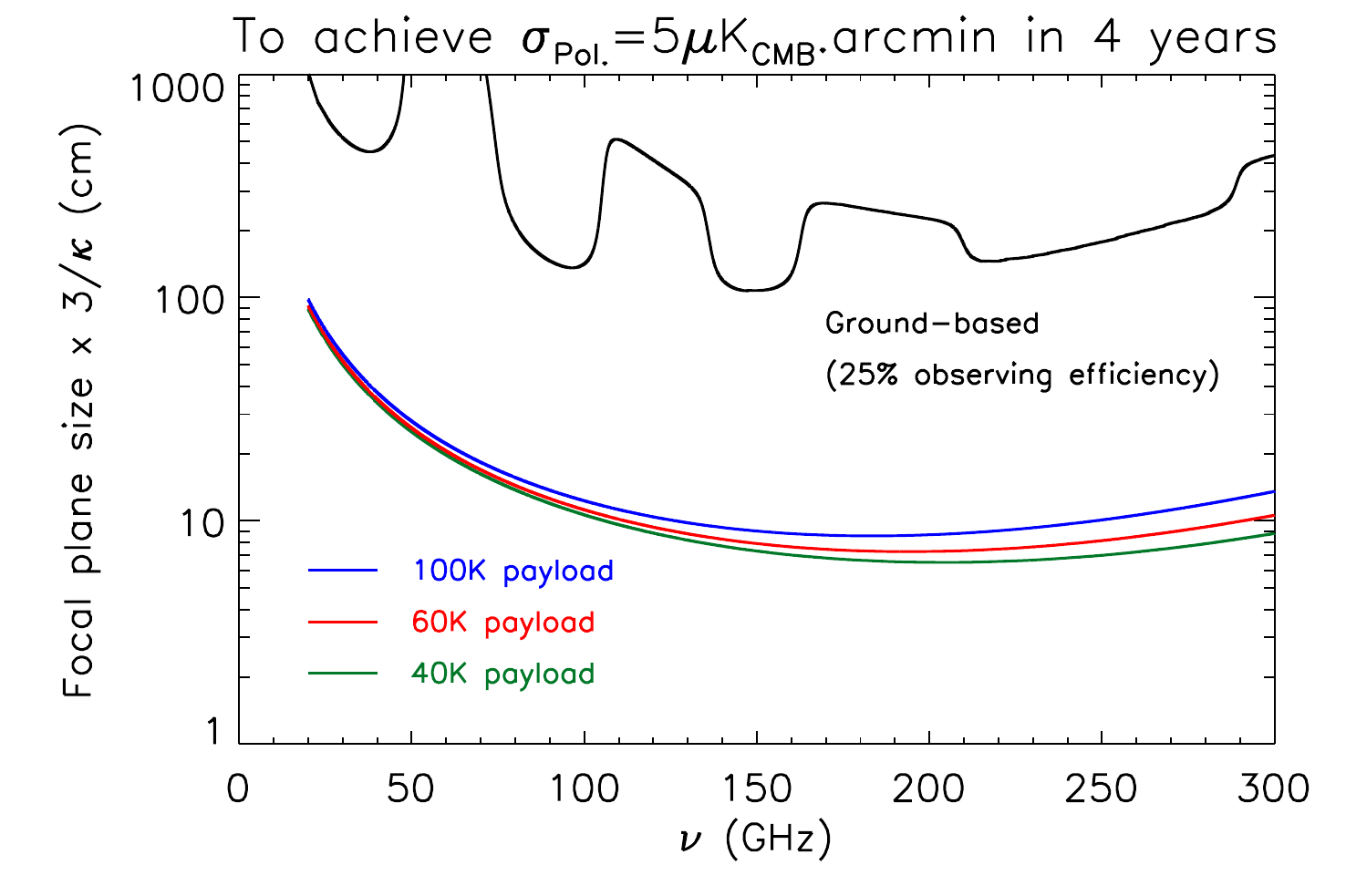}
\caption{\label{fig:cmb-mapping} {\it Left}: Observing time required
  to reach a CMB polarisation sensitivity of 5\,$\mu$K.arcmin (full
  sky equivalent) for a ground-based experiment and for a space
  mission for different assumed payload temperatures. For an observing
  time of 4 years, about 200--300 detectors in the 100--150\,GHz
  frequency range are required from space, while on the ground about 2
  orders of magnitude more detectors are needed. {\it Right}: Typical
  focal-plane size required to reach the same sensitivity, assuming
  here that 9$\lambda^2$ of focal-plane area is required per
  detector.}
\end{figure}

We now assume, for the sake of discussion, that we want to map CMB
polarisation at the level of $\sigma=5\,\mu$K.arcmin over the entire sky
(i.e., a sky fraction $f_{\rm sky}=1$).  Figure~\ref{fig:cmb-mapping}
gives the number of detector-years of observation required to reach
that sensitivity in polarisation, as a function of frequency. The
required order of magnitude is 500--1000 detector-years for a space
mission, and about $10^5$ detector-years on the ground. This number
scales proportionally to $1/\sigma^2$ and $f_{\rm sky}$, so results
can straightforwardly be scaled to any required noise level and any
observed sky fraction.

Figure~\ref{fig:cmb-mapping} also gives an estimate of the focal-plane
area $A$ required to reach this full-sky sensitivity after 4 years of
observation. For a single-mode pixel observing at wavelength
$\lambda$, the focal-plane area required is roughly proportional to
$\lambda^2$, i.e., $A=\kappa^2 \lambda^2$ where $\kappa^2$ is a
filling factor that depends on focal-plane technology and on the
$F$-number of the telescope. Assuming a filling factor of 9
($\kappa=3$) a focal-plane dimension of order $10\,$cm is required in
the focal plane of a space mission, while a focal plane of dimension
around 1\,m is required on the ground to reach about the same
performance. Multi-frequency, dual-polarisation detectors can help
reduce the size of the focal plane (e.g., by a factor $\sqrt{6}$ for
dual-polarisation, trichroic detectors), as long as their optical
efficiency is not worse than that of single frequency, single
polarisation detectors.

From space, the most efficient frequency range (in terms of
focal-plane area required to reach a given CMB sensitivity) is between
170 and 200$\,$GHz, depending somewhat on the payload temperature (see
right panel of figure~\ref{fig:cmb-mapping}). From the ground, the
best observing frequency (from a sensitivity per focal-plane area
point of view) is 150$\,$GHz.

\subsection{Atmospheric emission fluctuations}
\label{appendix:atmospheric-fluctuations}

In addition to emitting photons that contribute to the total photon
noise of the observations, atmospheric emission varies in time and
over the sky. These variations are mostly due to inhomogeneities of
the precipitable water vapour (clouds) and of the temperature. For an
atmosphere at around 300$\,$K with 2\% emissivity, the total emission
is of order 6$\,$K$_{\rm RJ}$. Temperature inhomogeneities of 0.1\%,
for instance, generate fluctuations of the signal of the order of
6$\,$mK$_{\rm RJ}$. Such fluctuations on the timescale of 1\,s are
more than an order of magnitude larger than the 150-GHz detector
white noise in typical observing conditions from the Atacama plateau,
shown in the bottom left panel of figure~\ref{fig:det_sensitivity},
and hence are the dominant source of low-frequency noise in the data
timestreams. In addition, since such inhomogeneities are correlated over
patches of several degrees, the large-scale noise they generate is
correlated between the detectors in the focal plane of a single
instrument, and does not average down as $\sqrt{N_{\rm det}}$. 
However, atmospheric emission is only very weakly
polarised. A polarisation modulator, such as a continuously rotating
half-wave plate (HWP), typically reduces these fluctuations of
atmospheric emission by a factor of around $10^3$, so they impact
mostly the measurement of temperature anisotropies, and much less
polarisation. However, because $B$ modes are more than 3 orders of
magnitude below temperature anisotropies, the atmospheric noise cannot
be ignored for CMB polarisation measurements on large scales.

From the ground, observations at frequencies up to 300\,GHz are
theoretically possible from sites with very low precipitable water
vapour and stable observing conditions, such as Antarctica or the
Atacama plateau. However, even from such excellent observing sites,
broad-band observations at $\nu > 300\,$GHz are very challenging in
practice, by reason of these fluctuations of atmospheric emission. Since
stratospheric balloons, which can avoid the noise excess due to the
atmosphere, are limited to short observing times (e.g., from a few
days typically to a few weeks for ultra-long duration ballooning), a
space mission is the only viable option to achieve low-noise,
few arcminute-angular resolution, large sky area observations at
sub-millimetre wavelengths.

\section{Scan strategy optimisation}
\label{appendix:scanning}

\subsection{Main requirements and design drivers}

We assume a scanning strategy such as that described in
section~\ref{sec:scanning}, in which the satellite is spun around its
main symmetry axis with a period $T_{\rm spin}$. The spin axis
precesses around the anti-solar direction with a period $T_{\rm prec}$.
The precession angle is $\alpha$, while the line of sight (LOS) is offset
from the principal axis of symmetry by an angle $\beta$. Hence, while
the satellite scans the sky, the centre of the field of view (FOV)
scans a near-circle of angular radius $\beta$. In this appendix, we
discuss the optimisation of $\alpha$, $\beta$, $T_{\rm prec}$, and
$T_{\rm spin}$.

\begin{figure}[htbp]
\begin{center}
    \begin{tabular}{lr}
    {\includegraphics[trim=0cm 0cm 0cm 0cm, clip=true,width=0.49\textwidth]{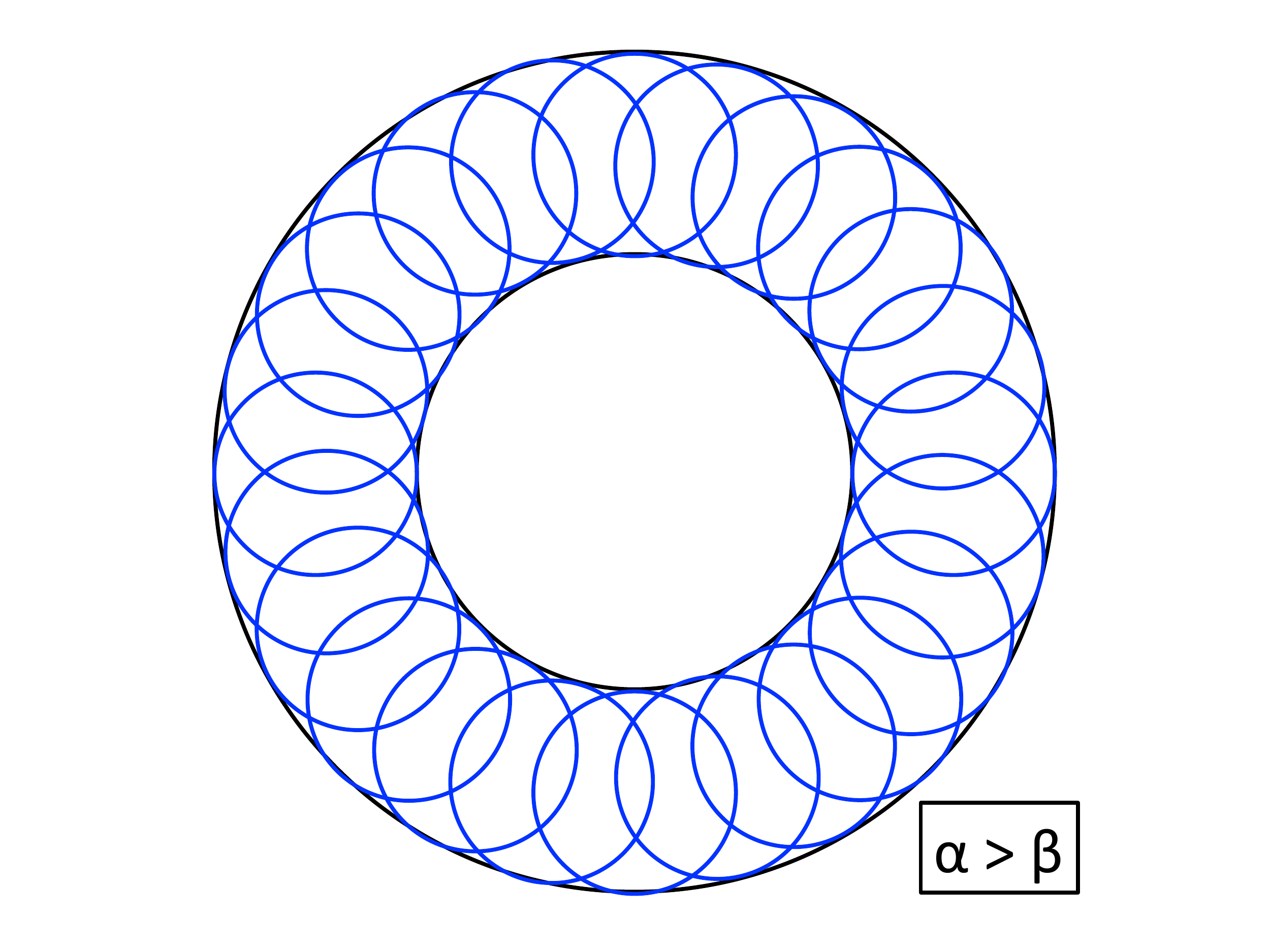}} 
    {\includegraphics[trim=0cm 0cm 0cm 0cm, clip=true,width=0.49\textwidth]{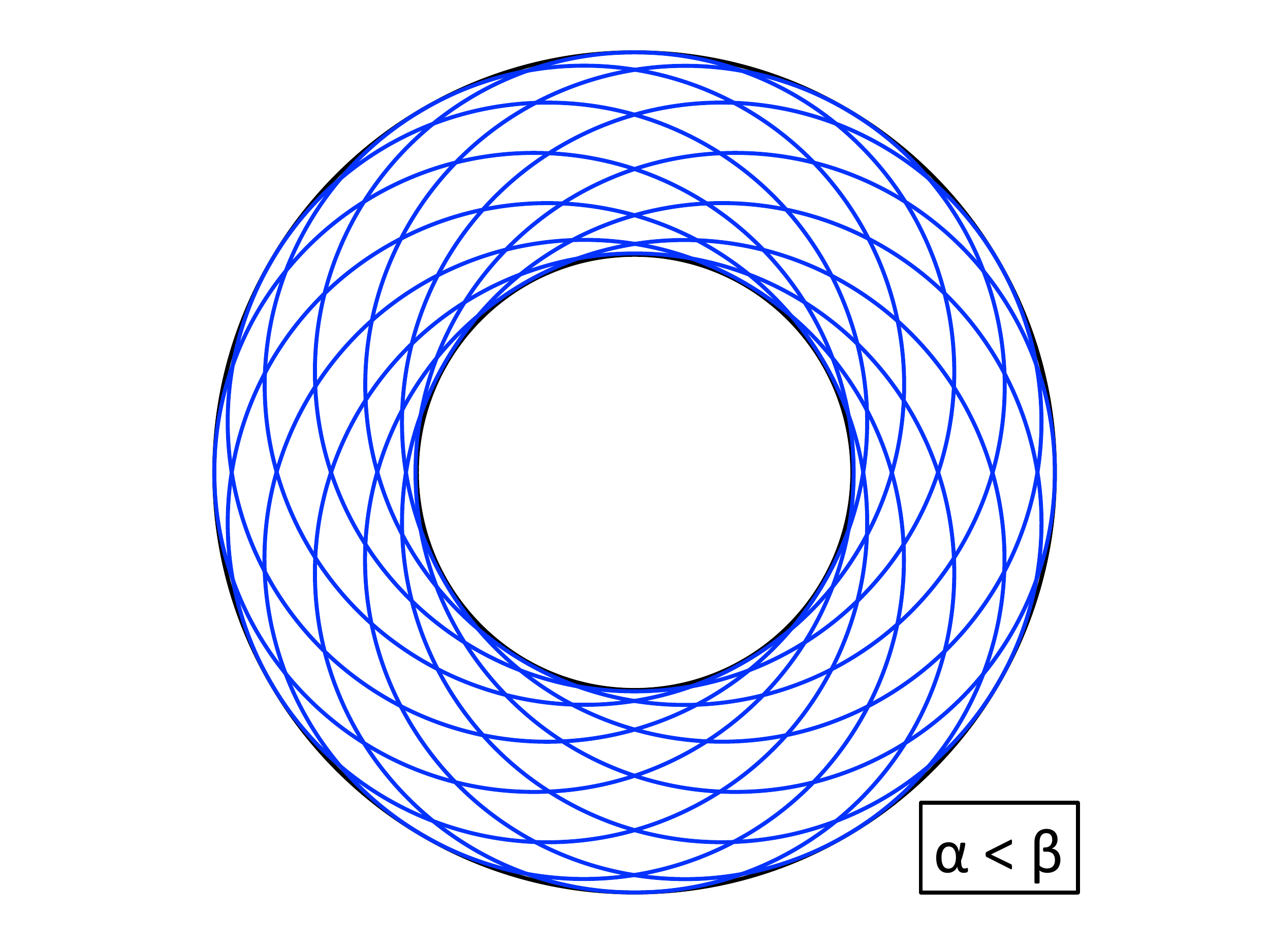}} 
    \end{tabular}
    \caption{ \label{fig:scans} For one complete precession, the FOV
      observes a ring of width $|\alpha+\beta| - |\alpha-\beta|$. {\it
        Left}: Case where $\alpha > \beta$. Small rings cross at large
      angles, for a good coverage of angles over timescales of order
      $\beta$ days, and good distribution of scanning directions for
      most pixels on the sky over the course of the mission. {\it
        Right}: Case where $\alpha < \beta$. Large rings cross at
      small angles, but connect very distant pixels on short periods.}
\end{center}
\end{figure}

Two general regimes of scanning can be considered. In the first one,
where the precession angle $\alpha$ is larger than the spin angle
$\beta$, the trajectory of the FOV covers, over the time of a
precession period, an annulus for which small circles intersect at large
crossing angles. In addition, annuli observed at times separated by
about $\beta$ (in degrees) days generate large-angle trajectory
crossings for most of the observed pixels. For such a scanning
strategy, single detectors observe each pixel at very different angles
that are almost evenly spread in $[0,2\pi]$. Hence, it should be
possible to make single-detector maps with good properties of the
final covariance of the reconstructed $I$, $Q$, $U$ maps: near-nominal
noise for both polarisation Stokes parameters; and low cross-correlation
between the errors in the three maps. In the second regime, where
$\alpha < \beta$, the trajectory of the FOV over one precession period
also covers a ring on the sky, but the trajectories cross at smaller
angles. Large-angle crossings occur only for data sets obtained at
very different times in the mission. It may be necessary to combine
several detectors to make polarisation maps, which requires mitigating
in the data-analysis step both beam asymmetries and possible bandpass
mismatches, i.e., the different response of different detectors to the
various astrophysical components present in the sky.
The scan geometries for the two scanning regimes are sketched in figure~\ref{fig:scans}.

For the Cosmic Origins Explorer version proposed in answer to the ESA
M4 call (\coreplus), the proposed scanning strategy used $\alpha = \beta =
45^\circ$. In practice, it would be desirable to slightly increase
these values (either $\alpha$, or $\beta$, or both), so that all the
detectors in the focal plane observe the complete sky. For the
\coremfive\ version proposed in answer to the M5 call, we set as a
baseline $\alpha=30^\circ$ and $\beta=65^\circ$, so that $\alpha +
\beta = 95^\circ$, which ensures that all the detectors within a focal
plane of angular radius of less than $5^\circ$ do cover the full
sky. We now discuss the rationale for deciding between these options,
and how they impact the global design of the payload and the mission.

\subsection{Practical constraints}

\paragraph{Thermal stability:}
The temperature of the payload depends on the incidence angle of solar
radiation on the \coremfive\ spacecraft. To ensure the thermal stability of
the payload, the precession axis will be maintained anti-solar (and
the payload symmetrical), so that the integrated solar illumination on
the solar panels at the bottom of the service module and on the outer
``V-groove'' screen remains constant.

\paragraph{Sun, Earth and Moon screening:}
The Sun should never shine on the inner V-groove screens nor on the
payload itself. This requirement constrains the precession angle
$\alpha$ to be lower than a limiting value $\alpha_{\rm max}$ set by the
geometry of the shields.
Similarly the Earth and Moon should never shine directly on the detector array.

\paragraph{Making full-sky maps:}
The requirement that a given detector $d$ be able to scan the full sky
imposes, for an anti-solar precession axis,
\begin{equation}
\alpha+\beta_d \geq 90^\circ,
\end{equation}
where $\beta_d = \beta+\Delta\beta_d$ is the offset angle of detector
$d$ with respect to the spin axis.  The angle $\Delta\beta_d$ is set
by the location of detector $d$ in the focal-plane array and by the
optical setup. To make full-sky maps with all detectors, we need
\begin{equation}
\alpha+\beta \geq 90^\circ + \theta_{\rm FP},
\label{eq:req-fullsky}
\end{equation}
where $\theta_{\rm FP}$ is the angular radius of the imprint of the
focal plane on the sky. Full sky coverage for each individual detector
is preferable. For \coremfive, the field of view is $\lsim 5^\circ$ in
radius, so $\alpha+\beta=95^\circ$ is appropriate. For much higher
values of $\alpha+\beta$, the Sun, Earth, and Moon would come closer to
the line of sight, which should be avoided.

\paragraph{Data transfer:}
Downloading the data to Earth requires pointing a steerable antenna
towards the Earth, and hence to cancel the spin and precession of the
spacecraft. Either a phased array such as that of {\it Gaia\/}
\citep{Gaia-antenna-array}, or a mechanical pointing
system 
are options. Possible Earth aspect angle $\theta_{\rm Earth}$ and
scanning speed depends on the specifications of the antenna
(maximum boresight angle and tracking
speed). 
The maximum Earth aspect angle during a precession is:
\begin{equation}
\theta_{\rm Earth} = \alpha + \arctan \left( \frac{r_{\rm orbit}^{\rm max}}{{1.5\times 10^6}\,{\rm km}} \right),
\end{equation}
where $r_{\rm orbit}^{\rm max}$ is the major axis of the \coremfive\
orbit around L2. For example, if we assume $\theta_{\rm Earth}^{\rm
  max}=60^\circ$ and $\alpha=30^\circ$, the maximum orbital radius is
about 860{,}000\,km, compatible with a large Lissajous orbit (see
Table~\ref{tab:orbit-injection}). For $\alpha=45^\circ$ the maximum
orbit radius is 400{,}000\,km (medium Lissajous), and for
$\alpha=50^\circ$ we get a maximum orbit radius of 260{,}000\,km
(small Lissajous). Injection into a large Lissajous orbit requires lower
$\Delta v$ and hence less propellant. This, however, is not a major
driver for the mission, since the Ariane~6.2 launcher is designed to
carry several tonnes (significantly more than the mass of \coremfive)
to an Earth-escape orbit.

\subsection{Sampling}

\paragraph{Co-scan sampling:} 
Denoting as $\theta_\parallel$ the sampling angle along the scan (in
arcminutes), we have the following equation between the sampling
period $T_{\rm sampling}$ and the spin period $T_{\rm spin}$:
\begin{equation}
\frac{(360 \times 60) \sin \beta}{\theta_\parallel} = \frac{T_{\rm spin}}{T_{\rm sampling}}.
\label{eq:co-scan-sampling-1}
\end{equation}
For $N_{\rm s}=4$ samples per beam length along the line of sight,
$\beta=65^\circ$, and $T_{\rm spin}=120\,$s (\coremfive\ proposal
baseline), the total number of samples per second for a detector with
a $4^\prime$ beam is 199, corresponding to a sampling period of about
$5\,$ms.

\paragraph{Cross-scan sampling:}
For each precession, the trajectory of the spin axis has an angular
length of about $(360 \times 60)\sin \alpha$ arcmin. Denoting as
$\theta_\perp$ the ``cross-scan sampling angle,'' i.e., the maximum
distance (in arcminutes) between two consecutive scan paths, as
measured along the trajectory of the spin axis, we have
\begin{equation}
\frac{(360 \times 60) \sin \alpha}{\theta_\perp} = \frac{T_{\rm prec}}{T_{\rm spin}} 
\label{eq:Tprec-Tspin-for-Xscan}
\end{equation}
and a cross-scan sampling step of
\begin{equation}
\left ( \frac{\theta_\perp}{1^\prime} \right ) = 3.75 \, \sin \alpha \, \left( \frac{T_{\rm spin}}{60 \, {\rm s}} \right) \, \left (\frac{96 \, {\rm hr}}{T_{\rm prec}} \right ) .
\label{eq:value-Tspin}
\end{equation}
Ideally, this cross-scan sampling step should also be $\simeq 1/4$ of the beam size.
For the baseline parameters proposed in response to the M5 call ($T_{\rm
  spin}=120\,$s, $T_{\rm prec}=4\,$day, and $\alpha=30^\circ$) we obtain
$\theta_\perp = 3.75^\prime$, which is not quite good enough for the highest 
frequency channels, calling for an increase of the spin rate (1 RPM would be better).
We note however that for most of the sky covered
by one precession the cross-scan sampling is actually smaller than
this maximum value (which is what we obtain for pixels on the trajectory
of the spin axis, along the circle of radius $\alpha=30^\circ$ centred
on the precession axis). 

\paragraph{Sky area covered during one precession:}
The sky area covered by one detector during one precession period is
the area of the sphere located between colatitudes $| \alpha-\beta |$
and $(\alpha+\beta)$. The sky fraction $f_{\rm sky}$ is
\begin{equation}
f_{\rm sky} = \frac{1}{2} \left ( \cos | \alpha \! - \! \beta | - \cos| \alpha \! + \! \beta | \right ).
\label{eq:fsky-one-precession}
\end{equation}
For $\alpha = 30^\circ$ and $\beta = 65^\circ$ (proposed baseline),
45\% of the sky is covered for each precession period.

\paragraph{Data samples per precession:}
For each precession period, the number of data points for a single
detector is
\begin{equation}
N_{\rm pts/prec.} \simeq 4.67 \times 10^{8} \, \sin \alpha \, \sin \beta \, \left ( \frac{1^\prime}{\theta_{\parallel}} \right ) \left ( \frac{1^\prime}{\theta_{\perp}} \right ).
\end{equation}
The average density of data points per square arcminute for one single
precession period is
\begin{equation}
N_{\rm pts/arcmin^2} \simeq 6.28 \, \left ( \frac{ \sin \alpha \, \sin \beta}{\cos | \alpha \! - \! \beta | - \cos| \alpha \! + \! \beta |} \right ) \left ( \frac{1^\prime}{\theta_{\parallel}} \right ) \left ( \frac{1^\prime}{\theta_{\perp}} \right ).
\end{equation}

\begin{figure}[htbp]
\begin{center}
    \includegraphics[trim=0cm 5cm 0cm 4cm, clip=true,width=\textwidth]{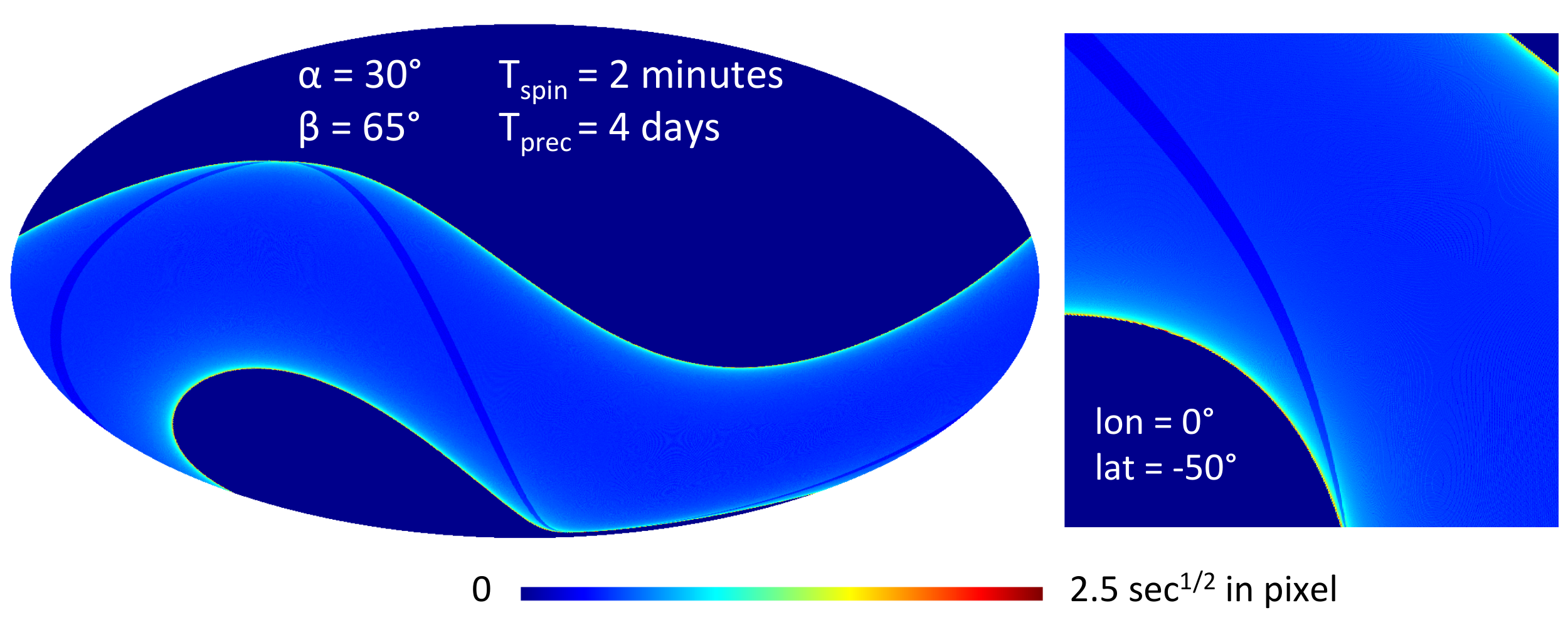}
    \includegraphics[trim=0cm 5cm 0cm 4cm, clip=true,width=\textwidth]{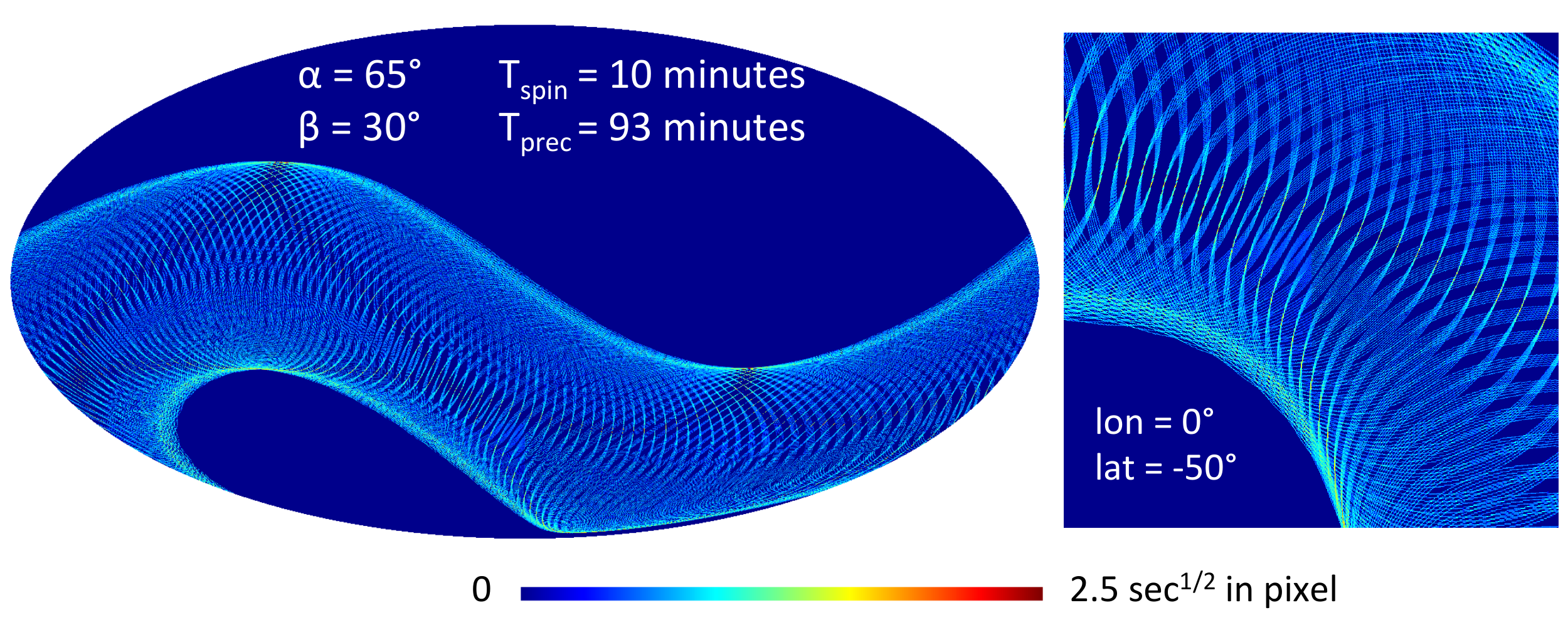}
    \caption{Sky coverage (in units of square root of time in {\tt HEALPix}
      map pixels for $N_{\rm side}=512$) for a single detector after 4 days
      of continuous observation for two choices of parameters for the
      scanning strategy. {\it Top}: Case A, \coremfive\ baseline parameters,
      with precession angle $\alpha=30^\circ$, scanning angle
      $\beta=65^\circ$, spin period 2 minutes, and precession period 4
      days. {\it Bottom}: Case B, LiteBIRD-like scan strategy, with
      precession angle $\alpha=65^\circ$, scanning angle
      $\beta=30^\circ$, spin period 10 minutes, and precession period 93
      minutes. Square maps on the right show a detail in the gnomonic
      projection centred at Galactic coordinates
      $(0^\circ,-50^\circ)$.}
    \label{fig:4day_coverage}
\end{center}
\end{figure}

\subsection{Optimisation}

The optimisation of the scanning strategy must be a compromise
between conflicting requirements. We consider as a starting point two
concrete examples.  The first case that is considered is the
\coremfive\ baseline (Case A: $\alpha=30^\circ$, $\beta=65^\circ$,
$T_{\rm spin}=120$\,s, $T_{\rm prec}=4$\,days), and corresponds to
scanning according to the pattern in the right panel of
figure~\ref{fig:scans}, while the second case (Case B:
$\alpha=65^\circ$, $\beta=30^\circ$, $T_{\rm spin}=600$\,s, $T_{\rm
  prec}=93$\,minutes) is representative of an option that is
considered for LiteBIRD, and corresponds to scanning according to the
left panel of figure~\ref{fig:scans}. As discussed in
Ref.~\citep{2017MNRAS.466..425W}, scan strategy B is nearly ideal in terms of
distribution of scan angles on the sky for single-detector map making.

According to eq.~(\ref{eq:fsky-one-precession}), and as illustrated
in figure~\ref{fig:scans}, both sets of angles allow the mission to
probe the same sky area over one precession period.
The sky-coverage as a function of time, however, also depends on the
choice that is made for $T_{\rm prec}/T_{\rm spin}$. In the present
example, scan strategy A fulfills the requirement to obtain a cross-scan
sampling angle small enough compared to the pixel size, while scan
strategy B does not.

Figure~\ref{fig:4day_coverage} shows the sky coverage for one single
detector after 4 days of observation for both scan strategies, in units of
$\sqrt{{\rm s}}$ per pixel in a {\tt HEALPix} $N_{\rm side}=512$ map in
Galactic coordinates. In scan strategy B, large gaps are left between the
scans; although they gradually fill up as precessions are accumulated,
the distribution of observing time is very inhomogeneous, as clearly
seen in the bottom panel of figure~\ref{fig:4day_coverage}.

To match the \coremfive\ cross-scan sampling of $3.75^\prime$ when
$\alpha=65^\circ$ and $\beta=30^\circ$, we need $T_{\rm prec}/T_{\rm
  spin} \simeq 5220$, which for $T_{\rm spin} = 10$~minutes results in
$T_{\rm prec} = 36$~days. If we require a precession period of 4 days,
the spin period must be $\simeq 1.1$\,minutes instead of 10.

In principle, it is also possible to adjust the ratio of $T_{\rm
  prec}/T_{\rm spin}$ so that the gaps between the scans fill up
optimally as precessions are being accumulated. Still, the total
number of spin periods needed to fill in the wide ring is no less than
36 days if $T_{\rm spin}=10$~minutes. This is not optimal for
cross-comparison of maps obtained at different times during the course
of the mission. In addition, in 36 days the precession axis moves by about
$35.5^\circ$ to follow the yearly motion of the Sun, leaving gaps in the 
sky coverage. To recover proper cross-scan sampling, $T_{\rm spin}$
must be reduced. We note however that the cross-scan sampling 
requirement is less stringent for LiteBIRD than for \coremfive\ because 
of its coarser angular resolution.

Optimally, for \coremfive, we may wish to retain the angles used in scan strategy B, but modify
$T_{\rm spin}$ and $T_{\rm prec}$. However, as discussed in
section~\ref{sec:payload-geometry}, $\alpha=65^\circ$ requires
additional solar panels and re-defining of the payload V-grooves, and
$T_{\rm spin}=1.1$\,min is more demanding on the attitude control
system. The parameters chosen for \coremfive\ are the result of a compromise 
between these different constraints.

\bibliographystyle{plainnat}
\bibliography{biblio}{}

\begin{thebibliography}{98}
\providecommand{\natexlab}[1]{#1}
\providecommand{\url}[1]{\texttt{#1}}
\expandafter\ifx\csname urlstyle\endcsname\relax
  \providecommand{\doi}[1]{doi: #1}\else
  \providecommand{\doi}{doi: \begingroup \urlstyle{rm}\Url}\fi

\bibitem[{Abazajian} et~al.(2016){Abazajian}, {Adshead}, {Ahmed}, {Allen},
  {Alonso}, {Arnold}, {Baccigalupi}, {Bartlett}, {Battaglia}, {Benson},
  {Bischoff}, {Borrill}, {Buza}, {Calabrese}, {Caldwell}, {Carlstrom}, {Chang},
  {Crawford}, {Cyr-Racine}, {De Bernardis}, {de Haan}, {di Serego Alighieri},
  {Dunkley}, {Dvorkin}, {Errard}, {Fabbian}, {Feeney}, {Ferraro}, {Filippini},
  {Flauger}, {Fuller}, {Gluscevic}, {Green}, {Grin}, {Grohs}, {Henning},
  {Hill}, {Hlozek}, {Holder}, {Holzapfel}, {Hu}, {Huffenberger}, {Keskitalo},
  {Knox}, {Kosowsky}, {Kovac}, {Kovetz}, {Kuo}, {Kusaka}, {Le Jeune}, {Lee},
  {Lilley}, {Loverde}, {Madhavacheril}, {Mantz}, {Marsh}, {McMahon},
  {Meerburg}, {Meyers}, {Miller}, {Munoz}, {Nguyen}, {Niemack}, {Peloso},
  {Peloton}, {Pogosian}, {Pryke}, {Raveri}, {Reichardt}, {Rocha}, {Rotti},
  {Schaan}, {Schmittfull}, {Scott}, {Sehgal}, {Shandera}, {Sherwin}, {Smith},
  {Sorbo}, {Starkman}, {Story}, {van Engelen}, {Vieira}, {Watson}, {Whitehorn},
  and {Kimmy Wu}]{2016arXiv161002743A}
K.~N. {Abazajian}, P.~{Adshead}, Z.~{Ahmed}, S.~W. {Allen}, D.~{Alonso}, K.~S.
  {Arnold}, C.~{Baccigalupi}, J.~G. {Bartlett}, N.~{Battaglia}, B.~A. {Benson},
  C.~A. {Bischoff}, J.~{Borrill}, V.~{Buza}, E.~{Calabrese}, R.~{Caldwell},
  J.~E. {Carlstrom}, C.~L. {Chang}, T.~M. {Crawford}, F.-Y. {Cyr-Racine},
  F.~{De Bernardis}, T.~{de Haan}, S.~{di Serego Alighieri}, J.~{Dunkley},
  C.~{Dvorkin}, J.~{Errard}, G.~{Fabbian}, S.~{Feeney}, S.~{Ferraro}, J.~P.
  {Filippini}, R.~{Flauger}, G.~M. {Fuller}, V.~{Gluscevic}, D.~{Green},
  D.~{Grin}, E.~{Grohs}, J.~W. {Henning}, J.~C. {Hill}, R.~{Hlozek},
  G.~{Holder}, W.~{Holzapfel}, W.~{Hu}, K.~M. {Huffenberger}, R.~{Keskitalo},
  L.~{Knox}, A.~{Kosowsky}, J.~{Kovac}, E.~D. {Kovetz}, C.-L. {Kuo},
  A.~{Kusaka}, M.~{Le Jeune}, A.~T. {Lee}, M.~{Lilley}, M.~{Loverde}, M.~S.
  {Madhavacheril}, A.~{Mantz}, D.~J.~E. {Marsh}, J.~{McMahon}, P.~D.
  {Meerburg}, J.~{Meyers}, A.~D. {Miller}, J.~B. {Munoz}, H.~N. {Nguyen}, M.~D.
  {Niemack}, M.~{Peloso}, J.~{Peloton}, L.~{Pogosian}, C.~{Pryke}, M.~{Raveri},
  C.~L. {Reichardt}, G.~{Rocha}, A.~{Rotti}, E.~{Schaan}, M.~M. {Schmittfull},
  D.~{Scott}, N.~{Sehgal}, S.~{Shandera}, B.~D. {Sherwin}, T.~L. {Smith},
  L.~{Sorbo}, G.~D. {Starkman}, K.~T. {Story}, A.~{van Engelen}, J.~D.
  {Vieira}, S.~{Watson}, N.~{Whitehorn}, and W.~L. {Kimmy Wu}.
\newblock {CMB-S4 Science Book, First Edition}.
\newblock \emph{ArXiv e-prints}, October 2016.

\bibitem[{Addison} et~al.(2016){Addison}, {Huang}, {Watts}, {Bennett},
  {Halpern}, {Hinshaw}, and {Weiland}]{2016ApJ...818..132A}
G.~E. {Addison}, Y.~{Huang}, D.~J. {Watts}, C.~L. {Bennett}, M.~{Halpern},
  G.~{Hinshaw}, and J.~L. {Weiland}.
\newblock {Quantifying Discordance in the 2015 Planck CMB Spectrum}.
\newblock \emph{\apj}, 818:\penalty0 132, February 2016.
\newblock \doi{10.3847/0004-637X/818/2/132}.

\bibitem[Allica et~al.(2010)Allica, Alonso, Amado, Baz‡n, Casares, de~Witte,
  Garc'a, McConnell, Montesano, Santilario, and Serrano]{Gaia-antenna-array}
J.~C. Allica, E.~Alonso, M.~Amado, A.~Baz‡n, F.~Casares, E.~de~Witte,
  Q.~Garc'a, T.~McConnell, A.~Montesano, J.~Santilario, and J.~L. Serrano.
\newblock Architecture of gaia satellite phased array antenna.
\newblock In \emph{Proceedings of the Fourth European Conference on Antennas
  and Propagation}, pages 1--4, April 2010.

\bibitem[{Andr{\'e}} et~al.(2014){Andr{\'e}}, {Baccigalupi}, {Banday},
  {Barbosa}, {Barreiro}, {Bartlett}, {Bartolo}, {Battistelli}, {Battye},
  {Bendo}, {Beno{\^i}t}, {Bernard}, {Bersanelli}, {B{\'e}thermin}, {Bielewicz},
  {Bonaldi}, {Bouchet}, {Boulanger}, {Brand}, {Bucher}, {Burigana}, {Cai},
  {Camus}, {Casas}, {Casasola}, {Castex}, {Challinor}, {Chluba}, {Chon},
  {Colafrancesco}, {Comis}, {Cuttaia}, {D'Alessandro}, {Da Silva}, {Davis}, {de
  Avillez}, {de Bernardis}, {de Petris}, {de Rosa}, {de Zotti}, {Delabrouille},
  {D{\'e}sert}, {Dickinson}, {Diego}, {Dunkley}, {En{\ss}lin}, {Errard},
  {Falgarone}, {Ferreira}, {Ferri{\`e}re}, {Finelli}, {Fletcher}, {Fosalba},
  {Fuller}, {Galli}, {Ganga}, {Garc{\'{\i}}a-Bellido}, {Ghribi}, {Giard},
  {Giraud-H{\'e}raud}, {Gonzalez-Nuevo}, {Grainge}, {Gruppuso}, {Hall},
  {Hamilton}, {Haverkorn}, {Hernandez-Monteagudo}, {Herranz}, {Jackson},
  {Jaffe}, {Khatri}, {Kunz}, {Lamagna}, {Lattanzi}, {Leahy}, {Lesgourgues},
  {Liguori}, {Liuzzo}, {Lopez-Caniego}, {Macias-Perez}, {Maffei}, {Maino},
  {Mangilli}, {Martinez-Gonzalez}, {Martins}, {Masi}, {Massardi}, {Matarrese},
  {Melchiorri}, {Melin}, {Mennella}, {Mignano}, {Miville-Desch{\^e}nes},
  {Monfardini}, {Murphy}, {Naselsky}, {Nati}, {Natoli}, {Negrello}, {Noviello},
  {O'Sullivan}, {Paci}, {Pagano}, {Paladino}, {Palanque-Delabrouille},
  {Paoletti}, {Peiris}, {Perrotta}, {Piacentini}, {Piat}, {Piccirillo},
  {Pisano}, {Polenta}, {Pollo}, {Ponthieu}, {Remazeilles}, {Ricciardi},
  {Roman}, {Rosset}, {Rubino-Martin}, {Salatino}, {Schillaci}, {Shellard},
  {Silk}, {Starobinsky}, {Stompor}, {Sunyaev}, {Tartari}, {Terenzi},
  {Toffolatti}, {Tomasi}, {Trappe}, {Tristram}, {Trombetti}, {Tucci}, {Van de
  Weijgaert}, {Van Tent}, {Verde}, {Vielva}, {Wandelt}, {Watson}, and
  {Withington}]{2014JCAP...02..006A}
P.~{Andr{\'e}}, C.~{Baccigalupi}, A.~{Banday}, D.~{Barbosa}, B.~{Barreiro},
  J.~{Bartlett}, N.~{Bartolo}, E.~{Battistelli}, R.~{Battye}, G.~{Bendo},
  A.~{Beno{\^i}t}, J.-P. {Bernard}, M.~{Bersanelli}, M.~{B{\'e}thermin},
  P.~{Bielewicz}, A.~{Bonaldi}, F.~{Bouchet}, F.~{Boulanger}, J.~{Brand},
  M.~{Bucher}, C.~{Burigana}, Z.-Y. {Cai}, P.~{Camus}, F.~{Casas},
  V.~{Casasola}, G.~{Castex}, A.~{Challinor}, J.~{Chluba}, G.~{Chon},
  S.~{Colafrancesco}, B.~{Comis}, F.~{Cuttaia}, G.~{D'Alessandro}, A.~{Da
  Silva}, R.~{Davis}, M.~{de Avillez}, P.~{de Bernardis}, M.~{de Petris},
  A.~{de Rosa}, G.~{de Zotti}, J.~{Delabrouille}, F.-X. {D{\'e}sert},
  C.~{Dickinson}, J.~M. {Diego}, J.~{Dunkley}, T.~{En{\ss}lin}, J.~{Errard},
  E.~{Falgarone}, P.~{Ferreira}, K.~{Ferri{\`e}re}, F.~{Finelli},
  A.~{Fletcher}, P.~{Fosalba}, G.~{Fuller}, S.~{Galli}, K.~{Ganga},
  J.~{Garc{\'{\i}}a-Bellido}, A.~{Ghribi}, M.~{Giard}, Y.~{Giraud-H{\'e}raud},
  J.~{Gonzalez-Nuevo}, K.~{Grainge}, A.~{Gruppuso}, A.~{Hall}, J.-C.
  {Hamilton}, M.~{Haverkorn}, C.~{Hernandez-Monteagudo}, D.~{Herranz},
  M.~{Jackson}, A.~{Jaffe}, R.~{Khatri}, M.~{Kunz}, L.~{Lamagna},
  M.~{Lattanzi}, P.~{Leahy}, J.~{Lesgourgues}, M.~{Liguori}, E.~{Liuzzo},
  M.~{Lopez-Caniego}, J.~{Macias-Perez}, B.~{Maffei}, D.~{Maino},
  A.~{Mangilli}, E.~{Martinez-Gonzalez}, C.~J.~A.~P. {Martins}, S.~{Masi},
  M.~{Massardi}, S.~{Matarrese}, A.~{Melchiorri}, J.-B. {Melin}, A.~{Mennella},
  A.~{Mignano}, M.-A. {Miville-Desch{\^e}nes}, A.~{Monfardini}, A.~{Murphy},
  P.~{Naselsky}, F.~{Nati}, P.~{Natoli}, M.~{Negrello}, F.~{Noviello},
  C.~{O'Sullivan}, F.~{Paci}, L.~{Pagano}, R.~{Paladino},
  N.~{Palanque-Delabrouille}, D.~{Paoletti}, H.~{Peiris}, F.~{Perrotta},
  F.~{Piacentini}, M.~{Piat}, L.~{Piccirillo}, G.~{Pisano}, G.~{Polenta},
  A.~{Pollo}, N.~{Ponthieu}, M.~{Remazeilles}, S.~{Ricciardi}, M.~{Roman},
  C.~{Rosset}, J.-A. {Rubino-Martin}, M.~{Salatino}, A.~{Schillaci},
  P.~{Shellard}, J.~{Silk}, A.~{Starobinsky}, R.~{Stompor}, R.~{Sunyaev},
  A.~{Tartari}, L.~{Terenzi}, L.~{Toffolatti}, M.~{Tomasi}, N.~{Trappe},
  M.~{Tristram}, T.~{Trombetti}, M.~{Tucci}, R.~{Van de Weijgaert}, B.~{Van
  Tent}, L.~{Verde}, P.~{Vielva}, B.~{Wandelt}, R.~{Watson}, and
  S.~{Withington}.
\newblock {PRISM (Polarized Radiation Imaging and Spectroscopy Mission): an
  extended white paper}.
\newblock \emph{\jcap}, 2:\penalty0 006, February 2014.
\newblock \doi{10.1088/1475-7516/2014/02/006}.

\bibitem[{Banerji et al.}(2017)]{Ranajoy.bandpass.paper}
R.~{Banerji et al.}
\newblock Correction of the effect of bandpass mismatch for future satellite
  cmb missions.
\newblock \emph{to be submitted}, 2017.

\bibitem[{Bennett} et~al.(2011){Bennett}, {Hill}, {Hinshaw}, {Larson}, {Smith},
  {Dunkley}, {Gold}, {Halpern}, {Jarosik}, {Kogut}, {Komatsu}, {Limon},
  {Meyer}, {Nolta}, {Odegard}, {Page}, {Spergel}, {Tucker}, {Weiland},
  {Wollack}, and {Wright}]{2011ApJS..192...17B}
C.~L. {Bennett}, R.~S. {Hill}, G.~{Hinshaw}, D.~{Larson}, K.~M. {Smith},
  J.~{Dunkley}, B.~{Gold}, M.~{Halpern}, N.~{Jarosik}, A.~{Kogut},
  E.~{Komatsu}, M.~{Limon}, S.~S. {Meyer}, M.~R. {Nolta}, N.~{Odegard},
  L.~{Page}, D.~N. {Spergel}, G.~S. {Tucker}, J.~L. {Weiland}, E.~{Wollack},
  and E.~L. {Wright}.
\newblock {Seven-year Wilkinson Microwave Anisotropy Probe (WMAP) Observations:
  Are There Cosmic Microwave Background Anomalies?}
\newblock \emph{\apjs}, 192:\penalty0 17, February 2011.
\newblock \doi{10.1088/0067-0049/192/2/17}.

\bibitem[{BICEP2 Collaboration} et~al.(2014){BICEP2 Collaboration}, {Ade},
  {Aikin}, {Amiri}, {Barkats}, {Benton}, {Bischoff}, {Bock}, {Brevik}, {Buder},
  {Bullock}, {Davis}, {Day}, {Dowell}, {Duband}, {Filippini}, {Fliescher},
  {Golwala}, {Halpern}, {Hasselfield}, {Hildebrandt}, {Hilton}, {Irwin},
  {Karkare}, {Kaufman}, {Keating}, {Kernasovskiy}, {Kovac}, {Kuo}, {Leitch},
  {Llombart}, {Lueker}, {Netterfield}, {Nguyen}, {O'Brient}, {Ogburn},
  {Orlando}, {Pryke}, {Reintsema}, {Richter}, {Schwarz}, {Sheehy},
  {Staniszewski}, {Story}, {Sudiwala}, {Teply}, {Tolan}, {Turner}, {Vieregg},
  {Wilson}, {Wong}, and {Yoon}]{2014ApJ...792...62B}
{BICEP2 Collaboration}, P.~A.~R. {Ade}, R.~W. {Aikin}, M.~{Amiri},
  D.~{Barkats}, S.~J. {Benton}, C.~A. {Bischoff}, J.~J. {Bock}, J.~A. {Brevik},
  I.~{Buder}, E.~{Bullock}, G.~{Davis}, P.~K. {Day}, C.~D. {Dowell},
  L.~{Duband}, J.~P. {Filippini}, S.~{Fliescher}, S.~R. {Golwala},
  M.~{Halpern}, M.~{Hasselfield}, S.~R. {Hildebrandt}, G.~C. {Hilton}, K.~D.
  {Irwin}, K.~S. {Karkare}, J.~P. {Kaufman}, B.~G. {Keating}, S.~A.
  {Kernasovskiy}, J.~M. {Kovac}, C.~L. {Kuo}, E.~M. {Leitch}, N.~{Llombart},
  M.~{Lueker}, C.~B. {Netterfield}, H.~T. {Nguyen}, R.~{O'Brient}, R.~W.
  {Ogburn}, IV, A.~{Orlando}, C.~{Pryke}, C.~D. {Reintsema}, S.~{Richter},
  R.~{Schwarz}, C.~D. {Sheehy}, Z.~K. {Staniszewski}, K.~T. {Story}, R.~V.
  {Sudiwala}, G.~P. {Teply}, J.~E. {Tolan}, A.~D. {Turner}, A.~G. {Vieregg},
  P.~{Wilson}, C.~L. {Wong}, and K.~W. {Yoon}.
\newblock {BICEP2. II. Experiment and three-year Data Set}.
\newblock \emph{\apj}, 792:\penalty0 62, September 2014.
\newblock \doi{10.1088/0004-637X/792/1/62}.

\bibitem[{BICEP2 Collaboration} et~al.(2016){BICEP2 Collaboration}, {Keck Array
  Collaboration}, {Ade}, {Ahmed}, {Aikin}, {Alexander}, {Barkats}, {Benton},
  {Bischoff}, {Bock}, {Bowens-Rubin}, {Brevik}, {Buder}, {Bullock}, {Buza},
  {Connors}, {Crill}, {Duband}, {Dvorkin}, {Filippini}, {Fliescher}, {Grayson},
  {Halpern}, {Harrison}, {Hilton}, {Hui}, {Irwin}, {Karkare}, {Karpel},
  {Kaufman}, {Keating}, {Kefeli}, {Kernasovskiy}, {Kovac}, {Kuo}, {Leitch},
  {Lueker}, {Megerian}, {Netterfield}, {Nguyen}, {O'Brient}, {Ogburn},
  {Orlando}, {Pryke}, {Richter}, {Schwarz}, {Sheehy}, {Staniszewski},
  {Steinbach}, {Sudiwala}, {Teply}, {Thompson}, {Tolan}, {Tucker}, {Turner},
  {Vieregg}, {Weber}, {Wiebe}, {Willmert}, {Wong}, {Wu}, and
  {Yoon}]{2016PhRvL.116c1302B}
{BICEP2 Collaboration}, {Keck Array Collaboration}, P.~A.~R. {Ade}, Z.~{Ahmed},
  R.~W. {Aikin}, K.~D. {Alexander}, D.~{Barkats}, S.~J. {Benton}, C.~A.
  {Bischoff}, J.~J. {Bock}, R.~{Bowens-Rubin}, J.~A. {Brevik}, I.~{Buder},
  E.~{Bullock}, V.~{Buza}, J.~{Connors}, B.~P. {Crill}, L.~{Duband},
  C.~{Dvorkin}, J.~P. {Filippini}, S.~{Fliescher}, J.~{Grayson}, M.~{Halpern},
  S.~{Harrison}, G.~C. {Hilton}, H.~{Hui}, K.~D. {Irwin}, K.~S. {Karkare},
  E.~{Karpel}, J.~P. {Kaufman}, B.~G. {Keating}, S.~{Kefeli}, S.~A.
  {Kernasovskiy}, J.~M. {Kovac}, C.~L. {Kuo}, E.~M. {Leitch}, M.~{Lueker},
  K.~G. {Megerian}, C.~B. {Netterfield}, H.~T. {Nguyen}, R.~{O'Brient}, R.~W.
  {Ogburn}, A.~{Orlando}, C.~{Pryke}, S.~{Richter}, R.~{Schwarz}, C.~D.
  {Sheehy}, Z.~K. {Staniszewski}, B.~{Steinbach}, R.~V. {Sudiwala}, G.~P.
  {Teply}, K.~L. {Thompson}, J.~E. {Tolan}, C.~{Tucker}, A.~D. {Turner}, A.~G.
  {Vieregg}, A.~C. {Weber}, D.~V. {Wiebe}, J.~{Willmert}, C.~L. {Wong},
  W.~L.~K. {Wu}, and K.~W. {Yoon}.
\newblock {Improved Constraints on Cosmology and Foregrounds from BICEP2 and
  Keck Array Cosmic Microwave Background Data with Inclusion of 95 GHz Band}.
\newblock \emph{Physical Review Letters}, 116\penalty0 (3):\penalty0 031302,
  January 2016.
\newblock \doi{10.1103/PhysRevLett.116.031302}.

\bibitem[{Bock} et~al.(2008){Bock}, {Cooray}, {Hanany}, {Keating}, {Lee},
  {Matsumura}, {Milligan}, {Ponthieu}, {Renbarger}, and
  {Tran}]{2008arXiv0805.4207B}
J.~{Bock}, A.~{Cooray}, S.~{Hanany}, B.~{Keating}, A.~{Lee}, T.~{Matsumura},
  M.~{Milligan}, N.~{Ponthieu}, T.~{Renbarger}, and H.~{Tran}.
\newblock {The Experimental Probe of Inflationary Cosmology (EPIC): A Mission
  Concept Study for NASA's Einstein Inflation Probe}.
\newblock \emph{ArXiv e-prints}, May 2008.

\bibitem[{Bock} et~al.(2009){Bock}, {Aljabri}, {Amblard}, {Baumann}, {Betoule},
  {Chui}, {Colombo}, {Cooray}, {Crumb}, {Day}, {Dickinson}, {Dowell},
  {Dragovan}, {Golwala}, {Gorski}, {Hanany}, {Holmes}, {Irwin}, {Johnson},
  {Keating}, {Kuo}, {Lee}, {Lange}, {Lawrence}, {Meyer}, {Miller}, {Nguyen},
  {Pierpaoli}, {Ponthieu}, {Puget}, {Raab}, {Richards}, {Satter}, {Seiffert},
  {Shimon}, {Tran}, {Williams}, and {Zmuidzinas}]{2009arXiv0906.1188B}
J.~{Bock}, A.~{Aljabri}, A.~{Amblard}, D.~{Baumann}, M.~{Betoule}, T.~{Chui},
  L.~{Colombo}, A.~{Cooray}, D.~{Crumb}, P.~{Day}, C.~{Dickinson}, D.~{Dowell},
  M.~{Dragovan}, S.~{Golwala}, K.~{Gorski}, S.~{Hanany}, W.~{Holmes},
  K.~{Irwin}, B.~{Johnson}, B.~{Keating}, C.-L. {Kuo}, A.~{Lee}, A.~{Lange},
  C.~{Lawrence}, S.~{Meyer}, N.~{Miller}, H.~{Nguyen}, E.~{Pierpaoli},
  N.~{Ponthieu}, J.-L. {Puget}, J.~{Raab}, P.~{Richards}, C.~{Satter},
  M.~{Seiffert}, M.~{Shimon}, H.~{Tran}, B.~{Williams}, and J.~{Zmuidzinas}.
\newblock {Study of the Experimental Probe of Inflationary Cosmology
  (EPIC)-Intemediate Mission for NASA's Einstein Inflation Probe}.
\newblock \emph{ArXiv e-prints}, June 2009.

\bibitem[{Bouchet} et~al.(2005){Bouchet}, {Beno{\^i}t}, {Camus}, {D{\'e}sert},
  {Piat}, and {Ponthieu}]{2005sf2a.conf..675B}
F.~R. {Bouchet}, A.~{Beno{\^i}t}, P.~{Camus}, F.~X. {D{\'e}sert}, M.~{Piat},
  and N.~{Ponthieu}.
\newblock {Charting the New Frontier of the Cosmic Microwave Background
  Polarization}.
\newblock In F.~{Casoli}, T.~{Contini}, J.~M. {Hameury}, and L.~{Pagani},
  editors, \emph{SF2A-2005: Semaine de l'Astrophysique Francaise}, page 675,
  December 2005.

\bibitem[{Branco} et~al.(2014){Branco}, {Charles}, and
  {Butterworth}]{2014SPIE.9144E..5VB}
M.~B.~C. {Branco}, I.~{Charles}, and J.~{Butterworth}.
\newblock {ATHENA X-IFU detector cooling chain}.
\newblock In \emph{Space Telescopes and Instrumentation 2014: Ultraviolet to
  Gamma Ray}, volume 9144 of \emph{\procspie}, page 91445V, July 2014.
\newblock \doi{10.1117/12.2056427}.

\bibitem[{Burigana} et~al.(2017){Burigana}, {Carvalho}, {Trombetti}, {Notari},
  {Quartin}, {De Gasperis}, {Buzzelli}, {Vittorio}, {De Zotti}, {de Bernardis},
  {Chluba}, {Bilicki}, {Danese}, {Delabrouille}, {Toffolatti}, {Lapi},
  {Negrello}, {Mazzotta}, {Scott}, {Contreras}, {Achucarro}, {Ade}, {Allison},
  {Ashdown}, {Ballardini}, {Banday}, {Banerji}, {Bartlett}, {Bartolo}, {Basak},
  {Bersanelli}, {Bonaldi}, {Bonato}, {Borrill}, {Bouchet}, {Boulanger},
  {Brinckmann}, {Bucher}, {Cabella}, {Cai}, {Calvo}, {Castellano}, {Challinor},
  {Clesse}, {Colantoni}, {Coppolecchia}, {Crook}, {D'Alessandro}, {Diego}, {Di
  Marco}, {Di Valentino}, {Errard}, {Feeney}, {Fernandez-Cobos}, {Ferraro},
  {Finelli}, {Forastieri}, {Galli}, {Genova-Santos}, {Gerbino},
  {Gonzalez-Nuevo}, {Grandis}, {Greenslade}, {Hagstotz}, {Hanany}, {Handley},
  {Hernandez-Monteagudo}, {Hervias-Caimapo}, {Hills}, {Hivon}, {Kiiveri},
  {Kisner}, {Kitching}, {Kunz}, {Kurki-Suonio}, {Lamagna}, {Lasenby},
  {Lattanzi}, {Lesgourgues}, {Liguori}, {Lindholm}, {Lopez-Caniego}, {Luzzi},
  {Maffei}, {Mandolesi}, {Martinez-Gonzalez}, {Martins}, {Masi}, {McCarthy},
  {Melchiorri}, {Melin}, {Molinari}, {Monfardini}, {Natoli}, {Paiella},
  {Paoletti}, {Patanchon}, {Piat}, {Pisano}, {Polastri}, {Polenta}, {Pollo},
  {Poulin}, {Remazeilles}, {Roman}, {Rubino-Martin}, {Salvati}, {Tartari},
  {Tomasi}, {Tramonte}, {Trappe}, {Tucker}, {Valiviita}, {Van de Weijgaert},
  {van Tent}, {Vennin}, {Vielva}, {Young}, {Zannoni}, and {for the CORE
  Collaboration}]{ECO.velocity.paper}
C.~{Burigana}, C.~S. {Carvalho}, T.~{Trombetti}, A.~{Notari}, M.~{Quartin},
  G.~{De Gasperis}, A.~{Buzzelli}, N.~{Vittorio}, G.~{De Zotti}, P.~{de
  Bernardis}, J.~{Chluba}, M.~{Bilicki}, L.~{Danese}, J.~{Delabrouille},
  L.~{Toffolatti}, A.~{Lapi}, M.~{Negrello}, P.~{Mazzotta}, D.~{Scott},
  D.~{Contreras}, A.~{Achucarro}, P.~{Ade}, R.~{Allison}, M.~{Ashdown},
  M.~{Ballardini}, A.~J. {Banday}, R.~{Banerji}, J.~{Bartlett}, N.~{Bartolo},
  S.~{Basak}, M.~{Bersanelli}, A.~{Bonaldi}, M.~{Bonato}, J.~{Borrill},
  F.~{Bouchet}, F.~{Boulanger}, T.~{Brinckmann}, M.~{Bucher}, P.~{Cabella},
  Z.-Y. {Cai}, M.~{Calvo}, G.~{Castellano}, A.~{Challinor}, S.~{Clesse},
  I.~{Colantoni}, A.~{Coppolecchia}, M.~{Crook}, G.~{D'Alessandro}, J.-M.
  {Diego}, A.~{Di Marco}, E.~{Di Valentino}, J.~{Errard}, S.~{Feeney},
  R.~{Fernandez-Cobos}, S.~{Ferraro}, F.~{Finelli}, F.~{Forastieri},
  S.~{Galli}, R.~{Genova-Santos}, M.~{Gerbino}, J.~{Gonzalez-Nuevo},
  S.~{Grandis}, J.~{Greenslade}, S.~{Hagstotz}, S.~{Hanany}, W.~{Handley},
  C.~{Hernandez-Monteagudo}, C.~{Hervias-Caimapo}, M.~{Hills}, E.~{Hivon},
  K.~{Kiiveri}, T.~{Kisner}, T.~{Kitching}, M.~{Kunz}, H.~{Kurki-Suonio},
  L.~{Lamagna}, A.~{Lasenby}, M.~{Lattanzi}, J.~{Lesgourgues}, M.~{Liguori},
  V.~{Lindholm}, M.~{Lopez-Caniego}, G.~{Luzzi}, B.~{Maffei}, N.~{Mandolesi},
  E.~{Martinez-Gonzalez}, C.~J.~A.~P. {Martins}, S.~{Masi}, D.~{McCarthy},
  A.~{Melchiorri}, J.-B. {Melin}, D.~{Molinari}, A.~{Monfardini}, P.~{Natoli},
  A.~{Paiella}, D.~{Paoletti}, G.~{Patanchon}, M.~{Piat}, G.~{Pisano},
  L.~{Polastri}, G.~{Polenta}, A.~{Pollo}, V.~{Poulin}, M.~{Remazeilles},
  M.~{Roman}, J.-A. {Rubino-Martin}, L.~{Salvati}, A.~{Tartari}, M.~{Tomasi},
  D.~{Tramonte}, N.~{Trappe}, C.~{Tucker}, J.~{Valiviita}, R.~{Van de
  Weijgaert}, B.~{van Tent}, V.~{Vennin}, P.~{Vielva}, K.~{Young},
  M.~{Zannoni}, and {for the CORE Collaboration}.
\newblock {Exploring cosmic origins with CORE: effects of observer peculiar
  motion}.
\newblock \emph{ArXiv e-prints}, April 2017.

\bibitem[{Carlstrom} et~al.(2002){Carlstrom}, {Holder}, and
  {Reese}]{2002ARA&A..40..643C}
J.~E. {Carlstrom}, G.~P. {Holder}, and E.~D. {Reese}.
\newblock {Cosmology with the Sunyaev-Zel'dovich Effect}.
\newblock \emph{\araa}, 40:\penalty0 643--680, 2002.
\newblock \doi{10.1146/annurev.astro.40.060401.093803}.

\bibitem[{Challinor et al.}(2017)]{ECO.lensing.paper}
A.~{Challinor et al.}
\newblock Exploring cosmic origins with \coremfive: Gravitational lensing of
  the cmb.
\newblock \emph{to be submitted}, 2017.

\bibitem[{Charles} et~al.(2016){Charles}, {Daniel}, {Andr{\'e}}, {Duband},
  {Duval}, {den Hartog}, {Mitsuda}, {Shinozaki}, {van Weers}, and
  {Yamasaki}]{2016SPIE.9905E..2JC}
I.~{Charles}, C.~{Daniel}, J.~{Andr{\'e}}, L.~{Duband}, J.-M. {Duval}, R.~{den
  Hartog}, K.~{Mitsuda}, K.~{Shinozaki}, H.~{van Weers}, and N.~Y. {Yamasaki}.
\newblock {Preliminary thermal architecture of the X-IFU instrument dewar}.
\newblock In \emph{Space Telescopes and Instrumentation 2016: Ultraviolet to
  Gamma Ray}, volume 9905 of \emph{\procspie}, page 99052J, July 2016.
\newblock \doi{10.1117/12.2232710}.

\bibitem[{Collaudin} et~al.(2010){Collaudin}, {Montet}, {Roche}, {Ilsen},
  {Schamberg}, {Cesa}, {Goodey}, {Rautakoski}, {Jewell}, {Idler}, {Koppe},
  {Sonn}, {Hendry}, {Hamer}, {Bauer}, {Feuchtgruber}, {Sawyer}, {Swinyard},
  {Sidher}, {Roelfsema}, {Dieleman}, and {Teyssier}]{2010AIPC.1218.1510C}
B.~{Collaudin}, D.~{Montet}, Y.~{Roche}, S.~{Ilsen}, C.~{Schamberg}, M.~{Cesa},
  K.~{Goodey}, J.~{Rautakoski}, C.~{Jewell}, S.~{Idler}, A.~{Koppe}, N.~{Sonn},
  D.~{Hendry}, S.~{Hamer}, O.~{Bauer}, H.~{Feuchtgruber}, E.~{Sawyer},
  B.~{Swinyard}, S.~{Sidher}, P.~{Roelfsema}, P.~{Dieleman}, and D.~{Teyssier}.
\newblock {Herschel: Testing of Cryogenics Instruments at Spacecraft Level and
  Early Flight Results}.
\newblock In J.~G. {Weisend}, editor, \emph{American Institute of Physics
  Conference Series}, volume 1218 of \emph{American Institute of Physics
  Conference Series}, pages 1510--1519, April 2010.
\newblock \doi{10.1063/1.3422331}.

\bibitem[{CORE Collaboration} et~al.(2016){CORE Collaboration}, {Finelli},
  {Bucher}, {Ach{\'u}carro}, {Ballardini}, {Bartolo}, {Baumann}, {Clesse},
  {Errard}, {Handley}, {Hindmarsh}, {Kiiveri}, {Kunz}, {Lasenby}, {Liguori},
  {Paoletti}, {Ringeval}, {V{\"a}liviita}, {van Tent}, {Vennin}, {Arroja},
  {Ashdown}, {Banday}, {Banerji}, {Baselmans}, {Bartlett}, {de Bernardis},
  {Bersanelli}, {Bonaldi}, {Borril}, {Bouchet}, {Boulanger}, {Brinckmann},
  {Cai}, {Calvo}, {Challinor}, {Chluba}, {D'Amico}, {Delabrouille},
  {Mar{\'{\i}}a Diego}, {De Zotti}, {Desjacques}, {Di Valentino}, {Feeney},
  {Fergusson}, {Ferraro}, {Forastieri}, {Galli}, {Garc{\'{\i}}a-Bellido},
  {G{\'e}nova-Santos}, {Gerbino}, {Gonz{\'a}lez-Nuevo}, {Grandis},
  {Greenslade}, {Hagstotz}, {Hanany}, {Hazra}, {Hern{\'a}ndez-Monteagudo},
  {Hivon}, {Hu}, {Kovetz}, {Kurki-Suonio}, {Lattanzi}, {Lesgourgues},
  {Lizarraga}, {L{\'o}pez-Caniego}, {Luzzi}, {Maffei}, {Martins},
  {Mart{\'{\i}}nez-Gonz{\'a}lez}, {McCarthy}, {Matarrese}, {Melchiorri},
  {Melin}, {Monfardini}, {Natoli}, {Negrello}, {Oppizzi}, {Pajer}, {Patil},
  {Piat}, {Pisano}, {Poulin}, {Ravenni}, {Remazeilles}, {Renzi}, {Roest},
  {Salvati}, {Tartari}, {Tasinato}, {Torrado}, {Trappe}, {Tucci}, {Urrestilla},
  {Vielva}, and {Van de Weygaert}]{2016arXiv161208270C}
{CORE Collaboration}, F.~{Finelli}, M.~{Bucher}, A.~{Ach{\'u}carro},
  M.~{Ballardini}, N.~{Bartolo}, D.~{Baumann}, S.~{Clesse}, J.~{Errard},
  W.~{Handley}, M.~{Hindmarsh}, K.~{Kiiveri}, M.~{Kunz}, A.~{Lasenby},
  M.~{Liguori}, D.~{Paoletti}, C.~{Ringeval}, J.~{V{\"a}liviita}, B.~{van
  Tent}, V.~{Vennin}, F.~{Arroja}, M.~{Ashdown}, A.~J. {Banday}, R.~{Banerji},
  J.~{Baselmans}, J.~G. {Bartlett}, P.~{de Bernardis}, M.~{Bersanelli},
  A.~{Bonaldi}, J.~{Borril}, F.~R. {Bouchet}, F.~{Boulanger}, T.~{Brinckmann},
  Z.-Y. {Cai}, M.~{Calvo}, A.~{Challinor}, J.~{Chluba}, G.~{D'Amico},
  J.~{Delabrouille}, J.~{Mar{\'{\i}}a Diego}, G.~{De Zotti}, V.~{Desjacques},
  E.~{Di Valentino}, S.~{Feeney}, J.~R. {Fergusson}, S.~{Ferraro},
  F.~{Forastieri}, S.~{Galli}, J.~{Garc{\'{\i}}a-Bellido}, R.~T.
  {G{\'e}nova-Santos}, M.~{Gerbino}, J.~{Gonz{\'a}lez-Nuevo}, S.~{Grandis},
  J.~{Greenslade}, S.~{Hagstotz}, S.~{Hanany}, D.~K. {Hazra},
  C.~{Hern{\'a}ndez-Monteagudo}, E.~{Hivon}, B.~{Hu}, E.~D. {Kovetz},
  H.~{Kurki-Suonio}, M.~{Lattanzi}, J.~{Lesgourgues}, J.~{Lizarraga},
  M.~{L{\'o}pez-Caniego}, G.~{Luzzi}, B.~{Maffei}, C.~J.~A.~P. {Martins},
  E.~{Mart{\'{\i}}nez-Gonz{\'a}lez}, D.~{McCarthy}, S.~{Matarrese},
  A.~{Melchiorri}, J.-B. {Melin}, A.~{Monfardini}, P.~{Natoli}, M.~{Negrello},
  F.~{Oppizzi}, E.~{Pajer}, S.~P. {Patil}, M.~{Piat}, G.~{Pisano}, V.~{Poulin},
  A.~{Ravenni}, M.~{Remazeilles}, A.~{Renzi}, D.~{Roest}, L.~{Salvati},
  A.~{Tartari}, G.~{Tasinato}, J.~{Torrado}, N.~{Trappe}, M.~{Tucci},
  J.~{Urrestilla}, P.~{Vielva}, and R.~{Van de Weygaert}.
\newblock {Exploring Cosmic Origins with CORE: Inflation}.
\newblock \emph{ArXiv e-prints}, December 2016.

\bibitem[{Couchot} et~al.(1999){Couchot}, {Delabrouille}, {Kaplan}, and
  {Revenu}]{1999A&AS..135..579C}
F.~{Couchot}, J.~{Delabrouille}, J.~{Kaplan}, and B.~{Revenu}.
\newblock {Optimised polarimeter configurations for measuring the Stokes
  parameters of the cosmic microwave background radiation}.
\newblock \emph{\aaps}, 135:\penalty0 579--584, March 1999.
\newblock \doi{10.1051/aas:1999191}.

\bibitem[{Couchot} et~al.(2017){Couchot}, {Henrot-Versill{\'e}}, {Perdereau},
  {Plaszczynski}, {Rouill{\'e} d'Orfeuil}, {Spinelli}, and
  {Tristram}]{2017A&A...597A.126C}
F.~{Couchot}, S.~{Henrot-Versill{\'e}}, O.~{Perdereau}, S.~{Plaszczynski},
  B.~{Rouill{\'e} d'Orfeuil}, M.~{Spinelli}, and M.~{Tristram}.
\newblock {Relieving tensions related to the lensing of the cosmic microwave
  background temperature power spectra}.
\newblock \emph{\aap}, 597:\penalty0 A126, January 2017.
\newblock \doi{10.1051/0004-6361/201527740}.

\bibitem[{De Bernardis} et~al.(2009){De Bernardis}, {Bucher}, {Burigana}, and
  {Piccirillo}]{2009ExA....23....5D}
P.~{De Bernardis}, M.~{Bucher}, C.~{Burigana}, and L.~{Piccirillo}.
\newblock {B-Pol: detecting primordial gravitational waves generated during
  inflation}.
\newblock \emph{Experimental Astronomy}, 23:\penalty0 5--16, March 2009.
\newblock \doi{10.1007/s10686-008-9120-y}.

\bibitem[{de Bernardis} et~al.(2017){de Bernardis}, {Ade}, {Baselmans},
  {Battistelli}, {Benoit}, {Bersanelli}, {Bideaud}, {Calvo}, {Casas},
  {Castellano}, {Catalano}, {Charles}, {Colantoni}, {Columbro}, {Coppolecchia},
  {Crook}, {D'Alessandro}, {De Petris}, {Delabrouille}, {Doyle}, {Franceschet},
  {Gomez}, {Goupy}, {Hanany}, {Hills}, {Lamagna}, {Macias-Perez}, {Maffei},
  {Martin}, {Martinez-Gonzalez}, {Masi}, {McCarthy}, {Mennella}, {Monfardini},
  {Noviello}, {Paiella}, {Piacentini}, {Piat}, {Pisano}, {Signorelli}, {Tan},
  {Tartari}, {Trappe}, {Triqueneaux}, {Tucker}, {Vermeulen}, {Young},
  {Zannoni}, {Ach{\'u}carro}, {Allison}, {Ashdown}, {Ballardini}, {Banday},
  {Banerji}, {Bartlett}, {Bartolo}, {Basak}, {Bonaldi}, {Bonato}, {Borrill},
  {Bouchet}, {Boulanger}, {Brinckmann}, {Bucher}, {Burigana}, {Buzzelli},
  {Cai}, {Carvalho}, {Challinor}, {Chluba}, {Clesse}, {De Gasperis}, {De
  Zotti}, {Di Valentino}, {Diego}, {Errard}, {Feeney}, {Fernandez-Cobos},
  {Finelli}, {Forastieri}, {Galli}, {G{\'e}nova-Santos}, {Gerbino},
  {Gonz{\'a}lez-Nuevo}, {Hagstotz}, {Greenslade}, {Handley},
  {Hern{\'a}ndez-Monteagudo}, {Hervias-Caimapo}, {Hivon}, {Kiiveri}, {Kisner},
  {Kitching}, {Kunz}, {Kurki-Suonio}, {Lasenby}, {Lattanzi}, {Lesgourgues},
  {Lewis}, {Liguori}, {Lindholm}, {Luzzi}, {Martins}, {Melchiorri}, {Melin},
  {Molinari}, {Natoli}, {Negrello}, {Notari}, {Paoletti}, {Patanchon},
  {Polastri}, {Polenta}, {Pollo}, {Poulin}, {Quartin}, {Remazeilles}, {Roman},
  {Rubi{\~n}o-Mart{\'{\i}}n}, {Salvati}, {Tomasi}, {Tramonte}, {Trombetti},
  {V{\"a}liviita}, {Van de Weijgaert}, {van Tent}, {Vennin}, {Vielva},
  {Vittorio}, and {for the CORE collaboration}]{ECO.instrument.paper}
P.~{de Bernardis}, P.~A.~R. {Ade}, J.~J.~A. {Baselmans}, E.~S. {Battistelli},
  A.~{Benoit}, M.~{Bersanelli}, A.~{Bideaud}, M.~{Calvo}, F.~J. {Casas},
  G.~{Castellano}, A.~{Catalano}, I.~{Charles}, I.~{Colantoni}, F.~{Columbro},
  A.~{Coppolecchia}, M.~{Crook}, G.~{D'Alessandro}, M.~{De Petris},
  J.~{Delabrouille}, S.~{Doyle}, C.~{Franceschet}, A.~{Gomez}, J.~{Goupy},
  S.~{Hanany}, M.~{Hills}, L.~{Lamagna}, J.~{Macias-Perez}, B.~{Maffei},
  S.~{Martin}, E.~{Martinez-Gonzalez}, S.~{Masi}, D.~{McCarthy}, A.~{Mennella},
  A.~{Monfardini}, F.~{Noviello}, A.~{Paiella}, F.~{Piacentini}, M.~{Piat},
  G.~{Pisano}, G.~{Signorelli}, C.~Y. {Tan}, A.~{Tartari}, N.~{Trappe},
  S.~{Triqueneaux}, C.~{Tucker}, G.~{Vermeulen}, K.~{Young}, M.~{Zannoni},
  A.~{Ach{\'u}carro}, R.~{Allison}, M.~{Ashdown}, M.~{Ballardini}, A.~J.
  {Banday}, R.~{Banerji}, J.~{Bartlett}, N.~{Bartolo}, S.~{Basak},
  A.~{Bonaldi}, M.~{Bonato}, J.~{Borrill}, F.~{Bouchet}, F.~{Boulanger},
  T.~{Brinckmann}, M.~{Bucher}, C.~{Burigana}, A.~{Buzzelli}, Z.~Y. {Cai},
  C.~S. {Carvalho}, A.~{Challinor}, J.~{Chluba}, S.~{Clesse}, G.~{De Gasperis},
  G.~{De Zotti}, E.~{Di Valentino}, J.~M. {Diego}, J.~{Errard}, S.~{Feeney},
  R.~{Fernandez-Cobos}, F.~{Finelli}, F.~{Forastieri}, S.~{Galli},
  R.~{G{\'e}nova-Santos}, M.~{Gerbino}, J.~{Gonz{\'a}lez-Nuevo}, S.~{Hagstotz},
  J.~{Greenslade}, W.~{Handley}, C.~{Hern{\'a}ndez-Monteagudo},
  C.~{Hervias-Caimapo}, E.~{Hivon}, K.~{Kiiveri}, T.~{Kisner}, T.~{Kitching},
  M.~{Kunz}, H.~{Kurki-Suonio}, A.~{Lasenby}, M.~{Lattanzi}, J.~{Lesgourgues},
  A.~{Lewis}, M.~{Liguori}, V.~{Lindholm}, G.~{Luzzi}, C.~J.~A.~P. {Martins},
  A.~{Melchiorri}, J.~B. {Melin}, D.~{Molinari}, P.~{Natoli}, M.~{Negrello},
  A.~{Notari}, D.~{Paoletti}, G.~{Patanchon}, L.~{Polastri}, G.~{Polenta},
  A.~{Pollo}, V.~{Poulin}, M.~{Quartin}, M.~{Remazeilles}, M.~{Roman}, J.~A.
  {Rubi{\~n}o-Mart{\'{\i}}n}, L.~{Salvati}, M.~{Tomasi}, D.~{Tramonte},
  T.~{Trombetti}, J.~{V{\"a}liviita}, R.~{Van de Weijgaert}, B.~{van Tent},
  V.~{Vennin}, P.~{Vielva}, N.~{Vittorio}, and {for the CORE collaboration}.
\newblock {Exploring Cosmic Origins with CORE: The Instrument}.
\newblock \emph{ArXiv e-prints}, May 2017.

\bibitem[{de Gasperis} et~al.(2005){de Gasperis}, {Balbi}, {Cabella}, {Natoli},
  and {Vittorio}]{2005A&A...436.1159D}
G.~{de Gasperis}, A.~{Balbi}, P.~{Cabella}, P.~{Natoli}, and N.~{Vittorio}.
\newblock {ROMA: A map-making algorithm for polarised CMB data sets}.
\newblock \emph{\aap}, 436:\penalty0 1159--1165, June 2005.
\newblock \doi{10.1051/0004-6361:20042512}.

\bibitem[{De Zotti} et~al.(2016){De Zotti}, {Gonzalez-Nuevo}, {Lopez-Caniego},
  {Negrello}, {Greenslade}, {Hernandez-Monteagudo}, {Delabrouille}, {Cai},
  {Biesiada}, {Bilicki}, {Bonaldi}, {Bonato}, {Burigana}, {Clements},
  {Colafrancesco}, {Diego}, {Le Brun}, {Massardi}, {Melin}, {Pollo}, {Roukema},
  {Serjeant}, {Toffolatti}, {Tucci}, and {the CORE
  collaboration}]{2016arXiv160907263D}
G.~{De Zotti}, J.~{Gonzalez-Nuevo}, M.~{Lopez-Caniego}, M.~{Negrello},
  J.~{Greenslade}, C.~{Hernandez-Monteagudo}, J.~{Delabrouille}, Z.-Y. {Cai},
  M.~{Biesiada}, M.~{Bilicki}, A.~{Bonaldi}, M.~{Bonato}, C.~{Burigana}, D.~L.
  {Clements}, S.~{Colafrancesco}, J.~M. {Diego}, A.~{Le Brun}, M.~{Massardi},
  J.~B. {Melin}, A.~{Pollo}, B.~{Roukema}, S.~{Serjeant}, L.~{Toffolatti},
  M.~{Tucci}, and {the CORE collaboration}.
\newblock {Exploring Cosmic Origins with CORE: Extragalactic sources in Cosmic
  Microwave Background maps}.
\newblock \emph{ArXiv e-prints}, September 2016.

\bibitem[{Delabrouille} and {Cardoso}(2009)]{2009LNP...665..159D}
J.~{Delabrouille} and J.-F. {Cardoso}.
\newblock {Diffuse Source Separation in CMB Observations}.
\newblock In V.~J. {Mart{\'{\i}}nez}, E.~{Saar},
  E.~{Mart{\'{\i}}nez-Gonz{\'a}lez}, and M.-J. {Pons-Border{\'{\i}}a}, editors,
  \emph{Data Analysis in Cosmology}, volume 665 of \emph{Lecture Notes in
  Physics, Berlin Springer Verlag}, pages 159--205, 2009.
\newblock \doi{10.1007/978-3-540-44767-2_6}.

\bibitem[{Delabrouille} et~al.(2002){Delabrouille}, {Patanchon}, and
  {Audit}]{2002MNRAS.330..807D}
J.~{Delabrouille}, G.~{Patanchon}, and E.~{Audit}.
\newblock {Separation of instrumental and astrophysical foregrounds for mapping
  cosmic microwave background anisotropies}.
\newblock \emph{\mnras}, 330:\penalty0 807--816, March 2002.
\newblock \doi{10.1046/j.1365-8711.2002.05200.x}.

\bibitem[{Di Valentino} et~al.(2016){Di Valentino}, {Brinckmann}, {Gerbino},
  {Poulin}, {Bouchet}, {Lesgourgues}, {Melchiorri}, {Chluba}, {Clesse},
  {Delabrouille}, {Dvorkin}, {Forastieri}, {Galli}, {Hooper}, {Lattanzi},
  {Martins}, {Salvati}, {Cabass}, {Caputo}, {Giusarma}, {Hivon}, {Natoli},
  {Pagano}, {Paradiso}, {Rubino-Martin}, {Achucarro}, {Ballardini}, {Bartolo},
  {Baumann}, {Bartlett}, {de Bernardis}, {Bonaldi}, {Bucher}, {Cai}, {De
  Zotti}, {Diego}, {Errard}, {Ferraro}, {Finelli}, {Genova-Santos},
  {Gonzalez-Nuevo}, {Grandis}, {Greenslade}, {Hagstotz}, {Handley},
  {Hindmarsh}, {Hernandez-Monteagudo}, {Kiiveri}, {Kunz}, {Lasenby}, {Liguori},
  {Lopez-Caniego}, {Luzzi}, {Melin}, {Mohr}, {Negrello}, {Paoletti},
  {Remazeilles}, {Ringeval}, {Valiviita}, {Van Tent}, {Vennin}, {Vittorio}, and
  {the CORE collaboration}]{2016arXiv161200021D}
E.~{Di Valentino}, T.~{Brinckmann}, M.~{Gerbino}, V.~{Poulin}, F.~R. {Bouchet},
  J.~{Lesgourgues}, A.~{Melchiorri}, J.~{Chluba}, S.~{Clesse},
  J.~{Delabrouille}, C.~{Dvorkin}, F.~{Forastieri}, S.~{Galli}, D.~C. {Hooper},
  M.~{Lattanzi}, C.~J.~A.~P. {Martins}, L.~{Salvati}, G.~{Cabass}, A.~{Caputo},
  E.~{Giusarma}, E.~{Hivon}, P.~{Natoli}, L.~{Pagano}, S.~{Paradiso}, J.~A.
  {Rubino-Martin}, A.~{Achucarro}, M.~{Ballardini}, N.~{Bartolo}, D.~{Baumann},
  J.~G. {Bartlett}, P.~{de Bernardis}, A.~{Bonaldi}, M.~{Bucher}, Z.-Y. {Cai},
  G.~{De Zotti}, J.~M. {Diego}, J.~{Errard}, S.~{Ferraro}, F.~{Finelli}, R.~T.
  {Genova-Santos}, J.~{Gonzalez-Nuevo}, S.~{Grandis}, J.~{Greenslade},
  S.~{Hagstotz}, W.~{Handley}, M.~{Hindmarsh}, C.~{Hernandez-Monteagudo},
  K.~{Kiiveri}, M.~{Kunz}, A.~{Lasenby}, M.~{Liguori}, M.~{Lopez-Caniego},
  G.~{Luzzi}, J.-B. {Melin}, J.~J. {Mohr}, M.~{Negrello}, D.~{Paoletti},
  M.~{Remazeilles}, C.~{Ringeval}, J.~{Valiviita}, B.~{Van Tent}, V.~{Vennin},
  N.~{Vittorio}, and {the CORE collaboration}.
\newblock {Exploring Cosmic Origins with CORE: Cosmological Parameters}.
\newblock \emph{ArXiv e-prints}, November 2016.

\bibitem[{Eriksen} et~al.(2004){Eriksen}, {Hansen}, {Banday}, {G{\'o}rski}, and
  {Lilje}]{2004ApJ...605...14E}
H.~K. {Eriksen}, F.~K. {Hansen}, A.~J. {Banday}, K.~M. {G{\'o}rski}, and P.~B.
  {Lilje}.
\newblock {Asymmetries in the Cosmic Microwave Background Anisotropy Field}.
\newblock \emph{\apj}, 605:\penalty0 14--20, April 2004.
\newblock \doi{10.1086/382267}.

\bibitem[{Errard} et~al.(2015){Errard}, {Ade}, {Akiba}, {Arnold}, {Atlas},
  {Baccigalupi}, {Barron}, {Boettger}, {Borrill}, {Chapman}, {Chinone},
  {Cukierman}, {Delabrouille}, {Dobbs}, {Ducout}, {Elleflot}, {Fabbian},
  {Feng}, {Feeney}, {Gilbert}, {Goeckner-Wald}, {Halverson}, {Hasegawa},
  {Hattori}, {Hazumi}, {Hill}, {Holzapfel}, {Hori}, {Inoue}, {Jaehnig},
  {Jaffe}, {Jeong}, {Katayama}, {Kaufman}, {Keating}, {Kermish}, {Keskitalo},
  {Kisner}, {Le Jeune}, {Lee}, {Leitch}, {Leon}, {Linder}, {Matsuda},
  {Matsumura}, {Miller}, {Myers}, {Navaroli}, {Nishino}, {Okamura}, {Paar},
  {Peloton}, {Poletti}, {Puglisi}, {Rebeiz}, {Reichardt}, {Richards}, {Ross},
  {Rotermund}, {Schenck}, {Sherwin}, {Siritanasak}, {Smecher}, {Stebor},
  {Steinbach}, {Stompor}, {Suzuki}, {Tajima}, {Takakura}, {Tikhomirov},
  {Tomaru}, {Whitehorn}, {Wilson}, {Yadav}, and {Zahn}]{2015ApJ...809...63E}
J.~{Errard}, P.~A.~R. {Ade}, Y.~{Akiba}, K.~{Arnold}, M.~{Atlas},
  C.~{Baccigalupi}, D.~{Barron}, D.~{Boettger}, J.~{Borrill}, S.~{Chapman},
  Y.~{Chinone}, A.~{Cukierman}, J.~{Delabrouille}, M.~{Dobbs}, A.~{Ducout},
  T.~{Elleflot}, G.~{Fabbian}, C.~{Feng}, S.~{Feeney}, A.~{Gilbert},
  N.~{Goeckner-Wald}, N.~W. {Halverson}, M.~{Hasegawa}, K.~{Hattori},
  M.~{Hazumi}, C.~{Hill}, W.~L. {Holzapfel}, Y.~{Hori}, Y.~{Inoue}, G.~C.
  {Jaehnig}, A.~H. {Jaffe}, O.~{Jeong}, N.~{Katayama}, J.~{Kaufman},
  B.~{Keating}, Z.~{Kermish}, R.~{Keskitalo}, T.~{Kisner}, M.~{Le Jeune}, A.~T.
  {Lee}, E.~M. {Leitch}, D.~{Leon}, E.~{Linder}, F.~{Matsuda}, T.~{Matsumura},
  N.~J. {Miller}, M.~J. {Myers}, M.~{Navaroli}, H.~{Nishino}, T.~{Okamura},
  H.~{Paar}, J.~{Peloton}, D.~{Poletti}, G.~{Puglisi}, G.~{Rebeiz}, C.~L.
  {Reichardt}, P.~L. {Richards}, C.~{Ross}, K.~M. {Rotermund}, D.~E. {Schenck},
  B.~D. {Sherwin}, P.~{Siritanasak}, G.~{Smecher}, N.~{Stebor}, B.~{Steinbach},
  R.~{Stompor}, A.~{Suzuki}, O.~{Tajima}, S.~{Takakura}, A.~{Tikhomirov},
  T.~{Tomaru}, N.~{Whitehorn}, B.~{Wilson}, A.~{Yadav}, and O.~{Zahn}.
\newblock {Modeling Atmospheric Emission for CMB Ground-based Observations}.
\newblock \emph{\apj}, 809:\penalty0 63, August 2015.
\newblock \doi{10.1088/0004-637X/809/1/63}.

\bibitem[{Essinger-Hileman} et~al.(2016){Essinger-Hileman}, {Kusaka}, {Appel},
  {Choi}, {Crowley}, {Ho}, {Jarosik}, {Page}, {Parker}, {Raghunathan}, {Simon},
  {Staggs}, and {Visnjic}]{2016RScI...87i4503E}
T.~{Essinger-Hileman}, A.~{Kusaka}, J.~W. {Appel}, S.~K. {Choi}, K.~{Crowley},
  S.~P. {Ho}, N.~{Jarosik}, L.~A. {Page}, L.~P. {Parker}, S.~{Raghunathan},
  S.~M. {Simon}, S.~T. {Staggs}, and K.~{Visnjic}.
\newblock {Systematic effects from an ambient-temperature, continuously
  rotating half-wave plate}.
\newblock \emph{Review of Scientific Instruments}, 87\penalty0 (9):\penalty0
  094503, September 2016.
\newblock \doi{10.1063/1.4962023}.

\bibitem[{Galli} et~al.(2014){Galli}, {Benabed}, {Bouchet}, {Cardoso},
  {Elsner}, {Hivon}, {Mangilli}, {Prunet}, and {Wandelt}]{2014PhRvD..90f3504G}
S.~{Galli}, K.~{Benabed}, F.~{Bouchet}, J.-F. {Cardoso}, F.~{Elsner},
  E.~{Hivon}, A.~{Mangilli}, S.~{Prunet}, and B.~{Wandelt}.
\newblock {CMB polarization can constrain cosmology better than CMB
  temperature}.
\newblock \emph{\prd}, 90\penalty0 (6):\penalty0 063504, September 2014.
\newblock \doi{10.1103/PhysRevD.90.063504}.

\bibitem[{G{\'o}rski} et~al.(2005){G{\'o}rski}, {Hivon}, {Banday}, {Wandelt},
  {Hansen}, {Reinecke}, and {Bartelmann}]{2005ApJ...622..759G}
K.~M. {G{\'o}rski}, E.~{Hivon}, A.~J. {Banday}, B.~D. {Wandelt}, F.~K.
  {Hansen}, M.~{Reinecke}, and M.~{Bartelmann}.
\newblock {HEALPix: A Framework for High-Resolution Discretization and Fast
  Analysis of Data Distributed on the Sphere}.
\newblock \emph{\apj}, 622:\penalty0 759--771, April 2005.
\newblock \doi{10.1086/427976}.

\bibitem[{Guo} et~al.(2010){Guo}, {White}, {Li}, and
  {Boylan-Kolchin}]{2010MNRAS.404.1111G}
Q.~{Guo}, S.~{White}, C.~{Li}, and M.~{Boylan-Kolchin}.
\newblock {How do galaxies populate dark matter haloes?}
\newblock \emph{\mnras}, 404:\penalty0 1111--1120, May 2010.
\newblock \doi{10.1111/j.1365-2966.2010.16341.x}.

\bibitem[{Hill} et~al.(2016){Hill}, {Beckman}, {Chinone}, {Goeckner-Wald},
  {Hazumi}, {Keating}, {Kusaka}, {Lee}, {Matsuda}, {Plambeck}, {Suzuki}, and
  {Takakura}]{2016SPIE.9914E..2UH}
C.~A. {Hill}, S.~{Beckman}, Y.~{Chinone}, N.~{Goeckner-Wald}, M.~{Hazumi},
  B.~{Keating}, A.~{Kusaka}, A.~T. {Lee}, F.~{Matsuda}, R.~{Plambeck},
  A.~{Suzuki}, and S.~{Takakura}.
\newblock {Design and development of an ambient-temperature
  continuously-rotating achromatic half-wave plate for CMB polarization
  modulation on the POLARBEAR-2 experiment}.
\newblock In \emph{Millimeter, Submillimeter, and Far-Infrared Detectors and
  Instrumentation for Astronomy VIII}, volume 9914 of \emph{\procspie}, page
  99142U, July 2016.
\newblock \doi{10.1117/12.2232280}.

\bibitem[{Hu} and {Okamoto}(2002)]{2002ApJ...574..566H}
W.~{Hu} and T.~{Okamoto}.
\newblock {Mass Reconstruction with Cosmic Microwave Background Polarization}.
\newblock \emph{\apj}, 574:\penalty0 566--574, August 2002.
\newblock \doi{10.1086/341110}.

\bibitem[{Ishino} et~al.(2016){Ishino}, {Akiba}, {Arnold}, {Barron}, {Borrill},
  {Chendra}, {Chinone}, {Cho}, {Cukierman}, {de Haan}, {Dobbs}, {Dominjon},
  {Dotani}, {Elleflot}, {Errard}, {Fujino}, {Fuke}, {Funaki}, {Goeckner-Wald},
  {Halverson}, {Harvey}, {Hasebe}, {Hasegawa}, {Hattori}, {Hattori}, {Hazumi},
  {Hidehira}, {Hill}, {Hilton}, {Holzapfel}, {Hori}, {Hubmayr}, {Ichiki},
  {Imada}, {Inatani}, {Inoue}, {Inoue}, {Irie}, {Irwin}, {Ishitsuka}, {Jeong},
  {Kanai}, {Karatsu}, {Kashima}, {Katayama}, {Kawano}, {Kawasaki}, {Keating},
  {Kernasovskiy}, {Keskitalo}, {Kibayashi}, {Kida}, {Kimura}, {Kimura},
  {Kisner}, {Kohri}, {Komatsu}, {Komatsu}, {Kuo}, {Kuromiya}, {Kusaka}, {Lee},
  {Li}, {Linder}, {Maki}, {Matsuhara}, {Matsumura}, {Matsuoka}, {Matsuura},
  {Mima}, {Minami}, {Mitsuda}, {Nagai}, {Nagasaki}, {Nagata}, {Nakajima},
  {Nakamura}, {Namikawa}, {Naruse}, {Nishibori}, {Nishijo}, {Nishino}, {Noda},
  {Noguchi}, {Ogawa}, {Ogburn}, {Oguri}, {Ohta}, {Okada}, {Okamoto}, {Okamura},
  {Otani}, {Pisano}, {Rebeiz}, {Richards}, {Sakai}, {Sakurai}, {Sato}, {Sato},
  {Segawa}, {Sekiguchi}, {Sekimoto}, {Sekine}, {Seljak}, {Sherwin}, {Shimizu},
  {Shinozaki}, {Shu}, {Stompor}, {Sugai}, {Sugita}, {Suzuki}, {Suzuki},
  {Suzuki}, {Tajima}, {Takada}, {Takakura}, {Takano}, {Takatori}, {Takei},
  {Tanabe}, {Tomaru}, {Tomita}, {Turin}, {Uozumi}, {Utsunomiya}, {Uzawa},
  {Wada}, {Watanabe}, {Westbrook}, {Whitehorn}, {Yamada}, {Yamamoto},
  {Yamasaki}, {Yamashita}, {Yoshida}, {Yoshida}, and
  {Yotsumoto}]{2016SPIE.9904E..0XI}
H.~{Ishino}, Y.~{Akiba}, K.~{Arnold}, D.~{Barron}, J.~{Borrill}, R.~{Chendra},
  Y.~{Chinone}, S.~{Cho}, A.~{Cukierman}, T.~{de Haan}, M.~{Dobbs},
  A.~{Dominjon}, T.~{Dotani}, T.~{Elleflot}, J.~{Errard}, T.~{Fujino},
  H.~{Fuke}, T.~{Funaki}, N.~{Goeckner-Wald}, N.~{Halverson}, P.~{Harvey},
  T.~{Hasebe}, M.~{Hasegawa}, K.~{Hattori}, M.~{Hattori}, M.~{Hazumi},
  N.~{Hidehira}, C.~{Hill}, G.~{Hilton}, W.~{Holzapfel}, Y.~{Hori},
  J.~{Hubmayr}, K.~{Ichiki}, H.~{Imada}, J.~{Inatani}, M.~{Inoue}, Y.~{Inoue},
  F.~{Irie}, K.~{Irwin}, H.~{Ishitsuka}, O.~{Jeong}, H.~{Kanai}, K.~{Karatsu},
  S.~{Kashima}, N.~{Katayama}, I.~{Kawano}, T.~{Kawasaki}, B.~{Keating},
  S.~{Kernasovskiy}, R.~{Keskitalo}, A.~{Kibayashi}, Y.~{Kida}, N.~{Kimura},
  K.~{Kimura}, T.~{Kisner}, K.~{Kohri}, E.~{Komatsu}, K.~{Komatsu}, C.-L.
  {Kuo}, S.~{Kuromiya}, A.~{Kusaka}, A.~{Lee}, D.~{Li}, E.~{Linder}, M.~{Maki},
  H.~{Matsuhara}, T.~{Matsumura}, S.~{Matsuoka}, S.~{Matsuura}, S.~{Mima},
  Y.~{Minami}, K.~{Mitsuda}, M.~{Nagai}, T.~{Nagasaki}, R.~{Nagata},
  M.~{Nakajima}, S.~{Nakamura}, T.~{Namikawa}, M.~{Naruse}, T.~{Nishibori},
  K.~{Nishijo}, H.~{Nishino}, A.~{Noda}, T.~{Noguchi}, H.~{Ogawa}, W.~{Ogburn},
  S.~{Oguri}, I.~{Ohta}, N.~{Okada}, A.~{Okamoto}, T.~{Okamura}, C.~{Otani},
  G.~{Pisano}, G.~{Rebeiz}, P.~{Richards}, S.~{Sakai}, Y.~{Sakurai}, Y.~{Sato},
  N.~{Sato}, Y.~{Segawa}, S.~{Sekiguchi}, Y.~{Sekimoto}, M.~{Sekine},
  U.~{Seljak}, B.~{Sherwin}, T.~{Shimizu}, K.~{Shinozaki}, S.~{Shu},
  R.~{Stompor}, H.~{Sugai}, H.~{Sugita}, J.~{Suzuki}, T.~{Suzuki}, A.~{Suzuki},
  O.~{Tajima}, S.~{Takada}, S.~{Takakura}, K.~{Takano}, S.~{Takatori},
  Y.~{Takei}, D.~{Tanabe}, T.~{Tomaru}, N.~{Tomita}, P.~{Turin}, S.~{Uozumi},
  S.~{Utsunomiya}, Y.~{Uzawa}, T.~{Wada}, H.~{Watanabe}, B.~{Westbrook},
  N.~{Whitehorn}, Y.~{Yamada}, R.~{Yamamoto}, N.~{Yamasaki}, T.~{Yamashita},
  T.~{Yoshida}, M.~{Yoshida}, and K.~{Yotsumoto}.
\newblock {LiteBIRD: lite satellite for the study of B-mode polarization and
  inflation from cosmic microwave background radiation detection}.
\newblock In \emph{Society of Photo-Optical Instrumentation Engineers (SPIE)
  Conference Series}, volume 9904 of \emph{\procspie}, page 99040X, July 2016.
\newblock \doi{10.1117/12.2231995}.

\bibitem[{Kamionkowski} and {Kovetz}(2016)]{2016ARA&A..54..227K}
M.~{Kamionkowski} and E.~D. {Kovetz}.
\newblock {The Quest for B Modes from Inflationary Gravitational Waves}.
\newblock \emph{\araa}, 54:\penalty0 227--269, September 2016.
\newblock \doi{10.1146/annurev-astro-081915-023433}.

\bibitem[{Kamionkowski} et~al.(1997){Kamionkowski}, {Kosowsky}, and
  {Stebbins}]{1997PhRvD..55.7368K}
M.~{Kamionkowski}, A.~{Kosowsky}, and A.~{Stebbins}.
\newblock {Statistics of cosmic microwave background polarization}.
\newblock \emph{\prd}, 55:\penalty0 7368--7388, June 1997.
\newblock \doi{10.1103/PhysRevD.55.7368}.

\bibitem[{Kaplinghat} et~al.(2003){Kaplinghat}, {Knox}, and
  {Song}]{2003PhRvL..91x1301K}
M.~{Kaplinghat}, L.~{Knox}, and Y.-S. {Song}.
\newblock {Determining Neutrino Mass from the Cosmic Microwave Background
  Alone}.
\newblock \emph{Physical Review Letters}, 91\penalty0 (24):\penalty0 241301,
  December 2003.
\newblock \doi{10.1103/PhysRevLett.91.241301}.

\bibitem[{Keisler} et~al.(2015){Keisler}, {Hoover}, {Harrington}, {Henning},
  {Ade}, {Aird}, {Austermann}, {Beall}, {Bender}, {Benson}, {Bleem},
  {Carlstrom}, {Chang}, {Chiang}, {Cho}, {Citron}, {Crawford}, {Crites}, {de
  Haan}, {Dobbs}, {Everett}, {Gallicchio}, {Gao}, {George}, {Gilbert},
  {Halverson}, {Hanson}, {Hilton}, {Holder}, {Holzapfel}, {Hou}, {Hrubes},
  {Huang}, {Hubmayr}, {Irwin}, {Knox}, {Lee}, {Leitch}, {Li}, {Luong-Van},
  {Marrone}, {McMahon}, {Mehl}, {Meyer}, {Mocanu}, {Natoli}, {Nibarger},
  {Novosad}, {Padin}, {Pryke}, {Reichardt}, {Ruhl}, {Saliwanchik}, {Sayre},
  {Schaffer}, {Shirokoff}, {Smecher}, {Stark}, {Story}, {Tucker},
  {Vanderlinde}, {Vieira}, {Wang}, {Whitehorn}, {Yefremenko}, and
  {Zahn}]{2015ApJ...807..151K}
R.~{Keisler}, S.~{Hoover}, N.~{Harrington}, J.~W. {Henning}, P.~A.~R. {Ade},
  K.~A. {Aird}, J.~E. {Austermann}, J.~A. {Beall}, A.~N. {Bender}, B.~A.
  {Benson}, L.~E. {Bleem}, J.~E. {Carlstrom}, C.~L. {Chang}, H.~C. {Chiang},
  H.-M. {Cho}, R.~{Citron}, T.~M. {Crawford}, A.~T. {Crites}, T.~{de Haan},
  M.~A. {Dobbs}, W.~{Everett}, J.~{Gallicchio}, J.~{Gao}, E.~M. {George},
  A.~{Gilbert}, N.~W. {Halverson}, D.~{Hanson}, G.~C. {Hilton}, G.~P. {Holder},
  W.~L. {Holzapfel}, Z.~{Hou}, J.~D. {Hrubes}, N.~{Huang}, J.~{Hubmayr}, K.~D.
  {Irwin}, L.~{Knox}, A.~T. {Lee}, E.~M. {Leitch}, D.~{Li}, D.~{Luong-Van},
  D.~P. {Marrone}, J.~J. {McMahon}, J.~{Mehl}, S.~S. {Meyer}, L.~{Mocanu},
  T.~{Natoli}, J.~P. {Nibarger}, V.~{Novosad}, S.~{Padin}, C.~{Pryke}, C.~L.
  {Reichardt}, J.~E. {Ruhl}, B.~R. {Saliwanchik}, J.~T. {Sayre}, K.~K.
  {Schaffer}, E.~{Shirokoff}, G.~{Smecher}, A.~A. {Stark}, K.~T. {Story},
  C.~{Tucker}, K.~{Vanderlinde}, J.~D. {Vieira}, G.~{Wang}, N.~{Whitehorn},
  V.~{Yefremenko}, and O.~{Zahn}.
\newblock {Measurements of Sub-degree B-mode Polarization in the Cosmic
  Microwave Background from 100 Square Degrees of SPTpol Data}.
\newblock \emph{\apj}, 807:\penalty0 151, July 2015.
\newblock \doi{10.1088/0004-637X/807/2/151}.

\bibitem[{Kogut} et~al.(2011){Kogut}, {Fixsen}, {Chuss}, {Dotson}, {Dwek},
  {Halpern}, {Hinshaw}, {Meyer}, {Moseley}, {Seiffert}, {Spergel}, and
  {Wollack}]{2011JCAP...07..025K}
A.~{Kogut}, D.~J. {Fixsen}, D.~T. {Chuss}, J.~{Dotson}, E.~{Dwek},
  M.~{Halpern}, G.~F. {Hinshaw}, S.~M. {Meyer}, S.~H. {Moseley}, M.~D.
  {Seiffert}, D.~N. {Spergel}, and E.~J. {Wollack}.
\newblock {The Primordial Inflation Explorer (PIXIE): a nulling polarimeter for
  cosmic microwave background observations}.
\newblock \emph{\jcap}, 7:\penalty0 025, July 2011.
\newblock \doi{10.1088/1475-7516/2011/07/025}.

\bibitem[{Kogut} et~al.(2016){Kogut}, {Chluba}, {Fixsen}, {Meyer}, and
  {Spergel}]{2016SPIE.9904E..0WK}
A.~{Kogut}, J.~{Chluba}, D.~J. {Fixsen}, S.~{Meyer}, and D.~{Spergel}.
\newblock {The Primordial Inflation Explorer (PIXIE)}.
\newblock In \emph{Society of Photo-Optical Instrumentation Engineers (SPIE)
  Conference Series}, volume 9904 of \emph{\procspie}, page 99040W, July 2016.
\newblock \doi{10.1117/12.2231090}.

\bibitem[{Kurki-Suonio} et~al.(2009){Kurki-Suonio}, {Keih{\"a}nen},
  {Keskitalo}, {Poutanen}, {Sirvi{\"o}}, {Maino}, and
  {Burigana}]{2009A&A...506.1511K}
H.~{Kurki-Suonio}, E.~{Keih{\"a}nen}, R.~{Keskitalo}, T.~{Poutanen}, A.-S.
  {Sirvi{\"o}}, D.~{Maino}, and C.~{Burigana}.
\newblock {Destriping CMB temperature and polarization maps}.
\newblock \emph{\aap}, 506:\penalty0 1511--1539, November 2009.
\newblock \doi{10.1051/0004-6361/200912361}.

\bibitem[{Kusaka} et~al.(2014){Kusaka}, {Essinger-Hileman}, {Appel},
  {Gallardo}, {Irwin}, {Jarosik}, {Nolta}, {Page}, {Parker}, {Raghunathan},
  {Sievers}, {Simon}, {Staggs}, and {Visnjic}]{2014RScI...85b4501K}
A.~{Kusaka}, T.~{Essinger-Hileman}, J.~W. {Appel}, P.~{Gallardo}, K.~D.
  {Irwin}, N.~{Jarosik}, M.~R. {Nolta}, L.~A. {Page}, L.~P. {Parker},
  S.~{Raghunathan}, J.~L. {Sievers}, S.~M. {Simon}, S.~T. {Staggs}, and
  K.~{Visnjic}.
\newblock {Modulation of cosmic microwave background polarization with a warm
  rapidly rotating half-wave plate on the Atacama B-Mode Search instrument}.
\newblock \emph{Review of Scientific Instruments}, 85\penalty0 (2):\penalty0
  024501, February 2014.
\newblock \doi{10.1063/1.4862058}.

\bibitem[{Lewis} and {Challinor}(2006)]{2006PhR...429....1L}
A.~{Lewis} and A.~{Challinor}.
\newblock {Weak gravitational lensing of the CMB}.
\newblock \emph{\physrep}, 429:\penalty0 1--65, June 2006.
\newblock \doi{10.1016/j.physrep.2006.03.002}.

\bibitem[{Lyth}(1997)]{1997PhRvL..78.1861L}
D.~H. {Lyth}.
\newblock {What Would We Learn by Detecting a Gravitational Wave Signal in the
  Cosmic Microwave Background Anisotropy?}
\newblock \emph{Physical Review Letters}, 78:\penalty0 1861--1863, March 1997.
\newblock \doi{10.1103/PhysRevLett.78.1861}.

\bibitem[{Matsumura} et~al.(2014){Matsumura}, {Akiba}, {Borrill}, {Chinone},
  {Dobbs}, {Fuke}, {Ghribi}, {Hasegawa}, {Hattori}, {Hattori}, {Hazumi},
  {Holzapfel}, {Inoue}, {Ishidoshiro}, {Ishino}, {Ishitsuka}, {Karatsu},
  {Katayama}, {Kawano}, {Kibayashi}, {Kibe}, {Kimura}, {Kimura}, {Koga},
  {Kozu}, {Komatsu}, {Lee}, {Matsuhara}, {Mima}, {Mitsuda}, {Mizukami},
  {Morii}, {Morishima}, {Murayama}, {Nagai}, {Nagata}, {Nakamura}, {Naruse},
  {Natsume}, {Nishibori}, {Nishino}, {Noda}, {Noguchi}, {Ogawa}, {Oguri},
  {Ohta}, {Otani}, {Richards}, {Sakai}, {Sato}, {Sato}, {Sekimoto}, {Shimizu},
  {Shinozaki}, {Sugita}, {Suzuki}, {Suzuki}, {Tajima}, {Takada}, {Takakura},
  {Takei}, {Tomaru}, {Uzawa}, {Wada}, {Watanabe}, {Yoshida}, {Yamasaki},
  {Yoshida}, and {Yotsumoto}]{2014JLTP..176..733M}
T.~{Matsumura}, Y.~{Akiba}, J.~{Borrill}, Y.~{Chinone}, M.~{Dobbs}, H.~{Fuke},
  A.~{Ghribi}, M.~{Hasegawa}, K.~{Hattori}, M.~{Hattori}, M.~{Hazumi},
  W.~{Holzapfel}, Y.~{Inoue}, K.~{Ishidoshiro}, H.~{Ishino}, H.~{Ishitsuka},
  K.~{Karatsu}, N.~{Katayama}, I.~{Kawano}, A.~{Kibayashi}, Y.~{Kibe},
  K.~{Kimura}, N.~{Kimura}, K.~{Koga}, M.~{Kozu}, E.~{Komatsu}, A.~{Lee},
  H.~{Matsuhara}, S.~{Mima}, K.~{Mitsuda}, K.~{Mizukami}, H.~{Morii},
  T.~{Morishima}, S.~{Murayama}, M.~{Nagai}, R.~{Nagata}, S.~{Nakamura},
  M.~{Naruse}, K.~{Natsume}, T.~{Nishibori}, H.~{Nishino}, A.~{Noda},
  T.~{Noguchi}, H.~{Ogawa}, S.~{Oguri}, I.~{Ohta}, C.~{Otani}, P.~{Richards},
  S.~{Sakai}, N.~{Sato}, Y.~{Sato}, Y.~{Sekimoto}, A.~{Shimizu},
  K.~{Shinozaki}, H.~{Sugita}, T.~{Suzuki}, A.~{Suzuki}, O.~{Tajima},
  S.~{Takada}, S.~{Takakura}, Y.~{Takei}, T.~{Tomaru}, Y.~{Uzawa}, T.~{Wada},
  H.~{Watanabe}, M.~{Yoshida}, N.~{Yamasaki}, T.~{Yoshida}, and K.~{Yotsumoto}.
\newblock {Mission Design of LiteBIRD}.
\newblock \emph{Journal of Low Temperature Physics}, 176:\penalty0 733--740,
  September 2014.
\newblock \doi{10.1007/s10909-013-0996-1}.

\bibitem[{Melin} et~al.(2017){Melin}, {Bonaldi}, {Remazeilles}, {Hagstotz},
  {Diego}, {Hern{\'a}ndez-Monteagudo}, {G{\'e}nova-Santos}, {Luzzi}, {Martins},
  {Grandis}, {Mohr}, {Bartlett}, {Delabrouille}, {Ferraro}, {Tramonte},
  {Rubi{\~n}o-Mart{\'{\i}}n}, {Mac{\`i}as-P{\'e}rez}, {Ach{\'u}carro}, {Ade},
  {Allison}, {Ashdown}, {Ballardini}, {Banday}, {Banerji}, {Bartolo}, {Basak},
  {Baselmans}, {Basu}, {Battye}, {Baumann}, {Bersanelli}, {Bonato}, {Borrill},
  {Bouchet}, {Boulanger}, {Brinckmann}, {Bucher}, {Burigana}, {Buzzelli},
  {Cai}, {Calvo}, {Carvalho}, {Castellano}, {Challinor}, {Chluba}, {Clesse},
  {Colafrancesco}, {Colantoni}, {Coppolecchia}, {Crook}, {D'Alessandro}, {de
  Bernardis}, {de Gasperis}, {De Petris}, {De Zotti}, {Di Valentino}, {Errard},
  {Feeney}, {Fern{\'a}ndez-Cobos}, {Finelli}, {Forastieri}, {Galli}, {Gerbino},
  {Gonz{\'a}lez-Nuevo}, {Greenslade}, {Hanany}, {Handley}, {Hervias-Caimapo},
  {Hills}, {Hivon}, {Kiiveri}, {Kisner}, {Kitching}, {Kunz}, {Kurki-Suonio},
  {Lamagna}, {Lasenby}, {Lattanzi}, {Le Brun}, {Lesgourgues}, {Lewis},
  {Liguori}, {Lindholm}, {Lopez-Caniego}, {Maffei}, {Martinez-Gonzalez},
  {Masi}, {McCarthy}, {Melchiorri}, {Molinari}, {Monfardini}, {Natoli},
  {Negrello}, {Notari}, {Paiella}, {Paoletti}, {Patanchon}, {Piat}, {Pisano},
  {Polastri}, {Polenta}, {Pollo}, {Poulin}, {Quartin}, {Roman}, {Salvati},
  {Tartari}, {Tomasi}, {Trappe}, {Triqueneaux}, {Trombetti}, {Tucker},
  {V{\"a}liviita}, {van de Weygaert}, {Van Tent}, {Vennin}, {Vielva},
  {Vittorio}, {Weller}, {Young}, {Zannoni}, and {for the CORE
  collaboration}]{2017arXiv170310456M}
J.-B. {Melin}, A.~{Bonaldi}, M.~{Remazeilles}, S.~{Hagstotz}, J.~M. {Diego},
  C.~{Hern{\'a}ndez-Monteagudo}, R.~T. {G{\'e}nova-Santos}, G.~{Luzzi},
  C.~J.~A.~P. {Martins}, S.~{Grandis}, J.~J. {Mohr}, J.~G. {Bartlett},
  J.~{Delabrouille}, S.~{Ferraro}, D.~{Tramonte}, J.~A.
  {Rubi{\~n}o-Mart{\'{\i}}n}, J.~F. {Mac{\`i}as-P{\'e}rez}, A.~{Ach{\'u}carro},
  P.~{Ade}, R.~{Allison}, M.~{Ashdown}, M.~{Ballardini}, A.~J. {Banday},
  R.~{Banerji}, N.~{Bartolo}, S.~{Basak}, J.~{Baselmans}, K.~{Basu}, R.~A.
  {Battye}, D.~{Baumann}, M.~{Bersanelli}, M.~{Bonato}, J.~{Borrill},
  F.~{Bouchet}, F.~{Boulanger}, T.~{Brinckmann}, M.~{Bucher}, C.~{Burigana},
  A.~{Buzzelli}, Z.-Y. {Cai}, M.~{Calvo}, C.~S. {Carvalho}, M.~G. {Castellano},
  A.~{Challinor}, J.~{Chluba}, S.~{Clesse}, S.~{Colafrancesco}, I.~{Colantoni},
  A.~{Coppolecchia}, M.~{Crook}, G.~{D'Alessandro}, P.~{de Bernardis}, G.~{de
  Gasperis}, M.~{De Petris}, G.~{De Zotti}, E.~{Di Valentino}, J.~{Errard},
  S.~M. {Feeney}, R.~{Fern{\'a}ndez-Cobos}, F.~{Finelli}, F.~{Forastieri},
  S.~{Galli}, M.~{Gerbino}, J.~{Gonz{\'a}lez-Nuevo}, J.~{Greenslade},
  S.~{Hanany}, W.~{Handley}, C.~{Hervias-Caimapo}, M.~{Hills}, E.~{Hivon},
  K.~{Kiiveri}, T.~{Kisner}, T.~{Kitching}, M.~{Kunz}, H.~{Kurki-Suonio},
  L.~{Lamagna}, A.~{Lasenby}, M.~{Lattanzi}, A.~M.~C. {Le Brun},
  J.~{Lesgourgues}, A.~{Lewis}, M.~{Liguori}, V.~{Lindholm},
  M.~{Lopez-Caniego}, B.~{Maffei}, E.~{Martinez-Gonzalez}, S.~{Masi},
  D.~{McCarthy}, A.~{Melchiorri}, D.~{Molinari}, A.~{Monfardini}, P.~{Natoli},
  M.~{Negrello}, A.~{Notari}, A.~{Paiella}, D.~{Paoletti}, G.~{Patanchon},
  M.~{Piat}, G.~{Pisano}, L.~{Polastri}, G.~{Polenta}, A.~{Pollo}, V.~{Poulin},
  M.~{Quartin}, M.~{Roman}, L.~{Salvati}, A.~{Tartari}, M.~{Tomasi},
  N.~{Trappe}, S.~{Triqueneaux}, T.~{Trombetti}, C.~{Tucker},
  J.~{V{\"a}liviita}, R.~{van de Weygaert}, B.~{Van Tent}, V.~{Vennin},
  P.~{Vielva}, N.~{Vittorio}, J.~{Weller}, K.~{Young}, M.~{Zannoni}, and {for
  the CORE collaboration}.
\newblock {Exploring Cosmic Origins with CORE: Cluster Science}.
\newblock \emph{ArXiv e-prints}, March 2017.

\bibitem[{Moncelsi} et~al.(2014){Moncelsi}, {Ade}, {Angil{\`e}}, {Benton},
  {Devlin}, {Fissel}, {Gandilo}, {Gundersen}, {Matthews}, {Netterfield},
  {Novak}, {Nutter}, {Pascale}, {Poidevin}, {Savini}, {Scott}, {Soler},
  {Spencer}, {Truch}, {Tucker}, and {Zhang}]{2014MNRAS.437.2772M}
L.~{Moncelsi}, P.~A.~R. {Ade}, F.~E. {Angil{\`e}}, S.~J. {Benton}, M.~J.
  {Devlin}, L.~M. {Fissel}, N.~N. {Gandilo}, J.~O. {Gundersen}, T.~G.
  {Matthews}, C.~B. {Netterfield}, G.~{Novak}, D.~{Nutter}, E.~{Pascale},
  F.~{Poidevin}, G.~{Savini}, D.~{Scott}, J.~D. {Soler}, L.~D. {Spencer},
  M.~D.~P. {Truch}, G.~S. {Tucker}, and J.~{Zhang}.
\newblock {Empirical modelling of the BLASTPol achromatic half-wave plate for
  precision submillimetre polarimetry}.
\newblock \emph{\mnras}, 437:\penalty0 2772--2789, January 2014.
\newblock \doi{10.1093/mnras/stt2090}.

\bibitem[{Nakamura} and {Petcov}(2016)]{NakamuraPetcov2016}
K.~{Nakamura} and S.~T. {Petcov}.
\newblock {Neutrino Mass, Mixing, and Oscillations}.
\newblock In C.~{Patrignani}, editor, \emph{{Review of Particle Physics}},
  volume~40 of \emph{{Chin. Phys. C}}, page 100001, 2016.
\newblock \doi{10.1088/1674-1137/40/10/090001}.

\bibitem[{Natoli et al.}(2017)]{ECO.systematics.paper}
P.~{Natoli et al.}
\newblock Exploring cosmic origins with \coremfive: Mitigation of systematic
  effects.
\newblock \emph{to be submitted}, 2017.

\bibitem[{Niemack}(2016)]{2016ApOpt..55.1686N}
M.~D. {Niemack}.
\newblock {Designs for a large-aperture telescope to map the CMB 10 times
  faster}.
\newblock \emph{\ao}, 55:\penalty0 1686, March 2016.
\newblock \doi{10.1364/AO.55.001686}.

\bibitem[{Parker}(1979)]{1979cmft.book.....P}
E.~N. {Parker}.
\newblock \emph{{Cosmical magnetic fields: Their origin and their activity}}.
\newblock {Clarendon Press, Oxford}, 1979.

\bibitem[{Patanchon} et~al.(2008){Patanchon}, {Ade}, {Bock}, {Chapin},
  {Devlin}, {Dicker}, {Griffin}, {Gundersen}, {Halpern}, {Hargrave}, {Hughes},
  {Klein}, {Marsden}, {Martin}, {Mauskopf}, {Netterfield}, {Olmi}, {Pascale},
  {Rex}, {Scott}, {Semisch}, {Truch}, {Tucker}, {Tucker}, {Viero}, and
  {Wiebe}]{2008ApJ...681..708P}
G.~{Patanchon}, P.~A.~R. {Ade}, J.~J. {Bock}, E.~L. {Chapin}, M.~J. {Devlin},
  S.~{Dicker}, M.~{Griffin}, J.~O. {Gundersen}, M.~{Halpern}, P.~C. {Hargrave},
  D.~H. {Hughes}, J.~{Klein}, G.~{Marsden}, P.~G. {Martin}, P.~{Mauskopf},
  C.~B. {Netterfield}, L.~{Olmi}, E.~{Pascale}, M.~{Rex}, D.~{Scott},
  C.~{Semisch}, M.~D.~P. {Truch}, C.~{Tucker}, G.~S. {Tucker}, M.~P. {Viero},
  and D.~V. {Wiebe}.
\newblock {SANEPIC: A Mapmaking Method for Time Stream Data from Large Arrays}.
\newblock \emph{\apj}, 681:\penalty0 708-725, July 2008.
\newblock \doi{10.1086/588543}.

\bibitem[{Planck Collaboration}(2016{\natexlab{a}})]{2016A&A...586A.136P}
{Planck Collaboration}.
\newblock {Planck intermediate results. XXXIII. Signature of the magnetic field
  geometry of interstellar filaments in dust polarization maps}.
\newblock \emph{\aap}, 586:\penalty0 A136, February 2016{\natexlab{a}}.
\newblock \doi{10.1051/0004-6361/201425305}.

\bibitem[{Planck Collaboration}(2016{\natexlab{b}})]{2016A&A...586A.137P}
{Planck Collaboration}.
\newblock {Planck intermediate results. XXXIV. The magnetic field structure in
  the Rosette Nebula}.
\newblock \emph{\aap}, 586:\penalty0 A137, February 2016{\natexlab{b}}.
\newblock \doi{10.1051/0004-6361/201525616}.

\bibitem[{Planck Collaboration}(2016{\natexlab{c}})]{2016A&A...586A.138P}
{Planck Collaboration}.
\newblock {Planck intermediate results. XXXV. Probing the role of the magnetic
  field in the formation of structure in molecular clouds}.
\newblock \emph{\aap}, 586:\penalty0 A138, February 2016{\natexlab{c}}.
\newblock \doi{10.1051/0004-6361/201525896}.

\bibitem[{Planck Collaboration}(2016{\natexlab{d}})]{2016A&A...586A.141P}
{Planck Collaboration}.
\newblock {Planck intermediate results. XXXVIII. E- and B-modes of dust
  polarization from the magnetized filamentary structure of the interstellar
  medium}.
\newblock \emph{\aap}, 586:\penalty0 A141, February 2016{\natexlab{d}}.
\newblock \doi{10.1051/0004-6361/201526506}.

\bibitem[{Planck Collaboration}(2016{\natexlab{e}})]{2016A&A...596A.103P}
{Planck Collaboration}.
\newblock {Planck intermediate results. XLII. Large-scale Galactic magnetic
  fields}.
\newblock \emph{\aap}, 596:\penalty0 A103, December 2016{\natexlab{e}}.
\newblock \doi{10.1051/0004-6361/201528033}.

\bibitem[{Planck Collaboration}(2016{\natexlab{f}})]{2016A&A...596A.105P}
{Planck Collaboration}.
\newblock {Planck intermediate results. XLIV. Structure of the Galactic
  magnetic field from dust polarization maps of the southern Galactic cap}.
\newblock \emph{\aap}, 596:\penalty0 A105, December 2016{\natexlab{f}}.
\newblock \doi{10.1051/0004-6361/201628636}.

\bibitem[{Planck Collaboration}(2016{\natexlab{g}})]{2016arXiv160802487P}
{Planck Collaboration}.
\newblock {Planck 2016 intermediate results. LI. Features in the cosmic
  microwave background temperature power spectrum and shifts in cosmological
  parameters}.
\newblock \emph{ArXiv e-prints}, August 2016{\natexlab{g}}.

\bibitem[{Planck Collaboration} et~al.(2011){Planck Collaboration}, {Ade},
  {Aghanim}, {Arnaud}, {Ashdown}, {Aumont}, {Baccigalupi}, {Baker}, {Balbi},
  {Banday}, and et~al.]{2011A&A...536A...2P}
{Planck Collaboration}, P.~A.~R. {Ade}, N.~{Aghanim}, M.~{Arnaud},
  M.~{Ashdown}, J.~{Aumont}, C.~{Baccigalupi}, M.~{Baker}, A.~{Balbi}, A.~J.
  {Banday}, and et~al.
\newblock {Planck early results. II. The thermal performance of Planck}.
\newblock \emph{\aap}, 536:\penalty0 A2, December 2011.
\newblock \doi{10.1051/0004-6361/201116486}.

\bibitem[{Planck Collaboration} et~al.(2014{\natexlab{a}}){Planck
  Collaboration}, {Ade}, {Aghanim}, {Alves}, {Armitage-Caplan}, {Arnaud},
  {Ashdown}, {Atrio-Barandela}, {Aumont}, {Baccigalupi}, and
  et~al.]{2014A&A...571A..13P}
{Planck Collaboration}, P.~A.~R. {Ade}, N.~{Aghanim}, M.~I.~R. {Alves},
  C.~{Armitage-Caplan}, M.~{Arnaud}, M.~{Ashdown}, F.~{Atrio-Barandela},
  J.~{Aumont}, C.~{Baccigalupi}, and et~al.
\newblock {Planck 2013 results. XIII. Galactic CO emission}.
\newblock \emph{\aap}, 571:\penalty0 A13, November 2014{\natexlab{a}}.
\newblock \doi{10.1051/0004-6361/201321553}.

\bibitem[{Planck Collaboration} et~al.(2014{\natexlab{b}}){Planck
  Collaboration}, {Ade}, {Aghanim}, {Armitage-Caplan}, {Arnaud}, {Ashdown},
  {Atrio-Barandela}, {Aumont}, {Baccigalupi}, {Banday}, and
  et~al.]{2014A&A...571A..10P}
{Planck Collaboration}, P.~A.~R. {Ade}, N.~{Aghanim}, C.~{Armitage-Caplan},
  M.~{Arnaud}, M.~{Ashdown}, F.~{Atrio-Barandela}, J.~{Aumont},
  C.~{Baccigalupi}, A.~J. {Banday}, and et~al.
\newblock {Planck 2013 results. X. HFI energetic particle effects:
  characterization, removal, and simulation}.
\newblock \emph{\aap}, 571:\penalty0 A10, November 2014{\natexlab{b}}.
\newblock \doi{10.1051/0004-6361/201321577}.

\bibitem[{Planck Collaboration} et~al.(2014{\natexlab{c}}){Planck
  Collaboration}, {Ade}, {Aghanim}, {Armitage-Caplan}, {Arnaud}, {Ashdown},
  {Atrio-Barandela}, {Aumont}, {Baccigalupi}, {Banday}, and
  et~al.]{2014A&A...571A..14P}
{Planck Collaboration}, P.~A.~R. {Ade}, N.~{Aghanim}, C.~{Armitage-Caplan},
  M.~{Arnaud}, M.~{Ashdown}, F.~{Atrio-Barandela}, J.~{Aumont},
  C.~{Baccigalupi}, A.~J. {Banday}, and et~al.
\newblock {Planck 2013 results. XIV. Zodiacal emission}.
\newblock \emph{\aap}, 571:\penalty0 A14, November 2014{\natexlab{c}}.
\newblock \doi{10.1051/0004-6361/201321562}.

\bibitem[{Planck Collaboration} et~al.(2014{\natexlab{d}}){Planck
  Collaboration}, {Ade}, {Aghanim}, {Armitage-Caplan}, {Arnaud}, {Ashdown},
  {Atrio-Barandela}, {Aumont}, {Baccigalupi}, {Banday}, and
  et~al.]{2014A&A...571A..16P}
{Planck Collaboration}, P.~A.~R. {Ade}, N.~{Aghanim}, C.~{Armitage-Caplan},
  M.~{Arnaud}, M.~{Ashdown}, F.~{Atrio-Barandela}, J.~{Aumont},
  C.~{Baccigalupi}, A.~J. {Banday}, and et~al.
\newblock {Planck 2013 results. XVI. Cosmological parameters}.
\newblock \emph{\aap}, 571:\penalty0 A16, November 2014{\natexlab{d}}.
\newblock \doi{10.1051/0004-6361/201321591}.

\bibitem[{Planck Collaboration} et~al.(2014{\natexlab{e}}){Planck
  Collaboration}, {Ade}, {Aghanim}, {Armitage-Caplan}, {Arnaud}, {Ashdown},
  {Atrio-Barandela}, {Aumont}, {Baccigalupi}, {Banday}, and
  et~al.]{2014A&A...571A..20P}
{Planck Collaboration}, P.~A.~R. {Ade}, N.~{Aghanim}, C.~{Armitage-Caplan},
  M.~{Arnaud}, M.~{Ashdown}, F.~{Atrio-Barandela}, J.~{Aumont},
  C.~{Baccigalupi}, A.~J. {Banday}, and et~al.
\newblock {Planck 2013 results. XX. Cosmology from Sunyaev-Zeldovich cluster
  counts}.
\newblock \emph{\aap}, 571:\penalty0 A20, November 2014{\natexlab{e}}.
\newblock \doi{10.1051/0004-6361/201321521}.

\bibitem[{Planck Collaboration} et~al.(2014{\natexlab{f}}){Planck
  Collaboration}, {Ade}, {Aghanim}, {Armitage-Caplan}, {Arnaud}, {Ashdown},
  {Atrio-Barandela}, {Aumont}, {Baccigalupi}, {Banday}, and
  et~al.]{2014A&A...571A..22P}
{Planck Collaboration}, P.~A.~R. {Ade}, N.~{Aghanim}, C.~{Armitage-Caplan},
  M.~{Arnaud}, M.~{Ashdown}, F.~{Atrio-Barandela}, J.~{Aumont},
  C.~{Baccigalupi}, A.~J. {Banday}, and et~al.
\newblock {Planck 2013 results. XXII. Constraints on inflation}.
\newblock \emph{\aap}, 571:\penalty0 A22, November 2014{\natexlab{f}}.
\newblock \doi{10.1051/0004-6361/201321569}.

\bibitem[{Planck Collaboration} et~al.(2014{\natexlab{g}}){Planck
  Collaboration}, {Ade}, {Aghanim}, {Armitage-Caplan}, {Arnaud}, {Ashdown},
  {Atrio-Barandela}, {Aumont}, {Baccigalupi}, {Banday}, and
  et~al.]{2014A&A...571A..23P}
{Planck Collaboration}, P.~A.~R. {Ade}, N.~{Aghanim}, C.~{Armitage-Caplan},
  M.~{Arnaud}, M.~{Ashdown}, F.~{Atrio-Barandela}, J.~{Aumont},
  C.~{Baccigalupi}, A.~J. {Banday}, and et~al.
\newblock {Planck 2013 results. XXIII. Isotropy and statistics of the CMB}.
\newblock \emph{\aap}, 571:\penalty0 A23, November 2014{\natexlab{g}}.
\newblock \doi{10.1051/0004-6361/201321534}.

\bibitem[{Planck Collaboration} et~al.(2016{\natexlab{a}}){Planck
  Collaboration}, {Adam}, {Ade}, {Aghanim}, {Akrami}, {Alves}, {Arg{\"u}eso},
  {Arnaud}, {Arroja}, {Ashdown}, and et~al.]{2016A&A...594A...1P}
{Planck Collaboration}, R.~{Adam}, P.~A.~R. {Ade}, N.~{Aghanim}, Y.~{Akrami},
  M.~I.~R. {Alves}, F.~{Arg{\"u}eso}, M.~{Arnaud}, F.~{Arroja}, M.~{Ashdown},
  and et~al.
\newblock {Planck 2015 results. I. Overview of products and scientific
  results}.
\newblock \emph{\aap}, 594:\penalty0 A1, September 2016{\natexlab{a}}.
\newblock \doi{10.1051/0004-6361/201527101}.

\bibitem[{Planck Collaboration} et~al.(2016{\natexlab{b}}){Planck
  Collaboration}, {Adam}, {Ade}, {Aghanim}, {Alves}, {Arnaud}, {Arzoumanian},
  {Ashdown}, {Aumont}, {Baccigalupi}, and et~al.]{2016A&A...586A.135P}
{Planck Collaboration}, R.~{Adam}, P.~A.~R. {Ade}, N.~{Aghanim}, M.~I.~R.
  {Alves}, M.~{Arnaud}, D.~{Arzoumanian}, M.~{Ashdown}, J.~{Aumont},
  C.~{Baccigalupi}, and et~al.
\newblock {Planck intermediate results. XXXII. The relative orientation between
  the magnetic field and structures traced by interstellar dust}.
\newblock \emph{\aap}, 586:\penalty0 A135, February 2016{\natexlab{b}}.
\newblock \doi{10.1051/0004-6361/201425044}.

\bibitem[{Planck Collaboration} et~al.(2016{\natexlab{c}}){Planck
  Collaboration}, {Adam}, {Ade}, {Aghanim}, {Arnaud}, {Aumont}, {Baccigalupi},
  {Banday}, {Barreiro}, {Bartlett}, and et~al.]{2016A&A...586A.133P}
{Planck Collaboration}, R.~{Adam}, P.~A.~R. {Ade}, N.~{Aghanim}, M.~{Arnaud},
  J.~{Aumont}, C.~{Baccigalupi}, A.~J. {Banday}, R.~B. {Barreiro}, J.~G.
  {Bartlett}, and et~al.
\newblock {Planck intermediate results. XXX. The angular power spectrum of
  polarized dust emission at intermediate and high Galactic latitudes}.
\newblock \emph{\aap}, 586:\penalty0 A133, February 2016{\natexlab{c}}.
\newblock \doi{10.1051/0004-6361/201425034}.

\bibitem[{Planck Collaboration} et~al.(2016{\natexlab{d}}){Planck
  Collaboration}, {Ade}, {Aghanim}, {Arnaud}, {Arroja}, {Ashdown}, {Aumont},
  {Baccigalupi}, {Ballardini}, {Banday}, and et~al.]{2016A&A...594A..20P}
{Planck Collaboration}, P.~A.~R. {Ade}, N.~{Aghanim}, M.~{Arnaud}, F.~{Arroja},
  M.~{Ashdown}, J.~{Aumont}, C.~{Baccigalupi}, M.~{Ballardini}, A.~J. {Banday},
  and et~al.
\newblock {Planck 2015 results. XX. Constraints on inflation}.
\newblock \emph{\aap}, 594:\penalty0 A20, September 2016{\natexlab{d}}.
\newblock \doi{10.1051/0004-6361/201525898}.

\bibitem[{Planck Collaboration} et~al.(2016{\natexlab{e}}){Planck
  Collaboration}, {Ade}, {Aghanim}, {Arnaud}, {Ashdown}, {Aumont},
  {Baccigalupi}, {Banday}, {Barreiro}, {Bartlett}, and
  et~al.]{2016A&A...594A..13P}
{Planck Collaboration}, P.~A.~R. {Ade}, N.~{Aghanim}, M.~{Arnaud},
  M.~{Ashdown}, J.~{Aumont}, C.~{Baccigalupi}, A.~J. {Banday}, R.~B.
  {Barreiro}, J.~G. {Bartlett}, and et~al.
\newblock {Planck 2015 results. XIII. Cosmological parameters}.
\newblock \emph{\aap}, 594:\penalty0 A13, September 2016{\natexlab{e}}.
\newblock \doi{10.1051/0004-6361/201525830}.

\bibitem[{Planck Collaboration} et~al.(2016{\natexlab{f}}){Planck
  Collaboration}, {Ade}, {Aghanim}, {Arnaud}, {Ashdown}, {Aumont},
  {Baccigalupi}, {Banday}, {Barreiro}, {Bartlett}, and
  et~al.]{2016A&A...594A..15P}
{Planck Collaboration}, P.~A.~R. {Ade}, N.~{Aghanim}, M.~{Arnaud},
  M.~{Ashdown}, J.~{Aumont}, C.~{Baccigalupi}, A.~J. {Banday}, R.~B.
  {Barreiro}, J.~G. {Bartlett}, and et~al.
\newblock {Planck 2015 results. XV. Gravitational lensing}.
\newblock \emph{\aap}, 594:\penalty0 A15, September 2016{\natexlab{f}}.
\newblock \doi{10.1051/0004-6361/201525941}.

\bibitem[{Planck Collaboration} et~al.(2016{\natexlab{g}}){Planck
  Collaboration}, {Ade}, {Aghanim}, {Arnaud}, {Ashdown}, {Aumont},
  {Baccigalupi}, {Banday}, {Barreiro}, {Bartlett}, and
  et~al.]{2016A&A...594A..24P}
{Planck Collaboration}, P.~A.~R. {Ade}, N.~{Aghanim}, M.~{Arnaud},
  M.~{Ashdown}, J.~{Aumont}, C.~{Baccigalupi}, A.~J. {Banday}, R.~B.
  {Barreiro}, J.~G. {Bartlett}, and et~al.
\newblock {Planck 2015 results. XXIV. Cosmology from Sunyaev-Zeldovich cluster
  counts}.
\newblock \emph{\aap}, 594:\penalty0 A24, September 2016{\natexlab{g}}.
\newblock \doi{10.1051/0004-6361/201525833}.

\bibitem[{Planck Collaboration} et~al.(2016{\natexlab{h}}){Planck
  Collaboration}, {Aghanim}, {Ashdown}, {Aumont}, {Baccigalupi}, {Ballardini},
  {Banday}, {Barreiro}, {Bartolo}, {Basak}, {Benabed}, {Bernard}, {Bersanelli},
  {Bielewicz}, {Bonavera}, {Bond}, {Borrill}, {Bouchet}, {Boulanger},
  {Burigana}, {Calabrese}, {Cardoso}, {Carron}, {Chiang}, {Colombo}, {Comis},
  {Couchot}, {Coulais}, {Crill}, {Curto}, {Cuttaia}, {de Bernardis}, {de
  Zotti}, {Delabrouille}, {Di Valentino}, {Dickinson}, {Diego}, {Dor{\'e}},
  {Douspis}, {Ducout}, {Dupac}, {Dusini}, {Elsner}, {En{\ss}lin}, {Eriksen},
  {Falgarone}, {Fantaye}, {Finelli}, {Forastieri}, {Frailis}, {Fraisse},
  {Franceschi}, {Frolov}, {Galeotta}, {Galli}, {Ganga}, {G{\'e}nova-Santos},
  {Gerbino}, {Ghosh}, {Giraud-H{\'e}raud}, {Gonz{\'a}lez-Nuevo}, {G{\'o}rski},
  {Gruppuso}, {Gudmundsson}, {Hansen}, {Helou}, {Henrot-Versill{\'e}},
  {Herranz}, {Hivon}, {Huang}, {Jaffe}, {Jones}, {Keih{\"a}nen}, {Keskitalo},
  {Kiiveri}, {Kisner}, {Krachmalnicoff}, {Kunz}, {Kurki-Suonio}, {Lamarre},
  {Langer}, {Lasenby}, {Lattanzi}, {Lawrence}, {Le Jeune}, {Levrier}, {Lilje},
  {Lilley}, {Lindholm}, {L{\'o}pez-Caniego}, {Ma}, {Mac{\'{\i}}as-P{\'e}rez},
  {Maggio}, {Maino}, {Mandolesi}, {Mangilli}, {Maris}, {Martin},
  {Mart{\'{\i}}nez-Gonz{\'a}lez}, {Matarrese}, {Mauri}, {McEwen}, {Melchiorri},
  {Mennella}, {Migliaccio}, {Miville-Desch{\^e}nes}, {Molinari}, {Moneti},
  {Montier}, {Morgante}, {Moss}, {Natoli}, {Oxborrow}, {Pagano}, {Paoletti},
  {Patanchon}, {Perdereau}, {Perotto}, {Pettorino}, {Piacentini},
  {Plaszczynski}, {Polastri}, {Polenta}, {Puget}, {Rachen}, {Racine},
  {Reinecke}, {Remazeilles}, {Renzi}, {Rocha}, {Rosset}, {Rossetti}, {Roudier},
  {Rubi{\~n}o-Mart{\'{\i}}n}, {Ruiz-Granados}, {Salvati}, {Sandri},
  {Savelainen}, {Scott}, {Sirignano}, {Sirri}, {Soler}, {Spencer}, {Suur-Uski},
  {Tauber}, {Tavagnacco}, {Tenti}, {Toffolatti}, {Tomasi}, {Tristram},
  {Trombetti}, {Valiviita}, {Van Tent}, {Vielva}, {Villa}, {Vittorio},
  {Wandelt}, {Wehus}, {Zacchei}, and {Zonca}]{2016A&A...596A.109P}
{Planck Collaboration}, N.~{Aghanim}, M.~{Ashdown}, J.~{Aumont},
  C.~{Baccigalupi}, M.~{Ballardini}, A.~J. {Banday}, R.~B. {Barreiro},
  N.~{Bartolo}, S.~{Basak}, K.~{Benabed}, J.-P. {Bernard}, M.~{Bersanelli},
  P.~{Bielewicz}, L.~{Bonavera}, J.~R. {Bond}, J.~{Borrill}, F.~R. {Bouchet},
  F.~{Boulanger}, C.~{Burigana}, E.~{Calabrese}, J.-F. {Cardoso}, J.~{Carron},
  H.~C. {Chiang}, L.~P.~L. {Colombo}, B.~{Comis}, F.~{Couchot}, A.~{Coulais},
  B.~P. {Crill}, A.~{Curto}, F.~{Cuttaia}, P.~{de Bernardis}, G.~{de Zotti},
  J.~{Delabrouille}, E.~{Di Valentino}, C.~{Dickinson}, J.~M. {Diego},
  O.~{Dor{\'e}}, M.~{Douspis}, A.~{Ducout}, X.~{Dupac}, S.~{Dusini},
  F.~{Elsner}, T.~A. {En{\ss}lin}, H.~K. {Eriksen}, E.~{Falgarone},
  Y.~{Fantaye}, F.~{Finelli}, F.~{Forastieri}, M.~{Frailis}, A.~A. {Fraisse},
  E.~{Franceschi}, A.~{Frolov}, S.~{Galeotta}, S.~{Galli}, K.~{Ganga}, R.~T.
  {G{\'e}nova-Santos}, M.~{Gerbino}, T.~{Ghosh}, Y.~{Giraud-H{\'e}raud},
  J.~{Gonz{\'a}lez-Nuevo}, K.~M. {G{\'o}rski}, A.~{Gruppuso}, J.~E.
  {Gudmundsson}, F.~K. {Hansen}, G.~{Helou}, S.~{Henrot-Versill{\'e}},
  D.~{Herranz}, E.~{Hivon}, Z.~{Huang}, A.~H. {Jaffe}, W.~C. {Jones},
  E.~{Keih{\"a}nen}, R.~{Keskitalo}, K.~{Kiiveri}, T.~S. {Kisner},
  N.~{Krachmalnicoff}, M.~{Kunz}, H.~{Kurki-Suonio}, J.-M. {Lamarre},
  M.~{Langer}, A.~{Lasenby}, M.~{Lattanzi}, C.~R. {Lawrence}, M.~{Le Jeune},
  F.~{Levrier}, P.~B. {Lilje}, M.~{Lilley}, V.~{Lindholm},
  M.~{L{\'o}pez-Caniego}, Y.-Z. {Ma}, J.~F. {Mac{\'{\i}}as-P{\'e}rez},
  G.~{Maggio}, D.~{Maino}, N.~{Mandolesi}, A.~{Mangilli}, M.~{Maris}, P.~G.
  {Martin}, E.~{Mart{\'{\i}}nez-Gonz{\'a}lez}, S.~{Matarrese}, N.~{Mauri},
  J.~D. {McEwen}, A.~{Melchiorri}, A.~{Mennella}, M.~{Migliaccio}, M.-A.
  {Miville-Desch{\^e}nes}, D.~{Molinari}, A.~{Moneti}, L.~{Montier},
  G.~{Morgante}, A.~{Moss}, P.~{Natoli}, C.~A. {Oxborrow}, L.~{Pagano},
  D.~{Paoletti}, G.~{Patanchon}, O.~{Perdereau}, L.~{Perotto}, V.~{Pettorino},
  F.~{Piacentini}, S.~{Plaszczynski}, L.~{Polastri}, G.~{Polenta}, J.-L.
  {Puget}, J.~P. {Rachen}, B.~{Racine}, M.~{Reinecke}, M.~{Remazeilles},
  A.~{Renzi}, G.~{Rocha}, C.~{Rosset}, M.~{Rossetti}, G.~{Roudier}, J.~A.
  {Rubi{\~n}o-Mart{\'{\i}}n}, B.~{Ruiz-Granados}, L.~{Salvati}, M.~{Sandri},
  M.~{Savelainen}, D.~{Scott}, C.~{Sirignano}, G.~{Sirri}, J.~D. {Soler}, L.~D.
  {Spencer}, A.-S. {Suur-Uski}, J.~A. {Tauber}, D.~{Tavagnacco}, M.~{Tenti},
  L.~{Toffolatti}, M.~{Tomasi}, M.~{Tristram}, T.~{Trombetti}, J.~{Valiviita},
  F.~{Van Tent}, P.~{Vielva}, F.~{Villa}, N.~{Vittorio}, B.~D. {Wandelt}, I.~K.
  {Wehus}, A.~{Zacchei}, and A.~{Zonca}.
\newblock {Planck intermediate results. XLVIII. Disentangling Galactic dust
  emission and cosmic infrared background anisotropies}.
\newblock \emph{\aap}, 596:\penalty0 A109, December 2016{\natexlab{h}}.
\newblock \doi{10.1051/0004-6361/201629022}.

\bibitem[{Planck HFI Core Team} et~al.(2011){Planck HFI Core Team}, {Ade},
  {Aghanim}, {Ansari}, {Arnaud}, {Ashdown}, {Aumont}, {Banday}, {Bartelmann},
  {Bartlett}, {Battaner}, {Benabed}, {Beno{\^i}t}, {Bernard}, {Bersanelli},
  {Bhatia}, {Bock}, {Bond}, {Borrill}, {Bouchet}, {Boulanger}, {Bradshaw},
  {Br{\'e}elle}, {Bucher}, {Camus}, {Cardoso}, {Catalano}, {Challinor},
  {Chamballu}, {Charra}, {Charra}, {Chary}, {Chiang}, {Church}, {Clements},
  {Colombi}, {Couchot}, {Coulais}, {Cressiot}, {Crill}, {Crook}, {de
  Bernardis}, {Delabrouille}, {Delouis}, {D{\'e}sert}, {Dolag}, {Dole},
  {Dor{\'e}}, {Douspis}, {Efstathiou}, {Eng}, {Filliard}, {Forni}, {Fosalba},
  {Fourmond}, {Ganga}, {Giard}, {Girard}, {Giraud-H{\'e}raud}, {Gispert},
  {G{\'o}rski}, {Gratton}, {Griffin}, {Guyot}, {Haissinski}, {Harrison},
  {Helou}, {Henrot-Versill{\'e}}, {Hern{\'a}ndez-Monteagudo}, {Hildebrandt},
  {Hills}, {Hivon}, {Hobson}, {Holmes}, {Huffenberger}, {Jaffe}, {Jones},
  {Kaplan}, {Kneissl}, {Knox}, {Lagache}, {Lamarre}, {Lami}, {Lange},
  {Lasenby}, {Lavabre}, {Lawrence}, {Leriche}, {Leroy}, {Longval},
  {Mac{\'{\i}}as-P{\'e}rez}, {Maciaszek}, {MacTavish}, {Maffei}, {Mandolesi},
  {Mann}, {Mansoux}, {Masi}, {Matsumura}, {McGehee}, {Melin}, {Mercier},
  {Miville-Desch{\^e}nes}, {Moneti}, {Montier}, {Mortlock}, {Murphy}, {Nati},
  {Netterfield}, {N{\o}rgaard-Nielsen}, {North}, {Noviello}, {Novikov},
  {Osborne}, {Paine}, {Pajot}, {Patanchon}, {Peacocke}, {Pearson}, {Perdereau},
  {Perotto}, {Piacentini}, {Piat}, {Plaszczynski}, {Pointecouteau}, {Pons},
  {Ponthieu}, {Pr{\'e}zeau}, {Prunet}, {Puget}, {Reach}, {Renault},
  {Ristorcelli}, {Rocha}, {Rosset}, {Roudier}, {Rowan-Robinson}, {Rusholme},
  {Santos}, {Savini}, {Schaefer}, {Shellard}, {Spencer}, {Starck}, {Stassi},
  {Stolyarov}, {Stompor}, {Sudiwala}, {Sunyaev}, {Sygnet}, {Tauber}, {Thum},
  {Torre}, {Touze}, {Tristram}, {van Leeuwen}, {Vibert}, {Vibert}, {Wade},
  {Wandelt}, {White}, {Wiesemeyer}, {Woodcraft}, {Yurchenko}, {Yvon}, and
  {Zacchei}]{2011A&A...536A...4P}
{Planck HFI Core Team}, P.~A.~R. {Ade}, N.~{Aghanim}, R.~{Ansari}, M.~{Arnaud},
  M.~{Ashdown}, J.~{Aumont}, A.~J. {Banday}, M.~{Bartelmann}, J.~G. {Bartlett},
  E.~{Battaner}, K.~{Benabed}, A.~{Beno{\^i}t}, J.-P. {Bernard},
  M.~{Bersanelli}, R.~{Bhatia}, J.~J. {Bock}, J.~R. {Bond}, J.~{Borrill}, F.~R.
  {Bouchet}, F.~{Boulanger}, T.~{Bradshaw}, E.~{Br{\'e}elle}, M.~{Bucher},
  P.~{Camus}, J.-F. {Cardoso}, A.~{Catalano}, A.~{Challinor}, A.~{Chamballu},
  J.~{Charra}, M.~{Charra}, R.-R. {Chary}, C.~{Chiang}, S.~{Church}, D.~L.
  {Clements}, S.~{Colombi}, F.~{Couchot}, A.~{Coulais}, C.~{Cressiot}, B.~P.
  {Crill}, M.~{Crook}, P.~{de Bernardis}, J.~{Delabrouille}, J.-M. {Delouis},
  F.-X. {D{\'e}sert}, K.~{Dolag}, H.~{Dole}, O.~{Dor{\'e}}, M.~{Douspis},
  G.~{Efstathiou}, P.~{Eng}, C.~{Filliard}, O.~{Forni}, P.~{Fosalba}, J.-J.
  {Fourmond}, K.~{Ganga}, M.~{Giard}, D.~{Girard}, Y.~{Giraud-H{\'e}raud},
  R.~{Gispert}, K.~M. {G{\'o}rski}, S.~{Gratton}, M.~{Griffin}, G.~{Guyot},
  J.~{Haissinski}, D.~{Harrison}, G.~{Helou}, S.~{Henrot-Versill{\'e}},
  C.~{Hern{\'a}ndez-Monteagudo}, S.~R. {Hildebrandt}, R.~{Hills}, E.~{Hivon},
  M.~{Hobson}, W.~A. {Holmes}, K.~M. {Huffenberger}, A.~H. {Jaffe}, W.~C.
  {Jones}, J.~{Kaplan}, R.~{Kneissl}, L.~{Knox}, G.~{Lagache}, J.-M. {Lamarre},
  P.~{Lami}, A.~E. {Lange}, A.~{Lasenby}, A.~{Lavabre}, C.~R. {Lawrence},
  B.~{Leriche}, C.~{Leroy}, Y.~{Longval}, J.~F. {Mac{\'{\i}}as-P{\'e}rez},
  T.~{Maciaszek}, C.~J. {MacTavish}, B.~{Maffei}, N.~{Mandolesi}, R.~{Mann},
  B.~{Mansoux}, S.~{Masi}, T.~{Matsumura}, P.~{McGehee}, J.-B. {Melin},
  C.~{Mercier}, M.-A. {Miville-Desch{\^e}nes}, A.~{Moneti}, L.~{Montier},
  D.~{Mortlock}, A.~{Murphy}, F.~{Nati}, C.~B. {Netterfield}, H.~U.
  {N{\o}rgaard-Nielsen}, C.~{North}, F.~{Noviello}, D.~{Novikov}, S.~{Osborne},
  C.~{Paine}, F.~{Pajot}, G.~{Patanchon}, T.~{Peacocke}, T.~J. {Pearson},
  O.~{Perdereau}, L.~{Perotto}, F.~{Piacentini}, M.~{Piat}, S.~{Plaszczynski},
  E.~{Pointecouteau}, R.~{Pons}, N.~{Ponthieu}, G.~{Pr{\'e}zeau}, S.~{Prunet},
  J.-L. {Puget}, W.~T. {Reach}, C.~{Renault}, I.~{Ristorcelli}, G.~{Rocha},
  C.~{Rosset}, G.~{Roudier}, M.~{Rowan-Robinson}, B.~{Rusholme}, D.~{Santos},
  G.~{Savini}, B.~M. {Schaefer}, P.~{Shellard}, L.~{Spencer}, J.-L. {Starck},
  P.~{Stassi}, V.~{Stolyarov}, R.~{Stompor}, R.~{Sudiwala}, R.~{Sunyaev}, J.-F.
  {Sygnet}, J.~A. {Tauber}, C.~{Thum}, J.-P. {Torre}, F.~{Touze},
  M.~{Tristram}, F.~{van Leeuwen}, L.~{Vibert}, D.~{Vibert}, L.~A. {Wade},
  B.~D. {Wandelt}, S.~D.~M. {White}, H.~{Wiesemeyer}, A.~{Woodcraft},
  V.~{Yurchenko}, D.~{Yvon}, and A.~{Zacchei}.
\newblock {Planck early results. IV. First assessment of the High Frequency
  Instrument in-flight performance}.
\newblock \emph{\aap}, 536:\penalty0 A4, December 2011.
\newblock \doi{10.1051/0004-6361/201116487}.

\bibitem[{PRISM Collaboration} et~al.(2013){PRISM Collaboration}, {Andre},
  {Baccigalupi}, {Barbosa}, {Bartlett}, {Bartolo}, {Battistelli}, {Battye},
  {Bendo}, {Bernard}, {Bersanelli}, {Bethermin}, {Bielewicz}, {Bonaldi},
  {Bouchet}, {Boulanger}, {Brand}, {Bucher}, {Burigana}, {Cai}, {Casasola},
  {Castex}, {Challinor}, {Chluba}, {Colafrancesco}, {Cuttaia}, {D'Alessandro},
  {Davis}, {de Avillez}, {de Bernardis}, {de Petris}, {de Rosa}, {de Zotti},
  {Delabrouille}, {Dickinson}, {Diego}, {Falgarone}, {Ferreira}, {Ferriere},
  {Finelli}, {Fletcher}, {Fuller}, {Galli}, {Ganga}, {Garcia-Bellido},
  {Ghribi}, {Gonzalez-Nuevo}, {Grainge}, {Gruppuso}, {Hall},
  {Hernandez-Monteagudo}, {Jackson}, {Jaffe}, {Khatri}, {Lamagna}, {Lattanzi},
  {Leahy}, {Liguori}, {Liuzzo}, {Lopez-Caniego}, {Macias-Perez}, {Maffei},
  {Maino}, {Masi}, {Mangilli}, {Massardi}, {Matarrese}, {Melchiorri}, {Melin},
  {Mennella}, {Mignano}, {Miville-Deschenes}, {Nati}, {Natoli}, {Negrello},
  {Noviello}, {Paci}, {Paladino}, {Paoletti}, {Perrotta}, {Piacentini}, {Piat},
  {Piccirillo}, {Pisano}, {Polenta}, {Ricciardi}, {Roman}, {Rubino-Martin},
  {Salatino}, {Schillaci}, {Shellard}, {Silk}, {Stompor}, {Sunyaev}, {Tartari},
  {Terenzi}, {Toffolatti}, {Tomasi}, {Trombetti}, {Tucci}, {Van Tent}, {Verde},
  {Wandelt}, and {Withington}]{2013arXiv1306.2259P}
{PRISM Collaboration}, P.~{Andre}, C.~{Baccigalupi}, D.~{Barbosa},
  J.~{Bartlett}, N.~{Bartolo}, E.~{Battistelli}, R.~{Battye}, G.~{Bendo}, J.-P.
  {Bernard}, M.~{Bersanelli}, M.~{Bethermin}, P.~{Bielewicz}, A.~{Bonaldi},
  F.~{Bouchet}, F.~{Boulanger}, J.~{Brand}, M.~{Bucher}, C.~{Burigana}, Z.-Y.
  {Cai}, V.~{Casasola}, G.~{Castex}, A.~{Challinor}, J.~{Chluba},
  S.~{Colafrancesco}, F.~{Cuttaia}, G.~{D'Alessandro}, R.~{Davis}, M.~{de
  Avillez}, P.~{de Bernardis}, M.~{de Petris}, A.~{de Rosa}, G.~{de Zotti},
  J.~{Delabrouille}, C.~{Dickinson}, J.~M. {Diego}, E.~{Falgarone},
  P.~{Ferreira}, K.~{Ferriere}, F.~{Finelli}, A.~{Fletcher}, G.~{Fuller},
  S.~{Galli}, K.~{Ganga}, J.~{Garcia-Bellido}, A.~{Ghribi},
  J.~{Gonzalez-Nuevo}, K.~{Grainge}, A.~{Gruppuso}, A.~{Hall},
  C.~{Hernandez-Monteagudo}, M.~{Jackson}, A.~{Jaffe}, R.~{Khatri},
  L.~{Lamagna}, M.~{Lattanzi}, P.~{Leahy}, M.~{Liguori}, E.~{Liuzzo},
  M.~{Lopez-Caniego}, J.~{Macias-Perez}, B.~{Maffei}, D.~{Maino}, S.~{Masi},
  A.~{Mangilli}, M.~{Massardi}, S.~{Matarrese}, A.~{Melchiorri}, J.-B. {Melin},
  A.~{Mennella}, A.~{Mignano}, M.-A. {Miville-Deschenes}, F.~{Nati},
  P.~{Natoli}, M.~{Negrello}, F.~{Noviello}, F.~{Paci}, R.~{Paladino},
  D.~{Paoletti}, F.~{Perrotta}, F.~{Piacentini}, M.~{Piat}, L.~{Piccirillo},
  G.~{Pisano}, G.~{Polenta}, S.~{Ricciardi}, M.~{Roman}, J.-A. {Rubino-Martin},
  M.~{Salatino}, A.~{Schillaci}, P.~{Shellard}, J.~{Silk}, R.~{Stompor},
  R.~{Sunyaev}, A.~{Tartari}, L.~{Terenzi}, L.~{Toffolatti}, M.~{Tomasi},
  T.~{Trombetti}, M.~{Tucci}, B.~{Van Tent}, L.~{Verde}, B.~{Wandelt}, and
  S.~{Withington}.
\newblock {PRISM (Polarized Radiation Imaging and Spectroscopy Mission): A
  White Paper on the Ultimate Polarimetric Spectro-Imaging of the Microwave and
  Far-Infrared Sky}.
\newblock \emph{ArXiv e-prints}, June 2013.

\bibitem[{Remazeilles} et~al.(2011){Remazeilles}, {Delabrouille}, and
  {Cardoso}]{2011MNRAS.418..467R}
M.~{Remazeilles}, J.~{Delabrouille}, and J.-F. {Cardoso}.
\newblock {Foreground component separation with generalized Internal Linear
  Combination}.
\newblock \emph{\mnras}, 418:\penalty0 467--476, November 2011.
\newblock \doi{10.1111/j.1365-2966.2011.19497.x}.

\bibitem[{Remazeilles} et~al.(2017){Remazeilles}, {Banday}, {Baccigalupi},
  {Basak}, {Bonaldi}, {De Zotti}, {Delabrouille}, {Dickinson}, {Eriksen},
  {Errard}, {Fernandez-Cobos}, {Fuskeland}, {Herv{\'{\i}}as-Caimapo},
  {L{\'o}pez-Caniego}, {Martinez-Gonz{\'a}lez}, {Roman}, {Vielva}, {Wehus},
  {Achucarro}, {Ade}, {Allison}, {Ashdown}, {Ballardini}, {Banerji}, {Bartolo},
  {Bartlett}, {Baumann}, {Bersanelli}, {Bonato}, {Borrill}, {Bouchet},
  {Boulanger}, {Brinckmann}, {Bucher}, {Burigana}, {Buzzelli}, {Cai}, {Calvo},
  {Carvalho}, {Castellano}, {Challinor}, {Chluba}, {Clesse}, {Colantoni},
  {Coppolecchia}, {Crook}, {D'Alessandro}, {de Bernardis}, {de Gasperis},
  {Diego}, {Di Valentino}, {Feeney}, {Ferraro}, {Finelli}, {Forastieri},
  {Galli}, {Genova-Santos}, {Gerbino}, {Gonz{\'a}lez-Nuevo}, {Grandis},
  {Greenslade}, {Hagstotz}, {Hanany}, {Handley}, {Hernandez-Monteagudo},
  {Hills}, {Hivon}, {Kiiveri}, {Kisner}, {Kitching}, {Kunz}, {Kurki-Suonio},
  {Lamagna}, {Lasenby}, {Lattanzi}, {Lesgourgues}, {Lewis}, {Liguori},
  {Lindholm}, {Luzzi}, {Maffei}, {Martins}, {Masi}, {McCarthy}, {Melin},
  {Melchiorri}, {Molinari}, {Monfardini}, {Natoli}, {Negrello}, {Notari},
  {Paiella}, {Paoletti}, {Patanchon}, {Piat}, {Pisano}, {Polastri}, {Polenta},
  {Pollo}, {Poulin}, {Quartin}, {Rubino-Martin}, {Salvati}, {Tartari},
  {Tomasi}, {Tramonte}, {Trappe}, {Trombetti}, {Tucker}, {Valiviita}, {Van de
  Weijgaert}, {van Tent}, {Vennin}, {Vittorio}, {Young}, and {for the CORE
  collaboration}]{ECO.foregrounds.paper}
M.~{Remazeilles}, A.~J. {Banday}, C.~{Baccigalupi}, S.~{Basak}, A.~{Bonaldi},
  G.~{De Zotti}, J.~{Delabrouille}, C.~{Dickinson}, H.~K. {Eriksen},
  J.~{Errard}, R.~{Fernandez-Cobos}, U.~{Fuskeland},
  C.~{Herv{\'{\i}}as-Caimapo}, M.~{L{\'o}pez-Caniego},
  E.~{Martinez-Gonz{\'a}lez}, M.~{Roman}, P.~{Vielva}, I.~{Wehus},
  A.~{Achucarro}, P.~{Ade}, R.~{Allison}, M.~{Ashdown}, M.~{Ballardini},
  R.~{Banerji}, N.~{Bartolo}, J.~{Bartlett}, D.~{Baumann}, M.~{Bersanelli},
  M.~{Bonato}, J.~{Borrill}, F.~{Bouchet}, F.~{Boulanger}, T.~{Brinckmann},
  M.~{Bucher}, C.~{Burigana}, A.~{Buzzelli}, Z.-Y. {Cai}, M.~{Calvo}, C.-S.
  {Carvalho}, G.~{Castellano}, A.~{Challinor}, J.~{Chluba}, S.~{Clesse},
  I.~{Colantoni}, A.~{Coppolecchia}, M.~{Crook}, G.~{D'Alessandro}, P.~{de
  Bernardis}, G.~{de Gasperis}, J.-M. {Diego}, E.~{Di Valentino}, S.~{Feeney},
  S.~{Ferraro}, F.~{Finelli}, F.~{Forastieri}, S.~{Galli}, R.~{Genova-Santos},
  M.~{Gerbino}, J.~{Gonz{\'a}lez-Nuevo}, S.~{Grandis}, J.~{Greenslade},
  S.~{Hagstotz}, S.~{Hanany}, W.~{Handley}, C.~{Hernandez-Monteagudo},
  M.~{Hills}, E.~{Hivon}, K.~{Kiiveri}, T.~{Kisner}, T.~{Kitching}, M.~{Kunz},
  H.~{Kurki-Suonio}, L.~{Lamagna}, A.~{Lasenby}, M.~{Lattanzi},
  J.~{Lesgourgues}, A.~{Lewis}, M.~{Liguori}, V.~{Lindholm}, G.~{Luzzi},
  B.~{Maffei}, C.~J.~A.~P. {Martins}, S.~{Masi}, D.~{McCarthy}, J.-B. {Melin},
  A.~{Melchiorri}, D.~{Molinari}, A.~{Monfardini}, P.~{Natoli}, M.~{Negrello},
  A.~{Notari}, A.~{Paiella}, D.~{Paoletti}, G.~{Patanchon}, M.~{Piat},
  G.~{Pisano}, L.~{Polastri}, G.~{Polenta}, A.~{Pollo}, V.~{Poulin},
  M.~{Quartin}, J.-A. {Rubino-Martin}, L.~{Salvati}, A.~{Tartari}, M.~{Tomasi},
  D.~{Tramonte}, N.~{Trappe}, T.~{Trombetti}, C.~{Tucker}, J.~{Valiviita},
  R.~{Van de Weijgaert}, B.~{van Tent}, V.~{Vennin}, N.~{Vittorio}, K.~{Young},
  and {for the CORE collaboration}.
\newblock {Exploring Cosmic Origins with CORE: B-mode Component Separation}.
\newblock \emph{ArXiv e-prints}, April 2017.

\bibitem[{Revenu} et~al.(2000){Revenu}, {Kim}, {Ansari}, {Couchot},
  {Delabrouille}, and {Kaplan}]{2000A&AS..142..499R}
B.~{Revenu}, A.~{Kim}, R.~{Ansari}, F.~{Couchot}, J.~{Delabrouille}, and
  J.~{Kaplan}.
\newblock {Destriping of polarized data in a CMB mission with a circular
  scanning strategy}.
\newblock \emph{\aaps}, 142:\penalty0 499--509, March 2000.
\newblock \doi{10.1051/aas:2000308}.

\bibitem[{Riess} et~al.(2011){Riess}, {Macri}, {Casertano}, {Lampeitl},
  {Ferguson}, {Filippenko}, {Jha}, {Li}, and {Chornock}]{2011ApJ...730..119R}
A.~G. {Riess}, L.~{Macri}, S.~{Casertano}, H.~{Lampeitl}, H.~C. {Ferguson},
  A.~V. {Filippenko}, S.~W. {Jha}, W.~{Li}, and R.~{Chornock}.
\newblock {A 3\% Solution: Determination of the Hubble Constant with the Hubble
  Space Telescope and Wide Field Camera 3}.
\newblock \emph{\apj}, 730:\penalty0 119, April 2011.
\newblock \doi{10.1088/0004-637X/730/2/119}.

\bibitem[{Rosset} et~al.(2007){Rosset}, {Yurchenko}, {Delabrouille}, {Kaplan},
  {Giraud-H{\'e}raud}, {Lamarre}, and {Murphy}]{2007A&A...464..405R}
C.~{Rosset}, V.~B. {Yurchenko}, J.~{Delabrouille}, J.~{Kaplan},
  Y.~{Giraud-H{\'e}raud}, J.-M. {Lamarre}, and J.~A. {Murphy}.
\newblock {Beam mismatch effects in cosmic microwave background polarization
  measurements}.
\newblock \emph{\aap}, 464:\penalty0 405--415, March 2007.
\newblock \doi{10.1051/0004-6361:20042230}.

\bibitem[{Scott} et~al.(2016){Scott}, {Contreras}, {Narimani}, and
  {Ma}]{2016JCAP...06..046S}
D.~{Scott}, D.~{Contreras}, A.~{Narimani}, and Y.-Z. {Ma}.
\newblock {The information content of cosmic microwave background
  anisotropies}.
\newblock \emph{\jcap}, 6:\penalty0 046, June 2016.
\newblock \doi{10.1088/1475-7516/2016/06/046}.

\bibitem[{Sherwin} and {Schmittfull}(2015)]{2015PhRvD..92d3005S}
B.~D. {Sherwin} and M.~{Schmittfull}.
\newblock {Delensing the CMB with the cosmic infrared background}.
\newblock \emph{\prd}, 92\penalty0 (4):\penalty0 043005, August 2015.
\newblock \doi{10.1103/PhysRevD.92.043005}.

\bibitem[{Sneiderman} et~al.(2016){Sneiderman}, {Shirron}, {Fujimoto},
  {Bialas}, {Boyce}, {Chiao}, {DiPirro}, {Eckart}, {Hartz}, {Ishisaki},
  {Kelley}, {Kilbourne}, {Masters}, {McCammon}, {Mitsuda}, {Noda}, {Porter},
  {Szymkowiak}, {Takei}, {Tsujimoto}, and {Yoshida}]{2016SPIE.9905E..3NS}
G.~A. {Sneiderman}, P.~J. {Shirron}, R.~{Fujimoto}, T.~G. {Bialas}, K.~R.
  {Boyce}, M.~P. {Chiao}, M.~J. {DiPirro}, M.~E. {Eckart}, L.~{Hartz},
  Y.~{Ishisaki}, R.~L. {Kelley}, C.~A. {Kilbourne}, C.~{Masters},
  D.~{McCammon}, K.~{Mitsuda}, H.~{Noda}, F.~S. {Porter}, A.~E. {Szymkowiak},
  Y.~{Takei}, M.~{Tsujimoto}, and S.~{Yoshida}.
\newblock {Cryogen-free operation of the Soft X-ray Spectrometer instrument}.
\newblock In \emph{Society of Photo-Optical Instrumentation Engineers (SPIE)
  Conference Series}, volume 9905 of \emph{\procspie}, page 99053N, July 2016.
\newblock \doi{10.1117/12.2232045}.

\bibitem[{Starobinsky}(1980)]{1980PhLB...91...99S}
A.~A. {Starobinsky}.
\newblock {A new type of isotropic cosmological models without singularity}.
\newblock \emph{Physics Letters B}, 91:\penalty0 99--102, March 1980.
\newblock \doi{10.1016/0370-2693(80)90670-X}.

\bibitem[{Tauber} et~al.(2010{\natexlab{a}}){Tauber}, {Mandolesi}, {Puget},
  {Banos}, {Bersanelli}, {Bouchet}, {Butler}, {Charra}, {Crone}, {Dodsworth},
  and et~al.]{2010A&A...520A...1T}
J.~A. {Tauber}, N.~{Mandolesi}, J.-L. {Puget}, T.~{Banos}, M.~{Bersanelli},
  F.~R. {Bouchet}, R.~C. {Butler}, J.~{Charra}, G.~{Crone}, J.~{Dodsworth}, and
  et~al.
\newblock {Planck pre-launch status: The Planck mission}.
\newblock \emph{\aap}, 520:\penalty0 A1, September 2010{\natexlab{a}}.
\newblock \doi{10.1051/0004-6361/200912983}.

\bibitem[{Tauber} et~al.(2010{\natexlab{b}}){Tauber}, {Norgaard-Nielsen},
  {Ade}, {Amiri Parian}, {Banos}, {Bersanelli}, {Burigana}, {Chamballu}, {de
  Chambure}, {Christensen}, {Corre}, {Cozzani}, {Crill}, {Crone},
  {D'Arcangelo}, {Daddato}, {Doyle}, {Dubruel}, {Forma}, {Hills},
  {Huffenberger}, {Jaffe}, {Jessen}, {Kletzkine}, {Lamarre}, {Leahy},
  {Longval}, {de Maagt}, {Maffei}, {Mandolesi}, {Mart{\'{\i}}-Canales},
  {Mart{\'{\i}}n-Polegre}, {Martin}, {Mendes}, {Murphy}, {Nielsen}, {Noviello},
  {Paquay}, {Peacocke}, {Ponthieu}, {Pontoppidan}, {Ristorcelli}, {Riti},
  {Rolo}, {Rosset}, {Sandri}, {Savini}, {Sudiwala}, {Tristram}, {Valenziano},
  {van der Vorst}, {van't Klooster}, {Villa}, and
  {Yurchenko}]{2010A&A...520A...2T}
J.~A. {Tauber}, H.~U. {Norgaard-Nielsen}, P.~A.~R. {Ade}, J.~{Amiri Parian},
  T.~{Banos}, M.~{Bersanelli}, C.~{Burigana}, A.~{Chamballu}, D.~{de Chambure},
  P.~R. {Christensen}, O.~{Corre}, A.~{Cozzani}, B.~{Crill}, G.~{Crone},
  O.~{D'Arcangelo}, R.~{Daddato}, D.~{Doyle}, D.~{Dubruel}, G.~{Forma},
  R.~{Hills}, K.~{Huffenberger}, A.~H. {Jaffe}, N.~{Jessen}, P.~{Kletzkine},
  J.~M. {Lamarre}, J.~P. {Leahy}, Y.~{Longval}, P.~{de Maagt}, B.~{Maffei},
  N.~{Mandolesi}, J.~{Mart{\'{\i}}-Canales}, A.~{Mart{\'{\i}}n-Polegre},
  P.~{Martin}, L.~{Mendes}, J.~A. {Murphy}, P.~{Nielsen}, F.~{Noviello},
  M.~{Paquay}, T.~{Peacocke}, N.~{Ponthieu}, K.~{Pontoppidan},
  I.~{Ristorcelli}, J.-B. {Riti}, L.~{Rolo}, C.~{Rosset}, M.~{Sandri},
  G.~{Savini}, R.~{Sudiwala}, M.~{Tristram}, L.~{Valenziano}, M.~{van der
  Vorst}, K.~{van't Klooster}, F.~{Villa}, and V.~{Yurchenko}.
\newblock {Planck pre-launch status: The optical system}.
\newblock \emph{\aap}, 520:\penalty0 A2, September 2010{\natexlab{b}}.
\newblock \doi{10.1051/0004-6361/200912911}.

\bibitem[{The COrE Collaboration} et~al.(2011){The COrE Collaboration},
  {Armitage-Caplan}, {Avillez}, {Barbosa}, {Banday}, {Bartolo}, {Battye},
  {Bernard}, {de Bernardis}, {Basak}, {Bersanelli}, {Bielewicz}, {Bonaldi},
  {Bucher}, {Bouchet}, {Boulanger}, {Burigana}, {Camus}, {Challinor},
  {Chongchitnan}, {Clements}, {Colafrancesco}, {Delabrouille}, {De Petris}, {De
  Zotti}, {Dickinson}, {Dunkley}, {Ensslin}, {Fergusson}, {Ferreira},
  {Ferriere}, {Finelli}, {Galli}, {Garcia-Bellido}, {Gauthier}, {Haverkorn},
  {Hindmarsh}, {Jaffe}, {Kunz}, {Lesgourgues}, {Liddle}, {Liguori},
  {Lopez-Caniego}, {Maffei}, {Marchegiani}, {Martinez-Gonzalez}, {Masi},
  {Mauskopf}, {Matarrese}, {Melchiorri}, {Mukherjee}, {Nati}, {Natoli},
  {Negrello}, {Pagano}, {Paoletti}, {Peacocke}, {Peiris}, {Perroto},
  {Piacentini}, {Piat}, {Piccirillo}, {Pisano}, {Ponthieu}, {Rath},
  {Ricciardi}, {Rubino Martin}, {Salatino}, {Shellard}, {Stompor},
  {Urrestilla}, {Van Tent}, {Verde}, {Wandelt}, and
  {Withington}]{2011arXiv1102.2181T}
{The COrE Collaboration}, C.~{Armitage-Caplan}, M.~{Avillez}, D.~{Barbosa},
  A.~{Banday}, N.~{Bartolo}, R.~{Battye}, J.~{Bernard}, P.~{de Bernardis},
  S.~{Basak}, M.~{Bersanelli}, P.~{Bielewicz}, A.~{Bonaldi}, M.~{Bucher},
  F.~{Bouchet}, F.~{Boulanger}, C.~{Burigana}, P.~{Camus}, A.~{Challinor},
  S.~{Chongchitnan}, D.~{Clements}, S.~{Colafrancesco}, J.~{Delabrouille},
  M.~{De Petris}, G.~{De Zotti}, C.~{Dickinson}, J.~{Dunkley}, T.~{Ensslin},
  J.~{Fergusson}, P.~{Ferreira}, K.~{Ferriere}, F.~{Finelli}, S.~{Galli},
  J.~{Garcia-Bellido}, C.~{Gauthier}, M.~{Haverkorn}, M.~{Hindmarsh},
  A.~{Jaffe}, M.~{Kunz}, J.~{Lesgourgues}, A.~{Liddle}, M.~{Liguori},
  M.~{Lopez-Caniego}, B.~{Maffei}, P.~{Marchegiani}, E.~{Martinez-Gonzalez},
  S.~{Masi}, P.~{Mauskopf}, S.~{Matarrese}, A.~{Melchiorri}, P.~{Mukherjee},
  F.~{Nati}, P.~{Natoli}, M.~{Negrello}, L.~{Pagano}, D.~{Paoletti},
  T.~{Peacocke}, H.~{Peiris}, L.~{Perroto}, F.~{Piacentini}, M.~{Piat},
  L.~{Piccirillo}, G.~{Pisano}, N.~{Ponthieu}, C.~{Rath}, S.~{Ricciardi},
  J.~{Rubino Martin}, M.~{Salatino}, P.~{Shellard}, R.~{Stompor}, L.~T.~J.
  {Urrestilla}, B.~{Van Tent}, L.~{Verde}, B.~{Wandelt}, and S.~{Withington}.
\newblock {COrE (Cosmic Origins Explorer) A White Paper}.
\newblock \emph{ArXiv e-prints}, February 2011.

\bibitem[{The EBEX Collaboration} et~al.(2017){The EBEX Collaboration},
  {Aboobaker}, {Ade}, {Araujo}, {Aubin}, {Baccigalupi}, {Bao}, {Chapman},
  {Didier}, {Dobbs}, {Geach}, {Grainger}, {Hanany}, {Helson}, {Hillbrand},
  {Hubmayr}, {Jaffe}, {Johnson}, {Jones}, {Klein}, {Korotkov}, {Lee},
  {Levinson}, {Limon}, {MacDermid}, {Matsumura}, {Miller}, {Milligan}, {Raach},
  {Reichborn-Kjennerud}, {Sagiv}, {Savini}, {Spencer}, {Tucker}, {Tucker},
  {Westbrook}, {Young}, and {Zilic}]{2017arXiv170303847T}
{The EBEX Collaboration}, A.~M. {Aboobaker}, P.~{Ade}, D.~{Araujo}, F.~{Aubin},
  C.~{Baccigalupi}, C.~{Bao}, D.~{Chapman}, J.~{Didier}, M.~{Dobbs},
  C.~{Geach}, W.~{Grainger}, S.~{Hanany}, K.~{Helson}, S.~{Hillbrand},
  J.~{Hubmayr}, A.~{Jaffe}, B.~{Johnson}, T.~{Jones}, J.~{Klein},
  A.~{Korotkov}, A.~{Lee}, L.~{Levinson}, M.~{Limon}, K.~{MacDermid},
  T.~{Matsumura}, A.~D. {Miller}, M.~{Milligan}, K.~{Raach},
  B.~{Reichborn-Kjennerud}, I.~{Sagiv}, G.~{Savini}, L.~{Spencer}, C.~{Tucker},
  G.~S. {Tucker}, B.~{Westbrook}, K.~{Young}, and K.~{Zilic}.
\newblock {The EBEX Balloon Borne Experiment - Optics, Receiver, and
  Polarimetry}.
\newblock \emph{ArXiv e-prints}, March 2017.

\bibitem[{The Polarbear Collaboration: P.~A.~R.~Ade} et~al.(2014){The Polarbear
  Collaboration: P.~A.~R.~Ade}, {Akiba}, {Anthony}, {Arnold}, {Atlas},
  {Barron}, {Boettger}, {Borrill}, {Chapman}, {Chinone}, {Dobbs}, {Elleflot},
  {Errard}, {Fabbian}, {Feng}, {Flanigan}, {Gilbert}, {Grainger}, {Halverson},
  {Hasegawa}, {Hattori}, {Hazumi}, {Holzapfel}, {Hori}, {Howard}, {Hyland},
  {Inoue}, {Jaehnig}, {Jaffe}, {Keating}, {Kermish}, {Keskitalo}, {Kisner}, {Le
  Jeune}, {Lee}, {Leitch}, {Linder}, {Lungu}, {Matsuda}, {Matsumura}, {Meng},
  {Miller}, {Morii}, {Moyerman}, {Myers}, {Navaroli}, {Nishino}, {Orlando},
  {Paar}, {Peloton}, {Poletti}, {Quealy}, {Rebeiz}, {Reichardt}, {Richards},
  {Ross}, {Schanning}, {Schenck}, {Sherwin}, {Shimizu}, {Shimmin}, {Shimon},
  {Siritanasak}, {Smecher}, {Spieler}, {Stebor}, {Steinbach}, {Stompor},
  {Suzuki}, {Takakura}, {Tomaru}, {Wilson}, {Yadav}, and
  {Zahn}]{2014ApJ...794..171T}
{The Polarbear Collaboration: P.~A.~R.~Ade}, Y.~{Akiba}, A.~E. {Anthony},
  K.~{Arnold}, M.~{Atlas}, D.~{Barron}, D.~{Boettger}, J.~{Borrill},
  S.~{Chapman}, Y.~{Chinone}, M.~{Dobbs}, T.~{Elleflot}, J.~{Errard},
  G.~{Fabbian}, C.~{Feng}, D.~{Flanigan}, A.~{Gilbert}, W.~{Grainger}, N.~W.
  {Halverson}, M.~{Hasegawa}, K.~{Hattori}, M.~{Hazumi}, W.~L. {Holzapfel},
  Y.~{Hori}, J.~{Howard}, P.~{Hyland}, Y.~{Inoue}, G.~C. {Jaehnig}, A.~H.
  {Jaffe}, B.~{Keating}, Z.~{Kermish}, R.~{Keskitalo}, T.~{Kisner}, M.~{Le
  Jeune}, A.~T. {Lee}, E.~M. {Leitch}, E.~{Linder}, M.~{Lungu}, F.~{Matsuda},
  T.~{Matsumura}, X.~{Meng}, N.~J. {Miller}, H.~{Morii}, S.~{Moyerman}, M.~J.
  {Myers}, M.~{Navaroli}, H.~{Nishino}, A.~{Orlando}, H.~{Paar}, J.~{Peloton},
  D.~{Poletti}, E.~{Quealy}, G.~{Rebeiz}, C.~L. {Reichardt}, P.~L. {Richards},
  C.~{Ross}, I.~{Schanning}, D.~E. {Schenck}, B.~D. {Sherwin}, A.~{Shimizu},
  C.~{Shimmin}, M.~{Shimon}, P.~{Siritanasak}, G.~{Smecher}, H.~{Spieler},
  N.~{Stebor}, B.~{Steinbach}, R.~{Stompor}, A.~{Suzuki}, S.~{Takakura},
  T.~{Tomaru}, B.~{Wilson}, A.~{Yadav}, and O.~{Zahn}.
\newblock {A Measurement of the Cosmic Microwave Background B-mode Polarization
  Power Spectrum at Sub-degree Scales with POLARBEAR}.
\newblock \emph{\apj}, 794:\penalty0 171, October 2014.
\newblock \doi{10.1088/0004-637X/794/2/171}.

\bibitem[{Tristram} et~al.(2011){Tristram}, {Filliard}, {Perdereau},
  {Plaszczynski}, {Stompor}, and {Touze}]{2011A&A...534A..88T}
M.~{Tristram}, C.~{Filliard}, O.~{Perdereau}, S.~{Plaszczynski}, R.~{Stompor},
  and F.~{Touze}.
\newblock {Iterative destriping and photometric calibration for Planck-HFI,
  polarized, multi-detector map-making}.
\newblock \emph{\aap}, 534:\penalty0 A88, October 2011.
\newblock \doi{10.1051/0004-6361/201116871}.

\bibitem[{Van Eck} et~al.(2017){Van Eck}, {Haverkorn}, {Alves}, {Beck}, {de
  Bruyn}, {En{\ss}lin}, {Farnes}, {Ferri{\`e}re}, {Heald}, {Horellou},
  {Horneffer}, {Iacobelli}, {Jeli{\'c}}, {Mart{\'{\i}}-Vidal}, {Mulcahy},
  {Reich}, {R{\"o}ttgering}, {Scaife}, {Schnitzeler}, {Sobey}, and
  {Sridhar}]{2017A&A...597A..98V}
C.~L. {Van Eck}, M.~{Haverkorn}, M.~I.~R. {Alves}, R.~{Beck}, A.~G. {de Bruyn},
  T.~{En{\ss}lin}, J.~S. {Farnes}, K.~{Ferri{\`e}re}, G.~{Heald},
  C.~{Horellou}, A.~{Horneffer}, M.~{Iacobelli}, V.~{Jeli{\'c}},
  I.~{Mart{\'{\i}}-Vidal}, D.~D. {Mulcahy}, W.~{Reich}, H.~J.~A.
  {R{\"o}ttgering}, A.~M.~M. {Scaife}, D.~H.~F.~M. {Schnitzeler}, C.~{Sobey},
  and S.~S. {Sridhar}.
\newblock {Faraday tomography of the local interstellar medium with LOFAR:
  Galactic foregrounds towards IC 342}.
\newblock \emph{\aap}, 597:\penalty0 A98, January 2017.
\newblock \doi{10.1051/0004-6361/201629707}.

\bibitem[{Wallis} et~al.(2017){Wallis}, {Brown}, {Battye}, and
  {Delabrouille}]{2017MNRAS.466..425W}
C.~G.~R. {Wallis}, M.~L. {Brown}, R.~A. {Battye}, and J.~{Delabrouille}.
\newblock {Optimal scan strategies for future CMB satellite experiments}.
\newblock \emph{\mnras}, 466:\penalty0 425--442, April 2017.
\newblock \doi{10.1093/mnras/stw2577}.

\bibitem[{Wu} and {Dor{\'e}}(2016)]{2016arXiv161202474W}
H.-Y. {Wu} and O.~{Dor{\'e}}.
\newblock {Optimizing future experiments of cosmic far-infrared background: a
  principal component approach}.
\newblock \emph{ArXiv e-prints}, December 2016.

\bibitem[{Zaldarriaga} and {Seljak}(1997)]{1997PhRvD..55.1830Z}
M.~{Zaldarriaga} and U.~{Seljak}.
\newblock {All-sky analysis of polarization in the microwave background}.
\newblock \emph{\prd}, 55:\penalty0 1830--1840, February 1997.
\newblock \doi{10.1103/PhysRevD.55.1830}.

\end{thebibliography}

\end{document}